\documentclass[mathematics,article,submit,pdftex,moreauthors]{Definitions/mdpi}
\preto{\abstractkeywords}{\nolinenumbers}  % turn off line numbering
\firstpage{1}
\pubvolume{1}
\issuenum{1}
\articlenumber{0}
\pubyear{2024}
\copyrightyear{2024}
\datereceived{ } 
\daterevised{ } % Comment out if no revised date
\dateaccepted{ } 
\datepublished{ } 
\hreflink{https://doi.org/} % If needed use \linebreak
\usepackage{amsmath,amsthm,verbatim,amssymb,amsfonts,amscd,graphicx}

\usepackage[utf8]{inputenc}
\usepackage{longtable}
\usepackage{pgfplots}
\usepackage{multirow}
\usepackage{makecell}
\usepackage{enumitem}
\usepackage{xfrac}
\usepackage[title]{appendix}
\usepackage[english]{babel}
\usepackage{pdflscape}
\usepackage{booktabs}

\usepackage[tableposition=top]{caption}  % caption is on top of table
\usepackage{subcaption}
\usepackage{tikz}
\usetikzlibrary{arrows,shapes,positioning,shadows,trees,calc}

\tikzset{
    basic/.style  = {draw, text width=2cm, drop shadow, font=\sffamily, rectangle},
    root/.style   = {basic, rounded corners=2pt, thin, align=center,
                     fill=green!30},
    level 2/.style = {basic, rounded corners=6pt, thin,align=center, fill=green!60,
                     text width=8em},
    level 3/.style = {basic, thin, align=center, fill=pink!60, text width=9.5em}
}

\usepackage[linesnumbered,ruled]{algorithm2e}

\def\R{\mathbb R}
\def\E{\mathbb E}
\def\N{\mathbb N}
\def\D{\mathbb D}
\def\X{\mathbb X}
\def\T{\mathcal T}
\def\A{\mathcal A}
\def\H{\mathcal H}
\def\I{\mathcal I}
\def\O{\mathcal O}

\let\oldemptyset\emptyset
\let\emptyset\varnothing

\DeclareMathOperator*{\argmax}{arg\,max}
\DeclareMathOperator*{\argmin}{arg\,min}

\newcommand{\eps}{\varepsilon}

\newlength{\summtableheight}  % height of summary tables
\newlength{\tableheight}  % height of each table in appendix
\usepackage{array}
\tikzstyle{linesAll} = [thick, line join=round, mark phase=1, mark repeat=1]
\def\myLineWidth{0.75}
\def\myMarkerLineWidth{0.25}
\def\myMarkSize{1.6}
\def\myMarkSizeD{0.2}
\tikzstyle{markerCirc} = [mark size=\myMarkSize pt, mark=*, mark options={solid, draw=black, line width=\myMarkerLineWidth pt}]
\tikzstyle{markerAst} = [mark size=\myMarkSize pt, mark=asterisk, mark options={solid, draw=black, line width=\myMarkerLineWidth pt}]
\tikzstyle{markerTria} = [mark size=\myMarkSize+2*\myMarkSizeD pt, mark=triangle*, mark options={solid, draw=black, line width=\myMarkerLineWidth pt}]
\tikzstyle{markerSquare} = [mark size=\myMarkSize-\myMarkSizeD pt, mark=square*, mark options={solid, draw=black, line
width=\myMarkerLineWidth pt}]
\tikzstyle{markerDiam} = [mark size=\myMarkSize+\myMarkSizeD pt, mark=diamond*, mark options={solid, draw=black, line
width=\myMarkerLineWidth pt}]
\tikzstyle{markerO} = [mark size=\myMarkSize pt, mark=o, mark options={solid, draw=black, line width=\myMarkerLineWidth pt}]
\tikzstyle{markerMercedes} = [mark size=\myMarkSize+4*\myMarkSizeD pt, mark=Mercedes star, mark options={solid, draw=black, line width=\myMarkerLineWidth pt}]
\tikzstyle{markerStick} = [mark size=\myMarkSize+4*\myMarkSizeD pt, mark=|, mark options={solid, draw=black, line width=\myMarkerLineWidth pt}]
\tikzstyle{plotStyle1} = [linesAll, markerCirc, color=red, solid]
\tikzstyle{plotStyle2} = [linesAll, markerAst, color=green, dashed]
\tikzstyle{plotStyle3} = [linesAll, markerTria, color=blue, dotted]
\tikzstyle{plotStyle4} = [linesAll, markerSquare, color=magenta, dashdotted]
\tikzstyle{plotStyle5} = [linesAll, markerDiam, color=pink, solid]
\tikzstyle{plotStyle6} = [linesAll, markerO, color=teal, dashed]
\tikzstyle{plotStyle7} = [linesAll, markerMercedes, color=violet, dotted]
\tikzstyle{plotStyle8} = [linesAll, markerStick, color=cyan, dashdotted]

\pgfplotsset{compat=1.18}

\Title{High-Performance Hybrid Algorithm for Minimum Sum-of-Squares Clustering of Infinitely Tall Data}

\TitleCitation{Title}

\newcommand{\orcidauthorA}{0000-0003-1105-5990} % Add \orcidA{} behind the author's name
\newcommand{\orcidauthorB}{0000-0001-7283-5144} % Add \orcidB{} behind the author's name

\Author{Ravil Mussabayev $^{1,2,\ddagger}$\orcidA{}, Rustam Mussabayev $^{2,3,\ddagger}$\orcidB{}}

\AuthorNames{Ravil Mussabayev, Rustam Mussabayev}

\AuthorCitation{Mussabayev, R.; Mussabayev, R.}
\address{%
	$^{1}$ \quad Department of Mathematics, University of Washington, Padelford Hall C-138, Seattle, 98195-4350, WA, USA; ravmus@uw.edu\\
	$^{2}$ \quad Satbayev University, Satbaev str. 22, Almaty, 050013, Kazakhstan; r.mussabayev@satbayev.university (Ravil M.); ru.mussabayev@satbayev.university (Rustam M.)\\
	$^{3}$ \quad Laboratory for Analysis and Modeling of Information Processes, Institute of Information and Computational Technologies, Pushkin str. 125, Almaty, 050010, Kazakhstan; rustam@iict.kz}

\corres{Correspondence: ravmus@uw.edu}

\secondnote{These authors contributed equally to this work.}
\abstract{This paper introduces a novel formulation of the clustering problem, namely the Minimum Sum-of-Squares Clustering of Infinitely Tall Data (MSSC-ITD), and presents HPClust, an innovative set of hybrid parallel approaches for its effective solution. By utilizing modern high-performance computing techniques, HPClust enhances key clustering metrics: effectiveness, computational efficiency, and scalability. In contrast to vanilla data parallelism, which only accelerates processing time through the MapReduce framework, our approach unlocks superior performance by leveraging the multi-strategy competitive-cooperative parallelism and intricate properties of the objective function landscape. Unlike other available algorithms that struggle to scale, our algorithm is inherently parallel in nature, improving solution quality through increased scalability and parallelism, and outperforming even advanced algorithms designed for small and medium-sized datasets. Our evaluation of HPClust, featuring four parallel strategies, demonstrates its superiority over traditional and cutting-edge methods by offering better performance in the key metrics. These results also show that parallel processing not only enhances the clustering efficiency, but the accuracy as well. Additionally, we explore the balance between computational efficiency and clustering quality, providing insights into optimal parallel strategies based on dataset specifics and resource availability. This research advances our understanding of parallelism in clustering algorithms, demonstrating that a judicious hybridization of advanced parallel approaches yields optimal results for MSSC-ITD. Experiments on synthetic data further confirm HPClust's exceptional scalability and robustness to noise.}


\begin{document}

\section{Introduction}

Clustering is a critical task that involves the identification of similar objects within a given set. As digital data continues to grow at an unprecedented rate, this problem has become increasingly challenging and has applications in diverse domains. For instance, in the biological and medical domains, it has been used for gene expression analysis~\cite{Jiang2004}, enhancing medical diagnostics~\cite{Mittal2021}, and advancing bioinformatics research~\cite{Deridder2013}. In the realm of technology and data, clustering optimizes vector quantization and data compression techniques~\cite{Yin2015}, identifies anomalies~\cite{Tu2020}, aids in pattern recognition and classification~\cite{Deridder2013}, dissects time series data for forecasting~\cite{Rakthanmanon2012}, and forms the basis for the finance and blockchain sectors~\cite{Lejun2024-datadet,Lejun2024-mtxdecom}. Furthermore, in the context of consumer and media analytics, clustering helps in segmenting customers for targeted marketing~\cite{Chen2018}, analyzing images and videos for content extraction~\cite{Yeung1998}, and understanding social media trends~\cite{Zhao2011}. Lastly, in the information sciences, it refines information retrieval systems~\cite{Djenouri2021} and processes natural language for better human-computer interaction~\cite{Alguliyev2019}, alongside analyzing network and traffic patterns~\cite{Depaire2008}.

%For instance, clustering has been used in gene expression analysis~\cite{Jiang2004}, enhancing medical diagnostics~\cite{Mittal2021}, optimizing vector quantization and data compression techniques~\cite{Yin2015}, identifying anomalies~\cite{Tu2020}, segmenting customers for targeted marketing~\cite{Chen2018}, analyzing images and videos for content extraction~\cite{Yeung1998}, understanding social media trends~\cite{Zhao2011}, refining information retrieval systems~\cite{Djenouri2021}, aiding in pattern recognition and classification~\cite{Deridder2013}, analyzing network and traffic patterns~\cite{Depaire2008}, advancing bioinformatics research~\cite{Deridder2013}, processing natural language for better human-computer interaction~\cite{Alguliyev2019}, and dissecting time series data for forecasting~\cite{Rakthanmanon2012}.

The most fundamental and widely studied clustering model is the Minimum Sum-of-Squares Clustering (MSSC)~\cite{Aloise2009}. It can be formulated as follows. Consider a set of $m$ data points $X=\{x_1, \ldots, x_m\}$ in the Euclidean space $\R^n$. Then, MSSC is aimed at finding $k$ cluster centers (centroids) $C=\left(c_1,\ldots,c_k\right) \in \R^{n \times k}$ that minimize the sum of squared distances from each data point $x_i$ to its nearest cluster center $c_j$:

\begin{equation}
\min\limits_{C} \ \ \ f\left(C,X\right)=\sum\limits_{i=1}^m \min_{j=1,\ldots,k} \| x_i - c_j \|^2
\label{eq:mssc}
\end{equation}
where $\| \cdot \|$ denotes the Euclidean norm. Each collection of centroids $C$ uniquely defines the corresponding partition $X = X_1 \cup \ldots \cup X_k$, where each subset (cluster) $X_j$ consists of the points that are closest to $c_j$ than to any other centroid. Equation~\eqref{eq:mssc} represents the objective function measuring the total squared deviation of data points from their closest centroids. Its global optimization leads to the simultaneous maximization of the similarity between objects within the same cluster and minimization of the similarity between objects in different clusters.

When dealing with big data, where the number of data points is unbounded, i.e., $|X| = m = \infty$, formulation~\eqref{eq:mssc} gives rise to the Minimum Sum-of-Squares Clustering of Infinitely Tall Data (MSSC-ITD) problem, which is one of the key contributions of our work. This problem makes traditional clustering methods unfeasible. The MSSC-ITD problem is a novel formulation that we have introduced in this paper, and our proposed algorithm is the first to provide an efficient solution to this challenge. In particular, few clustering algorithms exist that can address this problem, and even fewer can perform a global search in such conditions. Our approach fills this gap, providing a robust and efficient solution to the MSSC-ITD problem.
%This is a significant conceptual innovation of our approach, and we emphasize that the MSSC-ITD problem and its solution are original contributions of this paper.
%When dealing with big data, we can set $|X| = m = \infty$ in equation \eqref{eq:mssc}, giving rise to the Minimum Sum-of-Squares Clustering of Infinitely Tall Data (MSSC-ITD) problem. The key novelty of our work and the distinguishing feature of our proposed algorithm is its unique ability to tackle the MSSC-ITD problem.

Research has shown that global minimizers provide the most accurate representation of the clustering structure of a given dataset~\cite{Gribel2019}. However, achieving global minimizers in MSSC is a challenging task due to the highly non-convex nature of the objective function. This non-convexity becomes even more pronounced as the dataset size increases, making the task of finding global minimizers even more complex.

To address this challenge, several approaches have been proposed in the literature to explore the solution space and locate global minimizers, such as gradient-based optimization techniques~\cite{Karmitsa2018}, stochastic optimization algorithms~\cite{Franti2019}, metaheuristic search strategies~\cite{Gribel2019,Hansen2001}, and hybrid approaches~\cite{Mansueto2021}. Each of these approaches has its strengths and weaknesses, and there is no all-around solution. As a result, further research is needed to develop more efficient and robust techniques for locating global minimizers in the context of the MSSC-ITD problem.

Apart from the above classification, parallel processing in big data clustering algorithms presents another critical and frequently overlooked aspect. Most approaches that have been discovered in the literature are limited to only data parallelism, which is usually implemented using the MapReduce model. Meanwhile, more sophisticated parallel strategies are either not investigated or not applicable to the big data clustering algorithms available in the literature.

For general $k$ and $m$, the MSSC algorithms are known to be computationally intensive due to their NP-hard complexity~\cite{Aloise2009}. The NP-hardness of MSSC is heavily exacerbated by big data contexts. High-Performance Computing (HPC) technologies, including supercomputers and computer clusters, offer a robust platform for tackling such complex problems. By distributing the data across multiple nodes, computers, or processors, parallel processing enables scalable and efficient handling of big data. This approach leverages the combined computational power of multiple computing resources, allowing for faster and more effective execution of MSSC algorithms on massive datasets.

In this work, we propose HPClust, a set of novel parallel approaches for the MSSC-ITD problem. The decomposition principle is at the heart of the HPClust algorithm. This principle not only serves as the algorithm's cornerstone but also facilitates efficient and effective parallel processing of big data. Parallel processing is one of the core approaches employed for big data clustering. In the current work, we endeavor to comprehensively explore this dimension with the goal of maximizing the performance of the HPClust algorithm in big data contexts.

Four parallel approaches --- inner, competitive, cooperative, and hybrid --- are proposed to tackle the MSSC-ITD problem. The inner parallel method involves parallelizing distance evaluations in the K-means local search applied within each sequential clustering subproblem, offering scalability in the subproblem size. The competitive strategy implements concurrency at the subproblem level, maximizing diversity in initial clustering solutions. The cooperative approach simultaneously processes clustering subproblems, maximizing exploration by continuously selecting the best solution and capitalizing on it. The hybrid strategy combines the last two into a multi-strategy competitive-cooperative approach, aiming for an optimal exploration-exploitation trade-off in MSSC-ITD solutions.

%Four parallel approaches --- inner, competitive, cooperative, and hybrid --- are proposed that specifically address the MSSC-ITD problem. The inner parallel method involves parallel design of distance evaluations inside solving each sequential clustering subproblem via the K-means local search. The competitive scheme implements concurrency on the level of subproblems, maximizing the diversity of initial clustering solutions. The cooperative approach also uses simultaneous processing of clustering subproblem, but maximizes exploration by picking the best solution among all subproblems and capitalizing on its as much as possible. The hybrid strategy combines the last two with the goal of achieving the optimal exploration-exploitation trade-off of MSSC-ILD solutions.

The name HPClust can be interpreted in two ways, both reflecting the algorithm's key strengths. Firstly, ``High-Performance Clustering'' highlights the algorithm's computational efficiency, speed, and ability to scale through parallelism, making it a high-performance solution for clustering tasks. Alternatively, ``Hybrid Parallel Clustering'' emphasizes the innovative combined parallel clustering strategy employed by HPClust, which leverages the strengths of different parallel approaches to achieve superior performance. This hybrid strategy sets HPClust apart as a winning solution in the field of parallel clustering algorithms.

Notably, our algorithm boasts a significant conceptual advantage as one of the few clustering algorithms that is inherently parallel in nature. This allows it to improve solution quality through increased scalability and parallelism, setting it apart from other algorithms that may struggle to scale. Moreover, our algorithm is capable of competing with advanced clustering algorithms designed for small and medium-sized datasets, demonstrating its versatility and robustness. Unlike other algorithms where parallelism is a forced add-on, our algorithm's parallel nature is an intrinsic property that enables seamless scalability.

While other approaches to clustering often rely solely on data parallelism, our approach utilizes a combination of more advanced and sophisticated parallelism types. Data parallelism involves dividing the dataset into smaller chunks and processing each chunk simultaneously on different processors, but only brings advantages in processing time. In contrast, task parallelism (functional parallelism) enables us to execute different tasks or functions of the clustering algorithm in parallel, allowing for more flexibility and effectiveness when merging their results. Furthermore, hybrid parallelism combines these approaches, allowing us to leverage the strengths of each to achieve better results. Unlike other parallel approaches that only focus on scaling clustering in the data space without guarantees on solution quality, our approach leverages the strengths of different parallelism types by combining data parallelism with task parallelism and hybrid parallelism, achieving better results. This integrated approach sets our method apart from others, which often rely on a single type of parallelism, and enables us to deliver higher-quality clustering solutions and scalability in big data clustering.

Also, we provide a comprehensive review of various parallel and high-performance computing techniques used for big data clustering and indicate their strengths and weaknesses. We pinpoint the intricacies involved in the process of applying these approaches to HPClust, as well as exhibit the obtained insights in form of a tutorial on applying parallel and high-performance computing technologies to the problem of big data clustering.

%Our paper is structured as follows. Section~\ref{sec:related_works} surveys the key developments and strategies in the field of parallel clustering algorithms. Section~\ref{sec:methodology} presents our study's contributions, including the HPClust's proposed parallel strategies, and reviews modern high-performance techniques for optimizing clustering algorithms, highlighting implementation nuances. This section also describes our experimental setup. Section~\ref{sec:results} provides a detailed analysis and interpretation of our experimental findings, along with insights into trade-offs. Section~\ref{sec:guidelines} offers practical guidelines for selecting the optimal parallel strategy for HPClust, aimed at big data clustering practitioners. Finally, Section~\ref{sec:conclusion} concludes our work and identifies promising future research directions.

Our paper is structured as follows. Section~\ref{sec:related_works} surveys the key developments and strategies in the field of parallel clustering algorithms. Section~\ref{sec:proposed_algo} presents the proposed HPClust algorithm, while Section~\ref{sec:paral_strategies} describes its various parallel strategies. Section~\ref{sec:techniques_nuances} provides an overview of modern high-performance techniques for optimizing big data clustering algorithms, highlighting key nuances and considerations in the implementation details of the HPClust algorithm. Section~\ref{sec:experiments} describes our experimental setup and its methodology. Section~\ref{sec:results} provides a detailed analysis and interpretation of our experimental findings, along with insights into trade-offs. Section~\ref{sec:guidelines} offers practical guidelines for selecting the optimal parallel strategy for HPClust, aimed at big data clustering practitioners. Finally, Section~\ref{sec:conclusion} concludes our work and identifies promising future research directions.
%while Section~\ref{sec:nuances} delves deeper into the implementation details of the HPClust algorithm, highlighting key nuances and considerations.

\section{Related Works} \label{sec:related_works}

In the field of big data clustering, many methods have been created that work in parallel and distribute the workload to handle the difficulties presented by the large size, complex dimensions, and real-time nature of big data. Parallelism and distributed computing appear as two prominent techniques for big data clustering.

Usually, parallel processing in clustering algorithms involves dividing the data into smaller subsets, clustering them simultaneously on multiple processors, and aggregating these partial results into a global solution. This helps in reducing the computation time and makes the clustering process much more efficient. It is usually used when the data is too large to fit into memory or the computation time is a bottleneck.

Distributed computing, on the other hand, involves the distribution of big data across multiple machines. Clustering is then performed in a distributed manner using frameworks like Apache Hadoop or Apache Spark~\cite{Mohebi2016}. By distributing the data and computations, processing time is reduced, and scalability is achieved. This approach is useful when a dataset is unacceptably large to be stored and processed on a single machine.

K-means~\cite{Lloyd1982} algorithm with the Forgy initialization~\cite{Pena1999} is a commonly used traditional clustering method due to its simplicity and effectiveness. However, its application to big data can pose problems due to its high time complexity, which is $\O(m \cdot n \cdot k)$ for a single iteration, and the need to store all data in memory. The pseudocode of the Forgy K-means clustering method is provided in Algorithm~\ref{alg:forgy_kmeans}.

%\begin{algorithm}
%	\SetAlgoLined
%	\KwResult{Optimize centroids $C$ and assign dataset $X$ to clusters $Y$ via K-means++.}
%	\textbf{Initialization:}\\
%	Select $c_1$ randomly from $X$\;
%	\For{$i \gets 2$ \KwTo $k$}{
%		Calculate $d(x)$ as the minimum distance from $x \in X$ to the nearest centroid in ${c_1,\ldots,c_{i-1}}$\;
%		Select $c_i$ from $X$ with probability $\frac{d(x)^2}{\sum_{x \in X} d(x)^2}$\;
%	}
%	Set $C = (c_1, \ldots, c_k)$\;
%	
%	\textbf{Iterative Optimization:}\\
%	\Repeat{centroids $C$ are unchanged or maximum iterations reached}{
%		Assign each $x \in X$ to the nearest centroid in $C$\;
%		Update each $c_i \in C$ to the mean of points assigned to $c_i$\;
%	}
%	
%	\textbf{Cluster Assignment:}\\
%	Assign each $x \in X$ to its closest centroid in $C$, forming $Y$\;
%	\caption{K-Means++ Clustering}
%	\label{alg:kmeans_pp}
%\end{algorithm}

\begin{algorithm}
	\SetAlgoLined
	\KwResult{Optimize centroids \(C\) and assign dataset \(X\) to clusters \(Y\) via Forgy K-means.}
	\textbf{Initialization:}\\
	Randomly select \(k\) instances from \(X\) to serve as the initial centroids \(C = (c_1, \ldots, c_k)\)\;
	
	\textbf{Iterative Optimization:}\\
	\Repeat{centroids \(C\) are unchanged or maximum iterations reached}{
		Assign each \(x \in X\) to the nearest centroid in \(C\)\;
		Update each \(c_i \in C\) to the mean of points assigned to \(c_i\)\;
	}
	
	\textbf{Cluster Assignment:}\\
	Assign each \(x \in X\) to its closest centroid in \(C\), forming \(Y\).
	\caption{Forgy K-Means Clustering}
	\label{alg:forgy_kmeans}
\end{algorithm}

To circumvent the time complexity limitations of traditional approaches, like Forgy K-means, some parallel and distributed clustering algorithms have been suggested in the literature. The MapReduce framework is by far the most popular approach to scale clustering in the data space~\cite{Dean2008}. Zhao et al.~\cite{Zhao2009} implemented a distributed version of K-means according to the MapReduce concept that led to a significant speed-up compared to the sequential version without any guarantees on the clustering solution quality.

A widely adopted method to handle large datasets that cannot be accommodated entirely in RAM is the Minibatch K-means algorithm~\cite{Sculley2010}. It is an online version of the K-means algorithm that employs random subsets, or minibatches, of a dataset during each iteration to update the current solution. While this technique significantly accelerates computation time, it sacrifices the clustering quality since it exerts no control over the solution updates across iterations. Also, Minibatch K-means is an inherently sequential algorithm, amenable to only data parallelism.

Bahmani et al.~\cite{Bahmani2012} developed a scalable version of K-means++ that merges the advantages of K-means++ and Mini-batch K-means. However, our experimental evaluation on large real-world datasets showed that K-means$||$, while being on par with K-means++ in speed, is significantly worse that K-means++ with respect to solution quality.

Alguliyev et al. proposed an innovative approach in their study, where they introduced the Parallel Batch K-means For Big Data Clustering (PBK-BDC) algorithm \cite{Alguliyev2021}. This algorithm partitions large datasets into smaller segments, clusters them with the help of K-means, and aggregates the resulting cluster centers into a final pool. The algorithm then clusters the pool using K-means again. The pseudocode for the PBK-BDC algorithm can be found in Algorithm~\ref{alg:pbk_bdc}. Notably, PBK-BDC is one of the most prominent partition-based clustering algorithms. In the original paper, the authors empirically evaluated PBK-BDC and found that it outperformed the classical K-means algorithm \cite{Alguliyev2021}. However, this evaluation did not compare PBK-BDC to other advanced algorithms for clustering large datasets, leaving room for further research.

\begin{algorithm}
	\SetAlgoLined
	\KwResult{Determine the final centroids $C$ and cluster assignments $Y$ for a dataset $X$ utilizing the Parallel Batch K-means For Big Data Clustering (PBK-BDC) strategy.}
	\textbf{Initialization:}\\
	Divide the dataset $X$ into segments, each containing $p$ elements\;
	\ForEach{segment $C_i$}{
		Apply K-means clustering to $C_i$ to derive new centroids $C_{i, \text{new}}$\;
		Incorporate $C_{i, \text{new}}$ into the cooperative centroid repository $P$\;
	}
	Execute K-means clustering on repository $P$ to secure the ultimate centroids $C_{\text{final}}$\;
	Map every data point in $X$ to its closest centroid in $C_{\text{final}}$, establishing the ultimate cluster mappings $Y$.
	\caption{PBK-BDC Clustering Method}
	\label{alg:pbk_bdc}
\end{algorithm}

%\begin{algorithm}
%	\SetAlgoLined
%	\KwResult{Compute the final centroids $C$ and cluster assignments $Y$ for a dataset $X$ using the PBK-BDC algorithm.}
%	\textbf{Initialization:}\\
%	Partition the dataset $X$ into chunks of size $p$\;
%	\For{each chunk $C_i$}{
%		Cluster $C_i$ using K-means to obtain centroids $C_{i, \text{new}}$\;
%		Add $C_{i, \text{new}}$ to the pool of centroids $P$\;
%	}
%	Cluster the pool $P$ using K-means to obtain final centroids $C_{\text{final}}$\;
%	Assign each point in $X$ to the nearest centroid in $C_{\text{final}}$ to obtain final cluster assignments $Y$\;
%	\caption{PBK-BDC Clustering}
%	\label{alg:pbk_bdc}
%\end{algorithm}

% Despite these developments, there is an ongoing need for efficient, scalable, and high-quality clustering algorithms suitable for big data.

Mohebi et al.~\cite{Mohebi2016} conducted a comprehensive review of various parallel algorithms and concluded that the field of big data clustering using parallel computing is still in its emergent stage and offers significant scope for further research. They observed that parallel data processing can potentially reduce the clustering time of large datasets, but it may also have an adverse impact on the quality and performance of clustering. Thus, the primary objective of research in this area should be to achieve an optimal balance between quality and speed of clustering for big data applications.

%All in all, based on the detailed discussion of a large number of parallel algorithms, Mohebi et al.~\cite{Mohebi2016} concluded that the ﬁeld of parallel big data clustering is still young and open for new research. They argued that parallel data processing can help improve the clustering time of large datasets, but it may degrade the quality and performance. Therefore, the main concern is to achieve a reasonable trade-off between quality and speed in the context of big data.

Our proposed HPClust algorithm, utilizing advanced parallel processing techniques and intelligent sample selection, seeks to fill the gaps in the field. HPClust proves that advanced parallel strategies and careful algorithm design may optimize both the efficiency and effectiveness of clustering algorithms simultaneously, while maintaining exceptional scalability across various data scales.

%Big data is often seen as a challenge to be overcome by most standard algorithms and alternative heuristics, rather than an advantage to be leveraged to improve clustering results. Therefore, there is a significant need for new big data clustering algorithms that are relatively simple, effective for big data processing and able to use big data as an advantage to enhance clustering results. These algorithms should balance simplicity, result quality, and convergence speed, and perform global search for the optimal solution without relying on known global optimization metaheuristics. Our proposed HPClust algorithm~\cite{Mussabayev2023}, utilizing parallel processing and intelligent sample selection, seeks to fill this gap.

% Добавить в обзор публикации по различным существующим стратегиям распараллеливания алгоритмов кластеризации

\section{Proposed Algorithm} \label{sec:proposed_algo}

We propose HPClust, an array of parallel heuristic approaches for solving the MSSC-ITD problem via high-performance computing techniques. The algorithm's main idea is to apply the problem decomposition technique, letting each parallel worker iteratively process a sequence of subproblems, and intelligently combine the obtained partial results into a single global clustering solution.

Each parallel worker $w$ operates by sequentially clustering incoming samples of a large dataset. It begins by randomly selecting a small sample $S$ of size $s$ from $X$, and uses the K-means++ algorithm to obtain the initial configuration of centroids $C$. The worker then clusters each new incoming sample by the K-means algorithm using the best set of centroids $C_w$ (or $C_{best}$) obtained from all previously processed samples for the current worker (or among all parallel workers), called the incumbent solution. The incumbent solution is chosen based on the objective function value~\eqref{eq:mssc} obtained on a sample. This ``keep the best'' principle allows the algorithm not to lose information about the best local minimum obtained so far, and more iterations can only lead to further improvements.

%Each sequential worker of HPClust starts by randomly creating a sample $S$ of size $s$ from the given dataset $X$, where $s$ is much smaller than the total number of feature vectors $m$. The initial configuration of centroids, denoted as $C$, is obtained by applying the K-means++ algorithm to the first drawn sample. Then, the algorithm iteratively clusters each new incoming sample by the K-means algorithm initialized with the best set of centroids that have been obtained across all the already processed samples so far. In each iteration, such a set of the best centroids among the processed samples is called the incumbent solution. The criterion for choosing the incumbent solution is based on the objective function~\eqref{eq:mssc} evaluated on a sample. This iterative approach follows a ``keep the best'' principle, ensuring that the best solution found so far is always prioritized. Also, it is important to note that the current best (incumbent) solution $C$ is replaced or kept intact in each iteration, without growing in size as the number of iterations increases.

HPClust solves the issue of degenerate clusters (also known as empty clusters) by reinitializing them with K-means++ when all data points are reassigned to other clusters. This introduces new cluster centers, enhancing the overall clustering solution and increasing opportunities to minimize the objective function. Also, introducing new samples in each iteration perturbs the incumbent solution, injecting variability into the clustering outcomes.

When a stop condition is reached by any parallel worker (e.g., a time limit or maximum number of processed samples), the algorithm selects the final centroid set $C$ by choosing the solution obtained by the worker with the lowest incumbent sample objective function. Then, HPClust assigns data points of the entire dataset to their closest cluster centers in the final centroid set $C$. However, this final assignment step may be omitted if only the final centroids or a limited set of assignments are required.

HPClust's iteration time complexity is $\O(s \cdot n \cdot k)$ (where $k$ is the number of clusters). The algorithm's scalability can be fine-tuned by selecting suitable sample sizes and counts. By processing smaller subsets of the data in each iteration, the computational demands are substantially reduced. Additionally, employing random subsets of the data during each iteration and periodically re-initializing the centroids of degenerate clusters prevents the algorithm from being trapped in suboptimal solutions. This allows the algorithm to explore different parts of the objective function's landscape, potentially finding better solutions than a single application of the K-means algorithm.

\section{Parallel Strategies for the HPClust Algorithm} \label{sec:paral_strategies}

The HPClust algorithm is designed to be highly parallel in nature. Four different parallel strategies can be employed:

\begin{enumerate}
\item Inner parallelism (HPClust-inner): Employs parallel clustering at the implementation level of K-means and K-means++, processing individual data samples sequentially while parallelizing the calculation of minimum distances;

\item Competitive parallelism (HPClust-competitive): Independent workers process individual data samples in parallel, each using its own previous best centroids $C_w$ for initialization, and the best solution is selected when the stopping criterion is met. A pseudocode of the HPClust-competitive algorithm is shown in Algorithm~\ref{alg:competitive_hpclust};

\item Cooperative parallelism (HPClust-cooperative): Workers share information on best solutions and use the best set of centroids $C_{best}$ obtained from all previous iterations across every worker, initializing each subsequent sample using the cooperative knowledge. A pseudocode of the HPClust-cooperative algorithm is provided in Algorithm~\ref{alg:cooperative_hpclust};

\item Hybrid or competitive-cooperative parallelism (HPClust-hybrid): Combines competitive and cooperative strategies, initially utilizing diversity through competitive parallelism for a duration of $T_1$ seconds or $N_1$ iterations, and then capitalizing on the most successful evolved form through cooperative parallelism for an additional $T_2$ seconds or $N_2$ iterations. A pseudocode of the HPClust-cooperative algorithm is presented in Algorithm~\ref{alg:hybrid_hpclust}.
\end{enumerate}

\begin{figure}[htb]
\centering
\begin{tikzpicture}[
    node distance=1.0cm and 2cm,
    worker/.style={rectangle, draw, minimum width=1.5cm, minimum height=1cm, align=center},
    centroid/.style={rectangle, draw, minimum width=2cm, minimum height=1cm, align=center},
    loop/.style={rectangle, draw, minimum width=2cm, minimum height=1cm, align=center},
    final/.style={rectangle, draw, minimum width=3.5cm, minimum height=1cm, align=center},
    data/.style={rectangle, draw, minimum width=1.5cm, minimum height=1cm},
    arrow/.style={-latex, thick},
    looking1/.style={-latex, dashed, thick, black, font=\footnotesize, sloped}
]

    \node[data] (d1) {Data Sample 1};
    \node[right=of d1, yshift=-0.2cm] (ellipsis1) {$\ldots$};
    \node[data, right=of ellipsis1] (dN) {Data Sample N};

    \node[centroid, below=of d1] (c1) {Centroids 1};
    \node[below=of ellipsis1, yshift=-0.2cm] (ellipsis2) {$\ldots$};
    \node[centroid, below=of dN] (cN) {Centroids N};

    \node[loop, below=of c1] (loop1) {HPClust Loop on \\ Data Sample 1 using \\ the best Centroids \\ among all Workers};
    \node[below=of ellipsis2, yshift=-0.2cm] (ellipsis3) {$\ldots$};
    \node[loop, below=of cN] (loopN) {HPClust Loop on \\ Data Sample N using \\ the best Centroids \\ among all Workers};

    \node[final, below=of ellipsis3, yshift=-2.6cm] (final1) {Choose Centroids with \\ smallest objective function value};
    \node[final, below=of final1, yshift=-0.1cm] (finalN) {Assign each point to nearest \\ centroid in full dataset};

    \node[above=of d1] (d) {Dataset};

    \draw[arrow] (d.south) -- ++(0,-0.3cm) -| node[midway, left=0.5cm, yshift=-0.3cm]{Worker 1} (d1.north);
    \draw[arrow] (d.south) -- ++(0,-0.3cm) -| node[midway, left=0.5cm, yshift=-0.3cm]{Worker N} (dN.north);
    
    \draw[arrow] (d1.south) -- (c1.north);
    \draw[arrow] (dN.south) -- (cN.north);
    
    \draw[arrow] (c1.south) -- (loop1.north);
    \draw[arrow] (loop1.west) -- ++(-0.7cm,0) |- (d1.west);
    \draw[arrow] (cN.south) -- (loopN.north);
    \draw[arrow] (loopN.west) -- ++(-0.7cm,0) |- (dN.west);

    \draw[arrow] (loop1.south) -- ++(0,-0.5cm) -| (final1.north);
    \draw[arrow] (loopN.south) -- ++(0,-0.3cm) -| (final1.north);
    \draw[arrow] (final1.south) -- (finalN.north);

    \draw[looking1, bend left=30] (c1.east) to node[midway, above, sloped]{Use if better} (cN.west);
    \draw[looking1, bend left=30] (cN.west) to node[midway, above, sloped]{Use if better} (c1.east);
\end{tikzpicture}
\caption{Flowchart of the HPClust algorithm using a cooperative parallel strategy}
\label{fig:coop_flowchart_simplified}
\end{figure}
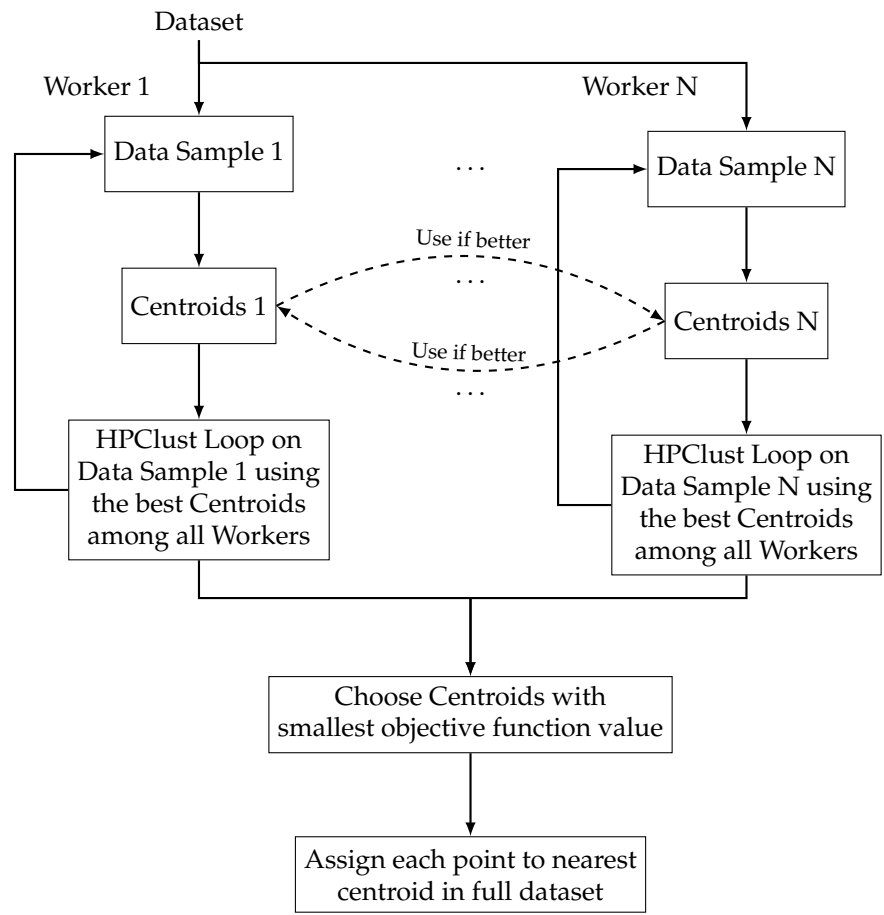

The goal of the hybrid mode is to leverage the advantages of both competitive and cooperative approaches, ensuring diversity and exploiting the most successful solutions. Flowcharts for the competitive and cooperative strategies are provided in Figures~\ref{fig:comp_flowchart_simplified} and \ref{fig:coop_flowchart_simplified}.

The HPClust algorithm source code, including implementations of various parallel strategies, is available at \href{https://github.com/rmusab/hpclust}{https://github.com/rmusab/hpclust}.

Our study focuses on the efficiency of parallel interaction strategies, assuming equal access to the full-sized dataset and independent sampling, without exploring distributed data storage optimizations, which are left for a separate study.

%\begin{algorithm}
%\SetAlgoLined
%\KwResult{Compute the final centroids $C$ and cluster assignments $Y$ for a dataset $X$ using the competitive HPClust algorithm.}
%\textbf{Initialization:}\\
%$C_w \leftarrow \text{Mark all } k \text{ centroids as degenerate for each worker } w$\;
%$\hat{f}_w \leftarrow \infty$ for each worker $w$\;
%$t_w \leftarrow 0$ for each worker $w$\;
%\While{$t_w < T$ for any worker $w$}{
%    \For{each parallel worker $w$}{
%        $S_{w} \leftarrow \text{Random sample of size } s \text{ from } X$\;
%        \For{each $c \in C_w$}{
%            \If{$c$ is the centroid associated with a degenerate cluster}{
%                \text{Reinitialize }$c \text{ using K-means++ on } S_{w}$\;
%            }
%        }
%        $C_{\text{new},w} \leftarrow \text{K-means clustering on } S_{w} \text{ with initial centroids } C_w$\;
%        \If{$f(C_{\text{new},w}, S_{w}) < \hat{f}_w$}{
%            $C_w \leftarrow C_{\text{new},w}$\;
%            $\hat{f}_w \leftarrow f(C_{\text{new},w}, S_{w})$\;
%        }
%        $t_w \leftarrow t_w + 1$\;
%    }
%}
%$C_{\text{best}} \leftarrow \text{Centroids of the worker with the smallest } \hat{f}_w \text{ value}$\;
%$Y \leftarrow \text{Assign each point in } X \text{ to nearest centroid in } C_{\text{best}}$\;
%\caption{Competitive HPClust Clustering}
%\label{alg:competitive_hpclust}
%\end{algorithm}

\begin{algorithm}
	\SetAlgoLined
	\KwResult{Determine the final centroids $C$ and cluster assignments $Y$ for a dataset $X$ using the competitive HPClust algorithm.}
	\textbf{Initialization:}\\
	$C_w \leftarrow \text{Set all } k \text{ centroids as degenerate for each worker } w$\;
	$\hat{f}_w \leftarrow \infty$ for each worker $w$\;
	$t_w \leftarrow 0$ for each worker $w$\;
	\While{there exists a worker $w$ with $t_w < T$}{
		\For{each worker $w$ in parallel}{
			$S_{w} \leftarrow \text{Select a random sample of size } s \text{ from } X$\;
			\For{each centroid $c$ in $C_w$}{
				\If{$c$ represents a degenerate cluster}{
					\text{Reinitialize }$c \text{ using K-means++ on } S_{w}$\;
				}
			}
			$C_{\text{new},w} \leftarrow \text{Perform K-means clustering on } S_{w} \text{ using } C_w \text{ as initial centroids}$\;
			\If{$f(C_{\text{new},w}, S_{w}) < \hat{f}_w$}{
				$C_w \leftarrow C_{\text{new},w}$\;
				$\hat{f}_w \leftarrow f(C_{\text{new},w}, S_{w})$\;
			}
			$t_w \leftarrow t_w + 1$\;
		}
	}
	$C_{\text{best}} \leftarrow \text{Identify centroids from the worker with the minimum } \hat{f}_w \text{ value}$\;
	$Y \leftarrow \text{Assign each data point in } X \text{ to the nearest centroid in } C_{\text{best}}$\;
	\caption{Competitive HPClust Clustering}
	\label{alg:competitive_hpclust}
\end{algorithm}

%\begin{algorithm}
%\SetAlgoLined
%\KwResult{Compute the final centroids $C$ and cluster assignments $Y$ for a dataset $X$ using the cooperative HPClust algorithm.}
%\textbf{Initialization:}\\
%$C_w \leftarrow \text{Mark all } k \text{ centroids as degenerate for each worker } w$\;
%$\hat{f}_w \leftarrow \infty$ for each worker $w$\;
%$t_w \leftarrow 0$ for each worker $w$\;
%\While{$t_w < T$ for any worker $w$}{
%    \For{each parallel worker $w$}{
%        $S_{w} \leftarrow \text{Random sample of size } s \text{ from } X$\;
%        $C_{\text{best}} \leftarrow \text{Centroids of the worker with the smallest } \hat{f}_w \text{ value}$\;
%        \For{each $c \in C_{\text{best}}$}{
%            \If{$c$ is the centroid associated with a degenerate cluster}{
%                \text{Reinitialize }$c \text{ using K-means++ on } S_{w}$\;
%            }
%        }
%        $C_{\text{new},w} \leftarrow \text{K-means clustering on } S_{w} \text{ with initial centroids } C_{\text{best}}$\;
%        \If{$f(C_{\text{new},w}, S_{w}) < \hat{f}_w$}{
%            $C_w \leftarrow C_{\text{new},w}$\;
%            $\hat{f}_w \leftarrow f(C_{\text{new},w}, S_{w})$\;
%        }
%        $t_w \leftarrow t_w + 1$\;
%    }
%}
%$C_{\text{best}} \leftarrow \text{Centroids of the worker with the smallest } \hat{f}_w \text{ value}$\;
%$Y \leftarrow \text{Assign each point in } X \text{ to nearest centroid in } C_{\text{best}}$\;
%\caption{cooperative HPClust Clustering}
%\label{alg:cooperative_hpclust}
%\end{algorithm}

\begin{algorithm}
	\SetAlgoLined
	\KwResult{Calculate the final centroids $C$ and cluster assignments $Y$ for a dataset $X$ using the cooperative HPClust algorithm.}
	\textbf{Initialization:}\\
	$C_w \leftarrow \text{Initialize all } k \text{ centroids as degenerate for each worker } w$\;
	$\hat{f}_w \leftarrow \infty$ for each worker $w$\;
	$t_w \leftarrow 0$ for each worker $w$\;
	\While{any worker $w$ has $t_w < T$}{
		\For{each worker $w$ in parallel}{
			$S_{w} \leftarrow \text{Take a random sample of size } s \text{ from } X$\;
			$C_{\text{best}} \leftarrow \text{Select centroids from the worker with the lowest } \hat{f}_w$\;
			\For{each centroid $c$ in $C_{\text{best}}$}{
				\If{$c$ represents a degenerate cluster}{
					\text{Reinitialize }$c \text{ using K-means++ based on } S_{w}$\;
				}
			}
			$C_{\text{new},w} \leftarrow \text{Apply K-means clustering to } S_{w} \text{ starting with } C_{\text{best}} \text{ as initial centroids}$\;
			\If{$f(C_{\text{new},w}, S_{w}) < \hat{f}_w$}{
				$C_w \leftarrow C_{\text{new},w}$\;
				$\hat{f}_w \leftarrow f(C_{\text{new},w}, S_{w})$\;
			}
			$t_w \leftarrow t_w + 1$\;
		}
	}
	$C_{\text{best}} \leftarrow \text{Retrieve centroids from the worker with the minimum } \hat{f}_w$\;
	$Y \leftarrow \text{Allocate each data point in } X \text{ to the closest centroid in } C_{\text{best}}$\;
	\caption{Cooperative HPClust Clustering}
	\label{alg:cooperative_hpclust}
\end{algorithm}

\begin{algorithm}
\SetAlgoLined
\KwResult{Compute the final centroids $C$ and cluster assignments $Y$ for a dataset $X$ using the hybrid HPClust algorithm.}
\textbf{Initialization:}\\
$C_w \leftarrow \text{Mark all } k \text{ centroids as degenerate for each worker } w$\;
$\hat{f}_w \leftarrow \infty$ for each worker $w$\;
$t_w \leftarrow 0$ for each worker $w$\;

\For{Phase in $(\text{Competitive}, \text{cooperative})$}{
    \While{$t_w < T_{\text{Phase}}$ for any worker $w$}{
        \For{each parallel worker $w$}{
            $S_{w} \leftarrow \text{Random sample of size } s \text{ from } X$\;
            \eIf{$\text{Phase} = \text{cooperative}$}{
                $C_{\text{base}} \leftarrow \text{Centroids of the worker with the smallest } \hat{f}_w \text{ value}$\;
            }{
                $C_{\text{base}} \leftarrow C_w$\;
            }
            \For{each $c \in C_{\text{base}}$}{
                \If{$c$ represents a degenerate cluster}{
                    \text{Reinitialize }$c \text{ using K-means++ on } S_{w}$\;
                }
            }
            $C_{\text{new},w} \leftarrow \text{K-means clustering on } S_{w} \text{ with initial centroids } C_{\text{base}}$\;
            \If{$f(C_{\text{new},w}, S_{w}) < \hat{f}_w$}{
                $C_w \leftarrow C_{\text{new},w}$\;
                $\hat{f}_w \leftarrow f(C_{\text{new},w}, S_{w})$\;
            }
            $t_w \leftarrow t_w + 1$\;
        }
    }
}

$C_{\text{best}} \leftarrow \text{Centroids of the worker with the smallest } \hat{f}_w \text{ value}$\;
$Y \leftarrow \text{Assign each point in } X \text{ to nearest centroid in } C_{\text{best}}$\;
\caption{Hybrid HPClust Clustering}
\label{alg:hybrid_hpclust}
\end{algorithm}

\begin{figure}[htb]
\centering
\begin{tikzpicture}[
    node distance=1.0cm and 2cm,
    worker/.style={rectangle, draw, minimum width=1.5cm, minimum height=1cm, align=center},
    centroid/.style={rectangle, draw, minimum width=2cm, minimum height=1cm, align=center},
    loop/.style={rectangle, draw, minimum width=2cm, minimum height=1cm, align=center},
    final/.style={rectangle, draw, minimum width=3.5cm, minimum height=1cm, align=center},
    data/.style={rectangle, draw, minimum width=1.5cm, minimum height=1cm},
    arrow/.style={-latex, thick}
]

    \node[data] (d1) {Data Sample 1};
    \node[right=of d1, yshift=-0.2cm] (ellipsis1) {$\ldots$};
    \node[data, right=of ellipsis1] (dN) {Data Sample N};

    \node[centroid, below=of d1] (c1) {Centroids 1};
    \node[below=of ellipsis1, yshift=-0.2cm] (ellipsis2) {$\ldots$};
    \node[centroid, below=of dN] (cN) {Centroids N};

    \node[loop, below=of c1] (loop1) {HPClust Loop \\ on Data Sample 1 \\ using Centroids 1};
    \node[below=of ellipsis2, yshift=-0.2cm] (ellipsis3) {$\ldots$};
    \node[loop, below=of cN] (loopN) {HPClust Loop \\ on Data Sample N \\ using Centroids N};

    \node[final, below=of ellipsis3, yshift=-2cm] (final1) {Choose Centroids with \\ smallest objective function value};
    \node[final, below=of final1, yshift=-0.1cm] (finalN) {Assign each point to nearest \\ centroid in full dataset};

    \node[above=of d1] (d) {Dataset};

    \draw[arrow] (d.south) -- ++(0,-0.3cm) -| node[midway, left=0.5cm, yshift=-0.3cm]{Worker 1} (d1.north);
    \draw[arrow] (d.south) -- ++(0,-0.3cm) -| node[midway, left=0.5cm, yshift=-0.3cm]{Worker N} (dN.north);
    
    \draw[arrow] (d1.south) -- (c1.north);
    \draw[arrow] (dN.south) -- (cN.north);
    
    \draw[arrow] (c1.south) -- (loop1.north);
    \draw[arrow] (loop1.west) -- ++(-0.7cm,0) |- (d1.west);
    \draw[arrow] (cN.south) -- (loopN.north);
    %\draw[arrow] (loopN.west) -- ++(-0.7cm,0) |- (dN.west);
    \draw[arrow] (loopN.west) -- ++(-0.7cm,0) |- (dN.west);

    \draw[arrow] (loop1.south) -- ++(0,-0.5cm) -| (final1.north);
    \draw[arrow] (loopN.south) -- ++(0,-0.3cm) -| (final1.north);
    \draw[arrow] (final1.south) -- (finalN.north);
\end{tikzpicture}
\caption{Flowchart of the HPClust algorithm with the competitive parallelism}
\label{fig:comp_flowchart_simplified}
\end{figure}
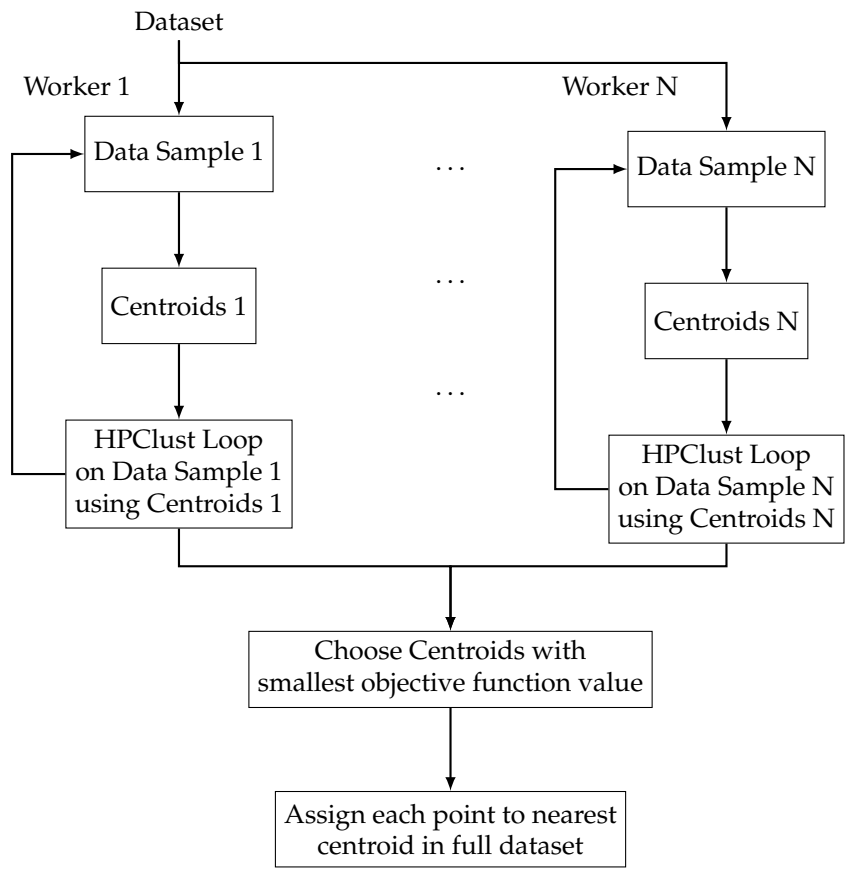

%The source code for the HPClust algorithm, which includes implementations of various parallelization strategies, is available at \href{https://github.com/R-Mussabayev/bigmeans/}{https://github.com/R-Mussabayev/bigmeans/}.

%In this article, we focus solely on researching the efficiency of various strategies of parallel interaction between individual workers. We assume that each worker has equal access to the full-sized dataset and can independently draw samples from it. For the sake of simplicity, in this study we are not exploring various available opportunities for further optimization of the algorithm, particularly those concerning distributed data storage across different nodes of the computing system. Such optimizations merit a separate study.

\section{High-Performance Techniques in HPClust} \label{sec:techniques_nuances}

\subsection{Analytical optimization}

In the analytical optimization of computational algorithms, several high-performance computing techniques are relevant. These techniques represent algorithmic improvements or theoretical advancements applied at the abstract level of the algorithm itself.

\begin{itemize}
	\item Parallel processing of iterations;
	\item Data sampling and partitioning;
	\item Tuning the level of parallelism;
	\item Optimizing inter-process communication.
\end{itemize}

Parallel processing of iterations allows for simultaneous processing of multiple iterations. This strategy employs the execution of various instances of the algorithm on different subsets of data, significantly reducing the time required for convergence~\cite{Crawford2000-softparopt}.

In relation to data management, HPClust can operate on subsets of data, allowing for a strategy of data partitioning. The initial dataset can be divided into smaller sections, each to be processed by an individual computing unit. This technique, known as data parallelism, proves particularly useful when handling datasets that exceed the memory capacity of a single machine~\cite{Dafir2021-parclust}.

The strategy of data sampling, wherein a random sample is selected from the dataset, can also be parallelized~\cite{He2015-parsampl}. Especially in cases of extensive datasets, scanning the complete dataset becomes time-consuming. By distributing the dataset across multiple processors, each can sample a section of the data independently. Then, the resultant samples can be combined.

Tuning the level of parallelism to the specifics of a dataset can lead to significant performance improvements~\cite{Sabne2013-adaptgpu}.

Optimizing inter-process communication by designing an algorithm to minimize data transfer between processes can improve performance. Techniques such as compression, delta encoding, or other forms of data reduction can also be utilized~\cite{Gupta1996-comm}.

\subsection{Nuances of parallelism in HPClust}

The HPClust algorithm, a partitioning-based clustering method, is well-suited for parallelism across its key processes. Within its inner parallel variant, HPClust-inner, two primary operations --- initialization and centroid updating --- can be executed concurrently. Initially, the algorithm leverages K-means++ on a subset of data, calculating distances from points to centroids, which can be done in parallel due to the independent nature of these calculations.

During each K-means iteration, the algorithm updates centroids (denoted as $C_{new}$) by measuring distances from all points in the sample to these new centroids, thereby redefining clusters. This centroid update phase shares the parallelizable characteristic of the initialization phase.

Despite the benefits of parallel processing in speeding up these tasks, it introduces certain challenges, such as the need for effective load balancing across cores or processors to avoid inefficiencies like idle processors, especially when the sample size $s$ is much smaller than the number of processors.

Moreover, implementing parallel computation in HPClust requires careful attention to concurrency control to avoid race conditions --- scenarios where the outcome depends on the order or timing of thread execution. In HPClust, threads may concurrently modify shared memory, such as updating centroid or data point memberships, potentially leading to inconsistent results.

To address these issues, synchronization mechanisms like locks, semaphores, or atomic operations are essential to ensure single-thread access to shared data, maintaining consistency and integrity. Optimizing the algorithm to reduce shared memory access can also help minimize race conditions. However, over-synchronization should be avoided as it can cause thread contention and decrease parallel efficiency.

For HPClust's parallel performance, it is important to achieve an optimal balance between data protection and computational speed. The aim is to improve computational speed through parallel processing without altering the clustering outcomes, maintaining consistency in results irrespective of the processor count. However, unlike other parallel clustering algorithms, this is not required for HPClust. Instead, HPClust can achieve higher accuracy by performing more iterations within a fixed time interval. This means that parallelism in HPClust improves not only efficiency but also accuracy.

Furthermore, the robustness of HPClust's parallel strategies is evident in centroid initialization, where allowing each worker to independently determine initial centroids helps overcome the challenges of poor initial selections, a known issue in K-means clustering. This feature emphasizes the importance of effective parallel design in maximizing HPClust's performance and accuracy.

\subsection{Implementation-level optimization}

To technically optimize the performance of HPClust on parallel or distributed computing systems, the following programmatic implementation-related techniques can be employed:

\begin{itemize}
	\item Vectorized operations;
	\item SIMD instructions;
	\item Concurrent data structures;
	\item Distributed computing;
	\item Load balancing;
	\item Parallel random number generation;
	\item Parallel input/output (I/O).
\end{itemize}

Further, the utilization of vectorized operations also contributes to the optimization process. Libraries such as NumPy in Python and Armadillo in C++ offer the capacity for vectorized operations. The use of these operations across entire arrays, rather than individual elements, can lead to substantial speed increases. This is due to the reduction in loop overhead and more efficient utilization of CPU features~\cite{Psarras2022-linalg}.

Simultaneously, modern CPUs provide support for SIMD (Single Instruction Multiple Data) instructions. With these, the same operation can be performed across multiple data points concurrently~\cite{Chhugani2008-simd}. Vectorizing computations, such as distance calculations in the HPClust algorithm, allows for the exploitation of these instructions, resulting in significant speed gains.

Modern programming languages and libraries offer concurrent data structures, which are designed for safe use across multiple threads or processes. These structures can prevent race conditions and synchronization issues, contributing to the efficiency of parallel algorithms~\cite{Saraswat2007-concur}.

For extremely large datasets that exceed the memory of a single machine, distributed computing frameworks such as Apache Hadoop or Apache Spark are beneficial. These frameworks facilitate the distribution of data and computation across several nodes in a cluster, accommodating larger datasets than would be possible on a single machine~\cite{Dafir2021-parclust}.

Load balancing is a strategy to efficiently use computational resources, ensuring an even distribution of work across all threads or processes. This strategy may include the dynamic assignment of tasks to processors based on their current workload. Alternatively, more sophisticated load balancing algorithms can be employed~\cite{JafarnejadGhomi2017-loadbal}.

The generation of random numbers, a function of the HPClust algorithm, can also be performed in parallel. Several techniques and libraries support parallel random number generation, maintaining independent and identically distributed numbers~\cite{Barash2014-gpurand}.

Finally, parallel I/O techniques can help alleviate the bottleneck caused by input/output operations such as reading data from disk or writing results back. A parallel file system or separate threads or processes performing I/O operations can facilitate this~\cite{May2001-pario}.

To implement these parallel strategies, various libraries and frameworks can be utilized. OpenMP or MPI in C/C++, and multiprocessing in Python offer traditional approaches. For GPU-accelerated parallel computation, CUDA or OpenCL are typically used. However, for a balance between functionality and simplicity, one might also consider employing modern libraries such as Numba. Numba provides a just-in-time compiler for Python that is easy to use yet powerful. Mojo is another notable option, providing simple and efficient parallelization solutions with a focus on high-level, user-friendly interfaces. To take full advantage of modern hardware architectures, one could use optimized numerical libraries, such as Intel's Math Kernel Library (MKL) or cuBLAS for GPUs. These libraries provide highly optimized implementations of common mathematical operations, which can lead to significant speedups.

Numba~\cite{Lam2015, Marowka2018} is a key instrument in high-performance computing, featuring optimization capabilities such as parallelization, multi-threading, and vectorization. These features are core strategies in performance optimization, transforming the execution speed of Python functions, loops, and numerical computations. Numba's dynamic generation of optimized machine code for both CPUs and GPUs further contributes to this performance boost, converging Python's usability and the speed of lower-level languages.

Numba's proficiency extends to CUDA support, facilitating the optimization of computational procedures through the use of NVIDIA GPUs. Moreover, it showcases seamless integration with Python's scientific stack, demonstrating compatibility with NumPy, SciPy, and Pandas, thereby optimizing Python's computational efficiency. In the context of distributed computing, Numba's interplay with Dask, a parallel computing library in Python, introduces an additional level of optimization, enabling efficient large-scale computations. Therefore, Numba serves as a potent tool in scientific computing, optimizing the bridge between Python's user-friendly nature and the computational efficiency of lower-level languages.

\subsection{Future optimization directions}

Future optimization of the HPClust algorithm can leverage the following high-performance techniques:

\begin{itemize}
	\item Dynamically adjusting the number of threads;
	\item Reducing communication overhead.
\end{itemize}

The number of threads can be adjusted dynamically, depending on the current system load and the size of the processed data subset, maximizing the use of CPU cores~\cite{Sabne2013-adaptgpu}.

The overhead in communication between different threads or processes is a major concern in parallel algorithms~\cite{Gupta1996-comm}. Designing the algorithm to allow each thread or process to operate independently, reducing the need for communication, can address this.

\section{Experimental Setup} \label{sec:experiments}

\subsection{Hardware and software}

Our experiments are conducted on an Ubuntu 22.04 64-bit system, equipped with an AMD EPYC 7663 Processor. The machine has 1.46 TB of RAM and runs Python 3.10.11, NumPy 1.24.3, and Numba 0.57.0. We utilize Numba to accelerate Python code through just-in-time compilation and also to enable parallel processing capabilities.

\subsection{Competitive algorithms}

We compare the performance of HPClust, equipped with different parallel strategies, to two benchmark algorithms: Forgy K-means~\cite{Pena1999} and PBK-BDC~\cite{Alguliyev2021}. Forgy K-means serves as a basic lower benchmark, representing a simple and straightforward approach. On the other hand, PBK-BDC is an advanced upper benchmark, which represents the most optimized big data clustering algorithm available in the literature~\cite{Alguliyev2021}.

\subsection{Datasets}

%The experiments were conducted on 23 datasets: 19 were publicly available (detailed in Table~\ref{tab:dataset_descr} and Table~\ref{tab:dataset_info}), and 4 were normalized.
The experiments are conducted on 23 datasets: 19 are publicly available (detailed in Table~\ref{tab:dataset_descr} and Table~\ref{tab:dataset_info}), and 4 are normalized. These datasets, which are numerically based and have no missing values, vary significantly in size, from 7,797 to 10,500,000 instances, and feature 2 to 5,000 attributes. This variety ensures testing of HPClust's adaptability across different data scales. Additionally, we align our methodology with Karmitsa et al.~\cite{Karmitsa2018} for comparative analysis.

\subsection{Experimental design and evaluation metrics}

Each dataset undergoes clustering $n_{exec}$ times into $k$ clusters of varying sizes. Each execution of an algorithm on some pair $(X, k)$ is considered an experiment. The total number of conducted experiments reaches $22,098$. We assess each experiment by measuring the resulting relative error ($\eps$), CPU time ($t$), and baseline convergence time ($\overline{t}$). The relative error reveals the performance relative to historical bests: $\eps = 100 \cdot (f - f^*) / f^*$. Sometimes, a relative error may yield a negative value, which actually indicates an even more impressive performance by the algorithm, surpassing expectations.

For HPClust, the clustering time $t$ represents the time until the last solution update of the fastest worker. Also, we employ a special baseline convergence time metric, $\overline{t}$, to more accurately measure clustering time, avoiding bias from minor late-stage improvements. More specifically, for each pair $(X, k)$, the baseline convergence time $\overline{t}$ is calculated as the time to achieve a baseline sample objective value $\overline{f}_s$, which is the maximum (relative to the algorithms) median of the best sample objectives obtained across $n_{exec}$ runs. Then, the baseline convergence time $\overline{t}$ is defined as the time until any worker reaches this baseline sample objective value.

\subsection{Hyperparameter selection}

We set a maximum CPU time limit $T$ and stop the K-means clustering process if iterations exceed $300$ or the improvement between two consecutive objectives is less than $10^{-4}$. For K-means++, we consider three candidate points for sampling each new centroid.

Sample sizes are optimized based on preliminary tests to ensure no further adjustments improve performance. The specific values of $T$ and $n_{exec}$ can be found in the detailed tables of experimental results included in Appendix~\ref{sec:appendix}.

\subsection{Preliminary experiments}

% Initially, we determined the best number of CPUs (8) for balancing performance and time across all algorithms.
Preliminary experiments helped establish baselines and optimize parameters. Initially, we established that having 8 CPUs would be the optimal value for the subsequent experiments. In this context, the optimal selection means that this choice achieves the best balance between the solution quality and execution time simultaneously for all the considered algorithms, allowing for further fair comparison under equal conditions.

Subsequent preliminary experiments involved running parallelized HPClust versions to establish baseline sample objective values $\overline{f}_s$ and fine-tuning the hybrid parallel approach by experimenting with different time splits ($T_1$ and $T_2$).

\subsection{Main experiments}

The main experimental results are displayed using a special table format. Each algorithm and pair $(X, k)$ originate a series of $n_{exec}$ experiments. Each series has a minimum, median, and maximum resulting values of relative accuracy and time, which are calculated across $n_{exec}$ executions of the algorithm on configuration $(X, k)$. The means of these metrics across the values of $k$ for each dataset are displayed in the corresponding columns of the presented tables. Tables~\ref{tab:result_e_par_strategies} and \ref{tab:result_t_bar_par_strategies} provide a comparison of the proposed HPClust parallel strategies, while Tables~\ref{tab:result_e_compet_algos} and \ref{tab:result_t_compet_algos} compare the best HPClust parallel strategy with the selected competitive algorithms.

For instance, for a particular algorithm, we have the following entry in a table: ISOLET \#Succ = 6/7; Min = 0.01; Med = 0.24; Max = 0.59. In this case, the ratio 6/7 indicates that for each of the 7 different values of $k \in$ \{2, 3, 5, 10, 15, 20, 25\}, we performed a series of runs for each of the compared algorithms. For each fixed choice of $(X, k)$, the corresponding series consists of $n_{exec}$ = 15 independent runs of each algorithm. Thus, for each dataset, we have 7 series of runs for each of the compared algorithms, with each series containing 15 independent results. The number 6 in the \#Succ ratio 6/7 indicates that the median objective function values for 6 out of 7 series of runs of this algorithm were lower than the mean objective function values in the corresponding series of all other algorithms.

%The results of the main experiments, displayed in these tables, show the minimum, mean, and maximum metrics for each algorithm and dataset, with means in the final rows highlighting overall performance across datasets.
The means in the final rows of these tables highlight overall performance across datasets. The best results for each metric and dataset pair were bolded, indicating top algorithm performance. The highest accuracy values for each dataset are displayed in bold among the algorithm results. Success is indicated when an algorithm's median performance on a series of executions for a value of $k$ outperforms or matches the best result among all algorithms for this series.

\subsection{Scaling experiment}

%Also, we designed an additional experiment to showcase the scalability of the proposed HPClust schemes. We sampled a synthetic dataset containing $10$ Gaussian blobs uniformly scattered in the box $(-40, 40)$, each with a uniformly random standard deviation drawn in the range $(0, 10)$. Then, the number of points was varied according to the law $3^{i + 7}$, where $i = 0, \ldots, 8$. For each choice of $i$, $10$ repetitions of the execution for each algorithm were performed and recorded. For the HPClust and PBK-BDC algorithm, the sample size $s = \min \{5000, n_samples - 1000\}$ was used across all executions, and parameter $T = 3.0$ seconds to limit the processing time. For the HPClust-hybrid regime, the naive time split $T_1 = T_2 = T / 2$ was used to avoid optimizing this parameter. A uniform random noise of $500$ points drawn at random from the box $(-50.0, 50.0)$ was added to each generated synthetic dataset.

Additionally, we conducted an experiment to demonstrate the scalability of our proposed HPClust strategies. We generated a synthetic dataset with $10$ features comprising $10$ Gaussian blobs uniformly distributed within the box $(-40, 40)$, each with a randomly sampled standard deviation between $(0, 10)$. The number of points was varied according to the law $m = 3^{i + 7}$, where $i = 0, ..., 8$. For each $i$, we performed $10$ execution repetitions for each algorithm and recorded the results. We employed a sample size of $s = \min\{5000, m - 1000\}$ and a processing time limit of $T = 3.0$ seconds for the HPClust and PBK-BDC algorithms. For HPClust-hybrid, we used a naive time split of $T_1 = T_2 = T/2$ to avoid additional optimization. To introduce noise, we added $500$ random points uniformly distributed within the box $(-50.0, 50.0)$ to each synthetic dataset. This experiment allowed us to assess the scalability of our algorithms under varying dataset sizes.

%The experiments were performed on a system running Ubuntu 22.04 64-bit, powered by an AMD EPYC 7663 56-Core Processor, with up to 16 cores used in our experiments. The system was equipped with 1.46 TB of RAM. The software stack consisted of Python 3.10.11 along with NumPy 1.24.3 and Numba 0.57.0. The Numba~\cite{Marowka2018} package was used to accelerate Python code execution and facilitate parallelism. The use of Numba is particularly advantageous for these purposes due to its ability to compile Python code into machine code at runtime and its capabilities for executing code on multiple processors.

%The performance of various parallelization versions of HPClust was evaluated on 19 publicly available datasets. Descriptions of these datasets are provided in Table~\ref{tab:dataset_descr}, with more details available on their corresponding webpages listed in Table~\ref{tab:dataset_info}. In addition to these, four datasets were normalized, bringing the total to 23 datasets.

\begin{table}[!htbp]
\scriptsize
\centering
\caption{Brief description of the datasets}
\label{tab:dataset_descr}
\begin{tabular}{lllll}
\hline
\multicolumn{1}{|l|}{\multirow{2}{*}{Datasets}} & \multicolumn{1}{c|}{No. instances} & \multicolumn{1}{c|}{No. attributes} & \multicolumn{1}{c|}{Size}  & \multicolumn{1}{c|}{\multirow{2}{*}{File size}} \\
\multicolumn{1}{|l|}{} & \multicolumn{1}{c|}{$m$} & \multicolumn{1}{c|}{$n$} & \multicolumn{1}{c|}{$m \times n$} & \multicolumn{1}{c|}{} \\ \hline
CORD-19 Embeddings & 599616 & 768 & 460505088 & 8.84 GB \\
HEPMASS & 10500000 & 28 & 294000000 & 7.5 GB \\
US Census Data 1990 & 2458285 & 68 & 167163380 & 361 MB \\
Gisette & 13500 & 5000 & 67500000 & 152.5 MB \\
Music Analysis & 106574 & 518 & 55205332 & 951 MB \\
Protein Homology & 145751 & 74 & 10785574 & 69.6 MB \\
MiniBooNE Particle Identification & 130064 & 50 & 6503200 & 91.2 MB \\
MFCCs for Speech Emotion Recognition & 85134  & 58 & 4937772 & 95.2 MB \\ 
ISOLET & 7797 & 617 & 4810749 & 40.5 MB \\
Sensorless Drive Diagnosis & 58509 & 48 & 2808432 & 25.6 MB \\
Online News Popularity & 39644 & 58 & 2299352 & 24.3 MB \\
Gas Sensor Array Drift & 13910 & 128 & 1780480 & 23.54 MB \\
3D Road Network & 434874 & 3 & 1304622 & 20.7 MB \\
KEGG Metabolic Relation Network (Directed) & 53413 & 20 & 1068260 & 7.34 MB \\
Skin Segmentation & 245057 & 3 & 735171 & 3.4 MB \\
Shuttle Control & 58000 & 9 & 522000 & 1.55 MB \\
EEG Eye State & 14980 & 14 & 209720 & 1.7 MB \\
Pla85900 & 85900 & 2 & 171800 & 1.79 MB \\
D15112 & 15112 & 2 & 30224 & 247 kB \\
\hline
\end{tabular}
\end{table}

\begin{table}[!htbp]
\scriptsize
\centering
\caption{URLs for the used datasets}
\label{tab:dataset_info}
\begin{tabular}{lp{8cm}}
\hline
\multicolumn{1}{|l|}{Datasets} & \multicolumn{1}{|l|}{URLs} \\ \hline
CORD-19 Embeddings & \url{https://www.kaggle.com/allen-institute-for-ai/CORD-19-research-challenge} \\
HEPMASS & \url{https://archive.ics.uci.edu/ml/datasets/HEPMASS}  \\
US Census Data 1990 & \url{https://archive.ics.uci.edu/ml/datasets/US+Census+Data+(1990)} \\
Gisette & \url{https://archive.ics.uci.edu/ml/datasets/Gisette} \\
Music Analysis & \url{https://archive.ics.uci.edu/ml/datasets/FMA\%3A+A+Dataset+For+Music+Analysis} \\
Protein Homology & \url{https://www.kdd.org/kdd-cup/view/kdd-cup-2004/Data} \\
MiniBooNE Particle Identification & \url{https://archive.ics.uci.edu/ml/datasets/MiniBooNE+particle+identification} \\
MFCCs for Speech Emotion Recognition & \url{https://www.kaggle.com/cracc97/features} \\
ISOLET & \url{https://archive.ics.uci.edu/ml/datasets/isolet} \\
Sensorless Drive Diagnosis & \url{https://archive.ics.uci.edu/ml/datasets/dataset+for+sensorless+drive+diagnosis} \\
Online News Popularity & \url{https://archive.ics.uci.edu/ml/datasets/online+news+popularity} \\
Gas Sensor Array Drift & \url{https://archive.ics.uci.edu/ml/datasets/gas+sensor+array+drift+dataset} \\
3D Road Network & \url{https://archive.ics.uci.edu/ml/datasets/3D+Road+Network+(North+Jutland,+Denmark)} \\
KEGG Metabolic Relation Network (Directed) & \url{https://archive.ics.uci.edu/ml/datasets/KEGG+Metabolic+Relation+Network+(Directed)} \\
Skin Segmentation & \url{https://archive.ics.uci.edu/ml/datasets/skin+segmentation} \\
Shuttle Control &  \url{https://archive.ics.uci.edu/ml/datasets/Statlog+(Shuttle)} \\
%EEG Eye State & https://archive.ics.uci.edu/ml/datasets/EEG+Eye+State \\
Pla85900 & \url{http://softlib.rice.edu/pub/tsplib/tsp/pla85900.tsp.gz} \\
D15112 & \url{https://github.com/mastqe/tsplib/blob/master/d15112.tsp} \\
\hline
\end{tabular}
\end{table}

\section{Experimental Results and Discussion} \label{sec:results}

\subsection{Performance Evaluation}

\begin{figure}[htbp]
  % \begin{adjustwidth}{-\extralength}{0cm}
  \captionsetup[subfigure]{justification=centering}
  \centering
  \begin{subfigure}{0.65\textwidth}
    \includegraphics[width=\linewidth]{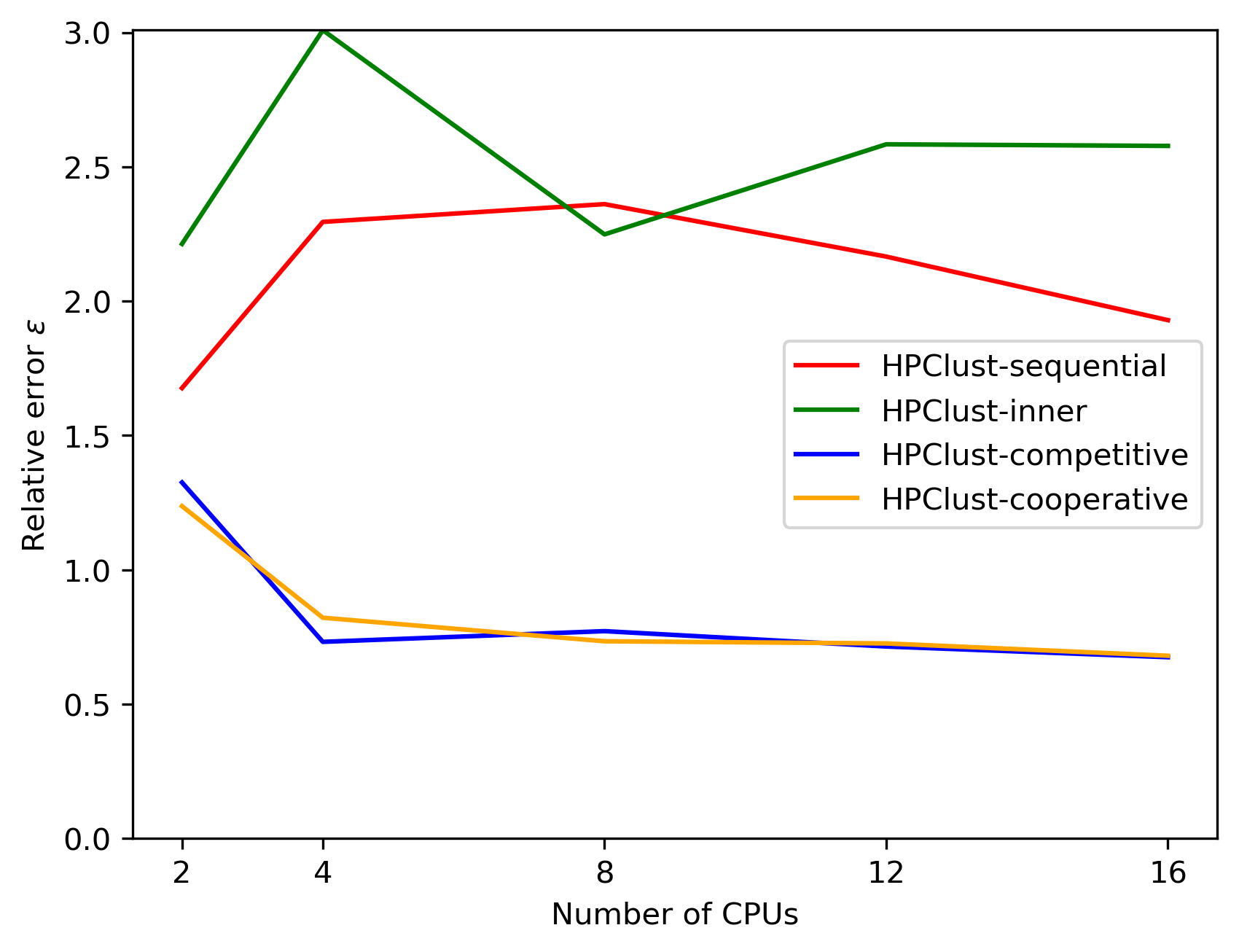}
    \caption{Median accuracy values}
    \label{fig:n_workers_e}
  \end{subfigure}
  %\hfill
  \par\bigskip
  \begin{subfigure}{0.65\textwidth}
    \includegraphics[width=\linewidth]{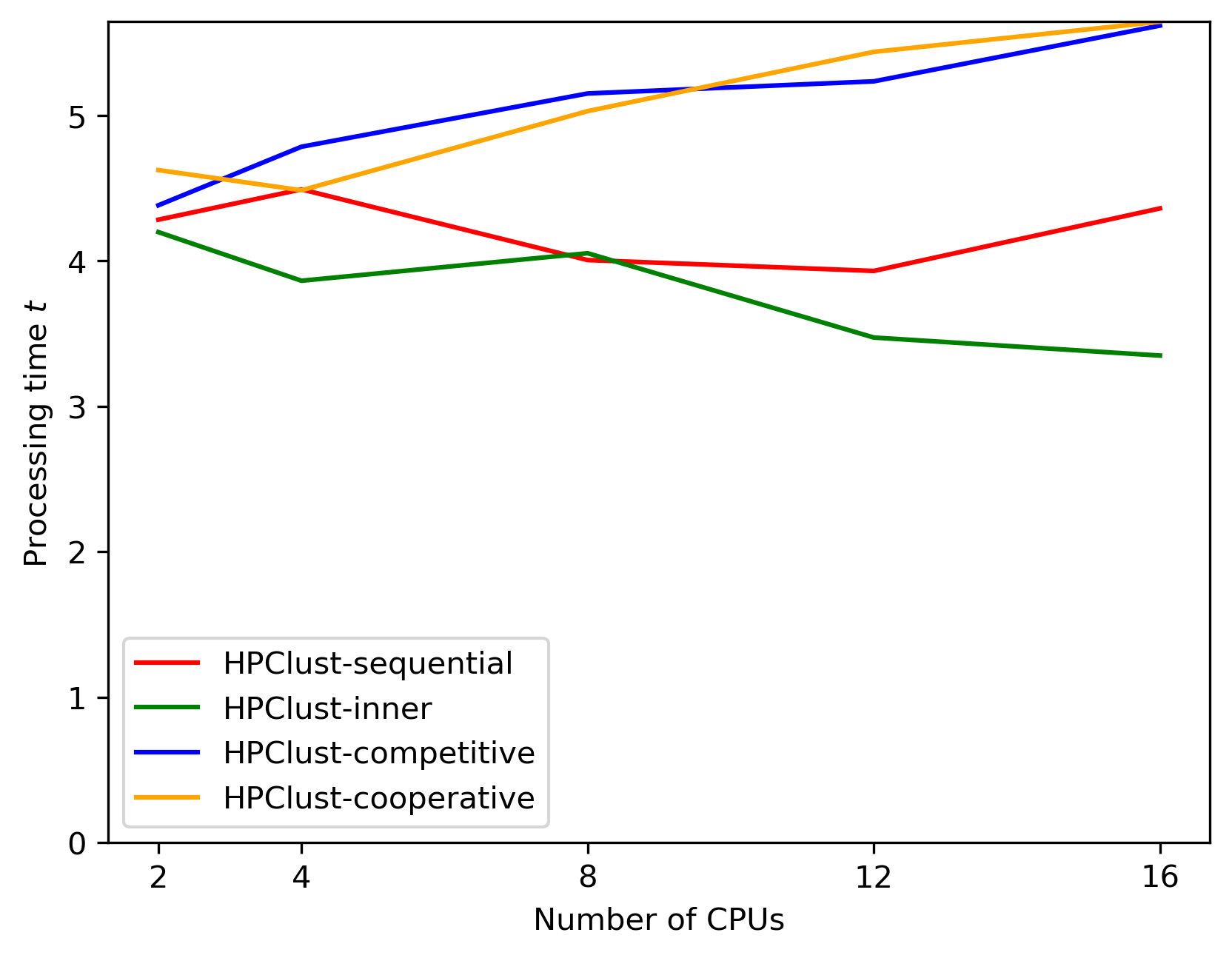}
    \caption{Median processing times}
    \label{fig:n_workers_t}
  \end{subfigure}
  \caption{Comparative results of the algorithms with respect to the number of employed CPUs averaged across all datasets}
  % \end{adjustwidth}
\end{figure}

The results of the first set of preliminary experiments, illustrated in Figures~\ref{fig:n_workers_e} and \ref{fig:n_workers_t}, determined the optimal number of CPUs for subsequent experiments, setting the stage for further investigation. As anticipated, the fully sequential strategy (HPClust-sequential) displayed no significant correlation with the number of parallel processors employed. The HPClust version with inner parallelism demonstrated a reduction in processing time with an increase in the number of CPUs, while the accuracy remained independent of the CPU count. In contrast, both the HPClust-competitive and HPClust-cooperative strategies exhibited an improvement in clustering accuracy as the number of CPUs increased. However, this accuracy gain came at the expense of increased processing time for these versions of HPClust. We attribute this observation to the need for coordination among multiple processors and the technical complexities introduced by Numba, such as parallel access to shared memory locations by multiple workers. Upon closer examination of the scores, we determined that utilizing 8 CPUs strikes the optimal balance between processing time and resulting accuracy across all algorithms on our machine. Thus, this choice of the CPU count was used in all the subsequent experiments.

Other preliminary experiments were straightforward. They allowed to obtain the necessary optimal values of the parameters for the main set of experiments.

A summary of the results of the main experiments are provided in Tables~\ref{tab:result_e_par_strategies}, \ref{tab:result_t_bar_par_strategies}, \ref{tab:result_e_compet_algos}, and \ref{tab:result_t_compet_algos}. Full details of the results of the main experiments are provided in Appendix~\ref{sec:appendix}.

\begin{table}[p]%
	\centering%
	\caption{Resulting relative clustering accuracies $\epsilon$ (\%) for the proposed parallel HPClust strategies.}%
	\label{tab:result_e_par_strategies}%
%	\begin{adjustwidth}{-\extralength}{0cm}
%		\newcolumntype{C}{>{\centering\arraybackslash}X}
	\resizebox{!}{\summtableheight}{%
		\begin{tabular}{l|cccc|cccc}
%		\begin{tabularx}{\fulllength}{p{5cm}|CCCC|CCCC}
			\hline
			\multirow{2}{*}{Dataset} & \multicolumn{4}{p{3cm}}{\mbox{HPClust-inner}}& \multicolumn{4}{p{3cm}}{\mbox{HPClust-competitive}} \\
			\cline{2-9}
			& \#Succ & Min & Median & Max & \#Succ & Min & Median & Max  \\
			\hline
			CORD-19 Embeddings & 0/7 & 0.07 & 0.21 & 0.34 & 3/7 & \textbf{0.0} & \textbf{0.07} & 0.18 \\
			HEPMASS & 0/7 & 0.08 & 0.22 & 0.66 & 3/7 & 0.03 & \textbf{0.07} & \textbf{0.19} \\
			US Census Data 1990 & 2/7 & 0.92 & 3.13 & 35.87 & 3/7 & 0.48 & \textbf{1.48} & \textbf{2.89} \\
			Gisette & 0/7 & -0.43 & -0.37 & -0.19 & 2/7 & -0.44 & -0.38 & -0.32 \\
			Music Analysis & 3/7 & 0.41 & 0.91 & 6.24 & 4/7 & 0.43 & \textbf{0.74} & \textbf{1.67} \\
			Protein Homology & 3/7 & \textbf{0.15} & 0.91 & 2.32 & 1/7 & 0.41 & \textbf{0.88} & 2.03 \\
			MiniBooNE Particle Identification & 2/7 & -0.03 & 0.51 & 402305.65 & 1/7 & -0.07 & \textbf{-0.0} & 719099.04 \\
			MiniBooNE Particle Identification (normalized) & 1/7 & 0.2 & 0.54 & 101.63 & 2/7 & 0.2 & 0.55 & \textbf{1.1} \\
			MFCCs for Speech Emotion Recognition & 2/7 & 0.14 & 0.64 & 1.95 & 1/7 & 0.11 & 0.34 & \textbf{0.76} \\
			ISOLET & 0/7 & 0.15 & 0.68 & 1.72 & 1/7 & 0.04 & \textbf{0.23} & 0.63 \\
			Sensorless Drive Diagnosis & 1/7 & -0.32 & 1.25 & 31.06 & 2/7 & \textbf{-0.43} & \textbf{-0.27} & 12.2 \\
			Sensorless Drive Diagnosis (normalized) & 1/7 & 0.4 & 3.03 & 9.69 & 4/7 & 0.31 & \textbf{1.06} & \textbf{3.26} \\
			Online News Popularity & 2/7 & 0.7 & 2.36 & 14.39 & 2/7 & 0.69 & 1.65 & \textbf{3.74} \\
			Gas Sensor Array Drift & 2/7 & 0.15 & 3.24 & 12.29 & 2/7 & \textbf{-0.05} & 1.78 & \textbf{3.77} \\
			3D Road Network & 2/7 & 0.04 & 0.4 & 1.24 & 2/7 & \textbf{0.03} & 0.22 & 1.06 \\
			Skin Segmentation & 1/7 & 0.04 & 2.91 & 9.72 & 2/7 & -0.05 & 1.05 & 4.36 \\
			KEGG Metabolic Relation Network (Directed) & 3/7 & -0.08 & 1.55 & 34.13 & 2/7 & \textbf{-0.42} & \textbf{0.24} & \textbf{2.5} \\
			Shuttle Control & 1/8 & 0.17 & 5.68 & 41.76 & 2/8 & -0.01 & 2.32 & 12.58 \\
			Shuttle Control (normalized) & 1/8 & 0.89 & 2.81 & 17.98 & 0/8 & \textbf{0.69} & 1.79 & \textbf{4.07} \\
			EEG Eye State & 3/8 & 0.54 & 0.79 & 7.15 & 2/8 & 0.53 & 0.56 & 119444.14 \\
			EEG Eye State (normalized) & 0/8 & \textbf{-0.06} & 2.4 & 31.49 & 6/8 & \textbf{-0.06} & \textbf{0.01} & 67.39 \\
			Pla85900 & 0/7 & \textbf{0.07} & 0.37 & 1.7 & 3/7 & \textbf{0.07} & \textbf{0.2} & 0.73 \\
			D15112 & 2/7 & 0.1 & 0.48 & 1.76 & 3/7 & 0.08 & \textbf{0.14} & \textbf{0.4} \\
			\hline
			Overall Results & 32/165 & 0.19 & 1.51 & 17507.42 & 53/165 & 0.11 & 0.64 & 36463.84 \\ \hline
		\end{tabular}%
	}

	\vspace{10pt}
%	\newpage
	
	\resizebox{!}{\summtableheight}{%
		\begin{tabular}{l|cccc|cccc}
%		\begin{tabularx}{\fulllength}{p{5cm}|XXXX|XXXX}
			\hline
			\multirow{2}{*}{Dataset} & \multicolumn{4}{p{3cm}}{\mbox{HPClust-cooperative}}& \multicolumn{4}{p{3cm}}{\mbox{HPClust-hybrid}} \\
			\cline{2-9}
			& \#Succ & Min & Median & Max & \#Succ & Min & Median & Max  \\
			\hline
			CORD-19 Embeddings & 3/7 & 0.04 & 0.08 & \textbf{0.16} & 1/7 & 0.02 & 0.08 & 0.24 \\
			HEPMASS & 0/7 & 0.04 & 0.16 & 0.57 & 4/7 & \textbf{-0.01} & 0.08 & 0.24 \\
			US Census Data 1990 & 1/7 & 0.45 & 1.64 & 4.37 & 1/7 & \textbf{0.32} & 1.72 & 3.17 \\
			Gisette & 2/7 & \textbf{-0.46} & -0.39 & -0.32 & 3/7 & -0.44 & \textbf{-0.4} & \textbf{-0.34} \\
			Music Analysis & 0/7 & 0.4 & 0.83 & 2.68 & 0/7 & \textbf{0.33} & 0.85 & 2.24 \\
			Protein Homology & 2/7 & 0.21 & 0.91 & \textbf{1.81} & 1/7 & 0.5 & 1.05 & 2.1 \\
			MiniBooNE Particle Identification & 2/7 & \textbf{-0.08} & \textbf{0.0} & 0.37 & 2/7 & -0.07 & \textbf{-0.0} & \textbf{0.15} \\
			MiniBooNE Particle Identification (normalized) & 1/7 & \textbf{0.19} & 0.56 & 1.43 & 3/7 & 0.23 & \textbf{0.51} & 1.29 \\
			MFCCs for Speech Emotion Recognition & 2/7 & \textbf{0.1} & 0.34 & 0.94 & 2/7 & 0.12 & \textbf{0.33} & 0.83 \\
			ISOLET & 2/7 & 0.03 & 0.25 & 0.68 & 4/7 & \textbf{0.01} & \textbf{0.23} & \textbf{0.59} \\
			Sensorless Drive Diagnosis & 2/7 & -0.41 & -0.21 & 11.82 & 2/7 & -0.42 & -0.21 & \textbf{8.18} \\
			Sensorless Drive Diagnosis (normalized) & 1/7 & \textbf{0.28} & 1.39 & 4.0 & 1/7 & 0.38 & 1.34 & 3.81 \\
			Online News Popularity & 2/7 & 0.56 & \textbf{1.6} & 7.79 & 1/7 & \textbf{0.47} & 1.69 & 7.86 \\
			Gas Sensor Array Drift & 1/7 & -0.04 & 0.91 & 4.05 & 2/7 & 0.06 & \textbf{0.79} & 3.99 \\
			3D Road Network & 2/7 & 0.04 & 0.22 & 1.04 & 1/7 & 0.04 & \textbf{0.21} & \textbf{0.88} \\
			Skin Segmentation & 2/7 & \textbf{-0.22} & 1.11 & 5.76 & 2/7 & -0.02 & \textbf{1.02} & \textbf{4.25} \\
			KEGG Metabolic Relation Network (Directed) & 1/7 & -0.3 & 0.35 & 6.26 & 1/7 & -0.29 & 0.25 & 23.7 \\
			Shuttle Control & 4/8 & \textbf{-0.14} & \textbf{1.55} & \textbf{4.76} & 1/8 & 0.08 & 1.86 & 9.13 \\
			Shuttle Control (normalized) & 2/8 & 0.81 & 2.22 & 4.97 & 5/8 & 0.71 & \textbf{1.61} & 4.49 \\
			EEG Eye State & 1/8 & 0.53 & 0.57 & \textbf{0.76} & 2/8 & \textbf{0.52} & \textbf{0.55} & 6.05 \\
			EEG Eye State (normalized) & 0/8 & \textbf{-0.06} & \textbf{0.01} & 56.9 & 2/8 & \textbf{-0.06} & 0.02 & \textbf{16.2} \\
			Pla85900 & 2/7 & \textbf{0.07} & 0.22 & 1.22 & 2/7 & \textbf{0.07} & \textbf{0.2} & \textbf{0.58} \\
			D15112 & 1/7 & 0.08 & 0.29 & 0.8 & 1/7 & \textbf{0.07} & 0.15 & 0.44 \\
			\hline
			Overall Results & 36/165 & \textbf{0.09} & 0.63 & 5.34 & 44/165 & 0.11 & \textbf{0.61} & \textbf{4.35} \\ \hline
		\end{tabular}%
	}
%	\end{adjustwidth}
\end{table}

\bigskip

\begin{table}[!htbp]%
	\centering%
	\caption{Baseline convergence times $\overline{t}$ (in seconds) of the HPClust parallel strategies.}%
	\label{tab:result_t_bar_par_strategies}%
	\resizebox{!}{\summtableheight}{%
		\begin{tabular}{l|cccc|cccc}
			\hline
			\multirow{2}{*}{Dataset} & \multicolumn{4}{p{3cm}}{\mbox{HPClust-inner}}& \multicolumn{4}{p{3cm}}{\mbox{HPClust-competitive}} \\
			\cline{2-9}
			& \#Succ & Min & Median & Max & \#Succ & Min & Median & Max  \\
			\hline
			CORD-19 Embeddings & 2/7 & \textbf{6.92} & 16.1 & 24.82 & 1/7 & 12.78 & 17.45 & 24.12 \\
			HEPMASS & 0/7 & 5.77 & 8.65 & 15.59 & 2/7 & \textbf{2.35} & 5.24 & 14.12 \\
			US Census Data 1990 & 0/7 & 0.24 & 0.63 & 2.07 & 1/7 & 0.19 & 0.53 & 1.63 \\
			Gisette & 4/7 & \textbf{3.27} & \textbf{4.4} & \textbf{6.38} & 0/7 & 17.06 & 19.38 & 23.72 \\
			Music Analysis & 0/7 & \textbf{0.58} & \textbf{3.22} & 7.19 & 0/7 & 1.44 & 4.02 & 7.89 \\
			Protein Homology & 2/7 & \textbf{0.79} & \textbf{1.71} & \textbf{3.18} & 1/7 & 1.63 & 2.45 & 4.03 \\
			MiniBooNE Particle Identification & 4/7 & \textbf{0.46} & \textbf{1.05} & \textbf{2.38} & 1/7 & 2.24 & 3.07 & 4.39 \\
			MiniBooNE Particle Identification (normalized) & 3/7 & \textbf{0.09} & \textbf{0.4} & \textbf{0.85} & 1/7 & 0.28 & 0.49 & 0.91 \\
			MFCCs for Speech Emotion Recognition & 1/7 & \textbf{0.12} & \textbf{0.39} & \textbf{0.83} & 1/7 & 0.29 & 0.57 & 0.96 \\
			ISOLET & 0/7 & \textbf{0.38} & \textbf{1.01} & 2.93 & 0/7 & 0.85 & 1.88 & 3.84 \\
			Sensorless Drive Diagnosis & 3/7 & \textbf{0.14} & \textbf{0.29} & \textbf{0.9} & 0/7 & 0.72 & 1.02 & 2.07 \\
			Sensorless Drive Diagnosis (normalized) & 0/7 & \textbf{0.02} & \textbf{0.09} & 0.28 & 0/7 & 0.04 & \textbf{0.09} & 0.26 \\
			Online News Popularity & 2/7 & \textbf{0.09} & \textbf{0.27} & 0.59 & 0/7 & 0.15 & 0.29 & 0.62 \\
			Gas Sensor Array Drift & 0/7 & \textbf{0.11} & \textbf{0.47} & 1.68 & 0/7 & 0.29 & 0.69 & 1.64 \\
			3D Road Network & 2/7 & \textbf{0.08} & \textbf{0.23} & \textbf{0.49} & 0/7 & 0.15 & 0.35 & 0.88 \\
			Skin Segmentation & 0/7 & 0.03 & 0.07 & 0.18 & 1/7 & \textbf{0.02} & 0.05 & \textbf{0.12} \\
			KEGG Metabolic Relation Network (Directed) & 0/7 & \textbf{0.1} & \textbf{0.3} & \textbf{0.82} & 1/7 & 0.26 & 0.46 & 0.97 \\
			Shuttle Control & 0/8 & 0.1 & 0.32 & 0.87 & 0/8 & 0.09 & 0.29 & 0.74 \\
			Shuttle Control (normalized) & 0/8 & 0.04 & 0.15 & 0.32 & 3/8 & \textbf{0.02} & 0.07 & \textbf{0.2} \\
			EEG Eye State & 0/8 & 0.13 & 0.43 & 1.11 & 0/8 & \textbf{0.06} & 0.31 & 0.78 \\
			EEG Eye State (normalized) & 0/8 & \textbf{0.04} & \textbf{0.11} & 0.74 & 0/8 & 0.06 & 0.12 & 0.34 \\
			Pla85900 & 0/7 & 0.07 & 0.64 & 1.42 & 1/7 & \textbf{0.05} & 0.35 & 1.14 \\
			D15112 & 0/7 & 0.06 & 0.42 & 1.11 & 3/7 & 0.06 & \textbf{0.21} & \textbf{0.77} \\
			\hline
			Overall Results & 23/165 & \textbf{0.85} & \textbf{1.8} & \textbf{3.34} & 17/165 & 1.79 & 2.58 & 4.18 \\ \hline
		\end{tabular}%
	}
		
	\vspace{10pt}
		
	\resizebox{!}{\summtableheight}{%
		\begin{tabular}{l|cccc|cccc}
			\hline
			\multirow{2}{*}{Dataset} & \multicolumn{4}{p{3cm}}{\mbox{HPClust-cooperative}}& \multicolumn{4}{p{3cm}}{\mbox{HPClust-hybrid}} \\
			\cline{2-9}
			& \#Succ & Min & Median & Max & \#Succ & Min & Median & Max  \\
			\hline
			CORD-19 Embeddings & 0/7 & 12.06 & 18.2 & 26.11 & 2/7 & 11.49 & \textbf{16.03} & \textbf{23.92} \\
			HEPMASS & 4/7 & 2.92 & \textbf{4.86} & \textbf{12.03} & 0/7 & 2.6 & 6.69 & 16.98 \\
			US Census Data 1990 & 1/7 & \textbf{0.14} & 0.46 & 1.47 & 3/7 & \textbf{0.14} & \textbf{0.45} & \textbf{1.41} \\
			Gisette & 0/7 & 16.95 & 18.99 & 23.07 & 0/7 & 16.99 & 19.33 & 23.71 \\
			Music Analysis & 2/7 & 1.58 & 3.33 & \textbf{6.99} & 0/7 & 1.35 & 3.88 & 8.19 \\
			Protein Homology & 2/7 & 1.73 & 2.52 & 4.23 & 0/7 & 1.83 & 2.91 & 4.43 \\
			MiniBooNE Particle Identification & 0/7 & 2.19 & 2.83 & 4.4 & 1/7 & 2.05 & 3.05 & 4.33 \\
			MiniBooNE Particle Identification (normalized) & 1/7 & 0.29 & 0.51 & 0.88 & 0/7 & 0.25 & 0.53 & 1.0 \\
			MFCCs for Speech Emotion Recognition & 1/7 & 0.22 & 0.49 & 0.99 & 1/7 & 0.26 & 0.55 & 1.06 \\
			ISOLET & 3/7 & 0.87 & 1.42 & \textbf{2.88} & 0/7 & 0.76 & 1.96 & 4.07 \\
			Sensorless Drive Diagnosis & 0/7 & 0.62 & 1.05 & 2.0 & 1/7 & 0.72 & 1.05 & 1.95 \\
			Sensorless Drive Diagnosis (normalized) & 3/7 & 0.04 & \textbf{0.09} & \textbf{0.25} & 2/7 & 0.04 & \textbf{0.09} & 0.27 \\
			Online News Popularity & 2/7 & 0.14 & 0.28 & \textbf{0.56} & 1/7 & 0.14 & 0.29 & 0.71 \\
			Gas Sensor Array Drift & 1/7 & 0.27 & 0.63 & \textbf{1.62} & 1/7 & 0.27 & 0.73 & 1.74 \\
			3D Road Network & 0/7 & 0.18 & 0.33 & 0.87 & 1/7 & 0.16 & 0.37 & 1.18 \\
			Skin Segmentation & 6/7 & \textbf{0.02} & \textbf{0.04} & 0.18 & 0/7 & \textbf{0.02} & \textbf{0.04} & 0.16 \\
			KEGG Metabolic Relation Network (Directed) & 2/7 & 0.24 & 0.42 & 0.98 & 1/7 & 0.25 & 0.44 & 0.97 \\
			Shuttle Control & 2/8 & 0.09 & \textbf{0.21} & \textbf{0.59} & 2/8 & \textbf{0.08} & \textbf{0.21} & 0.67 \\
			Shuttle Control (normalized) & 5/8 & \textbf{0.02} & \textbf{0.05} & 0.21 & 0/8 & \textbf{0.02} & 0.07 & 0.26 \\
			EEG Eye State & 2/8 & 0.07 & 0.23 & 0.89 & 3/8 & 0.08 & \textbf{0.22} & \textbf{0.77} \\
			EEG Eye State (normalized) & 2/8 & 0.06 & \textbf{0.11} & \textbf{0.33} & 0/8 & 0.06 & 0.15 & 0.46 \\
			Pla85900 & 6/7 & 0.06 & \textbf{0.24} & 1.16 & 0/7 & 0.06 & 0.35 & \textbf{1.11} \\
			D15112 & 3/7 & 0.05 & 0.24 & 0.83 & 1/7 & \textbf{0.04} & 0.24 & 0.89 \\
			\hline
			Overall Results & 48/165 & 1.77 & 2.5 & 4.07 & 20/165 & 1.72 & 2.59 & 4.36 \\ \hline
		\end{tabular}%
	}
\end{table}

\bigskip

\begin{table}[!htbp]%
	\centering%
	\caption{Relative clustering accuracies $\epsilon$ (in \%) resulting from the comparison of the hybrid HPClust strategy with the competitive algorithms.}%
	\label{tab:result_e_compet_algos}%
%	\resizebox{!}{\summtableheight}{%
	\begin{adjustwidth}{-\extralength}{0cm}
		\newcolumntype{C}{>{\centering\arraybackslash}X}
		\begin{tabularx}{\fulllength}{p{4cm}|CCCC|CCCC|CCCC}
			\toprule
			\multirow{2}{*}{Dataset} & \multicolumn{4}{p{3cm}}{\mbox{HPClust-hybrid}}& \multicolumn{4}{p{3cm}}{\mbox{Forgy K-means}}& \multicolumn{4}{p{3cm}}{\mbox{PBK-BDC}} \\
			\cline{2-13}
			& \#Succ & Min & Med & Max & \#Succ & Min & Med & Max & \#Succ & Min & Med & Max  \\
			\midrule
			CORD-19 Embeddings & 3/7 & 0.02 & \textbf{0.08} & \textbf{0.24} & 4/7 & \textbf{0.01} & 0.17 & 1.37 & 0/7 & 0.67 & 1.74 & 3.28 \\
			HEPMASS & 5/7 & \textbf{-0.01} & \textbf{0.08} & \textbf{0.24} & 2/7 & 0.02 & 0.18 & 0.63 & 0/7 & 0.63 & 1.45 & 3.21 \\
			US Census Data 1990 & 6/7 & \textbf{0.32} & \textbf{1.72} & \textbf{3.17} & 1/7 & 2.58 & 80.73 & 259.79 & 0/7 & 14.86 & 65.27 & 279.29 \\
			Gisette & 0/7 & -0.44 & -0.4 & -0.34 & 7/7 & \textbf{-0.52} & \textbf{-0.48} & \textbf{-0.39} & 0/7 & -0.47 & -0.42 & -0.32 \\
			Music Analysis & 1/7 & 0.33 & 0.85 & \textbf{2.24} & 6/7 & \textbf{-0.01} & \textbf{0.47} & 6.97 & 0/7 & 1.27 & 4.85 & 42.27 \\
			Protein Homology & 4/7 & \textbf{0.5} & \textbf{1.05} & \textbf{2.1} & 3/7 & 14.84 & 14.91 & 15.09 & 0/7 & 4.98 & 20.63 & 48.21 \\
			MiniBooNE Particle Identification & 4/7 & \textbf{-0.07} & \textbf{-0.0} & \textbf{0.15} & 3/7 & 2.62 & 19.52 & $111 \cdot 10^3$ & 0/7 & 2.61 & $41 \cdot 10^3$ & $111 \cdot 10^3$ \\
			MiniBooNE Particle Identification (normalized) & 2/7 & 0.23 & \textbf{0.51} & \textbf{1.29} & 5/7 & \textbf{-0.02} & 1.39 & 240.25 & 0/7 & 2.34 & 7.75 & 36.83 \\
			MFCCs for Speech Emotion Recognition & 4/7 & \textbf{0.12} & \textbf{0.33} & \textbf{0.83} & 3/7 & 0.22 & 1.49 & 2.92 & 0/7 & 1.97 & 10.1 & 40.56 \\
			ISOLET & 6/7 & \textbf{0.01} & \textbf{0.23} & \textbf{0.59} & 1/7 & 0.05 & 0.8 & 2.78 & 0/7 & 0.22 & 1.03 & 2.59 \\
			Sensorless Drive Diagnosis & 7/7 & \textbf{-0.42} & \textbf{-0.21} & \textbf{8.18} & 0/7 & 122.75 & 162.37 & 183.78 & 0/7 & 149.77 & 162.36 & 215.62 \\
			Sensorless Drive Diagnosis (normalized) & 6/7 & \textbf{0.38} & \textbf{1.34} & \textbf{3.81} & 1/7 & 1.3 & 6.21 & 26.96 & 0/7 & 4.49 & 11.24 & 48.1 \\
			Online News Popularity & 5/7 & \textbf{0.47} & \textbf{1.69} & \textbf{7.86} & 2/7 & 7.76 & 14.93 & 33.83 & 0/7 & 15.31 & 37.76 & 93.96 \\
			Gas Sensor Array Drift & 5/7 & \textbf{0.06} & \textbf{0.79} & \textbf{3.99} & 2/7 & 10.01 & 24.31 & 39.62 & 0/7 & 9.52 & 25.52 & 39.35 \\
			3D Road Network & 1/7 & 0.04 & \textbf{0.21} & 0.88 & 6/7 & \textbf{0.0} & 0.23 & \textbf{0.23} & 0/7 & 2.67 & 40.65 & 159.28 \\
			Skin Segmentation & 5/7 & \textbf{-0.02} & \textbf{1.02} & \textbf{4.25} & 2/7 & 2.17 & 9.02 & 21.32 & 0/7 & 7.46 & 20.55 & 71.1 \\
			KEGG Metabolic Relation Network (Directed) & 6/7 & \textbf{-0.29} & \textbf{0.25} & \textbf{23.7} & 1/7 & 94.27 & 95.67 & 108.63 & 0/7 & 94.26 & 94.92 & 107.54 \\
			Shuttle Control & 8/8 & \textbf{0.08} & \textbf{1.86} & \textbf{9.13} & 0/8 & 131.85 & 176.25 & 243.9 & 0/8 & 139.77 & 174.3 & 231.7 \\
			Shuttle Control (normalized) & 6/8 & \textbf{0.71} & \textbf{1.61} & \textbf{4.49} & 2/8 & 2.63 & 16.59 & 74.13 & 0/8 & 8.54 & 31.94 & 105.37 \\
			EEG Eye State & 7/8 & \textbf{0.52} & \textbf{0.55} & \textbf{6.05} & 1/8 & 27.46 & $876 \cdot 10^3$ & $102 \cdot 10^4$ & 0/8 & 3.81 & $804 \cdot 10^3$ & $102 \cdot 10^4$ \\
			EEG Eye State (normalized) & 8/8 & \textbf{-0.06} & \textbf{0.02} & \textbf{16.2} & 0/8 & 100.2 & 542.0 & 763.41 & 0/8 & 131.31 & 572.73 & 758.52 \\
			Pla85900 & 5/7 & 0.07 & \textbf{0.2} & \textbf{0.58} & 2/7 & \textbf{-0.02} & 0.39 & 1.99 & 0/7 & 2.55 & 10.5 & 39.62 \\
			D15112 & 4/7 & \textbf{0.07} & \textbf{0.15} & \textbf{0.44} & 3/7 & 0.11 & 1.15 & 5.82 & 0/7 & 0.31 & 1.41 & 6.39 \\
			\midrule
			Overall Results & 108/165 & \textbf{0.11} & \textbf{0.61} & \textbf{4.35} & 57/165 & 22.62 & {\scriptsize 38147.66} & {\scriptsize 49297.36} & 0/165 & 26.04 & {\scriptsize 36793.75} & {\scriptsize 49310.86} \\ \bottomrule
		\end{tabularx}%
	\end{adjustwidth}
%	}
\end{table}

\bigskip

\begin{table}[!htbp]%
	\centering%
	\caption{Total clustering times $t$ (in seconds) resulting from the comparison of the hybrid HPClust strategy with the competitive algorithms.}%
	\label{tab:result_t_compet_algos}%
%	\resizebox{!}{\summtableheight}{%
	\begin{adjustwidth}{-\extralength}{0cm}
		\newcolumntype{C}{>{\centering\arraybackslash}X}
		\begin{tabularx}{\fulllength}{p{5cm}|CCCC|CCCC|CCCC}
			\toprule
			\multirow{2}{*}{Dataset} & \multicolumn{4}{p{3cm}}{\mbox{HPClust-hybrid}}& \multicolumn{4}{p{3cm}}{\mbox{Forgy K-means}}& \multicolumn{4}{p{3cm}}{\mbox{PBK-BDC}} \\
			\cline{2-13}
			& \#Succ & Min & Med & Max & \#Succ & Min & Med & Max & \#Succ & Min & Med & Max  \\
			\midrule
			CORD-19 Embeddings & 4/7 & \textbf{14.71} & \textbf{27.15} & \textbf{36.39} & 0/7 & 419.46 & 704.62 & 1696.51 & 3/7 & 60.31 & 76.19 & 105.52 \\
			HEPMASS & 4/7 & \textbf{6.16} & \textbf{19.99} & \textbf{28.4} & 0/7 & 343.81 & 508.85 & 865.84 & 3/7 & 33.09 & 35.56 & 39.44 \\
			US Census Data 1990 & 5/7 & \textbf{0.34} & \textbf{2.08} & \textbf{2.96} & 0/7 & 29.55 & 61.8 & 120.46 & 2/7 & 4.18 & 4.72 & 5.46 \\
			Gisette & 6/7 & \textbf{18.01} & \textbf{21.12} & \textbf{26.19} & 1/7 & 28.7 & 52.93 & 97.55 & 0/7 & 21.53 & 33.18 & 63.21 \\
			Music Analysis & 4/7 & \textbf{1.51} & \textbf{5.61} & \textbf{8.53} & 0/7 & 49.88 & 86.6 & 145.67 & 3/7 & 5.34 & 7.32 & 10.42 \\
			Protein Homology & 4/7 & \textbf{1.82} & \textbf{3.37} & \textbf{5.21} & 0/7 & 13.77 & 19.31 & 31.43 & 3/7 & 5.56 & 7.9 & 11.86 \\
			MiniBooNE Particle Identification & 4/7 & \textbf{2.37} & \textbf{4.33} & \textbf{6.37} & 2/7 & 7.64 & 12.36 & 17.04 & 1/7 & 7.83 & 11.92 & 18.68 \\
			MiniBooNE Particle Identification (normalized) & 4/7 & \textbf{0.34} & \textbf{0.79} & \textbf{1.42} & 0/7 & 4.07 & 7.14 & 15.28 & 3/7 & 0.93 & 1.21 & 1.77 \\
			MFCCs for Speech Emotion Recognition & 3/7 & \textbf{0.28} & \textbf{0.72} & \textbf{1.26} & 0/7 & 2.99 & 4.91 & 8.07 & 4/7 & 0.67 & 0.94 & 1.3 \\
			ISOLET & 0/7 & 1.03 & 3.55 & 4.97 & 0/7 & 1.11 & 1.76 & 3.52 & 7/7 & \textbf{0.4} & \textbf{0.76} & \textbf{1.52} \\
			Sensorless Drive Diagnosis & 3/7 & \textbf{0.78} & \textbf{1.57} & \textbf{2.71} & 3/7 & 1.35 & 2.15 & 4.06 & 1/7 & 1.23 & 2.08 & 4.09 \\
			Sensorless Drive Diagnosis (normalized) & 2/7 & \textbf{0.05} & 0.22 & 0.33 & 0/7 & 0.4 & 0.76 & 1.9 & 5/7 & 0.1 & \textbf{0.15} & \textbf{0.21} \\
			Online News Popularity & 3/7 & \textbf{0.18} & \textbf{0.53} & \textbf{0.87} & 0/7 & 0.73 & 1.99 & 3.82 & 4/7 & 0.41 & 0.77 & 1.1 \\
			Gas Sensor Array Drift & 0/7 & 0.35 & 1.48 & 2.22 & 0/7 & 0.43 & 0.98 & 2.13 & 7/7 & \textbf{0.26} & \textbf{0.58} & \textbf{1.2} \\
			3D Road Network & 4/7 & \textbf{0.15} & \textbf{0.49} & \textbf{1.28} & 0/7 & 7.38 & 9.2 & 10.56 & 3/7 & 1.73 & 2.31 & 3.49 \\
			Skin Segmentation & 1/7 & \textbf{0.04} & 0.15 & 0.21 & 0/7 & 0.17 & 0.3 & 0.64 & 6/7 & 0.06 & \textbf{0.08} & \textbf{0.1} \\
			KEGG Metabolic Relation Network (Directed) & 3/7 & \textbf{0.34} & \textbf{0.85} & \textbf{1.28} & 0/7 & 1.14 & 1.61 & 2.23 & 4/7 & 1.2 & 1.64 & 2.09 \\
			Shuttle Control & 0/8 & 0.25 & 0.87 & 1.45 & 3/8 & \textbf{0.1} & 0.19 & 0.41 & 5/8 & 0.11 & \textbf{0.18} & \textbf{0.34} \\
			Shuttle Control (normalized) & 0/8 & 0.04 & 0.26 & 0.39 & 0/8 & 0.04 & 0.09 & 0.19 & 8/8 & \textbf{0.02} & \textbf{0.02} & \textbf{0.03} \\
			EEG Eye State & 0/8 & 0.21 & 0.98 & 1.43 & 4/8 & \textbf{0.07} & \textbf{0.13} & \textbf{0.22} & 4/8 & 0.08 & 0.14 & 0.23 \\
			EEG Eye State (normalized) & 0/8 & 0.11 & 0.66 & 0.99 & 2/8 & \textbf{0.06} & 0.14 & 0.33 & 6/8 & \textbf{0.06} & \textbf{0.11} & \textbf{0.23} \\
			Pla85900 & 0/7 & 0.11 & 0.93 & 1.47 & 0/7 & 0.13 & 0.26 & 0.58 & 7/7 & \textbf{0.05} & \textbf{0.07} & \textbf{0.14} \\
			D15112 & 0/7 & 0.2 & 0.9 & 1.43 & 0/7 & 0.02 & 0.03 & 0.06 & 7/7 & \textbf{0.01} & \textbf{0.01} & \textbf{0.02} \\
			\midrule
			Overall Results & 54/165 & \textbf{2.15} & \textbf{4.29} & \textbf{5.99} & 15/165 & 39.7 & 64.27 & 131.67 & 96/165 & 6.31 & 8.17 & 11.85 \\ \bottomrule
		\end{tabularx}%
	\end{adjustwidth}
%	}
\end{table}

As Table~\ref{tab:result_e_par_strategies} demonstrates, the HPClust-competitive, HPClust-cooperative, and HPClust-hybrid strategies markedly boost overall clustering quality, achieving results that are up to three times better than HPClust-inner.
% and several orders of magnitude superior to Forgy K-means and PBK-BDC.
%, showcasing the substantial improvement offered by these strategies.

%Between these, the HPClust-competitive approach exhibited slightly superior average clustering quality over the HPClust-cooperative method. This improvement is likely due to the comprehensive initializations at the outset, given the K-means algorithm's sensitivity to initial conditions. The choice between spending extensive time on local optimization with a single start point or executing multiple initializations emerged from the analysis. The experiments indicated that the latter, involving numerous distinct initializations, tends to be more beneficial. Unlike the HPClust-cooperative method, which eventually focuses on the outcomes of a single initialization, the competitive method persistently processes various K-means++ initializations, selecting the optimal one in the end.

The HPClust-competitive approach showed a slight edge in average clustering quality compared to HPClust-cooperative, likely due to comprehensive initializations that mitigate K-means' sensitivity to initial conditions. The analysis highlights a trade-off between extensive local optimization with a single start point and multiple initializations. The experiments suggest that multiple initializations, persistently processed by the competitive method, lead to better outcomes than the cooperative method's focus on a single initialization. This finding favors exploring diverse K-means++ initializations to select the optimal one in the end.

% The HPClust-hybrid showed the best performance with respect to the average clustering accuracy. This result is expected to some extent since the hybrid approach combines the advantages of both regimes. In the first stage, the competitive scheme provides a wide and rapid exploration of different K-means++ initializations on samples. In the second stage, the cooperative scheme provides a deep exploitation of the best solution returned by the first stage for the remaining time. However, the hybrid scheme requires an additional optimization regarding the parameter $T_1$, which defines the split between the competitive and cooperative regimes. This parameter is highly specific to the choice of a dataset and the number of clusters. In some applications, this might become an unbearable overhead, especially if the number of different datasets for clustering is quite large.

The HPClust-hybrid exhibited the highest average clustering accuracy among the tested methods. This outcome was anticipated to a certain extent, as the hybrid approach combines the strengths of both regimes. In the initial stage, the competitive strategy enables extensive and rapid exploration of various K-means++ initializations on samples. In the subsequent stage, the cooperative strategy facilitates a thorough exploitation of the best solution obtained from the first stage for the remaining time. However, the hybrid strategy necessitates an additional optimization concerning the parameter $T_1$, which determines the split between the competitive and cooperative regimes. This parameter is highly dependent on the specific dataset and the number of clusters. In certain scenarios, particularly when dealing with numerous diverse datasets for clustering, this might pose a significant overhead that could be challenging to handle.

In examining the baseline convergence times among various parallel strategies, it was evident that the HPClust-inner method achieved quicker baseline convergence than the alternatives for the majority of datasets. This disparity was especially notable in larger datasets, as shown at the beginning of Table~\ref{tab:result_t_bar_par_strategies}. For some datasets, to maintain high-quality clustering, substantial sample sizes were necessary, which were proportionate to the dataset sizes. The HPClust-inner strategy, by integrating parallelized K-means++ and K-means for each new sample, managed to expedite processing times relative to the sequential version in other parallel HPClust approaches. These findings highlight the crucial impact of algorithm selection and dataset characteristics on the delicate balance between computational efficiency and clustering accuracy. This underscores the importance of thoughtfully balancing sample size (which affects speed) with the quality of resulting clusters, as a careful trade-off is essential for achieving optimal outcomes.

Further analysis of the competitive, cooperative and hybrid HPClust strategies revealed an intricate interplay between the benefits of parallel processing and the resulting time costs. These methods did improve the solution quality, but the coordination required among multiple processors and the additional complexity from using the Numba library prolonged the convergence process, compared to the HPClust-inner method. Typically, with 8 CPUs, these strategies took up to twice as long to converge as the HPClust-inner method. This observation highlights the need to carefully weigh the trade-offs between exploiting computational resources to accelerate clustering and incurring additional overheads that may impact performance.

Table~\ref{tab:result_e_compet_algos} clearly demonstrates the superiority of the HPClust-hybrid algorithm over its competitors, exhibiting a significant lead in both the number of dominant series and average overall accuracy across all datasets. The HPClust-hybrid algorithm achieves an average accuracy that is a remarkable several orders of magnitude higher than its competitors.

As shown in Table~\ref{tab:result_t_compet_algos}, Forgy K-means, with its linear time complexity with respect to $m$, predictably exhibits a significant increase in time costs for the largest datasets, exceeding the fastest HPClust version by more than 20 times. While PBK-BDC is the quickest for small datasets, its average time costs for the largest datasets are triple those of HPClust, highlighting HPClust's efficiency advantage for large datasets.

%Nevertheless, where convergence speed is less critical, the HPClust-competitive, HPClust-cooperative, and HPClust-hybrid strategies markedly enhance the overall clustering quality, yielding results up to threefold better than other HPClust versions.

%The results of the scaling experiment are presented in Figures~\ref{fig:scaling_num_pts_eps} and \ref{fig:scaling_num_pts_t}. For each value on the $x$-axis the median of the obtained scores across $10$ repetitions are displayed. It can be clearly seen that all HPClust versions are highly robust and scalable with respect to the number of points. They were able to provide an optimal (within $0.2\%$ of the ground truth) clustering accuracy, while limiting their clustering time to under $3$ seconds irrespective of the number of points in the dataset. This is unlike the competitive algorithms: both Forgy K-means and PBK-BDC exhibited highly suboptimal clustering quality, with Forgy K-means incurring unacceptable linearly rising time costs with an increase in the number of points. For instance, on a dataset with around 43 million points, on average it took more than 2 hours for Forgy K-means to finish a single clustering execution. Although the PBK-BDC algorithm only slightly increased its time costs for the largest datasets, it was unable to provide an anywhere near the steadily optimal clustering solution at any data scale.

\begin{figure}[htbp]
	\captionsetup[subfigure]{justification=centering}
	\centering
	\begin{subfigure}{0.7\textwidth}
		\includegraphics[width=\linewidth]{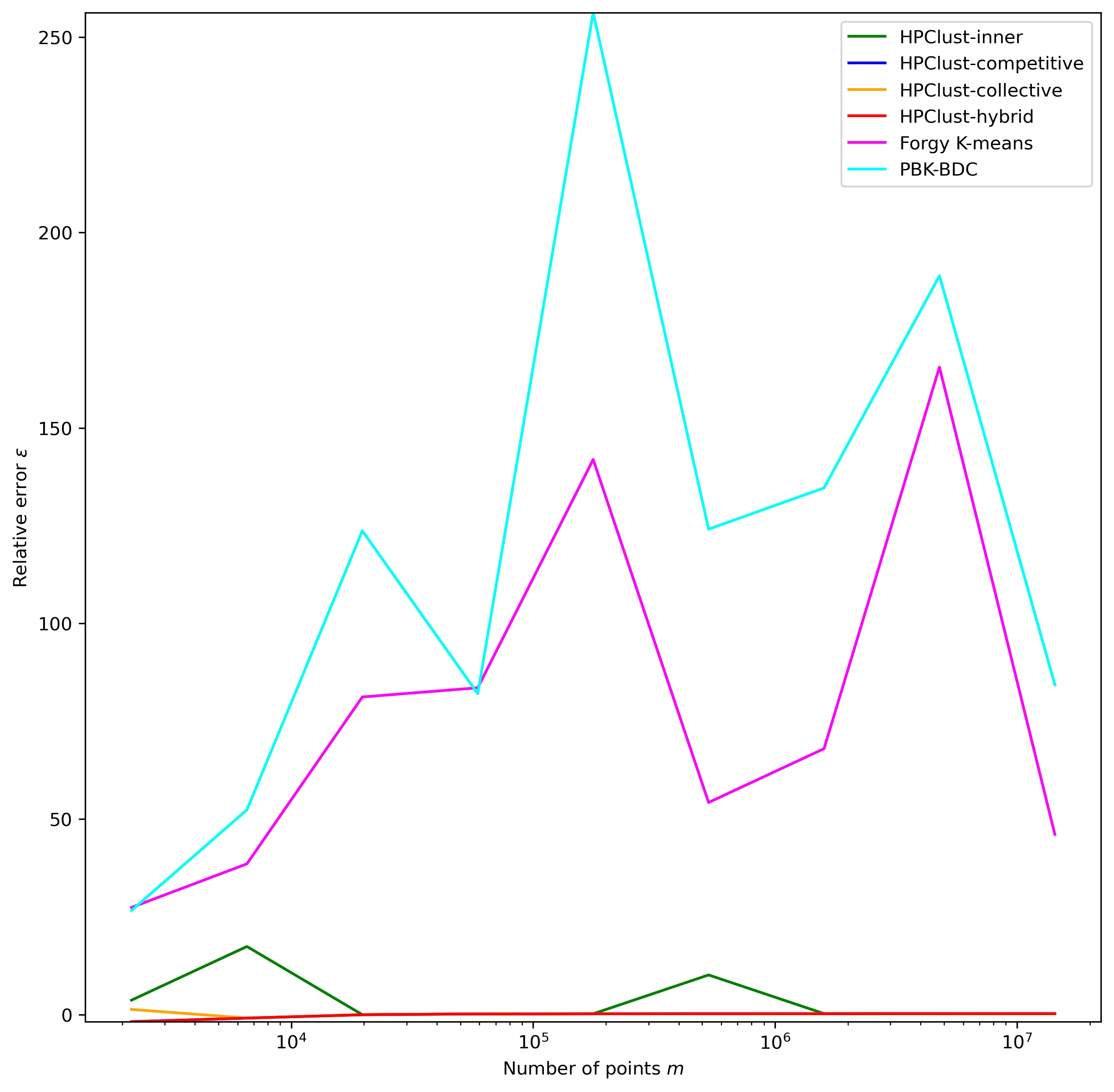}
		\caption{Median accuracy values}
		\label{fig:scaling_num_pts_eps}
	\end{subfigure}
	\par\bigskip
	\begin{subfigure}{0.7\textwidth}
		\includegraphics[width=\linewidth]{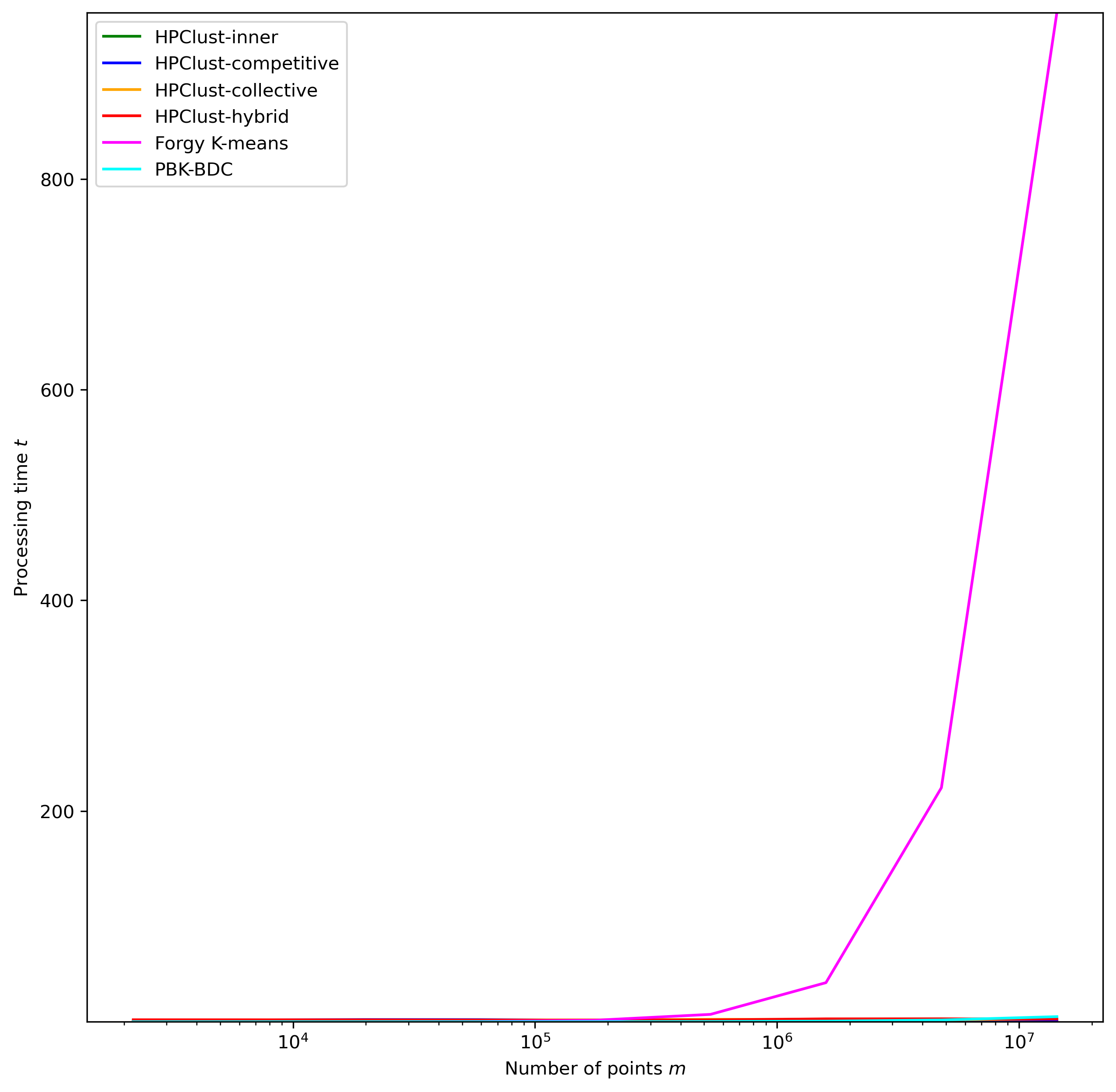}
		\caption{Median processing times}
		\label{fig:scaling_num_pts_t}
	\end{subfigure}
	\caption{Comparative results of the algorithms with respect to the number of points $m$ in a synthetic dataset}
\end{figure}

\begin{table}[!htbp]
	\centering
	\caption{Resulting relative clustering accuracies $\eps$ for the scaling experiment in the format $(median \ value, \pm standard \ deviation)$.}
	\label{tab:scaling_result_eps}
	\resizebox{\textwidth}{!}{
		\begin{tabular}{lllllll}
			\toprule
			$m$ & HPClust-inner & HPClust-competitive & HPClust-collective & HPClust-hybrid & Forgy K-means & PBK-BDC \\
			\midrule
			$3^{7}$ & 3.67 ($\pm$3.90) & -1.83 ($\pm$1.54) & 1.31 ($\pm$2.13) & -1.84 ($\pm$1.30) & 27.38 ($\pm$16.96) & 26.61 ($\pm$9.62) \\
			$3^{8}$ & 17.40 ($\pm$13.71) & -0.92 ($\pm$0.01) & -0.92 ($\pm$0.01) & -0.91 ($\pm$0.01) & 38.53 ($\pm$18.78) & 52.35 ($\pm$22.51) \\
			$3^{9}$ & -0.04 ($\pm$18.43) & -0.06 ($\pm$0.03) & -0.06 ($\pm$0.02) & -0.05 ($\pm$0.03) & 81.21 ($\pm$62.81) & 123.71 ($\pm$56.46) \\
			$3^{10}$ & 0.14 ($\pm$19.13) & 0.14 ($\pm$0.04) & 0.14 ($\pm$0.03) & 0.17 ($\pm$0.04) & 83.56 ($\pm$52.48) & 82.05 ($\pm$51.69) \\
			$3^{11}$ & 0.19 ($\pm$0.05) & 0.18 ($\pm$0.05) & 0.19 ($\pm$0.06) & 0.19 ($\pm$0.04) & 141.95 ($\pm$118.03) & 256.22 ($\pm$91.90) \\
			$3^{12}$ & 10.12 ($\pm$10.06) & 0.21 ($\pm$0.05) & 0.20 ($\pm$0.03) & 0.20 ($\pm$0.04) & 54.23 ($\pm$36.96) & 124.13 ($\pm$32.12) \\
			$3^{13}$ & 0.23 ($\pm$24.33) & 0.21 ($\pm$0.03) & 0.22 ($\pm$14.60) & 0.20 ($\pm$0.04) & 67.99 ($\pm$69.14) & 134.67 ($\pm$40.73) \\
			$3^{14}$ & 0.18 ($\pm$20.97) & 0.21 ($\pm$0.03) & 0.20 ($\pm$31.04) & 0.22 ($\pm$0.04) & 165.58 ($\pm$94.48) & 188.93 ($\pm$92.47) \\
			$3^{15}$ & 0.19 ($\pm$8.83) & 0.20 ($\pm$0.02) & 0.20 ($\pm$11.76) & 0.22 ($\pm$0.02) & 46.06 ($\pm$42.66) & 84.34 ($\pm$31.01) \\
			\bottomrule
		\end{tabular}
	}
	
	\bigskip
	
	\caption{Resulting clustering times $t$ for the scaling experiment in the format $(median \ value, \pm standard \ deviation)$.}
	\label{tab:scaling_result_t}
	\resizebox{\textwidth}{!}{
		\begin{tabular}{lllllll}
			\toprule
			$m$ & HPClust-inner & HPClust-competitive & HPClust-collective & HPClust-hybrid & Forgy K-means & PBK-BDC \\
			\midrule
			$3^{7}$ & 1.03 ($\pm$0.86) & 0.62 ($\pm$0.28) & 0.56 ($\pm$0.46) & 1.84 ($\pm$0.59) & 0.00 ($\pm$0.00) & 0.00 ($\pm$0.00) \\
			$3^{8}$ & 1.41 ($\pm$0.66) & 1.56 ($\pm$0.75) & 0.68 ($\pm$0.70) & 1.82 ($\pm$0.69) & 0.01 ($\pm$0.00) & 0.01 ($\pm$0.00) \\
			$3^{9}$ & 1.46 ($\pm$0.76) & 1.99 ($\pm$0.92) & 1.23 ($\pm$0.70) & 1.73 ($\pm$0.86) & 0.03 ($\pm$0.02) & 0.01 ($\pm$0.00) \\
			$3^{10}$ & 1.48 ($\pm$0.80) & 1.96 ($\pm$0.95) & 1.53 ($\pm$0.70) & 1.52 ($\pm$0.78) & 0.19 ($\pm$0.19) & 0.03 ($\pm$0.01) \\
			$3^{11}$ & 1.08 ($\pm$1.00) & 1.41 ($\pm$0.89) & 1.74 ($\pm$0.78) & 1.46 ($\pm$0.72) & 1.39 ($\pm$0.80) & 0.06 ($\pm$0.01) \\
			$3^{12}$ & 1.35 ($\pm$0.83) & 1.72 ($\pm$0.93) & 2.27 ($\pm$0.73) & 1.69 ($\pm$0.64) & 6.96 ($\pm$3.34) & 0.18 ($\pm$0.01) \\
			$3^{13}$ & 2.46 ($\pm$0.97) & 1.47 ($\pm$0.85) & 2.06 ($\pm$0.76) & 2.70 ($\pm$0.81) & 37.07 ($\pm$24.76) & 0.53 ($\pm$0.03) \\
			$3^{14}$ & 2.59 ($\pm$0.80) & 1.48 ($\pm$0.81) & 1.73 ($\pm$0.74) & 2.71 ($\pm$0.94) & 222.16 ($\pm$100.82) & 1.62 ($\pm$0.10) \\
			$3^{15}$ & 1.66 ($\pm$0.73) & 1.64 ($\pm$0.82) & 2.64 ($\pm$0.98) & 2.62 ($\pm$1.21) & 957.92 ($\pm$443.64) & 4.81 ($\pm$0.47) \\
			\bottomrule
		\end{tabular}
	}
\end{table}

The scaling experiment results are presented in Figures~\ref{fig:scaling_num_pts_eps} and \ref{fig:scaling_num_pts_t}. For each $x$-axis value, the median score across 10 repetitions is displayed. The figures clearly show that all HPClust versions are highly robust and scalable with respect to the number of points, achieving optimal clustering accuracy (within $0.2\%$ of ground truth) while keeping clustering time under $3$ seconds, regardless of dataset size. In contrast, competitive algorithms Forgy K-means and PBK-BDC exhibited substantially suboptimal clustering quality, with Forgy K-means incurring unacceptable linearly rising time costs with increasing points (e.g., over 2 hours for a single execution on a 43 million point dataset). Meanwhile, PBK-BDC failed to provide steadily optimal clustering solutions at any data scale, despite slightly increased time costs for larger datasets. The HPClust versions demonstrated superior performance and scalability. Detailed experimental results, showcasing median values across various data scales and algorithms, are presented in Tables~\ref{tab:scaling_result_eps} and \ref{tab:scaling_result_t} for a comprehensive understanding.

Surprisingly, the scaling experiment's results reveal an additional extraordinary property of HPClust: its iterative sampling processing with small samples renders it robust to noise and outliers, demonstrating a remarkable resilience to data perturbations and anomalies.

\subsection{Trade-offs Analysis}

Our experiments with the HPClust algorithm have revealed several key trade-offs. Here, we present an in-depth analysis of these trade-offs, which often involve intricate balancing acts between efficiency, accuracy, computation time, and dataset characteristics. The following are the primary trade-offs that practitioners might have to consider:

\begin{enumerate}
\item Accuracy vs. Computation Time: Our results showed that the choice of strategy significantly influences the balance between computation time and the resulting accuracy. For example, while HPClust-inner demonstrated faster convergence times, especially for large datasets, the HPClust-competitive, HPClust-cooperative, and HPClust-hybrid strategies offered improved clustering quality at the cost of slightly increased computation time. Thus, your choice should weigh the importance of quick results against the necessity of clustering precision;

\item Parallelism vs. Overhead: The level of parallelism used directly impacts the computation time and the overhead associated with managing multiple processors. While increasing the number of processors generally results in faster computation, it also introduces added overhead in coordinating these processors. This was particularly evident when using HPClust-competitive, HPClust-cooperative, and HPClust-hybrid strategies, which took nearly twice as long to converge as HPClust-inner, despite yielding superior solutions;

\item Sample Size vs. Quality of Clusters: The size of the sample used in the HPClust algorithm directly impacts the quality of clusters and the computation time. Larger samples often led to better approximations of the overall data distribution and improved final clustering quality. However, these benefits were offset by slower algorithmic performance, which is a crucial aspect to consider when dealing with large datasets;

\item Strategy Selection vs. Initialization Quality: In the context of HPClust, another critical trade-off lies in the choice of strategy and its influence on the quality of initializations. HPClust-competitive, which applies multiple initializations and continues clustering different K-means++ initializations to select the best one at the end, showed a slightly improved clustering quality over HPClust-cooperative. Meanwhile, the HPClust-hybrid strategy effectively amalgamated the comprehensive exploration capabilities of the competitive approach with the exploitation abilities of the cooperative approach. However, it should be noted that this comes with the requirement of additional optimization for the split parameter $T_1$. Therefore, the sensitivity of K-means to the quality of initial initialization is another critical factor to consider when choosing the strategy.
\end{enumerate}

In navigating these trade-offs, understanding the unique requirements of your task and the nature of your dataset is paramount. Each strategy presents its own advantages and disadvantages, which should be carefully considered in light of these trade-offs. With the correct approach, these trade-offs can be effectively managed to achieve optimal clustering results with the HPClust algorithm.

% \section{Guidelines for Choosing Parallelization Strategy}

% Based on the experimental results, we can suggest the following guidelines for choosing an appropriate parallelization strategy for the HPClust algorithm:
% \begin{enumerate}
% \item Choose HPClust-sequential when your main concern is accuracy and you have less constraints on computation time. This variant showed the most consistent performance across different datasets, achieving a good balance between the accuracy of the clustering and the computation time;

% \item Use HPClust-inner if memory usage is your primary constraint. This variant demonstrated the smallest memory footprint among all the parallelization strategies, making it the preferred option for memory-limited systems;

% \item Opt for HPClust-competitive when you have a high computational resource and need faster results. This variant delivered the fastest computation time on datasets with high dimensional data, making it suitable for time-sensitive tasks;

% \item Consider HPClust-cooperative if your task requires balancing speed and accuracy. This variant showed a relatively consistent performance in terms of both computation time and clustering accuracy.
% \end{enumerate}
% Remember, these are general guidelines and the specific choice of parallelization strategy would still depend on the nature of the dataset and the requirements of the task at hand.

\section{Guidelines for Choosing Parallel Strategy} \label{sec:guidelines}

Considering the outcomes of our research, we propose the following revised guidelines for selecting an appropriate parallel strategy for the HPClust algorithm:

\begin{enumerate}
\item If you are handling large datasets and have concerns over computation time, opt for the HPClust-inner strategy. This variant consistently showed faster convergence to baselines across most datasets, especially larger ones, as evidenced in the first rows of Table~\ref{tab:result_t_bar_par_strategies}. The employment of significant sample sizes, relative to the dataset sizes, along with parallelized K-means++ and K-means on each new sample, contributed to its accelerated processing times. However, remember that larger sample sizes often led to slower algorithmic performance, so balancing sample size with the quality of clusters remains crucial;

\item When computation time is less of a constraint and you aim for better clustering quality, choose between HPClust-competitive and HPClust-cooperative strategies. Both these strategies demonstrated an improved quality of final solutions compared to other versions of HPClust, on average three times better with 8 CPUs. However, due to the additional overhead of coordinating multiple processors and the complexities associated with the Numba library, they also exhibited longer convergence times, nearly twice as long as HPClust-inner with 8 CPUs;

\item If the clustering quality is your primary focus, HPClust-hybrid or HPClust-competitive should be the preferred choices. Our findings indicated a slightly improved clustering quality with HPClust-competitive compared to HPClust-cooperative. This improvement stems from the application of multiple initializations at the beginning, as K-means is highly sensitive to initial initialization quality. This strategy continues to cluster different K-means++ initializations, eventually selecting the best one at the end, leading to a superior solution. In the meantime, if you aim for superior clustering quality and willing to spend extra time on parameter optimization, opt for the HPClust-hybrid strategy. This choice demonstrated the best resulting clustering quality, while retaining the same degree of time efficiency as the competitive and cooperative approaches.
\end{enumerate}

These guidelines should assist researchers and practitioners in choosing an appropriate parallel strategy for their specific needs. However, keep in mind that these are general guidelines, and the choice of parallel strategy should be adapted to the specific requirements of your task and the nature of your dataset. This research strongly suggests that parallelism, when feasible, offers a significant enhancement in clustering accuracy and convergence time compared to the sequential variant.

Overall, the best strategy is likely to be one that strikes a balance between the need for accuracy, computation time, and the specific characteristics of the dataset at hand. The effectiveness of each strategy will inevitably depend on these factors, and the choice should be made accordingly.

\section{Conclusion and Future Works} \label{sec:conclusion}

%In this research, we proposed the HPClust algorithm and evaluated its three parallel strategies on a variety of datasets, both real-world and synthetic. Our comparative analysis was based on three key metrics: the resulting relative clustering accuracy $\eps$, the total runtime $t$, and the runtime with respect to the baseline $\overline{t}$. These metrics allowed us to assess both the effectiveness and efficiency of each parallelization strategy.

Our paper introduces the HPClust algorithm and explores its four parallel strategies on diverse datasets, including real-world and synthetic ones. Our comprehensive evaluation focuses on three essential metrics: relative clustering accuracy ($\eps$), total runtime ($t$), and baseline-normalized runtime ($\overline{t}$). These metrics provide a thorough assessment of each strategy's effectiveness and efficiency, enabling a well-rounded comparison.

The experimental results demonstrate HPClust's unrivaled effectiveness, efficiency, and scalability compared to baseline algorithms across a vast range of real-world datasets (spanning small to big sizes) and synthetic datasets. HPClust consistently outperforms its competitors, showcasing remarkable robustness to data scale and noise, as well as adaptability in various data settings.

%In conclusion, this study reveals that there is no universal parallelization strategy for the HPClust algorithm. Instead, the optimal strategy appears to be data-dependent, suggesting the need for adaptive techniques that can select the most suitable strategy based on the characteristics of the dataset. Nevertheless, in the majority of cases, we advise practitioners to utilize the competitive or competitive-cooperative (hybrid) parallelization strategies of the HPClust algorithm.

Also, this research demonstrates that no single parallel strategy universally optimizes the HPClust algorithm. Instead, the most effective approach depends on the dataset's characteristics, emphasizing the need for adaptive techniques that dynamically select the best strategy. However, in most cases, we recommend practitioners to employ either the competitive or hybrid (competitive-cooperative) parallel strategies of HPClust, which have shown superior performance and versatility.

Additionally, our work offers a comprehensive review of the primary high-performance techniques utilized for optimizing data clustering algorithms. We delve into the intricate aspects and nuances of applying parallel techniques, specifically analyzing the challenges and pitfalls associated with the HPClust algorithm. Through a detailed trade-off analysis, we provide practical guidelines to assist in selecting the most suitable parallel strategy for specific use cases. These guidelines aim to facilitate informed decision-making and provide actionable recommendations.

% Also, our work provided a detailed review of the main high-performance techniques for optimization of data clustering algorithms. We considered different intricate aspects and nuances of applying these technologies parallelization techniques, analyzing various pitfalls specific to the HPClust algorithm. Comprehensive trade-offs analysis was conducted, as well as summarized in actual hands-on guidelines for choosing the most appropriate parallelization strategy based on the use case.

%For future work, we plan to investigate adaptive techniques that can dynamically select the optimal parallelization strategy based on the dataset at hand. We also aim to delve deeper into the trade-offs observed in this study to gain a better understanding of their impacts on algorithmic performance and accuracy.

Future research will focus on developing adaptive methods that can intelligently choose the most suitable parallel strategy based on the specific dataset, optimizing performance and accuracy. Additionally, we will conduct a more in-depth analysis of the trade-offs revealed in this study, exploring their nuanced effects on algorithmic performance and accuracy, to uncover actionable insights for further improvement.

Another promising future research direction for the proposed HPClust algorithm is its potential adaptation for clustering streaming datasets or continuously growing datasets. This is particularly relevant in scenarios involving IoT sensors, financial transactions, social media feeds, and other real-time data sources, where data is constantly generated and requires efficient clustering techniques to uncover insights and patterns. By extending HPClust to handle streaming data, researchers can unlock new opportunities for real-time analytics and decision-making in various fields.

%The insights and observations gleaned from this study pave the way for further research into efficient and adaptive parallelization strategies for the HPClust algorithm and beyond. It is our hope that this research contributes significantly to the fields of data clustering and high-performance computing.

This study's findings and observations lay the groundwork for advancing efficient and adaptive parallel techniques for HPClust and beyond. Our goal is for this research to make a meaningful impact in the fields of data clustering and high-performance computing, driving innovation and improvement in these areas. By shedding light on the complex relationships between parallel strategies, dataset characteristics, and algorithmic performance, we aim to spark further discovery and progress.

\section*{Acknowledgements}

This research was funded by the Science Committee of the Ministry of Science and Higher Education of the Republic of Kazakhstan (grant no. BR21882268).

\bibliography{main}
%\printbibliography

\newpage

\appendix
\section{Extended experimental results} \label{sec:appendix}

Clustering details include the parameters and the following attributes of the clustering process:

\begin{itemize}
	\item $k$ is the number of clusters;
	\item $f^*$ is the best known objective function value multiplied by the number provided after the name of the dataset in the caption of each table;
	\item $s$ is the sample size;
	\item $n_{exec}$ is the number of executions for each choice of $k$;
	\item $n_{s}$ is the number of used samples;
	\item $T$ is the maximal CPU time allowed for the execution of an algorithm;
	\item $T_1$ and $T_2$ are the maximal CPU times allowed for the first and second phases of the HPClust-hybrid algorithm, respectively;
	\item $n_{d}$ is the number of distance function evaluations.
\end{itemize}

\begin{landscape}

%%%%%%%%%%%%%%%%%%%%%%%%%%%%%%%%%%%%%%%%%%%%%%%%%%%%%%%%%%%%%%%%%%%%%%%%%%%%%%%%%%%%%%%
%  START: CORD-19 Embeddings
%%%%%%%%%%%%%%%%%%%%%%%%%%%%%%%%%%%%%%%%%%%%%%%%%%%%%%%%%%%%%%%%%%%%%%%%%%%%%%%%%%%%%%%
\subsection{CORD-19 Embeddings}
Dimensions: $m$ = 599616, $n$ = 768.
\par
Description: COVID-19 Open Research Dataset (CORD-19) is a resource of more than half a million scholarly articles about COVID-19, SARS-CoV-2, and related coronaviruses represended as embeddings in vectorized form.

\begin{table}[!htbp]
	\centering
	
	\caption{Summary of the results with CORD-19 Embeddings ($\times10^{9}$)}
	\label{TabResultsD1}
	\small
	\resizebox{!}{\tableheight}{
		\begin{tabular}{|l|l|l|llllll|llllll|llllll|}
			\hline
			\multicolumn{1}{|c|}{\multirow{3}{*}{$k$}} & \multicolumn{1}{c|}{\multirow{3}{*}{$f^*$}} & \multicolumn{1}{c|}{\multirow{3}{*}{$\overline{f}$}} & \multicolumn{6}{c|}{HPClust-inner} & \multicolumn{6}{c|}{HPClust-competitive} & \multicolumn{6}{c|}{HPClust-cooperative} \\ \cline{4-21}
			\multicolumn{1}{|c|}{} & \multicolumn{1}{c|}{} & \multicolumn{1}{c|}{} & \multicolumn{2}{c|}{$\varepsilon$} & \multicolumn{2}{c|}{$\overline{t}$} & \multicolumn{2}{c|}{$t$} & \multicolumn{2}{c|}{$\varepsilon$} & \multicolumn{2}{c|}{$\overline{t}$} & \multicolumn{2}{c|}{$t$} & \multicolumn{2}{c|}{$\varepsilon$} & \multicolumn{2}{c|}{$\overline{t}$} & \multicolumn{2}{c|}{$t$} \\ \cline{4-21}
			\multicolumn{1}{|c|}{} & \multicolumn{1}{c|}{} & \multicolumn{1}{c|}{} & \multicolumn{1}{c|}{med} & \multicolumn{1}{c|}{std} & \multicolumn{1}{c|}{med} & \multicolumn{1}{c|}{std} & \multicolumn{1}{c|}{med} & \multicolumn{1}{c|}{std} & \multicolumn{1}{c|}{med} & \multicolumn{1}{c|}{std} & \multicolumn{1}{c|}{med} & \multicolumn{1}{c|}{std} & \multicolumn{1}{c|}{med} & \multicolumn{1}{c|}{std} & \multicolumn{1}{c|}{med} & \multicolumn{1}{c|}{std} & \multicolumn{1}{c|}{med} & \multicolumn{1}{c|}{std} & \multicolumn{1}{c|}{med} & \multicolumn{1}{c|}{std} \\ \hline
			2 & 2.03893$^*$ & 0.1082 & 0.006 & 0.001 & 15.834 & 0.0 & 15.834 & 6.423 & 0.006 & 0.001 & 15.225 & 3.105 & 20.295 & 9.046 & 0.006 & 0.0 & 15.556 & 15.461 & 27.333 & 14.943 \\
			3 & 1.9093$^*$ & 0.10141 & 0.019 & 0.007 & 23.69 & 13.913 & 33.566 & 13.481 & 0.011 & 0.006 & 10.312 & 4.304 & 16.699 & 11.92 & 0.019 & 0.004 & 13.461 & 7.223 & 15.598 & 10.088 \\
			5 & 1.77676$^*$ & 0.09433 & 0.145 & 0.066 & 22.952 & 10.562 & 27.398 & 11.366 & 0.015 & 0.003 & 11.529 & 2.162 & 12.459 & 7.228 & 0.015 & 0.002 & 21.706 & 4.474 & 27.337 & 5.387 \\
			10 & 1.62555$^*$ & 0.08679 & 0.463 & 0.253 & 3.419 & 2.783 & 29.185 & 8.184 & 0.067 & 0.057 & 7.935 & 1.012 & 34.535 & 10.962 & 0.063 & 0.045 & 7.959 & 1.433 & 21.939 & 10.646 \\
			15 & 1.55295$^*$ & 0.08276 & 0.282 & 0.104 & 12.552 & 7.758 & 31.399 & 8.814 & 0.073 & 0.124 & 19.717 & 5.888 & 34.974 & 7.468 & 0.128 & 0.061 & 14.53 & 1.741 & 20.63 & 8.791 \\
			20 & 1.49987$^*$ & 0.07991 & 0.414 & 0.075 & 17.89 & 8.654 & 38.208 & 13.756 & 0.196 & 0.146 & 25.292 & 4.827 & 28.762 & 2.341 & 0.143 & 0.101 & 21.548 & 1.319 & 32.616 & 4.514 \\
			25 & 1.46394$^*$ & 0.07789 & 0.166 & 0.195 & 16.347 & 7.16 & 31.45 & 9.114 & 0.092 & 0.07 & 32.139 & 5.446 & 36.157 & 4.77 & 0.153 & 0.097 & 32.654 & 4.919 & 33.721 & 6.338 \\
			\hline
			\multicolumn{3}{|c|}{Mean:} & \textbf{0.213} & & \textbf{16.098} & & \textbf{29.577} & & \textbf{0.066} & & \textbf{17.45} & & \textbf{26.269} & & \textbf{0.075} & & \textbf{18.202} & & \textbf{25.596} & \\ \hline
		\end{tabular}
	}
	
	\bigskip
	
	\small
	\resizebox{!}{\tableheight}{
		\begin{tabular}{|l|l|l|llllll|llllll|llllll|}
			\hline
			\multicolumn{1}{|c|}{\multirow{3}{*}{$k$}} & \multicolumn{1}{c|}{\multirow{3}{*}{$f^*$}} & \multicolumn{1}{c|}{\multirow{3}{*}{$\overline{f}$}} & \multicolumn{6}{c|}{HPClust-hybrid} & \multicolumn{6}{c|}{Forgy K-means} & \multicolumn{6}{c|}{PBK-BDC} \\ \cline{4-21}
			\multicolumn{1}{|c|}{} & \multicolumn{1}{c|}{} & \multicolumn{1}{c|}{} & \multicolumn{2}{c|}{$\varepsilon$} & \multicolumn{2}{c|}{$\overline{t}$} & \multicolumn{2}{c|}{$t$} & \multicolumn{2}{c|}{$\varepsilon$} & \multicolumn{2}{c|}{$\overline{t}$} & \multicolumn{2}{c|}{$t$} & \multicolumn{2}{c|}{$\varepsilon$} & \multicolumn{2}{c|}{$\overline{t}$} & \multicolumn{2}{c|}{$t$} \\ \cline{4-21}
			\multicolumn{1}{|c|}{} & \multicolumn{1}{c|}{} & \multicolumn{1}{c|}{} & \multicolumn{1}{c|}{med} & \multicolumn{1}{c|}{std} & \multicolumn{1}{c|}{med} & \multicolumn{1}{c|}{std} & \multicolumn{1}{c|}{med} & \multicolumn{1}{c|}{std} & \multicolumn{1}{c|}{med} & \multicolumn{1}{c|}{std} & \multicolumn{1}{c|}{med} & \multicolumn{1}{c|}{std} & \multicolumn{1}{c|}{med} & \multicolumn{1}{c|}{std} & \multicolumn{1}{c|}{med} & \multicolumn{1}{c|}{std} & \multicolumn{1}{c|}{med} & \multicolumn{1}{c|}{std} & \multicolumn{1}{c|}{med} & \multicolumn{1}{c|}{std} \\ \hline
			2 & 2.03893$^*$ & 0.1082 & 0.007 & 0.001 & 6.118 & 10.005 & 22.541 & 6.125 & 0.0 & 0.0 & -- & -- & 14.984 & 1.23 & 0.0 & 0.06 & -- & -- & 2.533 & 1.639 \\
			3 & 1.9093$^*$ & 0.10141 & 0.011 & 0.005 & 7.245 & 4.692 & 18.988 & 8.346 & 0.011 & 1.406 & -- & -- & 45.337 & 10.538 & 0.058 & 0.028 & -- & -- & 9.368 & 4.529 \\
			5 & 1.77676$^*$ & 0.09433 & 0.015 & 0.045 & 24.03 & 5.652 & 23.789 & 9.182 & -0.002 & 0.234 & -- & -- & 104.197 & 13.797 & 2.161 & 1.416 & -- & -- & 15.807 & 1.399 \\
			10 & 1.62555$^*$ & 0.08679 & 0.057 & 0.021 & 8.259 & 0.877 & 26.434 & 6.826 & 0.576 & 0.844 & -- & -- & 487.922 & 226.84 & 1.937 & 1.23 & -- & -- & 61.921 & 10.023 \\
			15 & 1.55295$^*$ & 0.08276 & 0.111 & 0.153 & 15.05 & 2.097 & 27.231 & 7.537 & 0.342 & 0.229 & -- & -- & 887.432 & 1046.333 & 2.333 & 1.291 & -- & -- & 110.804 & 27.208 \\
			20 & 1.49987$^*$ & 0.07991 & 0.161 & 0.184 & 25.027 & 3.976 & 38.075 & 7.841 & 0.233 & 0.319 & -- & -- & 1405.199 & 794.487 & 3.496 & 1.356 & -- & -- & 143.392 & 28.935 \\
			25 & 1.46394$^*$ & 0.07789 & 0.18 & 0.079 & 26.508 & 3.56 & 32.961 & 7.091 & 0.056 & 0.217 & -- & -- & 1987.234 & 837.859 & 2.21 & 0.842 & -- & -- & 189.484 & 31.495 \\
			\hline
			\multicolumn{3}{|c|}{Mean:} & \textbf{0.078} & & \textbf{16.034} & & \textbf{27.146} & & \textbf{0.174} & & \textbf{--} & & \textbf{704.615} & & \textbf{1.742} & & \textbf{--} & & \textbf{76.187} & \\ \hline
		\end{tabular}
	}
	
	\bigskip
	
	\caption{Clustering details with CORD-19 Embeddings}
	\label{TabDetailsD1}
%	\resizebox{\linewidth}{!}{
	\resizebox{!}{\tableheight}{
		\begin{tabular}{|l|l|llll|llll|llll|lllll|l|l|}
			\hline
			\multicolumn{1}{|c|}{\multirow{2}{*}{$k$}} & \multicolumn{1}{c|}{\multirow{2}{*}{$n_{exec}$}} & \multicolumn{4}{c|}{HPClust-inner} & \multicolumn{4}{c|}{HPClust-competitive} & \multicolumn{4}{c|}{HPClust-cooperative} & \multicolumn{5}{c|}{HPClust-hybrid} & \multicolumn{1}{c|}{Forgy K-means} & \multicolumn{1}{c|}{PBK-BDC} \\ \cline{3-21}
			\multicolumn{1}{|c|}{} & \multicolumn{1}{c|}{} & \multicolumn{1}{c|}{$s$} & \multicolumn{1}{c|}{$n_{s}$} & \multicolumn{1}{c|}{$T$} & \multicolumn{1}{c|}{$n_{d}$} & \multicolumn{1}{c|}{$s$} & \multicolumn{1}{c|}{$n_{s}$} & \multicolumn{1}{c|}{$T$} & \multicolumn{1}{c|}{$n_{d}$} & \multicolumn{1}{c|}{$s$} & \multicolumn{1}{c|}{$n_{s}$} & \multicolumn{1}{c|}{$T$} & \multicolumn{1}{c|}{$n_{d}$} & \multicolumn{1}{c|}{$s$} & \multicolumn{1}{c|}{$n_{s}$} & \multicolumn{1}{c|}{$T_1$} & \multicolumn{1}{c|}{$T_2$} & \multicolumn{1}{c|}{$n_{d}$} & \multicolumn{1}{c|}{$n_{d}$} & \multicolumn{1}{c|}{$n_{d}$} \\
			\hline
			2 & 7 & 32000 & 82 & 40.0 & 3.7E+07 & 32000 & 434 & 40.0 & 1.6E+08 & 32000 & 569 & 40.0 & 1.5E+08 & 32000 & 511 & 37.333 & 2.667 & 1.6E+08 & 1.4E+07 & 1.3E+07 \\
			3 & 7 & 32000 & 147 & 40.0 & 5.4E+07 & 32000 & 257 & 40.0 & 2.0E+08 & 32000 & 233 & 40.0 & 2.0E+08 & 32000 & 297 & 32.0 & 8.0 & 2.0E+08 & 5.6E+07 & 4.9E+07 \\
			5 & 7 & 32000 & 104 & 40.0 & 8.0E+07 & 32000 & 129 & 40.0 & 2.5E+08 & 32000 & 321 & 40.0 & 2.6E+08 & 32000 & 267 & 21.333 & 18.667 & 2.6E+08 & 1.3E+08 & 1.0E+08 \\
			10 & 7 & 32000 & 75 & 40.0 & 1.2E+08 & 32000 & 215 & 40.0 & 3.5E+08 & 32000 & 123 & 40.0 & 3.4E+08 & 32000 & 147 & 24.0 & 16.0 & 3.3E+08 & 6.9E+08 & 4.2E+08 \\
			15 & 7 & 32000 & 41 & 40.0 & 1.4E+08 & 32000 & 110 & 40.0 & 3.7E+08 & 32000 & 35 & 40.0 & 3.6E+08 & 32000 & 73 & 26.667 & 13.333 & 3.5E+08 & 1.3E+09 & 7.9E+08 \\
			20 & 7 & 32000 & 46 & 40.0 & 1.7E+08 & 32000 & 37 & 40.0 & 3.8E+08 & 32000 & 54 & 40.0 & 3.8E+08 & 32000 & 45 & 8.0 & 32.0 & 3.3E+08 & 2.1E+09 & 1.0E+09 \\
			25 & 7 & 32000 & 32 & 40.0 & 1.9E+08 & 32000 & 30 & 40.0 & 3.7E+08 & 32000 & 23 & 40.0 & 3.7E+08 & 32000 & 25 & 32.0 & 8.0 & 3.4E+08 & 2.9E+09 & 1.5E+09 \\
			\hline
		\end{tabular}
	}
	
\end{table}

\newpage

%%%%%%%%%%%%%%%%%%%%%%%%%%%%%%%%%%%%%%%%%%%%%%%%%%%%%%%%%%%%%%%%%%%%%%%%%%%%%%%%%%%%%%%
%  END: CORD-19 Embeddings
%%%%%%%%%%%%%%%%%%%%%%%%%%%%%%%%%%%%%%%%%%%%%%%%%%%%%%%%%%%%%%%%%%%%%%%%%%%%%%%%%%%%%%%

%%%%%%%%%%%%%%%%%%%%%%%%%%%%%%%%%%%%%%%%%%%%%%%%%%%%%%%%%%%%%%%%%%%%%%%%%%%%%%%%%%%%%%%
%  START: HEPMASS
%%%%%%%%%%%%%%%%%%%%%%%%%%%%%%%%%%%%%%%%%%%%%%%%%%%%%%%%%%%%%%%%%%%%%%%%%%%%%%%%%%%%%%%
\subsection{HEPMASS}
Dimensions: $m$ = 10500000, $n$ = 27.
\par
Description: The data set contains the 28 normalized features of physical particles that can be used for discovering the exotic ones in the field of high-energy physics.

\begin{table}[!htbp]
	\centering
	
	\caption{Summary of the results with HEPMASS ($\times10^{8}$)}
	\label{TabResultsD2}
	\small
	\resizebox{!}{\tableheight}{
		\begin{tabular}{|l|l|l|llllll|llllll|llllll|}
			\hline
			\multicolumn{1}{|c|}{\multirow{3}{*}{$k$}} & \multicolumn{1}{c|}{\multirow{3}{*}{$f^*$}} & \multicolumn{1}{c|}{\multirow{3}{*}{$\overline{f}$}} & \multicolumn{6}{c|}{HPClust-inner} & \multicolumn{6}{c|}{HPClust-competitive} & \multicolumn{6}{c|}{HPClust-cooperative} \\ \cline{4-21}
			\multicolumn{1}{|c|}{} & \multicolumn{1}{c|}{} & \multicolumn{1}{c|}{} & \multicolumn{2}{c|}{$\varepsilon$} & \multicolumn{2}{c|}{$\overline{t}$} & \multicolumn{2}{c|}{$t$} & \multicolumn{2}{c|}{$\varepsilon$} & \multicolumn{2}{c|}{$\overline{t}$} & \multicolumn{2}{c|}{$t$} & \multicolumn{2}{c|}{$\varepsilon$} & \multicolumn{2}{c|}{$\overline{t}$} & \multicolumn{2}{c|}{$t$} \\ \cline{4-21}
			\multicolumn{1}{|c|}{} & \multicolumn{1}{c|}{} & \multicolumn{1}{c|}{} & \multicolumn{1}{c|}{med} & \multicolumn{1}{c|}{std} & \multicolumn{1}{c|}{med} & \multicolumn{1}{c|}{std} & \multicolumn{1}{c|}{med} & \multicolumn{1}{c|}{std} & \multicolumn{1}{c|}{med} & \multicolumn{1}{c|}{std} & \multicolumn{1}{c|}{med} & \multicolumn{1}{c|}{std} & \multicolumn{1}{c|}{med} & \multicolumn{1}{c|}{std} & \multicolumn{1}{c|}{med} & \multicolumn{1}{c|}{std} & \multicolumn{1}{c|}{med} & \multicolumn{1}{c|}{std} & \multicolumn{1}{c|}{med} & \multicolumn{1}{c|}{std} \\ \hline
			2 & 2.48889$^*$ & 0.01512 & 0.004 & 0.001 & 8.694 & 6.958 & 19.003 & 8.514 & 0.004 & 0.0 & 8.529 & 5.832 & 26.287 & 3.089 & 0.003 & 0.001 & 11.257 & 5.634 & 18.327 & 7.876 \\
			3 & 2.36789$^*$ & 0.01439 & 0.009 & 0.62 & 16.367 & 2.089 & 19.541 & 6.92 & 0.005 & 0.003 & 7.582 & 3.877 & 13.209 & 6.614 & 0.008 & 0.436 & 4.764 & 3.203 & 16.956 & 8.51 \\
			5 & 2.21106$^*$ & 0.01349 & 0.341 & 0.437 & 5.451 & 2.961 & 20.678 & 10.902 & 0.012 & 0.161 & 1.352 & 4.625 & 17.709 & 7.415 & 0.333 & 0.378 & 2.665 & 0.943 & 14.077 & 7.155 \\
			10 & 2.00353$^*$ & 0.01223 & 0.289 & 0.078 & 5.034 & 6.332 & 18.004 & 8.518 & 0.086 & 0.069 & 2.335 & 0.651 & 16.864 & 10.671 & 0.122 & 0.066 & 1.619 & 0.369 & 13.306 & 5.406 \\
			15 & 1.89922$^*$ & 0.01157 & 0.397 & 0.191 & 5.965 & 0.0 & 12.862 & 7.789 & 0.094 & 0.068 & 6.256 & 6.402 & 23.113 & 5.153 & 0.155 & 0.12 & 4.364 & 1.089 & 16.703 & 6.439 \\
			20 & 1.82904$^*$ & 0.01114 & 0.322 & 0.051 & 15.461 & 4.451 & 17.015 & 8.134 & 0.156 & 0.087 & 6.688 & 5.447 & 20.223 & 7.415 & 0.209 & 0.089 & 3.768 & 5.851 & 20.596 & 4.425 \\
			25 & 1.77524$^*$ & 0.01082 & 0.189 & 0.179 & 3.61 & 4.257 & 22.461 & 6.812 & 0.111 & 0.032 & 3.925 & 1.402 & 24.055 & 4.933 & 0.279 & 0.151 & 5.554 & 4.163 & 19.474 & 7.38 \\
			\hline
			\multicolumn{3}{|c|}{Mean:} & \textbf{0.222} & & \textbf{8.655} & & \textbf{18.509} & & \textbf{0.067} & & \textbf{5.238} & & \textbf{20.209} & & \textbf{0.159} & & \textbf{4.856} & & \textbf{17.063} & \\ \hline
		\end{tabular}
	}
	
	\bigskip
	
	\small
	\resizebox{!}{\tableheight}{
		\begin{tabular}{|l|l|l|llllll|llllll|llllll|}
			\hline
			\multicolumn{1}{|c|}{\multirow{3}{*}{$k$}} & \multicolumn{1}{c|}{\multirow{3}{*}{$f^*$}} & \multicolumn{1}{c|}{\multirow{3}{*}{$\overline{f}$}} & \multicolumn{6}{c|}{HPClust-hybrid} & \multicolumn{6}{c|}{Forgy K-means} & \multicolumn{6}{c|}{PBK-BDC} \\ \cline{4-21}
			\multicolumn{1}{|c|}{} & \multicolumn{1}{c|}{} & \multicolumn{1}{c|}{} & \multicolumn{2}{c|}{$\varepsilon$} & \multicolumn{2}{c|}{$\overline{t}$} & \multicolumn{2}{c|}{$t$} & \multicolumn{2}{c|}{$\varepsilon$} & \multicolumn{2}{c|}{$\overline{t}$} & \multicolumn{2}{c|}{$t$} & \multicolumn{2}{c|}{$\varepsilon$} & \multicolumn{2}{c|}{$\overline{t}$} & \multicolumn{2}{c|}{$t$} \\ \cline{4-21}
			\multicolumn{1}{|c|}{} & \multicolumn{1}{c|}{} & \multicolumn{1}{c|}{} & \multicolumn{1}{c|}{med} & \multicolumn{1}{c|}{std} & \multicolumn{1}{c|}{med} & \multicolumn{1}{c|}{std} & \multicolumn{1}{c|}{med} & \multicolumn{1}{c|}{std} & \multicolumn{1}{c|}{med} & \multicolumn{1}{c|}{std} & \multicolumn{1}{c|}{med} & \multicolumn{1}{c|}{std} & \multicolumn{1}{c|}{med} & \multicolumn{1}{c|}{std} & \multicolumn{1}{c|}{med} & \multicolumn{1}{c|}{std} & \multicolumn{1}{c|}{med} & \multicolumn{1}{c|}{std} & \multicolumn{1}{c|}{med} & \multicolumn{1}{c|}{std} \\ \hline
			2 & 2.48889$^*$ & 0.01512 & 0.003 & 0.001 & 19.63 & 6.539 & 21.339 & 7.536 & 0.0 & 0.0 & -- & -- & 14.412 & 1.882 & 0.0 & 0.0 & -- & -- & 3.128 & 0.266 \\
			3 & 2.36789$^*$ & 0.01439 & 0.005 & 0.001 & 9.107 & 4.481 & 18.658 & 7.2 & 0.0 & 0.436 & -- & -- & 30.808 & 15.829 & 0.347 & 1.087 & -- & -- & 5.228 & 0.278 \\
			5 & 2.21106$^*$ & 0.01349 & 0.008 & 0.161 & 2.253 & 4.243 & 15.988 & 8.928 & 0.323 & 0.114 & -- & -- & 79.275 & 8.253 & 0.985 & 0.993 & -- & -- & 8.819 & 0.499 \\
			10 & 2.00353$^*$ & 0.01223 & 0.112 & 0.072 & 1.658 & 0.439 & 23.171 & 6.864 & 0.217 & 0.257 & -- & -- & 398.39 & 138.675 & 2.767 & 1.451 & -- & -- & 28.289 & 1.781 \\
			15 & 1.89922$^*$ & 0.01157 & 0.115 & 0.163 & 3.747 & 5.177 & 18.969 & 7.195 & 0.289 & 0.188 & -- & -- & 706.048 & 332.25 & 1.634 & 1.369 & -- & -- & 50.802 & 3.377 \\
			20 & 1.82904$^*$ & 0.01114 & 0.12 & 0.101 & 4.547 & 9.065 & 16.949 & 9.017 & 0.121 & 0.166 & -- & -- & 1103.684 & 294.319 & 2.351 & 0.863 & -- & -- & 67.597 & 2.791 \\
			25 & 1.77524$^*$ & 0.01082 & 0.186 & 0.139 & 5.863 & 4.647 & 24.828 & 7.007 & 0.344 & 0.269 & -- & -- & 1229.349 & 342.28 & 2.092 & 0.584 & -- & -- & 85.084 & 5.815 \\
			\hline
			\multicolumn{3}{|c|}{Mean:} & \textbf{0.078} & & \textbf{6.686} & & \textbf{19.986} & & \textbf{0.185} & & \textbf{--} & & \textbf{508.852} & & \textbf{1.454} & & \textbf{--} & & \textbf{35.564} & \\ \hline
		\end{tabular}
	}
	
	\bigskip
	
	\caption{Clustering details with HEPMASS}
	\label{TabDetailsD2}
	\resizebox{!}{\tableheight}{
		\begin{tabular}{|l|l|llll|llll|llll|lllll|l|l|}
			\hline
			\multicolumn{1}{|c|}{\multirow{2}{*}{$k$}} & \multicolumn{1}{c|}{\multirow{2}{*}{$n_{exec}$}} & \multicolumn{4}{c|}{HPClust-inner} & \multicolumn{4}{c|}{HPClust-competitive} & \multicolumn{4}{c|}{HPClust-cooperative} & \multicolumn{5}{c|}{HPClust-hybrid} & \multicolumn{1}{c|}{Forgy K-means} & \multicolumn{1}{c|}{PBK-BDC} \\ \cline{3-21}
			\multicolumn{1}{|c|}{} & \multicolumn{1}{c|}{} & \multicolumn{1}{c|}{$s$} & \multicolumn{1}{c|}{$n_{s}$} & \multicolumn{1}{c|}{$T$} & \multicolumn{1}{c|}{$n_{d}$} & \multicolumn{1}{c|}{$s$} & \multicolumn{1}{c|}{$n_{s}$} & \multicolumn{1}{c|}{$T$} & \multicolumn{1}{c|}{$n_{d}$} & \multicolumn{1}{c|}{$s$} & \multicolumn{1}{c|}{$n_{s}$} & \multicolumn{1}{c|}{$T$} & \multicolumn{1}{c|}{$n_{d}$} & \multicolumn{1}{c|}{$s$} & \multicolumn{1}{c|}{$n_{s}$} & \multicolumn{1}{c|}{$T_1$} & \multicolumn{1}{c|}{$T_2$} & \multicolumn{1}{c|}{$n_{d}$} & \multicolumn{1}{c|}{$n_{d}$} & \multicolumn{1}{c|}{$n_{d}$} \\
			\hline
			2 & 7 & 64000 & 25 & 30.0 & 3.2E+07 & 64000 & 256 & 30.0 & 1.1E+08 & 64000 & 172 & 30.0 & 1.0E+08 & 64000 & 188 & 17.0 & 13.0 & 1.0E+08 & 5.7E+08 & 4.0E+08 \\
			3 & 7 & 64000 & 27 & 30.0 & 5.6E+07 & 64000 & 116 & 30.0 & 2.0E+08 & 64000 & 155 & 30.0 & 1.9E+08 & 64000 & 164 & 7.0 & 23.0 & 2.0E+08 & 1.4E+09 & 1.1E+09 \\
			5 & 7 & 64000 & 26 & 30.0 & 9.6E+07 & 64000 & 166 & 30.0 & 3.8E+08 & 64000 & 121 & 30.0 & 3.6E+08 & 64000 & 135 & 7.0 & 23.0 & 3.7E+08 & 4.3E+09 & 2.7E+09 \\
			10 & 7 & 64000 & 23 & 30.0 & 2.0E+08 & 64000 & 135 & 30.0 & 8.0E+08 & 64000 & 112 & 30.0 & 7.4E+08 & 64000 & 189 & 16.0 & 14.0 & 7.7E+08 & 2.4E+10 & 1.1E+10 \\
			15 & 7 & 64000 & 16 & 30.0 & 3.1E+08 & 64000 & 189 & 30.0 & 1.3E+09 & 64000 & 142 & 30.0 & 1.2E+09 & 64000 & 148 & 9.0 & 21.0 & 1.3E+09 & 4.4E+10 & 2.2E+10 \\
			20 & 7 & 64000 & 21 & 30.0 & 4.4E+08 & 64000 & 154 & 30.0 & 1.9E+09 & 64000 & 162 & 30.0 & 1.8E+09 & 64000 & 129 & 28.0 & 2.0 & 1.8E+09 & 7.0E+10 & 2.9E+10 \\
			25 & 7 & 64000 & 26 & 30.0 & 5.7E+08 & 64000 & 175 & 30.0 & 2.4E+09 & 64000 & 140 & 30.0 & 2.2E+09 & 64000 & 170 & 22.0 & 8.0 & 2.3E+09 & 7.8E+10 & 3.7E+10 \\
			\hline
		\end{tabular}
	}
	
\end{table}

\newpage

%%%%%%%%%%%%%%%%%%%%%%%%%%%%%%%%%%%%%%%%%%%%%%%%%%%%%%%%%%%%%%%%%%%%%%%%%%%%%%%%%%%%%%%
%  END: HEPMASS
%%%%%%%%%%%%%%%%%%%%%%%%%%%%%%%%%%%%%%%%%%%%%%%%%%%%%%%%%%%%%%%%%%%%%%%%%%%%%%%%%%%%%%%

\newpage

%%%%%%%%%%%%%%%%%%%%%%%%%%%%%%%%%%%%%%%%%%%%%%%%%%%%%%%%%%%%%%%%%%%%%%%%%%%%%%%%%%%%%%%
%  START: US Census Data 1990
%%%%%%%%%%%%%%%%%%%%%%%%%%%%%%%%%%%%%%%%%%%%%%%%%%%%%%%%%%%%%%%%%%%%%%%%%%%%%%%%%%%%%%%
\subsection{US Census Data 1990}
Dimensions: $m$ = 2458285, $n$ = 68.
\par
Description: The data set was obtained from the (U.S. Department of Commerce) Census Bureau website and contains a one percent sample of the Public Use Microdata Samples (PUMS) person records drawn from the entire 1990 U.S. census sample.

\begin{table}[!htbp]
	\centering
	
	\caption{Summary of the results with US Census Data 1990 ($\times10^{8}$)}
	\label{TabResultsD3}
	\small
	\resizebox{!}{\tableheight}{
		\begin{tabular}{|l|l|l|llllll|llllll|llllll|}
			\hline
			\multicolumn{1}{|c|}{\multirow{3}{*}{$k$}} & \multicolumn{1}{c|}{\multirow{3}{*}{$f^*$}} & \multicolumn{1}{c|}{\multirow{3}{*}{$\overline{f}$}} & \multicolumn{6}{c|}{HPClust-inner} & \multicolumn{6}{c|}{HPClust-competitive} & \multicolumn{6}{c|}{HPClust-cooperative} \\ \cline{4-21}
			\multicolumn{1}{|c|}{} & \multicolumn{1}{c|}{} & \multicolumn{1}{c|}{} & \multicolumn{2}{c|}{$\varepsilon$} & \multicolumn{2}{c|}{$\overline{t}$} & \multicolumn{2}{c|}{$t$} & \multicolumn{2}{c|}{$\varepsilon$} & \multicolumn{2}{c|}{$\overline{t}$} & \multicolumn{2}{c|}{$t$} & \multicolumn{2}{c|}{$\varepsilon$} & \multicolumn{2}{c|}{$\overline{t}$} & \multicolumn{2}{c|}{$t$} \\ \cline{4-21}
			\multicolumn{1}{|c|}{} & \multicolumn{1}{c|}{} & \multicolumn{1}{c|}{} & \multicolumn{1}{c|}{med} & \multicolumn{1}{c|}{std} & \multicolumn{1}{c|}{med} & \multicolumn{1}{c|}{std} & \multicolumn{1}{c|}{med} & \multicolumn{1}{c|}{std} & \multicolumn{1}{c|}{med} & \multicolumn{1}{c|}{std} & \multicolumn{1}{c|}{med} & \multicolumn{1}{c|}{std} & \multicolumn{1}{c|}{med} & \multicolumn{1}{c|}{std} & \multicolumn{1}{c|}{med} & \multicolumn{1}{c|}{std} & \multicolumn{1}{c|}{med} & \multicolumn{1}{c|}{std} & \multicolumn{1}{c|}{med} & \multicolumn{1}{c|}{std} \\ \hline
			2 & 18.39812$^*$ & 0.04235 & 0.107 & 0.107 & 1.569 & 0.665 & 1.547 & 0.737 & 0.278 & 0.191 & 1.058 & 0.722 & 2.102 & 0.673 & 0.233 & 0.174 & 0.983 & 0.685 & 1.825 & 0.68 \\
			3 & 6.1591$^*$ & 0.01444 & 0.069 & 66.222 & 0.813 & 1.121 & 1.191 & 0.959 & 0.078 & 0.03 & 1.292 & 0.691 & 1.852 & 0.859 & 0.074 & 0.023 & 0.856 & 0.741 & 1.376 & 0.951 \\
			5 & 3.35214$^*$ & 0.00827 & 2.179 & 9.495 & 0.142 & 0.271 & 1.937 & 0.929 & 0.105 & 0.044 & 0.112 & 0.035 & 1.966 & 0.753 & 0.13 & 1.488 & 0.108 & 0.319 & 1.343 & 0.708 \\
			10 & 2.36352$^*$ & 0.00599 & 4.682 & 2.985 & 0.166 & 0.588 & 1.926 & 0.867 & 2.413 & 1.546 & 0.179 & 0.137 & 2.063 & 0.961 & 3.296 & 2.01 & 0.172 & 0.039 & 2.068 & 0.86 \\
			15 & 2.04097$^*$ & 0.00508 & 4.538 & 4.329 & 0.368 & 0.71 & 1.774 & 0.718 & 2.141 & 1.11 & 0.258 & 0.164 & 1.769 & 0.75 & 1.829 & 1.228 & 0.246 & 0.074 & 1.957 & 0.719 \\
			20 & 1.81278$^*$ & 0.00446 & 5.921 & 3.048 & 0.89 & 0.499 & 1.407 & 0.672 & 2.17 & 0.794 & 0.358 & 0.347 & 2.132 & 0.935 & 2.858 & 1.073 & 0.455 & 0.351 & 1.755 & 0.609 \\
			25 & 1.64602$^*$ & 0.00408 & 4.439 & 1.401 & 0.487 & 0.586 & 1.613 & 0.952 & 3.178 & 0.842 & 0.457 & 0.381 & 1.806 & 0.875 & 3.034 & 0.946 & 0.434 & 0.188 & 1.928 & 0.73 \\
			\hline
			\multicolumn{3}{|c|}{Mean:} & \textbf{3.134} & & \textbf{0.634} & & \textbf{1.628} & & \textbf{1.48} & & \textbf{0.531} & & \textbf{1.956} & & \textbf{1.636} & & \textbf{0.465} & & \textbf{1.75} & \\ \hline
		\end{tabular}
	}
	
	\bigskip
	
	\small
	\resizebox{!}{\tableheight}{
		\begin{tabular}{|l|l|l|llllll|llllll|llllll|}
			\hline
			\multicolumn{1}{|c|}{\multirow{3}{*}{$k$}} & \multicolumn{1}{c|}{\multirow{3}{*}{$f^*$}} & \multicolumn{1}{c|}{\multirow{3}{*}{$\overline{f}$}} & \multicolumn{6}{c|}{HPClust-hybrid} & \multicolumn{6}{c|}{Forgy K-means} & \multicolumn{6}{c|}{PBK-BDC} \\ \cline{4-21}
			\multicolumn{1}{|c|}{} & \multicolumn{1}{c|}{} & \multicolumn{1}{c|}{} & \multicolumn{2}{c|}{$\varepsilon$} & \multicolumn{2}{c|}{$\overline{t}$} & \multicolumn{2}{c|}{$t$} & \multicolumn{2}{c|}{$\varepsilon$} & \multicolumn{2}{c|}{$\overline{t}$} & \multicolumn{2}{c|}{$t$} & \multicolumn{2}{c|}{$\varepsilon$} & \multicolumn{2}{c|}{$\overline{t}$} & \multicolumn{2}{c|}{$t$} \\ \cline{4-21}
			\multicolumn{1}{|c|}{} & \multicolumn{1}{c|}{} & \multicolumn{1}{c|}{} & \multicolumn{1}{c|}{med} & \multicolumn{1}{c|}{std} & \multicolumn{1}{c|}{med} & \multicolumn{1}{c|}{std} & \multicolumn{1}{c|}{med} & \multicolumn{1}{c|}{std} & \multicolumn{1}{c|}{med} & \multicolumn{1}{c|}{std} & \multicolumn{1}{c|}{med} & \multicolumn{1}{c|}{std} & \multicolumn{1}{c|}{med} & \multicolumn{1}{c|}{std} & \multicolumn{1}{c|}{med} & \multicolumn{1}{c|}{std} & \multicolumn{1}{c|}{med} & \multicolumn{1}{c|}{std} & \multicolumn{1}{c|}{med} & \multicolumn{1}{c|}{std} \\ \hline
			2 & 18.39812$^*$ & 0.04235 & 0.193 & 0.101 & 0.529 & 0.722 & 2.371 & 0.787 & 0.0 & 0.0 & -- & -- & 0.824 & 0.235 & 0.0 & 0.0 & -- & -- & 0.334 & 0.019 \\
			3 & 6.1591$^*$ & 0.01444 & 0.083 & 0.027 & 1.114 & 0.72 & 1.452 & 0.882 & 162.972 & 61.576 & -- & -- & 1.837 & 1.254 & 170.038 & 55.912 & -- & -- & 0.652 & 0.036 \\
			5 & 3.35214$^*$ & 0.00827 & 0.117 & 0.037 & 0.106 & 0.037 & 1.563 & 0.7 & 356.299 & 149.498 & -- & -- & 16.932 & 7.178 & 216.673 & 185.79 & -- & -- & 1.675 & 0.106 \\
			10 & 2.36352$^*$ & 0.00599 & 3.426 & 1.645 & 0.168 & 0.042 & 2.665 & 1.017 & 12.78 & 250.618 & -- & -- & 41.385 & 15.171 & 21.403 & 168.656 & -- & -- & 4.014 & 0.287 \\
			15 & 2.04097$^*$ & 0.00508 & 2.063 & 1.189 & 0.279 & 0.159 & 2.463 & 0.757 & 9.039 & 185.544 & -- & -- & 74.563 & 18.755 & 17.079 & 191.017 & -- & -- & 6.282 & 0.398 \\
			20 & 1.81278$^*$ & 0.00446 & 3.548 & 1.232 & 0.572 & 0.434 & 1.873 & 0.766 & 12.895 & 6.605 & -- & -- & 117.544 & 73.095 & 16.362 & 10.313 & -- & -- & 9.041 & 0.82 \\
			25 & 1.64602$^*$ & 0.00408 & 2.641 & 1.094 & 0.41 & 0.262 & 2.147 & 0.78 & 11.136 & 5.757 & -- & -- & 179.535 & 60.011 & 15.337 & 6.458 & -- & -- & 11.032 & 0.508 \\
			\hline
			\multicolumn{3}{|c|}{Mean:} & \textbf{1.724} & & \textbf{0.454} & & \textbf{2.077} & & \textbf{80.732} & & \textbf{--} & & \textbf{61.803} & & \textbf{65.27} & & \textbf{--} & & \textbf{4.719} & \\ \hline
		\end{tabular}
	}
	
	\bigskip
	
	\caption{Clustering details with US Census Data 1990}
	\label{TabDetailsD3}
	\resizebox{!}{\tableheight}{
		\begin{tabular}{|l|l|llll|llll|llll|lllll|l|l|}
			\hline
			\multicolumn{1}{|c|}{\multirow{2}{*}{$k$}} & \multicolumn{1}{c|}{\multirow{2}{*}{$n_{exec}$}} & \multicolumn{4}{c|}{HPClust-inner} & \multicolumn{4}{c|}{HPClust-competitive} & \multicolumn{4}{c|}{HPClust-cooperative} & \multicolumn{5}{c|}{HPClust-hybrid} & \multicolumn{1}{c|}{Forgy K-means} & \multicolumn{1}{c|}{PBK-BDC} \\ \cline{3-21}
			\multicolumn{1}{|c|}{} & \multicolumn{1}{c|}{} & \multicolumn{1}{c|}{$s$} & \multicolumn{1}{c|}{$n_{s}$} & \multicolumn{1}{c|}{$T$} & \multicolumn{1}{c|}{$n_{d}$} & \multicolumn{1}{c|}{$s$} & \multicolumn{1}{c|}{$n_{s}$} & \multicolumn{1}{c|}{$T$} & \multicolumn{1}{c|}{$n_{d}$} & \multicolumn{1}{c|}{$s$} & \multicolumn{1}{c|}{$n_{s}$} & \multicolumn{1}{c|}{$T$} & \multicolumn{1}{c|}{$n_{d}$} & \multicolumn{1}{c|}{$s$} & \multicolumn{1}{c|}{$n_{s}$} & \multicolumn{1}{c|}{$T_1$} & \multicolumn{1}{c|}{$T_2$} & \multicolumn{1}{c|}{$n_{d}$} & \multicolumn{1}{c|}{$n_{d}$} & \multicolumn{1}{c|}{$n_{d}$} \\
			\hline
			2 & 20 & 6000 & 19 & 3.0 & 5.9E+06 & 6000 & 162 & 3.0 & 1.1E+07 & 6000 & 120 & 3.0 & 1.1E+07 & 6000 & 170 & 0.2 & 2.8 & 1.1E+07 & 1.5E+07 & 1.9E+07 \\
			3 & 20 & 6000 & 14 & 3.0 & 8.7E+06 & 6000 & 136 & 3.0 & 1.7E+07 & 6000 & 106 & 3.0 & 1.7E+07 & 6000 & 100 & 2.1 & 0.9 & 1.7E+07 & 3.7E+07 & 4.4E+07 \\
			5 & 20 & 6000 & 18 & 3.0 & 1.5E+07 & 6000 & 143 & 3.0 & 3.2E+07 & 6000 & 97 & 3.0 & 2.9E+07 & 6000 & 113 & 0.6 & 2.4 & 3.0E+07 & 3.6E+08 & 1.5E+08 \\
			10 & 20 & 6000 & 20 & 3.0 & 3.5E+07 & 6000 & 120 & 3.0 & 7.5E+07 & 6000 & 132 & 3.0 & 7.0E+07 & 6000 & 176 & 2.4 & 0.6 & 7.9E+07 & 9.5E+08 & 5.1E+08 \\
			15 & 20 & 6000 & 23 & 3.0 & 5.5E+07 & 6000 & 88 & 3.0 & 1.2E+08 & 6000 & 104 & 3.0 & 1.2E+08 & 6000 & 128 & 1.9 & 1.1 & 1.2E+08 & 1.7E+09 & 9.0E+08 \\
			20 & 20 & 6000 & 16 & 3.0 & 7.8E+07 & 6000 & 92 & 3.0 & 1.7E+08 & 6000 & 66 & 3.0 & 1.5E+08 & 6000 & 78 & 0.1 & 2.9 & 1.5E+08 & 2.7E+09 & 1.3E+09 \\
			25 & 20 & 6000 & 12 & 3.0 & 9.4E+07 & 6000 & 60 & 3.0 & 2.0E+08 & 6000 & 64 & 3.0 & 2.0E+08 & 6000 & 72 & 2.5 & 0.5 & 2.0E+08 & 4.1E+09 & 1.7E+09 \\
			\hline
		\end{tabular}
	}
	
\end{table}

\newpage

%%%%%%%%%%%%%%%%%%%%%%%%%%%%%%%%%%%%%%%%%%%%%%%%%%%%%%%%%%%%%%%%%%%%%%%%%%%%%%%%%%%%%%%
%  END: US Census Data 1990
%%%%%%%%%%%%%%%%%%%%%%%%%%%%%%%%%%%%%%%%%%%%%%%%%%%%%%%%%%%%%%%%%%%%%%%%%%%%%%%%%%%%%%%

%%%%%%%%%%%%%%%%%%%%%%%%%%%%%%%%%%%%%%%%%%%%%%%%%%%%%%%%%%%%%%%%%%%%%%%%%%%%%%%%%%%%%%%
%  START: Gisette
%%%%%%%%%%%%%%%%%%%%%%%%%%%%%%%%%%%%%%%%%%%%%%%%%%%%%%%%%%%%%%%%%%%%%%%%%%%%%%%%%%%%%%%
\subsection{Gisette}
Dimensions: $m$ = 13500, $n$ = 5000.
\par
Description: patterns for handwritten digit recognition problem.

\begin{table}[!htbp]
	\centering
	
	\caption{Summary of the results with Gisette ($\times10^{12}$)}
	\label{TabResultsD4}
	\small
	\resizebox{!}{\tableheight}{
		\begin{tabular}{|l|l|l|llllll|llllll|llllll|}
			\hline
			\multicolumn{1}{|c|}{\multirow{3}{*}{$k$}} & \multicolumn{1}{c|}{\multirow{3}{*}{$f^*$}} & \multicolumn{1}{c|}{\multirow{3}{*}{$\overline{f}$}} & \multicolumn{6}{c|}{HPClust-inner} & \multicolumn{6}{c|}{HPClust-competitive} & \multicolumn{6}{c|}{HPClust-cooperative} \\ \cline{4-21}
			\multicolumn{1}{|c|}{} & \multicolumn{1}{c|}{} & \multicolumn{1}{c|}{} & \multicolumn{2}{c|}{$\varepsilon$} & \multicolumn{2}{c|}{$\overline{t}$} & \multicolumn{2}{c|}{$t$} & \multicolumn{2}{c|}{$\varepsilon$} & \multicolumn{2}{c|}{$\overline{t}$} & \multicolumn{2}{c|}{$t$} & \multicolumn{2}{c|}{$\varepsilon$} & \multicolumn{2}{c|}{$\overline{t}$} & \multicolumn{2}{c|}{$t$} \\ \cline{4-21}
			\multicolumn{1}{|c|}{} & \multicolumn{1}{c|}{} & \multicolumn{1}{c|}{} & \multicolumn{1}{c|}{med} & \multicolumn{1}{c|}{std} & \multicolumn{1}{c|}{med} & \multicolumn{1}{c|}{std} & \multicolumn{1}{c|}{med} & \multicolumn{1}{c|}{std} & \multicolumn{1}{c|}{med} & \multicolumn{1}{c|}{std} & \multicolumn{1}{c|}{med} & \multicolumn{1}{c|}{std} & \multicolumn{1}{c|}{med} & \multicolumn{1}{c|}{std} & \multicolumn{1}{c|}{med} & \multicolumn{1}{c|}{std} & \multicolumn{1}{c|}{med} & \multicolumn{1}{c|}{std} & \multicolumn{1}{c|}{med} & \multicolumn{1}{c|}{std} \\ \hline
			2 & 4.19944 & 3.1048 & 0.009 & 0.005 & 2.673 & 1.476 & 3.135 & 1.445 & 0.007 & 0.006 & 3.941 & 0.683 & 4.994 & 0.75 & 0.009 & 0.005 & 4.327 & 0.746 & 4.623 & 0.759 \\
			3 & 4.11596 & 3.04579 & 0.036 & 0.148 & 1.315 & 1.4 & 2.646 & 1.244 & 0.029 & 0.013 & 5.224 & 0.563 & 5.991 & 0.964 & 0.023 & 0.017 & 5.001 & 0.426 & 5.271 & 0.749 \\
			5 & 4.02303 & 2.97834 & 0.077 & 0.03 & 3.049 & 0.836 & 3.353 & 0.894 & 0.064 & 0.037 & 8.468 & 0.901 & 8.529 & 0.875 & 0.081 & 0.037 & 8.397 & 1.456 & 8.484 & 1.568 \\
			10 & 3.87672 & 2.87532 & 0.165 & 0.099 & 4.147 & 0.874 & 4.501 & 0.82 & 0.156 & 0.045 & 16.714 & 0.814 & 18.757 & 2.217 & 0.132 & 0.061 & 16.53 & 1.032 & 18.299 & 1.802 \\
			15 & 3.81766 & 2.81586 & -0.282 & 0.051 & 5.514 & 0.535 & 5.47 & 0.646 & -0.297 & 0.048 & 25.545 & 3.208 & 26.532 & 5.26 & -0.315 & 0.048 & 25.196 & 2.045 & 28.983 & 2.846 \\
			20 & 3.81436 & 2.77677 & -1.6 & 0.048 & 6.593 & 0.908 & 6.593 & 0.713 & -1.628 & 0.045 & 32.543 & 3.114 & 38.244 & 4.044 & -1.627 & 0.066 & 32.486 & 1.482 & 34.75 & 2.311 \\
			25 & 3.74937 & 2.74501 & -1.002 & 0.072 & 7.495 & 0.988 & 6.831 & 1.347 & -1.022 & 0.053 & 43.202 & 3.331 & 46.72 & 3.559 & -1.055 & 0.055 & 40.976 & 4.052 & 45.528 & 3.191 \\
			\hline
			\multicolumn{3}{|c|}{Mean:} & \textbf{-0.371} & & \textbf{4.398} & & \textbf{4.647} & & \textbf{-0.384} & & \textbf{19.377} & & \textbf{21.395} & & \textbf{-0.393} & & \textbf{18.988} & & \textbf{20.848} & \\ \hline
		\end{tabular}
	}
	
	\bigskip
	
	\small
	\resizebox{!}{\tableheight}{
		\begin{tabular}{|l|l|l|llllll|llllll|llllll|}
			\hline
			\multicolumn{1}{|c|}{\multirow{3}{*}{$k$}} & \multicolumn{1}{c|}{\multirow{3}{*}{$f^*$}} & \multicolumn{1}{c|}{\multirow{3}{*}{$\overline{f}$}} & \multicolumn{6}{c|}{HPClust-hybrid} & \multicolumn{6}{c|}{Forgy K-means} & \multicolumn{6}{c|}{PBK-BDC} \\ \cline{4-21}
			\multicolumn{1}{|c|}{} & \multicolumn{1}{c|}{} & \multicolumn{1}{c|}{} & \multicolumn{2}{c|}{$\varepsilon$} & \multicolumn{2}{c|}{$\overline{t}$} & \multicolumn{2}{c|}{$t$} & \multicolumn{2}{c|}{$\varepsilon$} & \multicolumn{2}{c|}{$\overline{t}$} & \multicolumn{2}{c|}{$t$} & \multicolumn{2}{c|}{$\varepsilon$} & \multicolumn{2}{c|}{$\overline{t}$} & \multicolumn{2}{c|}{$t$} \\ \cline{4-21}
			\multicolumn{1}{|c|}{} & \multicolumn{1}{c|}{} & \multicolumn{1}{c|}{} & \multicolumn{1}{c|}{med} & \multicolumn{1}{c|}{std} & \multicolumn{1}{c|}{med} & \multicolumn{1}{c|}{std} & \multicolumn{1}{c|}{med} & \multicolumn{1}{c|}{std} & \multicolumn{1}{c|}{med} & \multicolumn{1}{c|}{std} & \multicolumn{1}{c|}{med} & \multicolumn{1}{c|}{std} & \multicolumn{1}{c|}{med} & \multicolumn{1}{c|}{std} & \multicolumn{1}{c|}{med} & \multicolumn{1}{c|}{std} & \multicolumn{1}{c|}{med} & \multicolumn{1}{c|}{std} & \multicolumn{1}{c|}{med} & \multicolumn{1}{c|}{std} \\ \hline
			2 & 4.19944 & 3.1048 & 0.008 & 0.004 & 4.459 & 0.432 & 4.499 & 0.702 & 0.0 & 0.0 & -- & -- & 3.691 & 1.633 & 0.006 & 0.003 & -- & -- & 3.726 & 1.139 \\
			3 & 4.11596 & 3.04579 & 0.023 & 0.016 & 5.623 & 1.684 & 5.777 & 1.243 & 0.0 & 0.0 & -- & -- & 7.901 & 2.493 & 0.01 & 0.002 & -- & -- & 8.518 & 2.308 \\
			5 & 4.02303 & 2.97834 & 0.071 & 0.032 & 8.021 & 0.722 & 8.617 & 1.411 & 0.011 & 0.027 & -- & -- & 32.88 & 20.091 & 0.037 & 0.041 & -- & -- & 16.183 & 8.116 \\
			10 & 3.87672 & 2.87532 & 0.127 & 0.046 & 16.899 & 1.893 & 19.183 & 2.955 & 0.038 & 0.049 & -- & -- & 45.665 & 21.177 & 0.116 & 0.042 & -- & -- & 31.801 & 17.762 \\
			15 & 3.81766 & 2.81586 & -0.301 & 0.026 & 25.974 & 2.107 & 27.735 & 2.35 & -0.442 & 0.046 & -- & -- & 59.415 & 23.158 & -0.332 & 0.069 & -- & -- & 43.361 & 10.205 \\
			20 & 3.81436 & 2.77677 & -1.658 & 0.044 & 32.582 & 2.488 & 36.983 & 4.158 & -1.782 & 0.045 & -- & -- & 106.076 & 27.759 & -1.69 & 0.058 & -- & -- & 60.231 & 25.017 \\
			25 & 3.74937 & 2.74501 & -1.049 & 0.042 & 41.765 & 2.526 & 45.036 & 3.52 & -1.215 & 0.082 & -- & -- & 114.912 & 27.943 & -1.11 & 0.079 & -- & -- & 68.469 & 23.838 \\
			\hline
			\multicolumn{3}{|c|}{Mean:} & \textbf{-0.397} & & \textbf{19.332} & & \textbf{21.119} & & \textbf{-0.484} & & \textbf{--} & & \textbf{52.934} & & \textbf{-0.423} & & \textbf{--} & & \textbf{33.184} & \\ \hline
		\end{tabular}
	}
	
	\bigskip
	
	\caption{Clustering details with Gisette}
	\label{TabDetailsD4}
	\resizebox{!}{\tableheight}{
		\begin{tabular}{|l|l|llll|llll|llll|lllll|l|l|}
			\hline
			\multicolumn{1}{|c|}{\multirow{2}{*}{$k$}} & \multicolumn{1}{c|}{\multirow{2}{*}{$n_{exec}$}} & \multicolumn{4}{c|}{HPClust-inner} & \multicolumn{4}{c|}{HPClust-competitive} & \multicolumn{4}{c|}{HPClust-cooperative} & \multicolumn{5}{c|}{HPClust-hybrid} & \multicolumn{1}{c|}{Forgy K-means} & \multicolumn{1}{c|}{PBK-BDC} \\ \cline{3-21}
			\multicolumn{1}{|c|}{} & \multicolumn{1}{c|}{} & \multicolumn{1}{c|}{$s$} & \multicolumn{1}{c|}{$n_{s}$} & \multicolumn{1}{c|}{$T$} & \multicolumn{1}{c|}{$n_{d}$} & \multicolumn{1}{c|}{$s$} & \multicolumn{1}{c|}{$n_{s}$} & \multicolumn{1}{c|}{$T$} & \multicolumn{1}{c|}{$n_{d}$} & \multicolumn{1}{c|}{$s$} & \multicolumn{1}{c|}{$n_{s}$} & \multicolumn{1}{c|}{$T$} & \multicolumn{1}{c|}{$n_{d}$} & \multicolumn{1}{c|}{$s$} & \multicolumn{1}{c|}{$n_{s}$} & \multicolumn{1}{c|}{$T_1$} & \multicolumn{1}{c|}{$T_2$} & \multicolumn{1}{c|}{$n_{d}$} & \multicolumn{1}{c|}{$n_{d}$} & \multicolumn{1}{c|}{$n_{d}$} \\
			\hline
			2 & 15 & 10000 & 8 & 5.0 & 1.1E+06 & 10000 & 20 & 5.0 & 3.4E+06 & 10000 & 18 & 5.0 & 3.4E+06 & 10000 & 15 & 4.5 & 0.5 & 3.3E+06 & 7.0E+05 & 6.7E+05 \\
			3 & 15 & 10000 & 5 & 5.0 & 1.5E+06 & 10000 & 7 & 5.0 & 4.0E+06 & 10000 & 4 & 5.0 & 4.1E+06 & 10000 & 5 & 1.833 & 3.167 & 3.9E+06 & 1.6E+06 & 1.6E+06 \\
			5 & 15 & 10000 & 5 & 5.0 & 2.0E+06 & 10000 & 4 & 5.0 & 6.3E+06 & 10000 & 6 & 5.0 & 6.4E+06 & 10000 & 5 & 3.167 & 1.833 & 6.4E+06 & 6.8E+06 & 3.2E+06 \\
			10 & 15 & 10000 & 1 & 5.0 & 2.9E+06 & 10000 & 6 & 5.0 & 1.6E+07 & 10000 & 6 & 5.0 & 1.6E+07 & 10000 & 5 & 3.833 & 1.167 & 1.7E+07 & 9.7E+06 & 6.8E+06 \\
			15 & 15 & 10000 & 1 & 5.0 & 3.9E+06 & 10000 & 5 & 5.0 & 2.5E+07 & 10000 & 7 & 5.0 & 2.5E+07 & 10000 & 6 & 3.0 & 2.0 & 2.4E+07 & 1.3E+07 & 9.5E+06 \\
			20 & 15 & 10000 & 1 & 5.0 & 4.4E+06 & 10000 & 6 & 5.0 & 3.5E+07 & 10000 & 3 & 5.0 & 3.3E+07 & 10000 & 6 & 2.333 & 2.667 & 3.5E+07 & 2.3E+07 & 1.3E+07 \\
			25 & 15 & 10000 & 1 & 5.0 & 6.1E+06 & 10000 & 6 & 5.0 & 4.4E+07 & 10000 & 6 & 5.0 & 4.4E+07 & 10000 & 5 & 4.667 & 0.333 & 4.4E+07 & 2.5E+07 & 1.5E+07 \\
			\hline
		\end{tabular}
	}
	
\end{table}

\newpage

%%%%%%%%%%%%%%%%%%%%%%%%%%%%%%%%%%%%%%%%%%%%%%%%%%%%%%%%%%%%%%%%%%%%%%%%%%%%%%%%%%%%%%%
%  END: Gisette
%%%%%%%%%%%%%%%%%%%%%%%%%%%%%%%%%%%%%%%%%%%%%%%%%%%%%%%%%%%%%%%%%%%%%%%%%%%%%%%%%%%%%%%

\newpage

%%%%%%%%%%%%%%%%%%%%%%%%%%%%%%%%%%%%%%%%%%%%%%%%%%%%%%%%%%%%%%%%%%%%%%%%%%%%%%%%%%%%%%%
%  START: Music Analysis
%%%%%%%%%%%%%%%%%%%%%%%%%%%%%%%%%%%%%%%%%%%%%%%%%%%%%%%%%%%%%%%%%%%%%%%%%%%%%%%%%%%%%%%
\subsection{Music Analysis}
Dimensions: $m$ = 106574, $n$ = 518.
\par
Description: a dataset for music analysis which contains different spectral and statistical attributes for each music track.

\begin{table}[!htbp]
	\centering
	
	\caption{Summary of the results with Music Analysis ($\times10^{11}$)}
	\label{TabResultsD5}
	\small
	\resizebox{!}{\tableheight}{
		\begin{tabular}{|l|l|l|llllll|llllll|llllll|}
			\hline
			\multicolumn{1}{|c|}{\multirow{3}{*}{$k$}} & \multicolumn{1}{c|}{\multirow{3}{*}{$f^*$}} & \multicolumn{1}{c|}{\multirow{3}{*}{$\overline{f}$}} & \multicolumn{6}{c|}{HPClust-inner} & \multicolumn{6}{c|}{HPClust-competitive} & \multicolumn{6}{c|}{HPClust-cooperative} \\ \cline{4-21}
			\multicolumn{1}{|c|}{} & \multicolumn{1}{c|}{} & \multicolumn{1}{c|}{} & \multicolumn{2}{c|}{$\varepsilon$} & \multicolumn{2}{c|}{$\overline{t}$} & \multicolumn{2}{c|}{$t$} & \multicolumn{2}{c|}{$\varepsilon$} & \multicolumn{2}{c|}{$\overline{t}$} & \multicolumn{2}{c|}{$t$} & \multicolumn{2}{c|}{$\varepsilon$} & \multicolumn{2}{c|}{$\overline{t}$} & \multicolumn{2}{c|}{$t$} \\ \cline{4-21}
			\multicolumn{1}{|c|}{} & \multicolumn{1}{c|}{} & \multicolumn{1}{c|}{} & \multicolumn{1}{c|}{med} & \multicolumn{1}{c|}{std} & \multicolumn{1}{c|}{med} & \multicolumn{1}{c|}{std} & \multicolumn{1}{c|}{med} & \multicolumn{1}{c|}{std} & \multicolumn{1}{c|}{med} & \multicolumn{1}{c|}{std} & \multicolumn{1}{c|}{med} & \multicolumn{1}{c|}{std} & \multicolumn{1}{c|}{med} & \multicolumn{1}{c|}{std} & \multicolumn{1}{c|}{med} & \multicolumn{1}{c|}{std} & \multicolumn{1}{c|}{med} & \multicolumn{1}{c|}{std} & \multicolumn{1}{c|}{med} & \multicolumn{1}{c|}{std} \\ \hline
			2 & 5.00474$^*$ & 0.26351 & 0.065 & 7.815 & 4.602 & 2.312 & 4.495 & 2.16 & 0.097 & 0.054 & 1.824 & 1.797 & 4.255 & 2.473 & 0.073 & 0.025 & 1.266 & 1.203 & 5.064 & 2.03 \\
			3 & 3.83748$^*$ & 0.20356 & 0.076 & 0.037 & 2.503 & 2.658 & 4.018 & 2.625 & 0.163 & 0.077 & 4.177 & 1.92 & 4.841 & 1.995 & 0.132 & 0.047 & 2.619 & 1.984 & 4.483 & 2.256 \\
			5 & 2.74249$^*$ & 0.14584 & 0.274 & 1.459 & 1.709 & 2.075 & 4.109 & 2.177 & 0.195 & 0.137 & 2.379 & 2.027 & 4.536 & 2.182 & 0.21 & 1.192 & 1.539 & 0.946 & 4.358 & 1.809 \\
			10 & 1.87296$^*$ & 0.10086 & 1.911 & 0.823 & 2.045 & 2.229 & 4.75 & 2.346 & 0.51 & 0.645 & 1.938 & 1.989 & 4.025 & 2.093 & 0.658 & 0.766 & 2.446 & 1.825 & 4.716 & 1.959 \\
			15 & 1.54422$^*$ & 0.08235 & 1.181 & 0.352 & 5.033 & 2.527 & 5.469 & 1.896 & 1.002 & 0.506 & 5.801 & 2.243 & 6.335 & 2.091 & 1.104 & 0.615 & 5.006 & 1.346 & 5.922 & 1.963 \\
			20 & 1.35315$^*$ & 0.07212 & 1.416 & 0.683 & 2.412 & 2.113 & 4.326 & 2.584 & 1.287 & 0.5 & 6.013 & 1.922 & 5.91 & 2.243 & 1.398 & 0.84 & 4.482 & 1.778 & 5.824 & 2.016 \\
			25 & 1.22622$^*$ & 0.06535 & 1.466 & 0.814 & 4.223 & 1.683 & 4.973 & 1.925 & 1.912 & 0.483 & 5.984 & 1.433 & 6.462 & 1.676 & 2.224 & 0.697 & 5.937 & 1.84 & 6.796 & 1.793 \\
			\hline
			\multicolumn{3}{|c|}{Mean:} & \textbf{0.913} & & \textbf{3.218} & & \textbf{4.591} & & \textbf{0.738} & & \textbf{4.017} & & \textbf{5.195} & & \textbf{0.829} & & \textbf{3.328} & & \textbf{5.309} & \\ \hline
		\end{tabular}
	}
	
	\bigskip
	
	\small
	\resizebox{!}{\tableheight}{
		\begin{tabular}{|l|l|l|llllll|llllll|llllll|}
			\hline
			\multicolumn{1}{|c|}{\multirow{3}{*}{$k$}} & \multicolumn{1}{c|}{\multirow{3}{*}{$f^*$}} & \multicolumn{1}{c|}{\multirow{3}{*}{$\overline{f}$}} & \multicolumn{6}{c|}{HPClust-hybrid} & \multicolumn{6}{c|}{Forgy K-means} & \multicolumn{6}{c|}{PBK-BDC} \\ \cline{4-21}
			\multicolumn{1}{|c|}{} & \multicolumn{1}{c|}{} & \multicolumn{1}{c|}{} & \multicolumn{2}{c|}{$\varepsilon$} & \multicolumn{2}{c|}{$\overline{t}$} & \multicolumn{2}{c|}{$t$} & \multicolumn{2}{c|}{$\varepsilon$} & \multicolumn{2}{c|}{$\overline{t}$} & \multicolumn{2}{c|}{$t$} & \multicolumn{2}{c|}{$\varepsilon$} & \multicolumn{2}{c|}{$\overline{t}$} & \multicolumn{2}{c|}{$t$} \\ \cline{4-21}
			\multicolumn{1}{|c|}{} & \multicolumn{1}{c|}{} & \multicolumn{1}{c|}{} & \multicolumn{1}{c|}{med} & \multicolumn{1}{c|}{std} & \multicolumn{1}{c|}{med} & \multicolumn{1}{c|}{std} & \multicolumn{1}{c|}{med} & \multicolumn{1}{c|}{std} & \multicolumn{1}{c|}{med} & \multicolumn{1}{c|}{std} & \multicolumn{1}{c|}{med} & \multicolumn{1}{c|}{std} & \multicolumn{1}{c|}{med} & \multicolumn{1}{c|}{std} & \multicolumn{1}{c|}{med} & \multicolumn{1}{c|}{std} & \multicolumn{1}{c|}{med} & \multicolumn{1}{c|}{std} & \multicolumn{1}{c|}{med} & \multicolumn{1}{c|}{std} \\ \hline
			2 & 5.00474$^*$ & 0.26351 & 0.081 & 0.042 & 2.83 & 1.904 & 3.613 & 2.307 & -0.0 & 9.283 & -- & -- & 2.863 & 0.593 & 1.266 & 6.592 & -- & -- & 0.446 & 0.06 \\
			3 & 3.83748$^*$ & 0.20356 & 0.142 & 0.048 & 3.071 & 1.871 & 4.941 & 1.695 & -0.0 & 4.505 & -- & -- & 4.797 & 2.883 & 2.19 & 13.482 & -- & -- & 1.174 & 0.314 \\
			5 & 2.74249$^*$ & 0.14584 & 0.208 & 0.494 & 2.046 & 1.981 & 5.308 & 2.497 & -0.001 & 1.834 & -- & -- & 11.907 & 2.354 & 1.223 & 29.104 & -- & -- & 2.204 & 0.38 \\
			10 & 1.87296$^*$ & 0.10086 & 0.604 & 0.786 & 2.652 & 1.685 & 5.809 & 2.564 & 1.448 & 1.175 & -- & -- & 56.994 & 26.9 & 9.839 & 8.146 & -- & -- & 6.313 & 0.994 \\
			15 & 1.54422$^*$ & 0.08235 & 1.565 & 0.612 & 4.07 & 2.336 & 6.531 & 2.011 & 0.649 & 0.425 & -- & -- & 138.407 & 34.68 & 5.648 & 4.717 & -- & -- & 10.277 & 1.354 \\
			20 & 1.35315$^*$ & 0.07212 & 1.405 & 0.658 & 5.708 & 2.343 & 6.508 & 2.543 & 0.597 & 0.594 & -- & -- & 151.525 & 44.902 & 7.064 & 3.501 & -- & -- & 13.586 & 2.985 \\
			25 & 1.22622$^*$ & 0.06535 & 1.945 & 0.795 & 6.795 & 2.159 & 6.584 & 2.231 & 0.611 & 0.548 & -- & -- & 239.709 & 65.327 & 6.692 & 4.415 & -- & -- & 17.212 & 2.663 \\
			\hline
			\multicolumn{3}{|c|}{Mean:} & \textbf{0.85} & & \textbf{3.882} & & \textbf{5.613} & & \textbf{0.472} & & \textbf{--} & & \textbf{86.6} & & \textbf{4.846} & & \textbf{--} & & \textbf{7.316} & \\ \hline
		\end{tabular}
	}
	
	\bigskip
	
	\caption{Clustering details with Music Analysis}
	\label{TabDetailsD5}
	\resizebox{!}{\tableheight}{
		\begin{tabular}{|l|l|llll|llll|llll|lllll|l|l|}
			\hline
			\multicolumn{1}{|c|}{\multirow{2}{*}{$k$}} & \multicolumn{1}{c|}{\multirow{2}{*}{$n_{exec}$}} & \multicolumn{4}{c|}{HPClust-inner} & \multicolumn{4}{c|}{HPClust-competitive} & \multicolumn{4}{c|}{HPClust-cooperative} & \multicolumn{5}{c|}{HPClust-hybrid} & \multicolumn{1}{c|}{Forgy K-means} & \multicolumn{1}{c|}{PBK-BDC} \\ \cline{3-21}
			\multicolumn{1}{|c|}{} & \multicolumn{1}{c|}{} & \multicolumn{1}{c|}{$s$} & \multicolumn{1}{c|}{$n_{s}$} & \multicolumn{1}{c|}{$T$} & \multicolumn{1}{c|}{$n_{d}$} & \multicolumn{1}{c|}{$s$} & \multicolumn{1}{c|}{$n_{s}$} & \multicolumn{1}{c|}{$T$} & \multicolumn{1}{c|}{$n_{d}$} & \multicolumn{1}{c|}{$s$} & \multicolumn{1}{c|}{$n_{s}$} & \multicolumn{1}{c|}{$T$} & \multicolumn{1}{c|}{$n_{d}$} & \multicolumn{1}{c|}{$s$} & \multicolumn{1}{c|}{$n_{s}$} & \multicolumn{1}{c|}{$T_1$} & \multicolumn{1}{c|}{$T_2$} & \multicolumn{1}{c|}{$n_{d}$} & \multicolumn{1}{c|}{$n_{d}$} & \multicolumn{1}{c|}{$n_{d}$} \\
			\hline
			2 & 20 & 6000 & 278 & 8.0 & 2.0E+07 & 6000 & 964 & 8.0 & 8.7E+07 & 6000 & 1249 & 8.0 & 8.6E+07 & 6000 & 834 & 1.333 & 6.667 & 8.5E+07 & 4.9E+06 & 3.3E+06 \\
			3 & 20 & 6000 & 157 & 8.0 & 2.7E+07 & 6000 & 644 & 8.0 & 9.7E+07 & 6000 & 629 & 8.0 & 9.7E+07 & 6000 & 680 & 1.6 & 6.4 & 9.6E+07 & 9.1E+06 & 8.0E+06 \\
			5 & 20 & 6000 & 116 & 8.0 & 3.8E+07 & 6000 & 318 & 8.0 & 1.1E+08 & 6000 & 282 & 8.0 & 1.1E+08 & 6000 & 345 & 1.333 & 6.667 & 1.1E+08 & 2.4E+07 & 1.7E+07 \\
			10 & 20 & 6000 & 50 & 8.0 & 5.5E+07 & 6000 & 59 & 8.0 & 1.2E+08 & 6000 & 84 & 8.0 & 1.2E+08 & 6000 & 99 & 6.133 & 1.867 & 1.1E+08 & 1.2E+08 & 6.3E+07 \\
			15 & 20 & 6000 & 34 & 8.0 & 6.0E+07 & 6000 & 59 & 8.0 & 1.2E+08 & 6000 & 46 & 8.0 & 1.2E+08 & 6000 & 46 & 4.533 & 3.467 & 1.1E+08 & 3.0E+08 & 1.0E+08 \\
			20 & 20 & 6000 & 14 & 8.0 & 6.4E+07 & 6000 & 29 & 8.0 & 1.2E+08 & 6000 & 30 & 8.0 & 1.2E+08 & 6000 & 21 & 0.533 & 7.467 & 1.0E+08 & 3.3E+08 & 1.5E+08 \\
			25 & 20 & 6000 & 16 & 8.0 & 6.6E+07 & 6000 & 21 & 8.0 & 1.2E+08 & 6000 & 23 & 8.0 & 1.2E+08 & 6000 & 10 & 0.267 & 7.733 & 8.8E+07 & 5.2E+08 & 1.9E+08 \\
			\hline
		\end{tabular}
	}
	
\end{table}

\newpage

%%%%%%%%%%%%%%%%%%%%%%%%%%%%%%%%%%%%%%%%%%%%%%%%%%%%%%%%%%%%%%%%%%%%%%%%%%%%%%%%%%%%%%%
%  END: Music Analysis
%%%%%%%%%%%%%%%%%%%%%%%%%%%%%%%%%%%%%%%%%%%%%%%%%%%%%%%%%%%%%%%%%%%%%%%%%%%%%%%%%%%%%%%

%%%%%%%%%%%%%%%%%%%%%%%%%%%%%%%%%%%%%%%%%%%%%%%%%%%%%%%%%%%%%%%%%%%%%%%%%%%%%%%%%%%%%%%
%  START: Protein Homology
%%%%%%%%%%%%%%%%%%%%%%%%%%%%%%%%%%%%%%%%%%%%%%%%%%%%%%%%%%%%%%%%%%%%%%%%%%%%%%%%%%%%%%%
\subsection{Protein Homology}
Dimensions: $m$ = 145751, $n$ = 74.
\par
Description: a data set for protein homology prediction which contains a features describing the match (e.g. the score of a sequence alignment) between the native protein sequence and the sequence that is tested for homology.

\begin{table}[!htbp]
	\centering
	
	\caption{Summary of the results with Protein Homology ($\times10^{11}$)}
	\label{TabResultsD6}
	\small
	\resizebox{!}{\tableheight}{
		\begin{tabular}{|l|l|l|llllll|llllll|llllll|}
			\hline
			\multicolumn{1}{|c|}{\multirow{3}{*}{$k$}} & \multicolumn{1}{c|}{\multirow{3}{*}{$f^*$}} & \multicolumn{1}{c|}{\multirow{3}{*}{$\overline{f}$}} & \multicolumn{6}{c|}{HPClust-inner} & \multicolumn{6}{c|}{HPClust-competitive} & \multicolumn{6}{c|}{HPClust-cooperative} \\ \cline{4-21}
			\multicolumn{1}{|c|}{} & \multicolumn{1}{c|}{} & \multicolumn{1}{c|}{} & \multicolumn{2}{c|}{$\varepsilon$} & \multicolumn{2}{c|}{$\overline{t}$} & \multicolumn{2}{c|}{$t$} & \multicolumn{2}{c|}{$\varepsilon$} & \multicolumn{2}{c|}{$\overline{t}$} & \multicolumn{2}{c|}{$t$} & \multicolumn{2}{c|}{$\varepsilon$} & \multicolumn{2}{c|}{$\overline{t}$} & \multicolumn{2}{c|}{$t$} \\ \cline{4-21}
			\multicolumn{1}{|c|}{} & \multicolumn{1}{c|}{} & \multicolumn{1}{c|}{} & \multicolumn{1}{c|}{med} & \multicolumn{1}{c|}{std} & \multicolumn{1}{c|}{med} & \multicolumn{1}{c|}{std} & \multicolumn{1}{c|}{med} & \multicolumn{1}{c|}{std} & \multicolumn{1}{c|}{med} & \multicolumn{1}{c|}{std} & \multicolumn{1}{c|}{med} & \multicolumn{1}{c|}{std} & \multicolumn{1}{c|}{med} & \multicolumn{1}{c|}{std} & \multicolumn{1}{c|}{med} & \multicolumn{1}{c|}{std} & \multicolumn{1}{c|}{med} & \multicolumn{1}{c|}{std} & \multicolumn{1}{c|}{med} & \multicolumn{1}{c|}{std} \\ \hline
			2 & 15.20433$^*$ & 4.88318 & 1.848 & 0.686 & 2.345 & 0.497 & 2.272 & 0.746 & 1.878 & 0.029 & 0.873 & 0.732 & 1.514 & 0.949 & 1.866 & 0.663 & 1.5 & 0.721 & 2.351 & 0.583 \\
			3 & 8.07129$^*$ & 2.89651 & 0.521 & 0.553 & 2.183 & 0.98 & 1.949 & 0.908 & 0.851 & 0.609 & 1.203 & 0.666 & 2.23 & 1.017 & 0.621 & 0.457 & 1.155 & 0.866 & 2.187 & 0.952 \\
			5 & 5.30537$^*$ & 1.86379 & 0.804 & 0.622 & 1.397 & 0.999 & 1.554 & 0.888 & 0.784 & 0.838 & 1.432 & 0.799 & 1.569 & 0.916 & 0.651 & 0.424 & 1.105 & 0.391 & 1.401 & 0.649 \\
			10 & 3.3767$^*$ & 1.26637 & 0.198 & 0.787 & 1.784 & 0.887 & 2.311 & 0.915 & 0.244 & 0.21 & 2.264 & 0.808 & 2.69 & 0.521 & 0.235 & 0.21 & 2.417 & 0.506 & 2.718 & 0.502 \\
			15 & 2.86473$^*$ & 1.08655 & 1.166 & 0.849 & 1.665 & 0.798 & 2.274 & 0.927 & 0.905 & 0.429 & 3.293 & 0.636 & 3.547 & 0.727 & 0.733 & 0.468 & 3.288 & 0.775 & 3.379 & 0.777 \\
			20 & 2.5732$^*$ & 0.98195 & 0.782 & 0.531 & 1.535 & 0.899 & 2.782 & 0.896 & 0.761 & 0.414 & 3.874 & 0.656 & 4.003 & 0.823 & 1.012 & 0.495 & 3.692 & 0.6 & 4.596 & 1.048 \\
			25 & 2.38539$^*$ & 0.90731 & 1.035 & 0.745 & 1.045 & 0.398 & 2.556 & 0.886 & 0.719 & 0.736 & 4.189 & 0.826 & 4.557 & 1.009 & 1.22 & 0.664 & 4.481 & 1.436 & 4.775 & 1.223 \\
			\hline
			\multicolumn{3}{|c|}{Mean:} & \textbf{0.908} & & \textbf{1.708} & & \textbf{2.242} & & \textbf{0.878} & & \textbf{2.447} & & \textbf{2.873} & & \textbf{0.906} & & \textbf{2.52} & & \textbf{3.058} & \\ \hline
		\end{tabular}
	}
	
	\bigskip
	
	\small
	\resizebox{!}{\tableheight}{
		\begin{tabular}{|l|l|l|llllll|llllll|llllll|}
			\hline
			\multicolumn{1}{|c|}{\multirow{3}{*}{$k$}} & \multicolumn{1}{c|}{\multirow{3}{*}{$f^*$}} & \multicolumn{1}{c|}{\multirow{3}{*}{$\overline{f}$}} & \multicolumn{6}{c|}{HPClust-hybrid} & \multicolumn{6}{c|}{Forgy K-means} & \multicolumn{6}{c|}{PBK-BDC} \\ \cline{4-21}
			\multicolumn{1}{|c|}{} & \multicolumn{1}{c|}{} & \multicolumn{1}{c|}{} & \multicolumn{2}{c|}{$\varepsilon$} & \multicolumn{2}{c|}{$\overline{t}$} & \multicolumn{2}{c|}{$t$} & \multicolumn{2}{c|}{$\varepsilon$} & \multicolumn{2}{c|}{$\overline{t}$} & \multicolumn{2}{c|}{$t$} & \multicolumn{2}{c|}{$\varepsilon$} & \multicolumn{2}{c|}{$\overline{t}$} & \multicolumn{2}{c|}{$t$} \\ \cline{4-21}
			\multicolumn{1}{|c|}{} & \multicolumn{1}{c|}{} & \multicolumn{1}{c|}{} & \multicolumn{1}{c|}{med} & \multicolumn{1}{c|}{std} & \multicolumn{1}{c|}{med} & \multicolumn{1}{c|}{std} & \multicolumn{1}{c|}{med} & \multicolumn{1}{c|}{std} & \multicolumn{1}{c|}{med} & \multicolumn{1}{c|}{std} & \multicolumn{1}{c|}{med} & \multicolumn{1}{c|}{std} & \multicolumn{1}{c|}{med} & \multicolumn{1}{c|}{std} & \multicolumn{1}{c|}{med} & \multicolumn{1}{c|}{std} & \multicolumn{1}{c|}{med} & \multicolumn{1}{c|}{std} & \multicolumn{1}{c|}{med} & \multicolumn{1}{c|}{std} \\ \hline
			2 & 15.20433$^*$ & 4.88318 & 1.874 & 0.025 & 0.995 & 1.107 & 2.419 & 0.768 & 1.824 & 0.0 & -- & -- & 0.44 & 0.048 & 1.825 & 0.004 & -- & -- & 0.307 & 0.037 \\
			3 & 8.07129$^*$ & 2.89651 & 0.811 & 0.617 & 2.182 & 1.037 & 2.114 & 1.027 & 0.0 & 0.0 & -- & -- & 1.241 & 0.216 & 0.017 & 53.129 & -- & -- & 0.899 & 0.154 \\
			5 & 5.30537$^*$ & 1.86379 & 1.689 & 0.508 & 1.242 & 0.849 & 2.808 & 0.666 & 0.001 & 0.0 & -- & -- & 3.087 & 0.22 & 15.471 & 10.927 & -- & -- & 1.797 & 0.297 \\
			10 & 3.3767$^*$ & 1.26637 & 0.387 & 0.211 & 2.967 & 0.52 & 3.022 & 0.539 & 18.12 & 0.0 & -- & -- & 12.253 & 2.843 & 25.368 & 10.08 & -- & -- & 5.637 & 2.004 \\
			15 & 2.86473$^*$ & 1.08655 & 0.949 & 0.462 & 3.528 & 1.02 & 3.837 & 1.13 & 23.941 & 0.064 & -- & -- & 31.292 & 7.263 & 31.217 & 10.86 & -- & -- & 11.102 & 1.696 \\
			20 & 2.5732$^*$ & 0.98195 & 0.623 & 0.542 & 4.412 & 0.955 & 4.681 & 1.218 & 28.605 & 0.245 & -- & -- & 35.658 & 11.432 & 33.088 & 11.515 & -- & -- & 15.362 & 3.452 \\
			25 & 2.38539$^*$ & 0.90731 & 0.982 & 0.808 & 5.021 & 0.448 & 4.72 & 1.107 & 31.848 & 0.149 & -- & -- & 51.196 & 12.361 & 37.4 & 7.777 & -- & -- & 20.165 & 4.787 \\
			\hline
			\multicolumn{3}{|c|}{Mean:} & \textbf{1.045} & & \textbf{2.907} & & \textbf{3.371} & & \textbf{14.906} & & \textbf{--} & & \textbf{19.31} & & \textbf{20.627} & & \textbf{--} & & \textbf{7.896} & \\ \hline
		\end{tabular}
	}
	
	\bigskip
	
	\caption{Clustering details with Protein Homology}
	\label{TabDetailsD6}
	\resizebox{!}{\tableheight}{
		\begin{tabular}{|l|l|llll|llll|llll|lllll|l|l|}
			\hline
			\multicolumn{1}{|c|}{\multirow{2}{*}{$k$}} & \multicolumn{1}{c|}{\multirow{2}{*}{$n_{exec}$}} & \multicolumn{4}{c|}{HPClust-inner} & \multicolumn{4}{c|}{HPClust-competitive} & \multicolumn{4}{c|}{HPClust-cooperative} & \multicolumn{5}{c|}{HPClust-hybrid} & \multicolumn{1}{c|}{Forgy K-means} & \multicolumn{1}{c|}{PBK-BDC} \\ \cline{3-21}
			\multicolumn{1}{|c|}{} & \multicolumn{1}{c|}{} & \multicolumn{1}{c|}{$s$} & \multicolumn{1}{c|}{$n_{s}$} & \multicolumn{1}{c|}{$T$} & \multicolumn{1}{c|}{$n_{d}$} & \multicolumn{1}{c|}{$s$} & \multicolumn{1}{c|}{$n_{s}$} & \multicolumn{1}{c|}{$T$} & \multicolumn{1}{c|}{$n_{d}$} & \multicolumn{1}{c|}{$s$} & \multicolumn{1}{c|}{$n_{s}$} & \multicolumn{1}{c|}{$T$} & \multicolumn{1}{c|}{$n_{d}$} & \multicolumn{1}{c|}{$s$} & \multicolumn{1}{c|}{$n_{s}$} & \multicolumn{1}{c|}{$T_1$} & \multicolumn{1}{c|}{$T_2$} & \multicolumn{1}{c|}{$n_{d}$} & \multicolumn{1}{c|}{$n_{d}$} & \multicolumn{1}{c|}{$n_{d}$} \\
			\hline
			2 & 15 & 56000 & 79 & 3.5 & 4.1E+07 & 56000 & 259 & 3.5 & 2.1E+08 & 56000 & 341 & 3.5 & 2.2E+08 & 56000 & 428 & 3.267 & 0.233 & 2.2E+08 & 6.7E+06 & 5.6E+06 \\
			3 & 15 & 56000 & 70 & 3.5 & 6.8E+07 & 56000 & 244 & 3.5 & 2.6E+08 & 56000 & 272 & 3.5 & 2.7E+08 & 56000 & 259 & 2.567 & 0.933 & 2.6E+08 & 2.0E+07 & 1.6E+07 \\
			5 & 15 & 56000 & 45 & 3.5 & 9.1E+07 & 56000 & 103 & 3.5 & 3.0E+08 & 56000 & 57 & 3.5 & 3.0E+08 & 56000 & 186 & 0.933 & 2.567 & 2.8E+08 & 5.8E+07 & 3.9E+07 \\
			10 & 15 & 56000 & 27 & 3.5 & 1.4E+08 & 56000 & 38 & 3.5 & 3.3E+08 & 56000 & 28 & 3.5 & 3.1E+08 & 56000 & 13 & 0.233 & 3.267 & 2.0E+08 & 2.5E+08 & 1.7E+08 \\
			15 & 15 & 56000 & 11 & 3.5 & 1.8E+08 & 56000 & 14 & 3.5 & 3.6E+08 & 56000 & 10 & 3.5 & 3.4E+08 & 56000 & 6 & 0.233 & 3.267 & 3.1E+08 & 6.5E+08 & 3.6E+08 \\
			20 & 15 & 56000 & 15 & 3.5 & 1.9E+08 & 56000 & 5 & 3.5 & 3.8E+08 & 56000 & 6 & 3.5 & 4.0E+08 & 56000 & 6 & 1.167 & 2.333 & 3.7E+08 & 7.5E+08 & 4.9E+08 \\
			25 & 15 & 56000 & 10 & 3.5 & 2.0E+08 & 56000 & 3 & 3.5 & 4.5E+08 & 56000 & 6 & 3.5 & 4.5E+08 & 56000 & 4 & 3.15 & 0.35 & 4.3E+08 & 1.0E+09 & 7.0E+08 \\
			\hline
		\end{tabular}
	}
	
\end{table}

\newpage

%%%%%%%%%%%%%%%%%%%%%%%%%%%%%%%%%%%%%%%%%%%%%%%%%%%%%%%%%%%%%%%%%%%%%%%%%%%%%%%%%%%%%%%
%  END: Protein Homology
%%%%%%%%%%%%%%%%%%%%%%%%%%%%%%%%%%%%%%%%%%%%%%%%%%%%%%%%%%%%%%%%%%%%%%%%%%%%%%%%%%%%%%%

\newpage

%%%%%%%%%%%%%%%%%%%%%%%%%%%%%%%%%%%%%%%%%%%%%%%%%%%%%%%%%%%%%%%%%%%%%%%%%%%%%%%%%%%%%%%
%  START: MiniBooNE Particle Identification
%%%%%%%%%%%%%%%%%%%%%%%%%%%%%%%%%%%%%%%%%%%%%%%%%%%%%%%%%%%%%%%%%%%%%%%%%%%%%%%%%%%%%%%
\subsection{MiniBooNE Particle Identification}
Dimensions: $m$ = 130064, $n$ = 50.
\par
Description: a data set for distinguishing electron neutrinos (signal) from muon neutrinos (background) which contains different particle variables for each event.

\begin{table}[!htbp]
	\centering
	
	\caption{Summary of the results with MiniBooNE Particle Identification ($\times10^{10}$)}
	\label{TabResultsD7}
	\small
	\resizebox{!}{\tableheight}{
		\begin{tabular}{|l|l|l|llllll|llllll|llllll|}
			\hline
			\multicolumn{1}{|c|}{\multirow{3}{*}{$k$}} & \multicolumn{1}{c|}{\multirow{3}{*}{$f^*$}} & \multicolumn{1}{c|}{\multirow{3}{*}{$\overline{f}$}} & \multicolumn{6}{c|}{HPClust-inner} & \multicolumn{6}{c|}{HPClust-competitive} & \multicolumn{6}{c|}{HPClust-cooperative} \\ \cline{4-21}
			\multicolumn{1}{|c|}{} & \multicolumn{1}{c|}{} & \multicolumn{1}{c|}{} & \multicolumn{2}{c|}{$\varepsilon$} & \multicolumn{2}{c|}{$\overline{t}$} & \multicolumn{2}{c|}{$t$} & \multicolumn{2}{c|}{$\varepsilon$} & \multicolumn{2}{c|}{$\overline{t}$} & \multicolumn{2}{c|}{$t$} & \multicolumn{2}{c|}{$\varepsilon$} & \multicolumn{2}{c|}{$\overline{t}$} & \multicolumn{2}{c|}{$t$} \\ \cline{4-21}
			\multicolumn{1}{|c|}{} & \multicolumn{1}{c|}{} & \multicolumn{1}{c|}{} & \multicolumn{1}{c|}{med} & \multicolumn{1}{c|}{std} & \multicolumn{1}{c|}{med} & \multicolumn{1}{c|}{std} & \multicolumn{1}{c|}{med} & \multicolumn{1}{c|}{std} & \multicolumn{1}{c|}{med} & \multicolumn{1}{c|}{std} & \multicolumn{1}{c|}{med} & \multicolumn{1}{c|}{std} & \multicolumn{1}{c|}{med} & \multicolumn{1}{c|}{std} & \multicolumn{1}{c|}{med} & \multicolumn{1}{c|}{std} & \multicolumn{1}{c|}{med} & \multicolumn{1}{c|}{std} & \multicolumn{1}{c|}{med} & \multicolumn{1}{c|}{std} \\ \hline
			2 & 8.92236 & 8.90824 & 0.0 & 0.0 & 0.961 & 0.493 & 1.719 & 0.593 & 0.0 & 0.0 & 0.546 & 0.757 & 1.931 & 0.836 & 0.0 & 0.0 & 0.624 & 0.462 & 1.507 & 0.735 \\
			3 & 5.22601 & 5.2178 & 0.0 & 5.409 & 0.881 & 0.86 & 1.598 & 0.83 & 0.0 & 0.0 & 0.497 & 0.317 & 1.844 & 0.773 & 0.0 & 0.001 & 0.559 & 0.448 & 2.014 & 0.656 \\
			5 & 1.82252 & 1.82055 & 0.005 & 29.133 & 0.542 & 1.107 & 1.646 & 1.114 & 0.005 & 0.006 & 1.349 & 0.526 & 1.784 & 0.63 & 0.003 & 0.003 & 1.362 & 0.464 & 2.068 & 0.43 \\
			10 & 0.9092 & 0.90911 & 0.094 & 702427.909 & 1.583 & 0.613 & 2.219 & 0.65 & 0.033 & 0.043 & 2.63 & 0.329 & 3.05 & 0.945 & 0.051 & 0.031 & 2.705 & 0.412 & 3.065 & 0.461 \\
			15 & 0.63506 & 0.64964 & 2.395 & 1.575 & 1.162 & 0.483 & 2.284 & 0.567 & 0.173 & 0.667 & 3.973 & 0.442 & 4.099 & 0.67 & 0.141 & 0.391 & 3.87 & 0.793 & 4.345 & 1.781 \\
			20 & 0.50863 & 0.54514 & 1.034 & 3.267 & 1.003 & 0.285 & 2.073 & 0.444 & 0.085 & 1255623.008 & 4.681 & 0.536 & 6.863 & 1.949 & 0.214 & 0.291 & 4.608 & 0.827 & 6.948 & 1.716 \\
			25 & 0.44425 & 0.44476 & 0.026 & 2.267 & 1.208 & 0.698 & 2.961 & 0.565 & -0.303 & 0.133 & 7.802 & 1.262 & 8.719 & 1.595 & -0.391 & 0.113 & 6.073 & 0.774 & 7.816 & 2.224 \\
			\hline
			\multicolumn{3}{|c|}{Mean:} & \textbf{0.508} & & \textbf{1.048} & & \textbf{2.071} & & \textbf{-0.001} & & \textbf{3.068} & & \textbf{4.041} & & \textbf{0.003} & & \textbf{2.829} & & \textbf{3.966} & \\ \hline
		\end{tabular}
	}
	
	\bigskip
	
	\small
	\resizebox{!}{\tableheight}{
		\begin{tabular}{|l|l|l|llllll|llllll|llllll|}
			\hline
			\multicolumn{1}{|c|}{\multirow{3}{*}{$k$}} & \multicolumn{1}{c|}{\multirow{3}{*}{$f^*$}} & \multicolumn{1}{c|}{\multirow{3}{*}{$\overline{f}$}} & \multicolumn{6}{c|}{HPClust-hybrid} & \multicolumn{6}{c|}{Forgy K-means} & \multicolumn{6}{c|}{PBK-BDC} \\ \cline{4-21}
			\multicolumn{1}{|c|}{} & \multicolumn{1}{c|}{} & \multicolumn{1}{c|}{} & \multicolumn{2}{c|}{$\varepsilon$} & \multicolumn{2}{c|}{$\overline{t}$} & \multicolumn{2}{c|}{$t$} & \multicolumn{2}{c|}{$\varepsilon$} & \multicolumn{2}{c|}{$\overline{t}$} & \multicolumn{2}{c|}{$t$} & \multicolumn{2}{c|}{$\varepsilon$} & \multicolumn{2}{c|}{$\overline{t}$} & \multicolumn{2}{c|}{$t$} \\ \cline{4-21}
			\multicolumn{1}{|c|}{} & \multicolumn{1}{c|}{} & \multicolumn{1}{c|}{} & \multicolumn{1}{c|}{med} & \multicolumn{1}{c|}{std} & \multicolumn{1}{c|}{med} & \multicolumn{1}{c|}{std} & \multicolumn{1}{c|}{med} & \multicolumn{1}{c|}{std} & \multicolumn{1}{c|}{med} & \multicolumn{1}{c|}{std} & \multicolumn{1}{c|}{med} & \multicolumn{1}{c|}{std} & \multicolumn{1}{c|}{med} & \multicolumn{1}{c|}{std} & \multicolumn{1}{c|}{med} & \multicolumn{1}{c|}{std} & \multicolumn{1}{c|}{med} & \multicolumn{1}{c|}{std} & \multicolumn{1}{c|}{med} & \multicolumn{1}{c|}{std} \\ \hline
			2 & 8.92236 & 8.90824 & 0.0 & 0.0 & 0.453 & 0.468 & 1.686 & 0.803 & 0.0 & 140555.682 & -- & -- & 0.135 & 0.109 & 286908.084 & 140555.682 & -- & -- & 0.197 & 0.069 \\
			3 & 5.22601 & 5.2178 & 0.0 & 0.0 & 0.817 & 0.383 & 1.932 & 0.66 & 0.0 & 166530.775 & -- & -- & 0.416 & 0.236 & 0.0 & 122199.794 & -- & -- & 0.343 & 0.125 \\
			5 & 1.82252 & 1.82055 & 0.008 & 0.005 & 2.417 & 0.64 & 2.724 & 0.639 & 116.777 & 55.608 & -- & -- & 1.563 & 0.473 & 116.777 & 57.841 & -- & -- & 1.636 & 0.439 \\
			10 & 0.9092 & 0.90911 & 0.03 & 0.053 & 2.787 & 1.204 & 4.616 & 0.816 & 0.002 & 0.0 & -- & -- & 13.383 & 3.122 & 0.002 & 0.001 & -- & -- & 12.268 & 2.158 \\
			15 & 0.63506 & 0.64964 & 0.123 & 0.048 & 3.514 & 0.685 & 5.845 & 1.681 & 3.883 & 0.764 & -- & -- & 17.098 & 4.58 & 3.883 & 0.777 & -- & -- & 15.296 & 4.864 \\
			20 & 0.50863 & 0.54514 & 0.091 & 0.169 & 5.003 & 0.552 & 5.87 & 1.558 & 7.051 & 0.439 & -- & -- & 24.047 & 4.968 & 7.052 & 0.603 & -- & -- & 24.811 & 6.02 \\
			25 & 0.44425 & 0.44476 & -0.267 & 0.182 & 6.377 & 0.932 & 7.623 & 1.661 & 8.936 & 0.202 & -- & -- & 29.885 & 4.702 & 8.936 & 0.38 & -- & -- & 28.882 & 7.385 \\
			\hline
			\multicolumn{3}{|c|}{Mean:} & \textbf{-0.002} & & \textbf{3.053} & & \textbf{4.328} & & \textbf{19.521} & & \textbf{--} & & \textbf{12.361} & & \textbf{41006.39} & & \textbf{--} & & \textbf{11.919} & \\ \hline
		\end{tabular}
	}
	
	\bigskip
	
	\caption{Clustering details with MiniBooNE Particle Identification}
	\label{TabDetailsD7}
	\resizebox{!}{\tableheight}{
		\begin{tabular}{|l|l|llll|llll|llll|lllll|l|l|}
			\hline
			\multicolumn{1}{|c|}{\multirow{2}{*}{$k$}} & \multicolumn{1}{c|}{\multirow{2}{*}{$n_{exec}$}} & \multicolumn{4}{c|}{HPClust-inner} & \multicolumn{4}{c|}{HPClust-competitive} & \multicolumn{4}{c|}{HPClust-cooperative} & \multicolumn{5}{c|}{HPClust-hybrid} & \multicolumn{1}{c|}{Forgy K-means} & \multicolumn{1}{c|}{PBK-BDC} \\ \cline{3-21}
			\multicolumn{1}{|c|}{} & \multicolumn{1}{c|}{} & \multicolumn{1}{c|}{$s$} & \multicolumn{1}{c|}{$n_{s}$} & \multicolumn{1}{c|}{$T$} & \multicolumn{1}{c|}{$n_{d}$} & \multicolumn{1}{c|}{$s$} & \multicolumn{1}{c|}{$n_{s}$} & \multicolumn{1}{c|}{$T$} & \multicolumn{1}{c|}{$n_{d}$} & \multicolumn{1}{c|}{$s$} & \multicolumn{1}{c|}{$n_{s}$} & \multicolumn{1}{c|}{$T$} & \multicolumn{1}{c|}{$n_{d}$} & \multicolumn{1}{c|}{$s$} & \multicolumn{1}{c|}{$n_{s}$} & \multicolumn{1}{c|}{$T_1$} & \multicolumn{1}{c|}{$T_2$} & \multicolumn{1}{c|}{$n_{d}$} & \multicolumn{1}{c|}{$n_{d}$} & \multicolumn{1}{c|}{$n_{d}$} \\
			\hline
			2 & 15 & 130000 & 59 & 3.0 & 5.5E+07 & 130000 & 201 & 3.0 & 2.1E+08 & 130000 & 182 & 3.0 & 2.1E+08 & 130000 & 203 & 2.5 & 0.5 & 1.9E+08 & 2.3E+06 & 4.7E+06 \\
			3 & 15 & 130000 & 42 & 3.0 & 7.2E+07 & 130000 & 161 & 3.0 & 2.5E+08 & 130000 & 173 & 3.0 & 2.6E+08 & 130000 & 159 & 2.5 & 0.5 & 2.5E+08 & 8.6E+06 & 8.2E+06 \\
			5 & 15 & 130000 & 39 & 3.0 & 1.1E+08 & 130000 & 74 & 3.0 & 3.1E+08 & 130000 & 98 & 3.0 & 3.1E+08 & 130000 & 74 & 0.2 & 2.8 & 2.2E+08 & 4.5E+07 & 4.8E+07 \\
			10 & 15 & 130000 & 29 & 3.0 & 1.7E+08 & 130000 & 8 & 3.0 & 3.8E+08 & 130000 & 12 & 3.0 & 3.7E+08 & 130000 & 11 & 2.0 & 1.0 & 3.4E+08 & 4.1E+08 & 3.8E+08 \\
			15 & 15 & 130000 & 21 & 3.0 & 2.1E+08 & 130000 & 4 & 3.0 & 5.8E+08 & 130000 & 4 & 3.0 & 5.9E+08 & 130000 & 6 & 0.5 & 2.5 & 6.2E+08 & 5.4E+08 & 4.9E+08 \\
			20 & 15 & 130000 & 14 & 3.0 & 2.3E+08 & 130000 & 5 & 3.0 & 8.2E+08 & 130000 & 5 & 3.0 & 8.5E+08 & 130000 & 4 & 2.9 & 0.1 & 8.5E+08 & 7.6E+08 & 7.8E+08 \\
			25 & 15 & 130000 & 16 & 3.0 & 2.5E+08 & 130000 & 5 & 3.0 & 1.1E+09 & 130000 & 5 & 3.0 & 9.9E+08 & 130000 & 3 & 1.7 & 1.3 & 1.1E+09 & 9.5E+08 & 9.3E+08 \\
			\hline
		\end{tabular}
	}
	
\end{table}

\newpage

%%%%%%%%%%%%%%%%%%%%%%%%%%%%%%%%%%%%%%%%%%%%%%%%%%%%%%%%%%%%%%%%%%%%%%%%%%%%%%%%%%%%%%%
%  END: MiniBooNE Particle Identification
%%%%%%%%%%%%%%%%%%%%%%%%%%%%%%%%%%%%%%%%%%%%%%%%%%%%%%%%%%%%%%%%%%%%%%%%%%%%%%%%%%%%%%%

%%%%%%%%%%%%%%%%%%%%%%%%%%%%%%%%%%%%%%%%%%%%%%%%%%%%%%%%%%%%%%%%%%%%%%%%%%%%%%%%%%%%%%%
%  START: MiniBooNE Particle Identification (normalized)
%%%%%%%%%%%%%%%%%%%%%%%%%%%%%%%%%%%%%%%%%%%%%%%%%%%%%%%%%%%%%%%%%%%%%%%%%%%%%%%%%%%%%%%
\subsection{MiniBooNE Particle Identification (normalized)}
Dimensions: $m$ = 130064, $n$ = 50.
\par
Description: a data set for distinguishing electron neutrinos (signal) from muon neutrinos (background) which contains different particle variables for each event. Min-max scaling was used for normalization of data set values for better clusterization.

\begin{table}[!htbp]
	\centering
	
	\caption{Summary of the results with MiniBooNE Particle Identification (normalized) ($\times10^{2}$)}
	\label{TabResultsD8}
	\small
	\resizebox{!}{\tableheight}{
		\begin{tabular}{|l|l|l|llllll|llllll|llllll|}
			\hline
			\multicolumn{1}{|c|}{\multirow{3}{*}{$k$}} & \multicolumn{1}{c|}{\multirow{3}{*}{$f^*$}} & \multicolumn{1}{c|}{\multirow{3}{*}{$\overline{f}$}} & \multicolumn{6}{c|}{HPClust-inner} & \multicolumn{6}{c|}{HPClust-competitive} & \multicolumn{6}{c|}{HPClust-cooperative} \\ \cline{4-21}
			\multicolumn{1}{|c|}{} & \multicolumn{1}{c|}{} & \multicolumn{1}{c|}{} & \multicolumn{2}{c|}{$\varepsilon$} & \multicolumn{2}{c|}{$\overline{t}$} & \multicolumn{2}{c|}{$t$} & \multicolumn{2}{c|}{$\varepsilon$} & \multicolumn{2}{c|}{$\overline{t}$} & \multicolumn{2}{c|}{$t$} & \multicolumn{2}{c|}{$\varepsilon$} & \multicolumn{2}{c|}{$\overline{t}$} & \multicolumn{2}{c|}{$t$} \\ \cline{4-21}
			\multicolumn{1}{|c|}{} & \multicolumn{1}{c|}{} & \multicolumn{1}{c|}{} & \multicolumn{1}{c|}{med} & \multicolumn{1}{c|}{std} & \multicolumn{1}{c|}{med} & \multicolumn{1}{c|}{std} & \multicolumn{1}{c|}{med} & \multicolumn{1}{c|}{std} & \multicolumn{1}{c|}{med} & \multicolumn{1}{c|}{std} & \multicolumn{1}{c|}{med} & \multicolumn{1}{c|}{std} & \multicolumn{1}{c|}{med} & \multicolumn{1}{c|}{std} & \multicolumn{1}{c|}{med} & \multicolumn{1}{c|}{std} & \multicolumn{1}{c|}{med} & \multicolumn{1}{c|}{std} & \multicolumn{1}{c|}{med} & \multicolumn{1}{c|}{std} \\ \hline
			2 & 28.01938$^*$ & 2.49407 & 0.014 & 150.663 & 0.37 & 0.279 & 0.417 & 0.303 & 0.019 & 0.01 & 0.142 & 0.209 & 0.546 & 0.302 & 0.027 & 0.009 & 0.219 & 0.156 & 0.625 & 0.321 \\
			3 & 19.85673$^*$ & 1.75033 & 0.031 & 3.034 & 0.352 & 0.262 & 0.492 & 0.231 & 0.026 & 0.014 & 0.15 & 0.168 & 0.534 & 0.293 & 0.031 & 0.014 & 0.152 & 0.135 & 0.52 & 0.303 \\
			5 & 12.10267$^*$ & 1.11597 & 0.12 & 1.745 & 0.023 & 0.013 & 0.647 & 0.301 & 0.089 & 0.028 & 0.066 & 0.016 & 0.604 & 0.299 & 0.087 & 0.043 & 0.064 & 0.027 & 0.459 & 0.249 \\
			10 & 8.57382$^*$ & 0.76679 & 0.668 & 0.528 & 0.612 & 0.378 & 0.479 & 0.322 & 0.471 & 0.33 & 0.444 & 0.196 & 0.627 & 0.263 & 0.647 & 0.564 & 0.692 & 0.202 & 0.837 & 0.209 \\
			15 & 7.24131$^*$ & 0.64941 & 0.619 & 0.26 & 0.27 & 0.294 & 0.467 & 0.26 & 0.75 & 0.287 & 0.55 & 0.184 & 0.76 & 0.222 & 0.772 & 0.445 & 0.66 & 0.197 & 0.763 & 0.221 \\
			20 & 6.30493$^*$ & 0.56979 & 1.164 & 0.703 & 0.463 & 0.253 & 0.586 & 0.298 & 1.282 & 0.747 & 1.0 & 0.273 & 1.045 & 0.208 & 0.963 & 0.577 & 0.848 & 0.209 & 0.951 & 0.222 \\
			25 & 5.71335$^*$ & 0.51724 & 1.147 & 0.47 & 0.693 & 0.269 & 0.738 & 0.253 & 1.209 & 0.447 & 1.108 & 0.231 & 1.171 & 0.232 & 1.363 & 0.605 & 0.915 & 0.219 & 1.157 & 0.303 \\
			\hline
			\multicolumn{3}{|c|}{Mean:} & \textbf{0.538} & & \textbf{0.398} & & \textbf{0.546} & & \textbf{0.549} & & \textbf{0.494} & & \textbf{0.755} & & \textbf{0.556} & & \textbf{0.507} & & \textbf{0.759} & \\ \hline
		\end{tabular}
	}
	
	\bigskip
	
	\small
	\resizebox{!}{\tableheight}{
		\begin{tabular}{|l|l|l|llllll|llllll|llllll|}
			\hline
			\multicolumn{1}{|c|}{\multirow{3}{*}{$k$}} & \multicolumn{1}{c|}{\multirow{3}{*}{$f^*$}} & \multicolumn{1}{c|}{\multirow{3}{*}{$\overline{f}$}} & \multicolumn{6}{c|}{HPClust-hybrid} & \multicolumn{6}{c|}{Forgy K-means} & \multicolumn{6}{c|}{PBK-BDC} \\ \cline{4-21}
			\multicolumn{1}{|c|}{} & \multicolumn{1}{c|}{} & \multicolumn{1}{c|}{} & \multicolumn{2}{c|}{$\varepsilon$} & \multicolumn{2}{c|}{$\overline{t}$} & \multicolumn{2}{c|}{$t$} & \multicolumn{2}{c|}{$\varepsilon$} & \multicolumn{2}{c|}{$\overline{t}$} & \multicolumn{2}{c|}{$t$} & \multicolumn{2}{c|}{$\varepsilon$} & \multicolumn{2}{c|}{$\overline{t}$} & \multicolumn{2}{c|}{$t$} \\ \cline{4-21}
			\multicolumn{1}{|c|}{} & \multicolumn{1}{c|}{} & \multicolumn{1}{c|}{} & \multicolumn{1}{c|}{med} & \multicolumn{1}{c|}{std} & \multicolumn{1}{c|}{med} & \multicolumn{1}{c|}{std} & \multicolumn{1}{c|}{med} & \multicolumn{1}{c|}{std} & \multicolumn{1}{c|}{med} & \multicolumn{1}{c|}{std} & \multicolumn{1}{c|}{med} & \multicolumn{1}{c|}{std} & \multicolumn{1}{c|}{med} & \multicolumn{1}{c|}{std} & \multicolumn{1}{c|}{med} & \multicolumn{1}{c|}{std} & \multicolumn{1}{c|}{med} & \multicolumn{1}{c|}{std} & \multicolumn{1}{c|}{med} & \multicolumn{1}{c|}{std} \\ \hline
			2 & 28.01938$^*$ & 2.49407 & 0.02 & 0.009 & 0.239 & 0.194 & 0.582 & 0.231 & 0.0 & 336.097 & -- & -- & 0.245 & 0.099 & 2.67 & 1.876 & -- & -- & 0.064 & 0.019 \\
			3 & 19.85673$^*$ & 1.75033 & 0.03 & 0.014 & 0.222 & 0.185 & 0.473 & 0.3 & 6.987 & 389.115 & -- & -- & 0.391 & 0.08 & 9.686 & 10.446 & -- & -- & 0.084 & 0.033 \\
			5 & 12.10267$^*$ & 1.11597 & 0.084 & 0.039 & 0.064 & 0.028 & 0.727 & 0.236 & -0.002 & 1.531 & -- & -- & 0.824 & 0.251 & 12.653 & 27.352 & -- & -- & 0.234 & 0.062 \\
			10 & 8.57382$^*$ & 0.76679 & 0.648 & 0.403 & 0.588 & 0.237 & 0.689 & 0.208 & 1.487 & 1.117 & -- & -- & 4.124 & 1.152 & 7.647 & 5.029 & -- & -- & 1.027 & 0.229 \\
			15 & 7.24131$^*$ & 0.64941 & 0.499 & 0.313 & 0.594 & 0.301 & 0.85 & 0.229 & 0.33 & 1.282 & -- & -- & 10.031 & 4.537 & 7.806 & 7.356 & -- & -- & 1.735 & 0.332 \\
			20 & 6.30493$^*$ & 0.56979 & 1.232 & 0.793 & 0.994 & 0.298 & 1.075 & 0.354 & 0.803 & 0.492 & -- & -- & 14.275 & 4.417 & 7.111 & 3.61 & -- & -- & 2.324 & 0.485 \\
			25 & 5.71335$^*$ & 0.51724 & 1.068 & 0.299 & 0.989 & 0.331 & 1.144 & 0.532 & 0.118 & 0.256 & -- & -- & 20.063 & 10.132 & 6.67 & 3.059 & -- & -- & 2.98 & 0.44 \\
			\hline
			\multicolumn{3}{|c|}{Mean:} & \textbf{0.512} & & \textbf{0.527} & & \textbf{0.791} & & \textbf{1.389} & & \textbf{--} & & \textbf{7.136} & & \textbf{7.749} & & \textbf{--} & & \textbf{1.207} & \\ \hline
		\end{tabular}
	}
	
	\bigskip
	
	\caption{Clustering details with MiniBooNE Particle Identification (normalized)}
	\label{TabDetailsD8}
	\resizebox{!}{\tableheight}{
		\begin{tabular}{|l|l|llll|llll|llll|lllll|l|l|}
			\hline
			\multicolumn{1}{|c|}{\multirow{2}{*}{$k$}} & \multicolumn{1}{c|}{\multirow{2}{*}{$n_{exec}$}} & \multicolumn{4}{c|}{HPClust-inner} & \multicolumn{4}{c|}{HPClust-competitive} & \multicolumn{4}{c|}{HPClust-cooperative} & \multicolumn{5}{c|}{HPClust-hybrid} & \multicolumn{1}{c|}{Forgy K-means} & \multicolumn{1}{c|}{PBK-BDC} \\ \cline{3-21}
			\multicolumn{1}{|c|}{} & \multicolumn{1}{c|}{} & \multicolumn{1}{c|}{$s$} & \multicolumn{1}{c|}{$n_{s}$} & \multicolumn{1}{c|}{$T$} & \multicolumn{1}{c|}{$n_{d}$} & \multicolumn{1}{c|}{$s$} & \multicolumn{1}{c|}{$n_{s}$} & \multicolumn{1}{c|}{$T$} & \multicolumn{1}{c|}{$n_{d}$} & \multicolumn{1}{c|}{$s$} & \multicolumn{1}{c|}{$n_{s}$} & \multicolumn{1}{c|}{$T$} & \multicolumn{1}{c|}{$n_{d}$} & \multicolumn{1}{c|}{$s$} & \multicolumn{1}{c|}{$n_{s}$} & \multicolumn{1}{c|}{$T_1$} & \multicolumn{1}{c|}{$T_2$} & \multicolumn{1}{c|}{$n_{d}$} & \multicolumn{1}{c|}{$n_{d}$} & \multicolumn{1}{c|}{$n_{d}$} \\
			\hline
			2 & 20 & 12000 & 54 & 1.0 & 7.5E+06 & 12000 & 437 & 1.0 & 4.3E+07 & 12000 & 594 & 1.0 & 4.5E+07 & 12000 & 488 & 0.033 & 0.967 & 4.3E+07 & 4.4E+06 & 4.1E+06 \\
			3 & 20 & 12000 & 56 & 1.0 & 1.6E+07 & 12000 & 282 & 1.0 & 6.7E+07 & 12000 & 291 & 1.0 & 7.0E+07 & 12000 & 260 & 0.033 & 0.967 & 6.9E+07 & 9.9E+06 & 7.2E+06 \\
			5 & 20 & 12000 & 56 & 1.0 & 2.3E+07 & 12000 & 194 & 1.0 & 9.0E+07 & 12000 & 164 & 1.0 & 9.1E+07 & 12000 & 264 & 0.167 & 0.833 & 8.9E+07 & 2.3E+07 & 1.8E+07 \\
			10 & 20 & 12000 & 26 & 1.0 & 5.5E+07 & 12000 & 50 & 1.0 & 1.2E+08 & 12000 & 72 & 1.0 & 1.3E+08 & 12000 & 50 & 0.667 & 0.333 & 1.1E+08 & 1.2E+08 & 8.6E+07 \\
			15 & 20 & 12000 & 14 & 1.0 & 7.5E+07 & 12000 & 20 & 1.0 & 1.3E+08 & 12000 & 24 & 1.0 & 1.3E+08 & 12000 & 23 & 0.867 & 0.133 & 1.2E+08 & 3.1E+08 & 1.5E+08 \\
			20 & 20 & 12000 & 12 & 1.0 & 8.3E+07 & 12000 & 14 & 1.0 & 1.3E+08 & 12000 & 14 & 1.0 & 1.3E+08 & 12000 & 10 & 0.233 & 0.767 & 1.0E+08 & 4.2E+08 & 2.3E+08 \\
			25 & 20 & 12000 & 11 & 1.0 & 8.8E+07 & 12000 & 10 & 1.0 & 1.4E+08 & 12000 & 12 & 1.0 & 1.4E+08 & 12000 & 10 & 0.733 & 0.267 & 1.4E+08 & 6.1E+08 & 3.0E+08 \\
			\hline
		\end{tabular}
	}
	
\end{table}

\newpage

%%%%%%%%%%%%%%%%%%%%%%%%%%%%%%%%%%%%%%%%%%%%%%%%%%%%%%%%%%%%%%%%%%%%%%%%%%%%%%%%%%%%%%%
%  END: MiniBooNE Particle Identification (normalized)
%%%%%%%%%%%%%%%%%%%%%%%%%%%%%%%%%%%%%%%%%%%%%%%%%%%%%%%%%%%%%%%%%%%%%%%%%%%%%%%%%%%%%%%

\newpage

%%%%%%%%%%%%%%%%%%%%%%%%%%%%%%%%%%%%%%%%%%%%%%%%%%%%%%%%%%%%%%%%%%%%%%%%%%%%%%%%%%%%%%%
%  START: MFCCs for Speech Emotion Recognition
%%%%%%%%%%%%%%%%%%%%%%%%%%%%%%%%%%%%%%%%%%%%%%%%%%%%%%%%%%%%%%%%%%%%%%%%%%%%%%%%%%%%%%%
\subsection{MFCCs for Speech Emotion Recognition}
Dimensions: $m$ = 85134, $n$ = 58.
\par
Description: a data set for predicting females and males speech emotions based on Mel Frequency Cepstral Coefficients (MFCCs) values.

\begin{table}[!htbp]
	\centering
	
	\caption{Summary of the results with MFCCs for Speech Emotion Recognition ($\times10^{9}$)}
	\label{TabResultsD9}
	\small
	\resizebox{!}{\tableheight}{
		\begin{tabular}{|l|l|l|llllll|llllll|llllll|}
			\hline
			\multicolumn{1}{|c|}{\multirow{3}{*}{$k$}} & \multicolumn{1}{c|}{\multirow{3}{*}{$f^*$}} & \multicolumn{1}{c|}{\multirow{3}{*}{$\overline{f}$}} & \multicolumn{6}{c|}{HPClust-inner} & \multicolumn{6}{c|}{HPClust-competitive} & \multicolumn{6}{c|}{HPClust-cooperative} \\ \cline{4-21}
			\multicolumn{1}{|c|}{} & \multicolumn{1}{c|}{} & \multicolumn{1}{c|}{} & \multicolumn{2}{c|}{$\varepsilon$} & \multicolumn{2}{c|}{$\overline{t}$} & \multicolumn{2}{c|}{$t$} & \multicolumn{2}{c|}{$\varepsilon$} & \multicolumn{2}{c|}{$\overline{t}$} & \multicolumn{2}{c|}{$t$} & \multicolumn{2}{c|}{$\varepsilon$} & \multicolumn{2}{c|}{$\overline{t}$} & \multicolumn{2}{c|}{$t$} \\ \cline{4-21}
			\multicolumn{1}{|c|}{} & \multicolumn{1}{c|}{} & \multicolumn{1}{c|}{} & \multicolumn{1}{c|}{med} & \multicolumn{1}{c|}{std} & \multicolumn{1}{c|}{med} & \multicolumn{1}{c|}{std} & \multicolumn{1}{c|}{med} & \multicolumn{1}{c|}{std} & \multicolumn{1}{c|}{med} & \multicolumn{1}{c|}{std} & \multicolumn{1}{c|}{med} & \multicolumn{1}{c|}{std} & \multicolumn{1}{c|}{med} & \multicolumn{1}{c|}{std} & \multicolumn{1}{c|}{med} & \multicolumn{1}{c|}{std} & \multicolumn{1}{c|}{med} & \multicolumn{1}{c|}{std} & \multicolumn{1}{c|}{med} & \multicolumn{1}{c|}{std} \\ \hline
			2 & 0.74513$^*$ & 0.10188 & 0.029 & 0.015 & 0.55 & 0.262 & 0.623 & 0.289 & 0.04 & 0.016 & 0.444 & 0.279 & 0.609 & 0.273 & 0.04 & 0.023 & 0.302 & 0.222 & 0.561 & 0.214 \\
			3 & 0.50215$^*$ & 0.06923 & 0.037 & 0.022 & 0.352 & 0.308 & 0.366 & 0.297 & 0.043 & 0.027 & 0.186 & 0.133 & 0.458 & 0.3 & 0.051 & 0.029 & 0.25 & 0.162 & 0.555 & 0.184 \\
			5 & 0.3456$^*$ & 0.04777 & 0.059 & 0.03 & 0.499 & 0.253 & 0.499 & 0.291 & 0.063 & 0.043 & 0.32 & 0.213 & 0.579 & 0.256 & 0.057 & 0.022 & 0.281 & 0.176 & 0.596 & 0.255 \\
			10 & 0.21763$^*$ & 0.03009 & 1.209 & 1.243 & 0.366 & 0.133 & 0.363 & 0.252 & 0.11 & 0.033 & 0.57 & 0.22 & 0.662 & 0.23 & 0.129 & 0.046 & 0.51 & 0.165 & 0.644 & 0.179 \\
			15 & 0.17608$^*$ & 0.02458 & 1.2 & 0.733 & 0.301 & 0.19 & 0.564 & 0.205 & 0.237 & 0.37 & 0.49 & 0.158 & 0.746 & 0.21 & 0.519 & 0.534 & 0.464 & 0.207 & 0.812 & 0.218 \\
			20 & 0.15383$^*$ & 0.0214 & 0.8 & 1.017 & 0.315 & 0.185 & 0.706 & 0.29 & 0.789 & 0.397 & 0.863 & 0.151 & 0.903 & 0.247 & 0.644 & 0.364 & 0.755 & 0.186 & 0.982 & 0.217 \\
			25 & 0.14109$^*$ & 0.01968 & 1.142 & 0.742 & 0.351 & 0.218 & 0.526 & 0.247 & 1.09 & 0.41 & 1.104 & 0.165 & 0.974 & 0.198 & 0.97 & 0.443 & 0.893 & 0.368 & 1.079 & 0.303 \\
			\hline
			\multicolumn{3}{|c|}{Mean:} & \textbf{0.639} & & \textbf{0.39} & & \textbf{0.521} & & \textbf{0.339} & & \textbf{0.568} & & \textbf{0.704} & & \textbf{0.344} & & \textbf{0.494} & & \textbf{0.747} & \\ \hline
		\end{tabular}
	}
	
	\bigskip
	
	\small
	\resizebox{!}{\tableheight}{
		\begin{tabular}{|l|l|l|llllll|llllll|llllll|}
			\hline
			\multicolumn{1}{|c|}{\multirow{3}{*}{$k$}} & \multicolumn{1}{c|}{\multirow{3}{*}{$f^*$}} & \multicolumn{1}{c|}{\multirow{3}{*}{$\overline{f}$}} & \multicolumn{6}{c|}{HPClust-hybrid} & \multicolumn{6}{c|}{Forgy K-means} & \multicolumn{6}{c|}{PBK-BDC} \\ \cline{4-21}
			\multicolumn{1}{|c|}{} & \multicolumn{1}{c|}{} & \multicolumn{1}{c|}{} & \multicolumn{2}{c|}{$\varepsilon$} & \multicolumn{2}{c|}{$\overline{t}$} & \multicolumn{2}{c|}{$t$} & \multicolumn{2}{c|}{$\varepsilon$} & \multicolumn{2}{c|}{$\overline{t}$} & \multicolumn{2}{c|}{$t$} & \multicolumn{2}{c|}{$\varepsilon$} & \multicolumn{2}{c|}{$\overline{t}$} & \multicolumn{2}{c|}{$t$} \\ \cline{4-21}
			\multicolumn{1}{|c|}{} & \multicolumn{1}{c|}{} & \multicolumn{1}{c|}{} & \multicolumn{1}{c|}{med} & \multicolumn{1}{c|}{std} & \multicolumn{1}{c|}{med} & \multicolumn{1}{c|}{std} & \multicolumn{1}{c|}{med} & \multicolumn{1}{c|}{std} & \multicolumn{1}{c|}{med} & \multicolumn{1}{c|}{std} & \multicolumn{1}{c|}{med} & \multicolumn{1}{c|}{std} & \multicolumn{1}{c|}{med} & \multicolumn{1}{c|}{std} & \multicolumn{1}{c|}{med} & \multicolumn{1}{c|}{std} & \multicolumn{1}{c|}{med} & \multicolumn{1}{c|}{std} & \multicolumn{1}{c|}{med} & \multicolumn{1}{c|}{std} \\ \hline
			2 & 0.74513$^*$ & 0.10188 & 0.039 & 0.02 & 0.275 & 0.278 & 0.345 & 0.294 & 0.001 & 0.0 & -- & -- & 0.174 & 0.065 & 0.001 & 0.0 & -- & -- & 0.049 & 0.007 \\
			3 & 0.50215$^*$ & 0.06923 & 0.044 & 0.032 & 0.242 & 0.25 & 0.643 & 0.261 & 0.001 & 0.0 & -- & -- & 0.236 & 0.058 & 0.001 & 34.911 & -- & -- & 0.069 & 0.008 \\
			5 & 0.3456$^*$ & 0.04777 & 0.057 & 0.028 & 0.331 & 0.22 & 0.538 & 0.242 & -0.002 & 0.0 & -- & -- & 0.774 & 0.129 & 25.789 & 19.059 & -- & -- & 0.2 & 0.03 \\
			10 & 0.21763$^*$ & 0.03009 & 0.102 & 0.024 & 0.724 & 0.222 & 0.78 & 0.209 & 3.278 & 1.286 & -- & -- & 2.347 & 0.54 & 11.788 & 8.119 & -- & -- & 0.693 & 0.145 \\
			15 & 0.17608$^*$ & 0.02458 & 0.251 & 0.263 & 0.644 & 0.235 & 0.855 & 0.228 & 1.7 & 1.843 & -- & -- & 5.663 & 2.193 & 12.054 & 8.998 & -- & -- & 1.202 & 0.215 \\
			20 & 0.15383$^*$ & 0.0214 & 0.791 & 0.603 & 0.832 & 0.25 & 0.91 & 0.344 & 2.096 & 1.593 & -- & -- & 10.052 & 2.014 & 11.16 & 4.233 & -- & -- & 2.035 & 0.301 \\
			25 & 0.14109$^*$ & 0.01968 & 1.017 & 0.327 & 0.801 & 0.166 & 0.989 & 0.366 & 3.385 & 1.68 & -- & -- & 15.119 & 5.08 & 9.933 & 5.571 & -- & -- & 2.359 & 0.407 \\
			\hline
			\multicolumn{3}{|c|}{Mean:} & \textbf{0.329} & & \textbf{0.55} & & \textbf{0.723} & & \textbf{1.494} & & \textbf{--} & & \textbf{4.909} & & \textbf{10.104} & & \textbf{--} & & \textbf{0.944} & \\ \hline
		\end{tabular}
	}
	
	\bigskip
	
	\caption{Clustering details with MFCCs for Speech Emotion Recognition}
	\label{TabDetailsD9}
	\resizebox{!}{\tableheight}{
		\begin{tabular}{|l|l|llll|llll|llll|lllll|l|l|}
			\hline
			\multicolumn{1}{|c|}{\multirow{2}{*}{$k$}} & \multicolumn{1}{c|}{\multirow{2}{*}{$n_{exec}$}} & \multicolumn{4}{c|}{HPClust-inner} & \multicolumn{4}{c|}{HPClust-competitive} & \multicolumn{4}{c|}{HPClust-cooperative} & \multicolumn{5}{c|}{HPClust-hybrid} & \multicolumn{1}{c|}{Forgy K-means} & \multicolumn{1}{c|}{PBK-BDC} \\ \cline{3-21}
			\multicolumn{1}{|c|}{} & \multicolumn{1}{c|}{} & \multicolumn{1}{c|}{$s$} & \multicolumn{1}{c|}{$n_{s}$} & \multicolumn{1}{c|}{$T$} & \multicolumn{1}{c|}{$n_{d}$} & \multicolumn{1}{c|}{$s$} & \multicolumn{1}{c|}{$n_{s}$} & \multicolumn{1}{c|}{$T$} & \multicolumn{1}{c|}{$n_{d}$} & \multicolumn{1}{c|}{$s$} & \multicolumn{1}{c|}{$n_{s}$} & \multicolumn{1}{c|}{$T$} & \multicolumn{1}{c|}{$n_{d}$} & \multicolumn{1}{c|}{$s$} & \multicolumn{1}{c|}{$n_{s}$} & \multicolumn{1}{c|}{$T_1$} & \multicolumn{1}{c|}{$T_2$} & \multicolumn{1}{c|}{$n_{d}$} & \multicolumn{1}{c|}{$n_{d}$} & \multicolumn{1}{c|}{$n_{d}$} \\
			\hline
			2 & 20 & 12000 & 109 & 1.0 & 1.4E+07 & 12000 & 469 & 1.0 & 6.1E+07 & 12000 & 386 & 1.0 & 5.7E+07 & 12000 & 230 & 0.767 & 0.233 & 5.6E+07 & 3.5E+06 & 3.1E+06 \\
			3 & 20 & 12000 & 60 & 1.0 & 1.9E+07 & 12000 & 252 & 1.0 & 7.2E+07 & 12000 & 312 & 1.0 & 7.6E+07 & 12000 & 350 & 0.2 & 0.8 & 7.3E+07 & 5.4E+06 & 4.8E+06 \\
			5 & 20 & 12000 & 54 & 1.0 & 2.7E+07 & 12000 & 186 & 1.0 & 9.1E+07 & 12000 & 172 & 1.0 & 8.8E+07 & 12000 & 150 & 0.833 & 0.167 & 8.9E+07 & 1.9E+07 & 1.5E+07 \\
			10 & 20 & 12000 & 26 & 1.0 & 5.0E+07 & 12000 & 67 & 1.0 & 1.0E+08 & 12000 & 60 & 1.0 & 1.1E+08 & 12000 & 78 & 0.967 & 0.033 & 1.1E+08 & 5.7E+07 & 4.8E+07 \\
			15 & 20 & 12000 & 21 & 1.0 & 6.0E+07 & 12000 & 24 & 1.0 & 1.1E+08 & 12000 & 30 & 1.0 & 1.1E+08 & 12000 & 19 & 0.033 & 0.967 & 7.9E+07 & 1.5E+08 & 9.9E+07 \\
			20 & 20 & 12000 & 20 & 1.0 & 6.6E+07 & 12000 & 17 & 1.0 & 1.2E+08 & 12000 & 18 & 1.0 & 1.1E+08 & 12000 & 13 & 0.9 & 0.1 & 1.1E+08 & 2.6E+08 & 1.6E+08 \\
			25 & 20 & 12000 & 8 & 1.0 & 7.1E+07 & 12000 & 10 & 1.0 & 1.2E+08 & 12000 & 13 & 1.0 & 1.2E+08 & 12000 & 9 & 0.833 & 0.167 & 1.2E+08 & 4.2E+08 & 2.1E+08 \\
			\hline
		\end{tabular}
	}
	
\end{table}

\newpage

%%%%%%%%%%%%%%%%%%%%%%%%%%%%%%%%%%%%%%%%%%%%%%%%%%%%%%%%%%%%%%%%%%%%%%%%%%%%%%%%%%%%%%%
%  END: MFCCs for Speech Emotion Recognition
%%%%%%%%%%%%%%%%%%%%%%%%%%%%%%%%%%%%%%%%%%%%%%%%%%%%%%%%%%%%%%%%%%%%%%%%%%%%%%%%%%%%%%%

%%%%%%%%%%%%%%%%%%%%%%%%%%%%%%%%%%%%%%%%%%%%%%%%%%%%%%%%%%%%%%%%%%%%%%%%%%%%%%%%%%%%%%%
%  START: ISOLET
%%%%%%%%%%%%%%%%%%%%%%%%%%%%%%%%%%%%%%%%%%%%%%%%%%%%%%%%%%%%%%%%%%%%%%%%%%%%%%%%%%%%%%%
\subsection{ISOLET}
Dimensions: $m$ = 7797, $n$ = 617.
\par
Description: data set of patterns for spoken letter recognition which contains the spectral coefficients and other additional features.

\begin{table}[!htbp]
	\centering
	
	\caption{Summary of the results with ISOLET ($\times10^{5}$)}
	\label{TabResultsD10}
	\small
	\resizebox{!}{\tableheight}{
		\begin{tabular}{|l|l|l|llllll|llllll|llllll|}
			\hline
			\multicolumn{1}{|c|}{\multirow{3}{*}{$k$}} & \multicolumn{1}{c|}{\multirow{3}{*}{$f^*$}} & \multicolumn{1}{c|}{\multirow{3}{*}{$\overline{f}$}} & \multicolumn{6}{c|}{HPClust-inner} & \multicolumn{6}{c|}{HPClust-competitive} & \multicolumn{6}{c|}{HPClust-cooperative} \\ \cline{4-21}
			\multicolumn{1}{|c|}{} & \multicolumn{1}{c|}{} & \multicolumn{1}{c|}{} & \multicolumn{2}{c|}{$\varepsilon$} & \multicolumn{2}{c|}{$\overline{t}$} & \multicolumn{2}{c|}{$t$} & \multicolumn{2}{c|}{$\varepsilon$} & \multicolumn{2}{c|}{$\overline{t}$} & \multicolumn{2}{c|}{$t$} & \multicolumn{2}{c|}{$\varepsilon$} & \multicolumn{2}{c|}{$\overline{t}$} & \multicolumn{2}{c|}{$t$} \\ \cline{4-21}
			\multicolumn{1}{|c|}{} & \multicolumn{1}{c|}{} & \multicolumn{1}{c|}{} & \multicolumn{1}{c|}{med} & \multicolumn{1}{c|}{std} & \multicolumn{1}{c|}{med} & \multicolumn{1}{c|}{std} & \multicolumn{1}{c|}{med} & \multicolumn{1}{c|}{std} & \multicolumn{1}{c|}{med} & \multicolumn{1}{c|}{std} & \multicolumn{1}{c|}{med} & \multicolumn{1}{c|}{std} & \multicolumn{1}{c|}{med} & \multicolumn{1}{c|}{std} & \multicolumn{1}{c|}{med} & \multicolumn{1}{c|}{std} & \multicolumn{1}{c|}{med} & \multicolumn{1}{c|}{std} & \multicolumn{1}{c|}{med} & \multicolumn{1}{c|}{std} \\ \hline
			2 & 7.2194 & 3.66767 & 0.033 & 0.008 & 1.796 & 1.551 & 2.954 & 1.115 & 0.033 & 0.007 & 1.503 & 1.185 & 2.82 & 1.431 & 0.032 & 0.007 & 0.459 & 0.692 & 1.448 & 1.457 \\
			3 & 6.78782 & 3.4509 & 0.054 & 0.279 & 0.969 & 0.694 & 2.918 & 1.369 & 0.044 & 0.008 & 2.079 & 1.157 & 3.222 & 0.959 & 0.043 & 0.006 & 0.793 & 0.573 & 2.038 & 1.324 \\
			5 & 6.13651 & 3.11969 & 0.456 & 0.41 & 0.624 & 0.341 & 2.13 & 1.492 & 0.066 & 0.135 & 1.094 & 1.663 & 3.945 & 1.463 & 0.071 & 0.098 & 0.889 & 0.608 & 2.951 & 1.373 \\
			10 & 5.28577 & 2.70109 & 0.622 & 0.502 & 0.82 & 1.018 & 2.565 & 1.323 & 0.189 & 0.087 & 1.256 & 0.335 & 2.976 & 1.041 & 0.343 & 0.236 & 0.805 & 0.342 & 3.197 & 1.258 \\
			15 & 4.87391 & 2.49236 & 1.4 & 0.56 & 0.313 & 1.402 & 3.013 & 1.381 & 0.647 & 0.373 & 1.625 & 0.748 & 3.45 & 1.206 & 0.552 & 0.321 & 1.563 & 0.437 & 2.674 & 0.99 \\
			20 & 4.60857 & 2.35574 & 1.162 & 0.868 & 1.516 & 1.047 & 2.941 & 1.138 & 0.357 & 0.365 & 2.369 & 0.887 & 4.128 & 0.785 & 0.391 & 0.376 & 2.135 & 0.599 & 3.588 & 0.914 \\
			25 & 4.44323 & 2.25735 & 1.0 & 0.372 & 1.028 & 0.139 & 2.428 & 1.433 & 0.28 & 0.329 & 3.2 & 0.591 & 4.233 & 1.024 & 0.332 & 0.224 & 3.263 & 0.714 & 4.174 & 0.931 \\
			\hline
			\multicolumn{3}{|c|}{Mean:} & \textbf{0.675} & & \textbf{1.01} & & \textbf{2.707} & & \textbf{0.231} & & \textbf{1.875} & & \textbf{3.539} & & \textbf{0.252} & & \textbf{1.415} & & \textbf{2.867} & \\ \hline
		\end{tabular}
	}
	
	\bigskip
	
	\small
	\resizebox{!}{\tableheight}{
		\begin{tabular}{|l|l|l|llllll|llllll|llllll|}
			\hline
			\multicolumn{1}{|c|}{\multirow{3}{*}{$k$}} & \multicolumn{1}{c|}{\multirow{3}{*}{$f^*$}} & \multicolumn{1}{c|}{\multirow{3}{*}{$\overline{f}$}} & \multicolumn{6}{c|}{HPClust-hybrid} & \multicolumn{6}{c|}{Forgy K-means} & \multicolumn{6}{c|}{PBK-BDC} \\ \cline{4-21}
			\multicolumn{1}{|c|}{} & \multicolumn{1}{c|}{} & \multicolumn{1}{c|}{} & \multicolumn{2}{c|}{$\varepsilon$} & \multicolumn{2}{c|}{$\overline{t}$} & \multicolumn{2}{c|}{$t$} & \multicolumn{2}{c|}{$\varepsilon$} & \multicolumn{2}{c|}{$\overline{t}$} & \multicolumn{2}{c|}{$t$} & \multicolumn{2}{c|}{$\varepsilon$} & \multicolumn{2}{c|}{$\overline{t}$} & \multicolumn{2}{c|}{$t$} \\ \cline{4-21}
			\multicolumn{1}{|c|}{} & \multicolumn{1}{c|}{} & \multicolumn{1}{c|}{} & \multicolumn{1}{c|}{med} & \multicolumn{1}{c|}{std} & \multicolumn{1}{c|}{med} & \multicolumn{1}{c|}{std} & \multicolumn{1}{c|}{med} & \multicolumn{1}{c|}{std} & \multicolumn{1}{c|}{med} & \multicolumn{1}{c|}{std} & \multicolumn{1}{c|}{med} & \multicolumn{1}{c|}{std} & \multicolumn{1}{c|}{med} & \multicolumn{1}{c|}{std} & \multicolumn{1}{c|}{med} & \multicolumn{1}{c|}{std} & \multicolumn{1}{c|}{med} & \multicolumn{1}{c|}{std} & \multicolumn{1}{c|}{med} & \multicolumn{1}{c|}{std} \\ \hline
			2 & 7.2194 & 3.66767 & 0.032 & 0.006 & 1.592 & 1.149 & 2.841 & 1.579 & -0.0 & 0.0 & -- & -- & 0.119 & 0.022 & 0.026 & 0.01 & -- & -- & 0.062 & 0.008 \\
			3 & 6.78782 & 3.4509 & 0.045 & 0.007 & 1.657 & 1.094 & 4.006 & 1.218 & 0.552 & 0.27 & -- & -- & 0.311 & 0.283 & 0.047 & 0.245 & -- & -- & 0.151 & 0.049 \\
			5 & 6.13651 & 3.11969 & 0.07 & 0.13 & 1.762 & 1.292 & 2.122 & 1.538 & 0.392 & 0.797 & -- & -- & 0.63 & 0.345 & 0.444 & 0.691 & -- & -- & 0.31 & 0.087 \\
			10 & 5.28577 & 2.70109 & 0.166 & 0.122 & 0.949 & 0.483 & 3.472 & 1.262 & 0.936 & 1.01 & -- & -- & 1.475 & 1.051 & 1.281 & 0.647 & -- & -- & 0.609 & 0.367 \\
			15 & 4.87391 & 2.49236 & 0.731 & 0.354 & 2.276 & 1.118 & 3.789 & 0.858 & 1.403 & 1.382 & -- & -- & 1.611 & 1.074 & 2.444 & 1.378 & -- & -- & 0.964 & 0.392 \\
			20 & 4.60857 & 2.35574 & 0.34 & 0.346 & 2.067 & 0.664 & 4.11 & 0.879 & 1.079 & 0.845 & -- & -- & 3.058 & 1.241 & 1.816 & 1.161 & -- & -- & 1.208 & 0.389 \\
			25 & 4.44323 & 2.25735 & 0.259 & 0.261 & 3.407 & 0.958 & 4.48 & 1.001 & 1.252 & 0.854 & -- & -- & 5.116 & 0.925 & 1.127 & 0.67 & -- & -- & 1.989 & 0.743 \\
			\hline
			\multicolumn{3}{|c|}{Mean:} & \textbf{0.235} & & \textbf{1.959} & & \textbf{3.546} & & \textbf{0.802} & & \textbf{--} & & \textbf{1.76} & & \textbf{1.026} & & \textbf{--} & & \textbf{0.756} & \\ \hline
		\end{tabular}
	}
	
	\bigskip
	
	\caption{Clustering details with ISOLET}
	\label{TabDetailsD10}
	\resizebox{!}{\tableheight}{
		\begin{tabular}{|l|l|llll|llll|llll|lllll|l|l|}
			\hline
			\multicolumn{1}{|c|}{\multirow{2}{*}{$k$}} & \multicolumn{1}{c|}{\multirow{2}{*}{$n_{exec}$}} & \multicolumn{4}{c|}{HPClust-inner} & \multicolumn{4}{c|}{HPClust-competitive} & \multicolumn{4}{c|}{HPClust-cooperative} & \multicolumn{5}{c|}{HPClust-hybrid} & \multicolumn{1}{c|}{Forgy K-means} & \multicolumn{1}{c|}{PBK-BDC} \\ \cline{3-21}
			\multicolumn{1}{|c|}{} & \multicolumn{1}{c|}{} & \multicolumn{1}{c|}{$s$} & \multicolumn{1}{c|}{$n_{s}$} & \multicolumn{1}{c|}{$T$} & \multicolumn{1}{c|}{$n_{d}$} & \multicolumn{1}{c|}{$s$} & \multicolumn{1}{c|}{$n_{s}$} & \multicolumn{1}{c|}{$T$} & \multicolumn{1}{c|}{$n_{d}$} & \multicolumn{1}{c|}{$s$} & \multicolumn{1}{c|}{$n_{s}$} & \multicolumn{1}{c|}{$T$} & \multicolumn{1}{c|}{$n_{d}$} & \multicolumn{1}{c|}{$s$} & \multicolumn{1}{c|}{$n_{s}$} & \multicolumn{1}{c|}{$T_1$} & \multicolumn{1}{c|}{$T_2$} & \multicolumn{1}{c|}{$n_{d}$} & \multicolumn{1}{c|}{$n_{d}$} & \multicolumn{1}{c|}{$n_{d}$} \\
			\hline
			2 & 15 & 4000 & 283 & 5.0 & 1.3E+07 & 4000 & 977 & 5.0 & 4.9E+07 & 4000 & 518 & 5.0 & 5.1E+07 & 4000 & 1120 & 1.167 & 3.833 & 5.0E+07 & 1.7E+05 & 1.1E+05 \\
			3 & 15 & 4000 & 240 & 5.0 & 1.4E+07 & 4000 & 879 & 5.0 & 5.4E+07 & 4000 & 523 & 5.0 & 5.4E+07 & 4000 & 1096 & 1.667 & 3.333 & 5.3E+07 & 4.9E+05 & 2.9E+05 \\
			5 & 15 & 4000 & 128 & 5.0 & 1.8E+07 & 4000 & 606 & 5.0 & 5.6E+07 & 4000 & 431 & 5.0 & 5.6E+07 & 4000 & 300 & 1.833 & 3.167 & 5.5E+07 & 1.1E+06 & 4.8E+05 \\
			10 & 15 & 4000 & 79 & 5.0 & 2.5E+07 & 4000 & 147 & 5.0 & 5.9E+07 & 4000 & 209 & 5.0 & 5.9E+07 & 4000 & 186 & 4.667 & 0.333 & 5.7E+07 & 2.5E+06 & 1.1E+06 \\
			15 & 15 & 4000 & 66 & 5.0 & 2.8E+07 & 4000 & 112 & 5.0 & 5.9E+07 & 4000 & 58 & 5.0 & 5.9E+07 & 4000 & 87 & 1.167 & 3.833 & 4.8E+07 & 2.9E+06 & 1.9E+06 \\
			20 & 15 & 4000 & 40 & 5.0 & 3.1E+07 & 4000 & 77 & 5.0 & 5.8E+07 & 4000 & 75 & 5.0 & 5.9E+07 & 4000 & 80 & 3.333 & 1.667 & 5.7E+07 & 5.6E+06 & 2.4E+06 \\
			25 & 15 & 4000 & 27 & 5.0 & 3.3E+07 & 4000 & 43 & 5.0 & 5.6E+07 & 4000 & 40 & 5.0 & 5.8E+07 & 4000 & 37 & 3.167 & 1.833 & 4.5E+07 & 7.8E+06 & 3.2E+06 \\
			\hline
		\end{tabular}
	}
	
\end{table}

\newpage

%%%%%%%%%%%%%%%%%%%%%%%%%%%%%%%%%%%%%%%%%%%%%%%%%%%%%%%%%%%%%%%%%%%%%%%%%%%%%%%%%%%%%%%
%  END: ISOLET
%%%%%%%%%%%%%%%%%%%%%%%%%%%%%%%%%%%%%%%%%%%%%%%%%%%%%%%%%%%%%%%%%%%%%%%%%%%%%%%%%%%%%%%

\newpage

%%%%%%%%%%%%%%%%%%%%%%%%%%%%%%%%%%%%%%%%%%%%%%%%%%%%%%%%%%%%%%%%%%%%%%%%%%%%%%%%%%%%%%%
%  START: Sensorless Drive Diagnosis
%%%%%%%%%%%%%%%%%%%%%%%%%%%%%%%%%%%%%%%%%%%%%%%%%%%%%%%%%%%%%%%%%%%%%%%%%%%%%%%%%%%%%%%
\subsection{Sensorless Drive Diagnosis}
Dimensions: $m$ = 58509, $n$ = 48.
\par
Description: a data set for sensorless drive diagnosis with features extracted from motor current.

\begin{table}[!htbp]
	\centering
	
	\caption{Summary of the results with Sensorless Drive Diagnosis ($\times10^{7}$)}
	\label{TabResultsD11}
	\small
	\resizebox{!}{\tableheight}{
		\begin{tabular}{|l|l|l|llllll|llllll|llllll|}
			\hline
			\multicolumn{1}{|c|}{\multirow{3}{*}{$k$}} & \multicolumn{1}{c|}{\multirow{3}{*}{$f^*$}} & \multicolumn{1}{c|}{\multirow{3}{*}{$\overline{f}$}} & \multicolumn{6}{c|}{HPClust-inner} & \multicolumn{6}{c|}{HPClust-competitive} & \multicolumn{6}{c|}{HPClust-cooperative} \\ \cline{4-21}
			\multicolumn{1}{|c|}{} & \multicolumn{1}{c|}{} & \multicolumn{1}{c|}{} & \multicolumn{2}{c|}{$\varepsilon$} & \multicolumn{2}{c|}{$\overline{t}$} & \multicolumn{2}{c|}{$t$} & \multicolumn{2}{c|}{$\varepsilon$} & \multicolumn{2}{c|}{$\overline{t}$} & \multicolumn{2}{c|}{$t$} & \multicolumn{2}{c|}{$\varepsilon$} & \multicolumn{2}{c|}{$\overline{t}$} & \multicolumn{2}{c|}{$t$} \\ \cline{4-21}
			\multicolumn{1}{|c|}{} & \multicolumn{1}{c|}{} & \multicolumn{1}{c|}{} & \multicolumn{1}{c|}{med} & \multicolumn{1}{c|}{std} & \multicolumn{1}{c|}{med} & \multicolumn{1}{c|}{std} & \multicolumn{1}{c|}{med} & \multicolumn{1}{c|}{std} & \multicolumn{1}{c|}{med} & \multicolumn{1}{c|}{std} & \multicolumn{1}{c|}{med} & \multicolumn{1}{c|}{std} & \multicolumn{1}{c|}{med} & \multicolumn{1}{c|}{std} & \multicolumn{1}{c|}{med} & \multicolumn{1}{c|}{std} & \multicolumn{1}{c|}{med} & \multicolumn{1}{c|}{std} & \multicolumn{1}{c|}{med} & \multicolumn{1}{c|}{std} \\ \hline
			2 & 3.88116 & 3.87915 & -0.0 & 15.678 & 0.305 & 0.296 & 0.392 & 0.313 & -0.0 & 2.101 & 0.226 & 0.151 & 0.544 & 0.234 & -0.0 & 1.438 & 0.26 & 0.169 & 0.716 & 0.257 \\
			3 & 2.91313 & 3.22719 & -0.0 & 5.899 & 0.022 & 0.161 & 0.516 & 0.227 & -0.0 & 0.55 & 0.077 & 0.008 & 0.578 & 0.247 & -0.0 & 0.869 & 0.082 & 0.013 & 0.659 & 0.267 \\
			5 & 1.93651 & 1.93613 & 0.022 & 8.618 & 0.307 & 0.219 & 0.653 & 0.285 & 0.015 & 7.434 & 0.559 & 0.187 & 0.764 & 0.219 & 0.011 & 1.235 & 0.48 & 0.147 & 0.805 & 0.184 \\
			10 & 0.98472 & 1.0394 & 5.588 & 8.042 & 0.177 & 0.279 & 0.58 & 0.257 & -2.401 & 1.407 & 0.74 & 0.203 & 1.017 & 0.15 & -2.394 & 1.676 & 0.717 & 0.179 & 1.018 & 0.183 \\
			15 & 0.62816 & 0.63072 & 0.481 & 4.002 & 0.291 & 0.251 & 0.681 & 0.196 & 0.034 & 0.858 & 1.28 & 0.448 & 1.616 & 0.661 & 0.028 & 7.247 & 1.475 & 0.412 & 1.731 & 0.566 \\
			20 & 0.49884 & 0.50187 & 0.486 & 1.649 & 0.413 & 0.135 & 0.734 & 0.203 & -0.557 & 1.962 & 1.78 & 0.45 & 2.104 & 0.596 & -0.053 & 1.871 & 1.966 & 0.425 & 2.326 & 0.584 \\
			25 & 0.42225 & 0.43197 & 2.193 & 1.768 & 0.508 & 0.139 & 0.811 & 0.191 & 1.049 & 0.546 & 2.509 & 0.452 & 2.867 & 0.808 & 0.94 & 0.502 & 2.384 & 0.74 & 2.826 & 0.794 \\
			\hline
			\multicolumn{3}{|c|}{Mean:} & \textbf{1.253} & & \textbf{0.289} & & \textbf{0.624} & & \textbf{-0.266} & & \textbf{1.024} & & \textbf{1.356} & & \textbf{-0.21} & & \textbf{1.052} & & \textbf{1.44} & \\ \hline
		\end{tabular}
	}
	
	\bigskip
	
	\small
	\resizebox{!}{\tableheight}{
		\begin{tabular}{|l|l|l|llllll|llllll|llllll|}
			\hline
			\multicolumn{1}{|c|}{\multirow{3}{*}{$k$}} & \multicolumn{1}{c|}{\multirow{3}{*}{$f^*$}} & \multicolumn{1}{c|}{\multirow{3}{*}{$\overline{f}$}} & \multicolumn{6}{c|}{HPClust-hybrid} & \multicolumn{6}{c|}{Forgy K-means} & \multicolumn{6}{c|}{PBK-BDC} \\ \cline{4-21}
			\multicolumn{1}{|c|}{} & \multicolumn{1}{c|}{} & \multicolumn{1}{c|}{} & \multicolumn{2}{c|}{$\varepsilon$} & \multicolumn{2}{c|}{$\overline{t}$} & \multicolumn{2}{c|}{$t$} & \multicolumn{2}{c|}{$\varepsilon$} & \multicolumn{2}{c|}{$\overline{t}$} & \multicolumn{2}{c|}{$t$} & \multicolumn{2}{c|}{$\varepsilon$} & \multicolumn{2}{c|}{$\overline{t}$} & \multicolumn{2}{c|}{$t$} \\ \cline{4-21}
			\multicolumn{1}{|c|}{} & \multicolumn{1}{c|}{} & \multicolumn{1}{c|}{} & \multicolumn{1}{c|}{med} & \multicolumn{1}{c|}{std} & \multicolumn{1}{c|}{med} & \multicolumn{1}{c|}{std} & \multicolumn{1}{c|}{med} & \multicolumn{1}{c|}{std} & \multicolumn{1}{c|}{med} & \multicolumn{1}{c|}{std} & \multicolumn{1}{c|}{med} & \multicolumn{1}{c|}{std} & \multicolumn{1}{c|}{med} & \multicolumn{1}{c|}{std} & \multicolumn{1}{c|}{med} & \multicolumn{1}{c|}{std} & \multicolumn{1}{c|}{med} & \multicolumn{1}{c|}{std} & \multicolumn{1}{c|}{med} & \multicolumn{1}{c|}{std} \\ \hline
			2 & 3.88116 & 3.87915 & -0.0 & 0.636 & 0.213 & 0.139 & 0.628 & 0.276 & 100.19 & 0.0 & -- & -- & 0.136 & 0.038 & 100.19 & 0.0 & -- & -- & 0.14 & 0.022 \\
			3 & 2.91313 & 3.22719 & -0.0 & 0.968 & 0.081 & 0.014 & 0.751 & 0.258 & 10.865 & 71.87 & -- & -- & 0.433 & 0.158 & 10.865 & 67.653 & -- & -- & 0.483 & 0.152 \\
			5 & 1.93651 & 1.93613 & 0.016 & 6.804 & 0.704 & 0.201 & 0.844 & 0.207 & 37.859 & 0.003 & -- & -- & 0.496 & 0.068 & 37.853 & 35.156 & -- & -- & 0.524 & 0.132 \\
			10 & 0.98472 & 1.0394 & -2.404 & 0.932 & 0.69 & 0.194 & 1.581 & 0.459 & 127.202 & 0.358 & -- & -- & 1.696 & 0.338 & 127.256 & 0.036 & -- & -- & 1.77 & 0.387 \\
			15 & 0.62816 & 0.63072 & 0.029 & 0.082 & 1.339 & 0.335 & 1.989 & 0.587 & 235.577 & 0.573 & -- & -- & 1.93 & 0.551 & 235.435 & 5.741 & -- & -- & 1.789 & 0.399 \\
			20 & 0.49884 & 0.50187 & -0.058 & 0.448 & 1.895 & 0.392 & 2.297 & 0.56 & 309.27 & 23.686 & -- & -- & 3.769 & 1.075 & 309.269 & 23.374 & -- & -- & 3.477 & 1.446 \\
			25 & 0.42225 & 0.43197 & 0.92 & 0.55 & 2.407 & 0.607 & 2.924 & 0.944 & 315.617 & 35.414 & -- & -- & 6.618 & 2.05 & 315.673 & 0.549 & -- & -- & 6.369 & 1.881 \\
			\hline
			\multicolumn{3}{|c|}{Mean:} & \textbf{-0.214} & & \textbf{1.047} & & \textbf{1.573} & & \textbf{162.369} & & \textbf{--} & & \textbf{2.154} & & \textbf{162.363} & & \textbf{--} & & \textbf{2.079} & \\ \hline
		\end{tabular}
	}
	
	\bigskip
	
	\caption{Clustering details with Sensorless Drive Diagnosis}
	\label{TabDetailsD11}
	\resizebox{!}{\tableheight}{
		\begin{tabular}{|l|l|llll|llll|llll|lllll|l|l|}
			\hline
			\multicolumn{1}{|c|}{\multirow{2}{*}{$k$}} & \multicolumn{1}{c|}{\multirow{2}{*}{$n_{exec}$}} & \multicolumn{4}{c|}{HPClust-inner} & \multicolumn{4}{c|}{HPClust-competitive} & \multicolumn{4}{c|}{HPClust-cooperative} & \multicolumn{5}{c|}{HPClust-hybrid} & \multicolumn{1}{c|}{Forgy K-means} & \multicolumn{1}{c|}{PBK-BDC} \\ \cline{3-21}
			\multicolumn{1}{|c|}{} & \multicolumn{1}{c|}{} & \multicolumn{1}{c|}{$s$} & \multicolumn{1}{c|}{$n_{s}$} & \multicolumn{1}{c|}{$T$} & \multicolumn{1}{c|}{$n_{d}$} & \multicolumn{1}{c|}{$s$} & \multicolumn{1}{c|}{$n_{s}$} & \multicolumn{1}{c|}{$T$} & \multicolumn{1}{c|}{$n_{d}$} & \multicolumn{1}{c|}{$s$} & \multicolumn{1}{c|}{$n_{s}$} & \multicolumn{1}{c|}{$T$} & \multicolumn{1}{c|}{$n_{d}$} & \multicolumn{1}{c|}{$s$} & \multicolumn{1}{c|}{$n_{s}$} & \multicolumn{1}{c|}{$T_1$} & \multicolumn{1}{c|}{$T_2$} & \multicolumn{1}{c|}{$n_{d}$} & \multicolumn{1}{c|}{$n_{d}$} & \multicolumn{1}{c|}{$n_{d}$} \\
			\hline
			2 & 40 & 58500 & 32 & 1.0 & 1.9E+07 & 58500 & 177 & 1.0 & 8.5E+07 & 58500 & 228 & 1.0 & 8.5E+07 & 58500 & 199 & 0.267 & 0.733 & 8.4E+07 & 3.9E+06 & 3.9E+06 \\
			3 & 40 & 58500 & 38 & 1.0 & 2.6E+07 & 58500 & 134 & 1.0 & 9.8E+07 & 58500 & 140 & 1.0 & 1.0E+08 & 58500 & 170 & 0.2 & 0.8 & 9.5E+07 & 1.4E+07 & 1.5E+07 \\
			5 & 40 & 58500 & 36 & 1.0 & 4.0E+07 & 58500 & 59 & 1.0 & 1.1E+08 & 58500 & 84 & 1.0 & 1.1E+08 & 58500 & 80 & 0.833 & 0.167 & 1.0E+08 & 1.6E+07 & 1.7E+07 \\
			10 & 40 & 58500 & 18 & 1.0 & 5.7E+07 & 58500 & 12 & 1.0 & 1.3E+08 & 58500 & 13 & 1.0 & 1.3E+08 & 58500 & 10 & 0.633 & 0.367 & 1.2E+08 & 5.7E+07 & 5.9E+07 \\
			15 & 40 & 58500 & 14 & 1.0 & 7.4E+07 & 58500 & 5 & 1.0 & 2.2E+08 & 58500 & 3 & 1.0 & 2.3E+08 & 58500 & 5 & 0.4 & 0.6 & 2.2E+08 & 6.6E+07 & 5.9E+07 \\
			20 & 40 & 58500 & 10 & 1.0 & 7.7E+07 & 58500 & 4 & 1.0 & 2.8E+08 & 58500 & 5 & 1.0 & 2.9E+08 & 58500 & 4 & 0.533 & 0.467 & 2.7E+08 & 1.3E+08 & 1.2E+08 \\
			25 & 40 & 58500 & 7 & 1.0 & 8.3E+07 & 58500 & 5 & 1.0 & 3.7E+08 & 58500 & 4 & 1.0 & 3.5E+08 & 58500 & 5 & 0.767 & 0.233 & 3.6E+08 & 2.2E+08 & 2.1E+08 \\
			\hline
		\end{tabular}
	}
	
\end{table}

\newpage

%%%%%%%%%%%%%%%%%%%%%%%%%%%%%%%%%%%%%%%%%%%%%%%%%%%%%%%%%%%%%%%%%%%%%%%%%%%%%%%%%%%%%%%
%  END: Sensorless Drive Diagnosis
%%%%%%%%%%%%%%%%%%%%%%%%%%%%%%%%%%%%%%%%%%%%%%%%%%%%%%%%%%%%%%%%%%%%%%%%%%%%%%%%%%%%%%%

%%%%%%%%%%%%%%%%%%%%%%%%%%%%%%%%%%%%%%%%%%%%%%%%%%%%%%%%%%%%%%%%%%%%%%%%%%%%%%%%%%%%%%%
%  START: Sensorless Drive Diagnosis (normalized)
%%%%%%%%%%%%%%%%%%%%%%%%%%%%%%%%%%%%%%%%%%%%%%%%%%%%%%%%%%%%%%%%%%%%%%%%%%%%%%%%%%%%%%%
\subsection{Sensorless Drive Diagnosis (normalized)}
Dimensions: $m$ = 58509, $n$ = 48.
\par
Description: a data set for sensorless drive diagnosis with features extracted from motor current. Min-max scaling was used for normalization of data set values for better clusterization.

\begin{table}[!htbp]
	\centering
	
	\caption{Summary of the results with Sensorless Drive Diagnosis (normalized) ($\times10^{3}$)}
	\label{TabResultsD12}
	\small
	\resizebox{!}{\tableheight}{
		\begin{tabular}{|l|l|l|llllll|llllll|llllll|}
			\hline
			\multicolumn{1}{|c|}{\multirow{3}{*}{$k$}} & \multicolumn{1}{c|}{\multirow{3}{*}{$f^*$}} & \multicolumn{1}{c|}{\multirow{3}{*}{$\overline{f}$}} & \multicolumn{6}{c|}{HPClust-inner} & \multicolumn{6}{c|}{HPClust-competitive} & \multicolumn{6}{c|}{HPClust-cooperative} \\ \cline{4-21}
			\multicolumn{1}{|c|}{} & \multicolumn{1}{c|}{} & \multicolumn{1}{c|}{} & \multicolumn{2}{c|}{$\varepsilon$} & \multicolumn{2}{c|}{$\overline{t}$} & \multicolumn{2}{c|}{$t$} & \multicolumn{2}{c|}{$\varepsilon$} & \multicolumn{2}{c|}{$\overline{t}$} & \multicolumn{2}{c|}{$t$} & \multicolumn{2}{c|}{$\varepsilon$} & \multicolumn{2}{c|}{$\overline{t}$} & \multicolumn{2}{c|}{$t$} \\ \cline{4-21}
			\multicolumn{1}{|c|}{} & \multicolumn{1}{c|}{} & \multicolumn{1}{c|}{} & \multicolumn{1}{c|}{med} & \multicolumn{1}{c|}{std} & \multicolumn{1}{c|}{med} & \multicolumn{1}{c|}{std} & \multicolumn{1}{c|}{med} & \multicolumn{1}{c|}{std} & \multicolumn{1}{c|}{med} & \multicolumn{1}{c|}{std} & \multicolumn{1}{c|}{med} & \multicolumn{1}{c|}{std} & \multicolumn{1}{c|}{med} & \multicolumn{1}{c|}{std} & \multicolumn{1}{c|}{med} & \multicolumn{1}{c|}{std} & \multicolumn{1}{c|}{med} & \multicolumn{1}{c|}{std} & \multicolumn{1}{c|}{med} & \multicolumn{1}{c|}{std} \\ \hline
			2 & 15.64798$^*$ & 0.89303 & 0.067 & 0.035 & 0.1 & 0.082 & 0.122 & 0.08 & 0.091 & 0.046 & 0.073 & 0.052 & 0.122 & 0.082 & 0.088 & 0.051 & 0.076 & 0.069 & 0.15 & 0.087 \\
			3 & 12.19375$^*$ & 0.70587 & 3.467 & 2.324 & 0.058 & 0.097 & 0.139 & 0.098 & 0.187 & 1.016 & 0.054 & 0.053 & 0.159 & 0.076 & 0.187 & 1.145 & 0.044 & 0.044 & 0.163 & 0.091 \\
			5 & 7.85054$^*$ & 0.45202 & 0.363 & 1.748 & 0.099 & 0.078 & 0.166 & 0.076 & 0.343 & 0.255 & 0.066 & 0.058 & 0.172 & 0.08 & 0.293 & 0.21 & 0.056 & 0.062 & 0.181 & 0.087 \\
			10 & 4.71275$^*$ & 0.28067 & 3.764 & 2.034 & 0.089 & 0.08 & 0.165 & 0.08 & 0.609 & 1.073 & 0.067 & 0.058 & 0.212 & 0.074 & 1.936 & 1.295 & 0.064 & 0.038 & 0.201 & 0.075 \\
			15 & 3.62541$^*$ & 0.21493 & 3.765 & 2.962 & 0.106 & 0.07 & 0.229 & 0.091 & 1.445 & 0.992 & 0.111 & 0.048 & 0.203 & 0.072 & 1.85 & 1.395 & 0.091 & 0.045 & 0.223 & 0.071 \\
			20 & 2.971$^*$ & 0.17797 & 4.762 & 2.238 & 0.059 & 0.068 & 0.169 & 0.087 & 2.142 & 0.786 & 0.101 & 0.051 & 0.23 & 0.065 & 2.391 & 1.266 & 0.099 & 0.034 & 0.233 & 0.066 \\
			25 & 2.60929$^*$ & 0.15364 & 5.017 & 2.274 & 0.111 & 0.065 & 0.2 & 0.086 & 2.629 & 1.204 & 0.155 & 0.078 & 0.25 & 0.076 & 2.993 & 1.446 & 0.185 & 0.07 & 0.246 & 0.061 \\
			\hline
			\multicolumn{3}{|c|}{Mean:} & \textbf{3.029} & & \textbf{0.089} & & \textbf{0.17} & & \textbf{1.064} & & \textbf{0.09} & & \textbf{0.193} & & \textbf{1.391} & & \textbf{0.088} & & \textbf{0.2} & \\ \hline
		\end{tabular}
	}
	
	\bigskip
	
	\small
	\resizebox{!}{\tableheight}{
		\begin{tabular}{|l|l|l|llllll|llllll|llllll|}
			\hline
			\multicolumn{1}{|c|}{\multirow{3}{*}{$k$}} & \multicolumn{1}{c|}{\multirow{3}{*}{$f^*$}} & \multicolumn{1}{c|}{\multirow{3}{*}{$\overline{f}$}} & \multicolumn{6}{c|}{HPClust-hybrid} & \multicolumn{6}{c|}{Forgy K-means} & \multicolumn{6}{c|}{PBK-BDC} \\ \cline{4-21}
			\multicolumn{1}{|c|}{} & \multicolumn{1}{c|}{} & \multicolumn{1}{c|}{} & \multicolumn{2}{c|}{$\varepsilon$} & \multicolumn{2}{c|}{$\overline{t}$} & \multicolumn{2}{c|}{$t$} & \multicolumn{2}{c|}{$\varepsilon$} & \multicolumn{2}{c|}{$\overline{t}$} & \multicolumn{2}{c|}{$t$} & \multicolumn{2}{c|}{$\varepsilon$} & \multicolumn{2}{c|}{$\overline{t}$} & \multicolumn{2}{c|}{$t$} \\ \cline{4-21}
			\multicolumn{1}{|c|}{} & \multicolumn{1}{c|}{} & \multicolumn{1}{c|}{} & \multicolumn{1}{c|}{med} & \multicolumn{1}{c|}{std} & \multicolumn{1}{c|}{med} & \multicolumn{1}{c|}{std} & \multicolumn{1}{c|}{med} & \multicolumn{1}{c|}{std} & \multicolumn{1}{c|}{med} & \multicolumn{1}{c|}{std} & \multicolumn{1}{c|}{med} & \multicolumn{1}{c|}{std} & \multicolumn{1}{c|}{med} & \multicolumn{1}{c|}{std} & \multicolumn{1}{c|}{med} & \multicolumn{1}{c|}{std} & \multicolumn{1}{c|}{med} & \multicolumn{1}{c|}{std} & \multicolumn{1}{c|}{med} & \multicolumn{1}{c|}{std} \\ \hline
			2 & 15.64798$^*$ & 0.89303 & 0.089 & 0.059 & 0.067 & 0.081 & 0.175 & 0.086 & 0.0 & 11.811 & -- & -- & 0.051 & 0.019 & 0.002 & 17.132 & -- & -- & 0.012 & 0.001 \\
			3 & 12.19375$^*$ & 0.70587 & 0.15 & 0.596 & 0.078 & 0.065 & 0.191 & 0.074 & 0.979 & 3.288 & -- & -- & 0.09 & 0.066 & 1.574 & 11.101 & -- & -- & 0.033 & 0.008 \\
			5 & 7.85054$^*$ & 0.45202 & 0.297 & 0.238 & 0.06 & 0.044 & 0.157 & 0.087 & 0.535 & 2.44 & -- & -- & 0.21 & 0.15 & 11.592 & 16.743 & -- & -- & 0.055 & 0.012 \\
			10 & 4.71275$^*$ & 0.28067 & 1.148 & 1.18 & 0.062 & 0.051 & 0.253 & 0.08 & 6.68 & 3.559 & -- & -- & 0.563 & 0.402 & 13.32 & 7.63 & -- & -- & 0.131 & 0.025 \\
			15 & 3.62541$^*$ & 0.21493 & 1.781 & 0.887 & 0.094 & 0.056 & 0.244 & 0.079 & 8.774 & 3.827 & -- & -- & 0.978 & 0.407 & 14.032 & 8.194 & -- & -- & 0.201 & 0.032 \\
			20 & 2.971$^*$ & 0.17797 & 3.125 & 1.012 & 0.101 & 0.063 & 0.251 & 0.073 & 12.594 & 5.298 & -- & -- & 1.644 & 0.589 & 18.915 & 6.425 & -- & -- & 0.265 & 0.051 \\
			25 & 2.60929$^*$ & 0.15364 & 2.768 & 1.106 & 0.161 & 0.072 & 0.275 & 0.065 & 13.879 & 6.179 & -- & -- & 1.806 & 0.741 & 19.277 & 4.994 & -- & -- & 0.318 & 0.05 \\
			\hline
			\multicolumn{3}{|c|}{Mean:} & \textbf{1.337} & & \textbf{0.089} & & \textbf{0.221} & & \textbf{6.206} & & \textbf{--} & & \textbf{0.763} & & \textbf{11.245} & & \textbf{--} & & \textbf{0.145} & \\ \hline
		\end{tabular}
	}
	
	\bigskip
	
	\caption{Clustering details with Sensorless Drive Diagnosis (normalized)}
	\label{TabDetailsD12}
	\resizebox{!}{\tableheight}{
		\begin{tabular}{|l|l|llll|llll|llll|lllll|l|l|}
			\hline
			\multicolumn{1}{|c|}{\multirow{2}{*}{$k$}} & \multicolumn{1}{c|}{\multirow{2}{*}{$n_{exec}$}} & \multicolumn{4}{c|}{HPClust-inner} & \multicolumn{4}{c|}{HPClust-competitive} & \multicolumn{4}{c|}{HPClust-cooperative} & \multicolumn{5}{c|}{HPClust-hybrid} & \multicolumn{1}{c|}{Forgy K-means} & \multicolumn{1}{c|}{PBK-BDC} \\ \cline{3-21}
			\multicolumn{1}{|c|}{} & \multicolumn{1}{c|}{} & \multicolumn{1}{c|}{$s$} & \multicolumn{1}{c|}{$n_{s}$} & \multicolumn{1}{c|}{$T$} & \multicolumn{1}{c|}{$n_{d}$} & \multicolumn{1}{c|}{$s$} & \multicolumn{1}{c|}{$n_{s}$} & \multicolumn{1}{c|}{$T$} & \multicolumn{1}{c|}{$n_{d}$} & \multicolumn{1}{c|}{$s$} & \multicolumn{1}{c|}{$n_{s}$} & \multicolumn{1}{c|}{$T$} & \multicolumn{1}{c|}{$n_{d}$} & \multicolumn{1}{c|}{$s$} & \multicolumn{1}{c|}{$n_{s}$} & \multicolumn{1}{c|}{$T_1$} & \multicolumn{1}{c|}{$T_2$} & \multicolumn{1}{c|}{$n_{d}$} & \multicolumn{1}{c|}{$n_{d}$} & \multicolumn{1}{c|}{$n_{d}$} \\
			\hline
			2 & 40 & 3500 & 38 & 0.3 & 2.2E+06 & 3500 & 252 & 0.3 & 1.4E+07 & 3500 & 298 & 0.3 & 1.4E+07 & 3500 & 362 & 0.16 & 0.14 & 1.4E+07 & 1.2E+06 & 1.1E+06 \\
			3 & 40 & 3500 & 42 & 0.3 & 3.4E+06 & 3500 & 248 & 0.3 & 1.9E+07 & 3500 & 293 & 0.3 & 2.0E+07 & 3500 & 321 & 0.13 & 0.17 & 1.9E+07 & 2.7E+06 & 2.6E+06 \\
			5 & 40 & 3500 & 36 & 0.3 & 5.4E+06 & 3500 & 174 & 0.3 & 2.6E+07 & 3500 & 194 & 0.3 & 2.6E+07 & 3500 & 158 & 0.01 & 0.29 & 2.5E+07 & 5.9E+06 & 5.4E+06 \\
			10 & 40 & 3500 & 27 & 0.3 & 9.7E+06 & 3500 & 110 & 0.3 & 3.4E+07 & 3500 & 114 & 0.3 & 3.2E+07 & 3500 & 134 & 0.23 & 0.07 & 3.3E+07 & 1.9E+07 & 1.4E+07 \\
			15 & 40 & 3500 & 31 & 0.3 & 1.5E+07 & 3500 & 52 & 0.3 & 3.7E+07 & 3500 & 60 & 0.3 & 3.7E+07 & 3500 & 70 & 0.26 & 0.04 & 3.6E+07 & 3.2E+07 & 2.3E+07 \\
			20 & 40 & 3500 & 20 & 0.3 & 1.8E+07 & 3500 & 39 & 0.3 & 3.9E+07 & 3500 & 40 & 0.3 & 3.8E+07 & 3500 & 34 & 0.1 & 0.2 & 3.4E+07 & 5.6E+07 & 2.9E+07 \\
			25 & 40 & 3500 & 20 & 0.3 & 2.1E+07 & 3500 & 30 & 0.3 & 4.0E+07 & 3500 & 28 & 0.3 & 3.9E+07 & 3500 & 28 & 0.1 & 0.2 & 3.6E+07 & 6.2E+07 & 3.7E+07 \\
			\hline
		\end{tabular}
	}
	
\end{table}

\newpage

%%%%%%%%%%%%%%%%%%%%%%%%%%%%%%%%%%%%%%%%%%%%%%%%%%%%%%%%%%%%%%%%%%%%%%%%%%%%%%%%%%%%%%%
%  END: Sensorless Drive Diagnosis (normalized)
%%%%%%%%%%%%%%%%%%%%%%%%%%%%%%%%%%%%%%%%%%%%%%%%%%%%%%%%%%%%%%%%%%%%%%%%%%%%%%%%%%%%%%%

\newpage

%%%%%%%%%%%%%%%%%%%%%%%%%%%%%%%%%%%%%%%%%%%%%%%%%%%%%%%%%%%%%%%%%%%%%%%%%%%%%%%%%%%%%%%
%  START: Online News Popularity
%%%%%%%%%%%%%%%%%%%%%%%%%%%%%%%%%%%%%%%%%%%%%%%%%%%%%%%%%%%%%%%%%%%%%%%%%%%%%%%%%%%%%%%
\subsection{Online News Popularity}
Dimensions: $m$ = 39644, $n$ = 58.
\par
Description: this dataset summarizes a heterogeneous set of features about articles published by Mashable in a period of two years for predicting the number of shares in social networks (popularity).

\begin{table}[!htbp]
	\centering
	
	\caption{Summary of the results with Online News Popularity ($\times10^{14}$)}
	\label{TabResultsD13}
	\small
	\resizebox{!}{\tableheight}{
		\begin{tabular}{|l|l|l|llllll|llllll|llllll|}
			\hline
			\multicolumn{1}{|c|}{\multirow{3}{*}{$k$}} & \multicolumn{1}{c|}{\multirow{3}{*}{$f^*$}} & \multicolumn{1}{c|}{\multirow{3}{*}{$\overline{f}$}} & \multicolumn{6}{c|}{HPClust-inner} & \multicolumn{6}{c|}{HPClust-competitive} & \multicolumn{6}{c|}{HPClust-cooperative} \\ \cline{4-21}
			\multicolumn{1}{|c|}{} & \multicolumn{1}{c|}{} & \multicolumn{1}{c|}{} & \multicolumn{2}{c|}{$\varepsilon$} & \multicolumn{2}{c|}{$\overline{t}$} & \multicolumn{2}{c|}{$t$} & \multicolumn{2}{c|}{$\varepsilon$} & \multicolumn{2}{c|}{$\overline{t}$} & \multicolumn{2}{c|}{$t$} & \multicolumn{2}{c|}{$\varepsilon$} & \multicolumn{2}{c|}{$\overline{t}$} & \multicolumn{2}{c|}{$t$} \\ \cline{4-21}
			\multicolumn{1}{|c|}{} & \multicolumn{1}{c|}{} & \multicolumn{1}{c|}{} & \multicolumn{1}{c|}{med} & \multicolumn{1}{c|}{std} & \multicolumn{1}{c|}{med} & \multicolumn{1}{c|}{std} & \multicolumn{1}{c|}{med} & \multicolumn{1}{c|}{std} & \multicolumn{1}{c|}{med} & \multicolumn{1}{c|}{std} & \multicolumn{1}{c|}{med} & \multicolumn{1}{c|}{std} & \multicolumn{1}{c|}{med} & \multicolumn{1}{c|}{std} & \multicolumn{1}{c|}{med} & \multicolumn{1}{c|}{std} & \multicolumn{1}{c|}{med} & \multicolumn{1}{c|}{std} & \multicolumn{1}{c|}{med} & \multicolumn{1}{c|}{std} \\ \hline
			2 & 9.53913 & 2.23789 & 0.012 & 0.011 & 0.328 & 0.146 & 0.345 & 0.177 & 0.023 & 0.015 & 0.134 & 0.131 & 0.254 & 0.164 & 0.019 & 0.009 & 0.064 & 0.098 & 0.187 & 0.132 \\
			3 & 5.91077 & 1.35797 & 0.05 & 7.212 & 0.403 & 0.156 & 0.357 & 0.174 & 0.061 & 0.035 & 0.211 & 0.175 & 0.392 & 0.219 & 0.072 & 0.031 & 0.256 & 0.146 & 0.415 & 0.164 \\
			5 & 3.09885 & 0.70224 & 0.08 & 6.232 & 0.473 & 0.147 & 0.483 & 0.162 & 0.068 & 0.018 & 0.193 & 0.101 & 0.34 & 0.186 & 0.08 & 0.034 & 0.146 & 0.093 & 0.374 & 0.181 \\
			10 & 1.17247 & 0.27667 & 3.005 & 5.598 & 0.198 & 0.172 & 0.359 & 0.188 & 1.531 & 0.834 & 0.265 & 0.131 & 0.571 & 0.178 & 1.001 & 1.526 & 0.267 & 0.128 & 0.424 & 0.126 \\
			15 & 0.77637 & 0.1913 & 2.99 & 5.129 & 0.149 & 0.162 & 0.468 & 0.167 & 2.225 & 1.195 & 0.265 & 0.156 & 0.473 & 0.171 & 1.863 & 1.21 & 0.259 & 0.123 & 0.564 & 0.161 \\
			20 & 0.59809 & 0.14447 & 4.752 & 2.196 & 0.156 & 0.147 & 0.441 & 0.226 & 2.587 & 1.268 & 0.406 & 0.122 & 0.568 & 0.152 & 3.388 & 1.168 & 0.418 & 0.15 & 0.552 & 0.167 \\
			25 & 0.49616 & 0.1202 & 5.599 & 1.786 & 0.205 & 0.161 & 0.262 & 0.24 & 5.083 & 2.276 & 0.551 & 0.147 & 0.62 & 0.149 & 4.767 & 7.225 & 0.529 & 0.149 & 0.594 & 0.16 \\
			\hline
			\multicolumn{3}{|c|}{Mean:} & \textbf{2.355} & & \textbf{0.273} & & \textbf{0.388} & & \textbf{1.654} & & \textbf{0.289} & & \textbf{0.46} & & \textbf{1.598} & & \textbf{0.277} & & \textbf{0.444} & \\ \hline
		\end{tabular}
	}
	
	\bigskip
	
	\small
	\resizebox{!}{\tableheight}{
		\begin{tabular}{|l|l|l|llllll|llllll|llllll|}
			\hline
			\multicolumn{1}{|c|}{\multirow{3}{*}{$k$}} & \multicolumn{1}{c|}{\multirow{3}{*}{$f^*$}} & \multicolumn{1}{c|}{\multirow{3}{*}{$\overline{f}$}} & \multicolumn{6}{c|}{HPClust-hybrid} & \multicolumn{6}{c|}{Forgy K-means} & \multicolumn{6}{c|}{PBK-BDC} \\ \cline{4-21}
			\multicolumn{1}{|c|}{} & \multicolumn{1}{c|}{} & \multicolumn{1}{c|}{} & \multicolumn{2}{c|}{$\varepsilon$} & \multicolumn{2}{c|}{$\overline{t}$} & \multicolumn{2}{c|}{$t$} & \multicolumn{2}{c|}{$\varepsilon$} & \multicolumn{2}{c|}{$\overline{t}$} & \multicolumn{2}{c|}{$t$} & \multicolumn{2}{c|}{$\varepsilon$} & \multicolumn{2}{c|}{$\overline{t}$} & \multicolumn{2}{c|}{$t$} \\ \cline{4-21}
			\multicolumn{1}{|c|}{} & \multicolumn{1}{c|}{} & \multicolumn{1}{c|}{} & \multicolumn{1}{c|}{med} & \multicolumn{1}{c|}{std} & \multicolumn{1}{c|}{med} & \multicolumn{1}{c|}{std} & \multicolumn{1}{c|}{med} & \multicolumn{1}{c|}{std} & \multicolumn{1}{c|}{med} & \multicolumn{1}{c|}{std} & \multicolumn{1}{c|}{med} & \multicolumn{1}{c|}{std} & \multicolumn{1}{c|}{med} & \multicolumn{1}{c|}{std} & \multicolumn{1}{c|}{med} & \multicolumn{1}{c|}{std} & \multicolumn{1}{c|}{med} & \multicolumn{1}{c|}{std} & \multicolumn{1}{c|}{med} & \multicolumn{1}{c|}{std} \\ \hline
			2 & 9.53913 & 2.23789 & 0.015 & 0.008 & 0.131 & 0.128 & 0.316 & 0.158 & -0.0 & 0.0 & -- & -- & 0.036 & 0.015 & 0.001 & 0.0 & -- & -- & 0.017 & 0.004 \\
			3 & 5.91077 & 1.35797 & 0.089 & 0.033 & 0.21 & 0.182 & 0.495 & 0.193 & 0.0 & 0.0 & -- & -- & 0.206 & 0.062 & 1.679 & 29.518 & -- & -- & 0.057 & 0.015 \\
			5 & 3.09885 & 0.70224 & 0.076 & 0.026 & 0.187 & 0.136 & 0.378 & 0.147 & 12.069 & 7.238 & -- & -- & 0.253 & 0.167 & 80.754 & 51.871 & -- & -- & 0.117 & 0.046 \\
			10 & 1.17247 & 0.27667 & 0.928 & 0.717 & 0.276 & 0.177 & 0.581 & 0.186 & 12.363 & 17.442 & -- & -- & 0.877 & 0.705 & 38.376 & 20.585 & -- & -- & 0.511 & 0.163 \\
			15 & 0.77637 & 0.1913 & 1.91 & 3.185 & 0.355 & 0.2 & 0.611 & 0.24 & 16.822 & 9.168 & -- & -- & 1.898 & 0.58 & 43.769 & 12.082 & -- & -- & 0.819 & 0.229 \\
			20 & 0.59809 & 0.14447 & 3.687 & 1.27 & 0.395 & 0.204 & 0.675 & 0.297 & 25.504 & 6.434 & -- & -- & 3.253 & 1.371 & 46.453 & 19.54 & -- & -- & 1.822 & 0.357 \\
			25 & 0.49616 & 0.1202 & 5.135 & 5.994 & 0.495 & 0.14 & 0.643 & 0.248 & 37.787 & 10.213 & -- & -- & 7.378 & 2.996 & 53.319 & 12.025 & -- & -- & 2.015 & 0.471 \\
			\hline
			\multicolumn{3}{|c|}{Mean:} & \textbf{1.692} & & \textbf{0.293} & & \textbf{0.529} & & \textbf{14.935} & & \textbf{--} & & \textbf{1.986} & & \textbf{37.764} & & \textbf{--} & & \textbf{0.765} & \\ \hline
		\end{tabular}
	}
	
	\bigskip
	
	\caption{Clustering details with Online News Popularity}
	\label{TabDetailsD13}
	\resizebox{!}{\tableheight}{
		\begin{tabular}{|l|l|llll|llll|llll|lllll|l|l|}
			\hline
			\multicolumn{1}{|c|}{\multirow{2}{*}{$k$}} & \multicolumn{1}{c|}{\multirow{2}{*}{$n_{exec}$}} & \multicolumn{4}{c|}{HPClust-inner} & \multicolumn{4}{c|}{HPClust-competitive} & \multicolumn{4}{c|}{HPClust-cooperative} & \multicolumn{5}{c|}{HPClust-hybrid} & \multicolumn{1}{c|}{Forgy K-means} & \multicolumn{1}{c|}{PBK-BDC} \\ \cline{3-21}
			\multicolumn{1}{|c|}{} & \multicolumn{1}{c|}{} & \multicolumn{1}{c|}{$s$} & \multicolumn{1}{c|}{$n_{s}$} & \multicolumn{1}{c|}{$T$} & \multicolumn{1}{c|}{$n_{d}$} & \multicolumn{1}{c|}{$s$} & \multicolumn{1}{c|}{$n_{s}$} & \multicolumn{1}{c|}{$T$} & \multicolumn{1}{c|}{$n_{d}$} & \multicolumn{1}{c|}{$s$} & \multicolumn{1}{c|}{$n_{s}$} & \multicolumn{1}{c|}{$T$} & \multicolumn{1}{c|}{$n_{d}$} & \multicolumn{1}{c|}{$s$} & \multicolumn{1}{c|}{$n_{s}$} & \multicolumn{1}{c|}{$T_1$} & \multicolumn{1}{c|}{$T_2$} & \multicolumn{1}{c|}{$n_{d}$} & \multicolumn{1}{c|}{$n_{d}$} & \multicolumn{1}{c|}{$n_{d}$} \\
			\hline
			2 & 20 & 10000 & 98 & 0.7 & 8.7E+06 & 10000 & 382 & 0.7 & 4.5E+07 & 10000 & 281 & 0.7 & 4.4E+07 & 10000 & 455 & 0.63 & 0.07 & 4.5E+07 & 5.6E+05 & 4.4E+05 \\
			3 & 20 & 10000 & 54 & 0.7 & 1.3E+07 & 10000 & 266 & 0.7 & 5.9E+07 & 10000 & 258 & 0.7 & 6.2E+07 & 10000 & 312 & 0.14 & 0.56 & 6.0E+07 & 4.9E+06 & 2.0E+06 \\
			5 & 20 & 10000 & 74 & 0.7 & 2.0E+07 & 10000 & 130 & 0.7 & 6.6E+07 & 10000 & 191 & 0.7 & 6.8E+07 & 10000 & 168 & 0.467 & 0.233 & 6.6E+07 & 6.0E+06 & 4.1E+06 \\
			10 & 20 & 10000 & 32 & 0.7 & 3.0E+07 & 10000 & 97 & 0.7 & 7.3E+07 & 10000 & 60 & 0.7 & 7.5E+07 & 10000 & 84 & 0.49 & 0.21 & 6.6E+07 & 2.4E+07 & 1.9E+07 \\
			15 & 20 & 10000 & 26 & 0.7 & 4.2E+07 & 10000 & 26 & 0.7 & 7.7E+07 & 10000 & 33 & 0.7 & 7.9E+07 & 10000 & 14 & 0.047 & 0.653 & 5.7E+07 & 4.8E+07 & 3.4E+07 \\
			20 & 20 & 10000 & 14 & 0.7 & 4.7E+07 & 10000 & 16 & 0.7 & 8.2E+07 & 10000 & 16 & 0.7 & 8.0E+07 & 10000 & 20 & 0.63 & 0.07 & 8.1E+07 & 9.1E+07 & 6.6E+07 \\
			25 & 20 & 10000 & 5 & 0.7 & 5.1E+07 & 10000 & 10 & 0.7 & 8.4E+07 & 10000 & 11 & 0.7 & 8.5E+07 & 10000 & 13 & 0.653 & 0.047 & 8.2E+07 & 2.1E+08 & 8.2E+07 \\
			\hline
		\end{tabular}
	}
	
\end{table}

\newpage

%%%%%%%%%%%%%%%%%%%%%%%%%%%%%%%%%%%%%%%%%%%%%%%%%%%%%%%%%%%%%%%%%%%%%%%%%%%%%%%%%%%%%%%
%  END: Online News Popularity
%%%%%%%%%%%%%%%%%%%%%%%%%%%%%%%%%%%%%%%%%%%%%%%%%%%%%%%%%%%%%%%%%%%%%%%%%%%%%%%%%%%%%%%

%%%%%%%%%%%%%%%%%%%%%%%%%%%%%%%%%%%%%%%%%%%%%%%%%%%%%%%%%%%%%%%%%%%%%%%%%%%%%%%%%%%%%%%
%  START: Gas Sensor Array Drift
%%%%%%%%%%%%%%%%%%%%%%%%%%%%%%%%%%%%%%%%%%%%%%%%%%%%%%%%%%%%%%%%%%%%%%%%%%%%%%%%%%%%%%%
\subsection{Gas Sensor Array Drift}
Dimensions: $m$ = 13910, $n$ = 128.
\par
Description: this data set contains measurements from chemical sensors utilized in simulations for drift compensation in a discrimination task of different gases at various levels of concentrations.

\begin{table}[!htbp]
	\centering
	
	\caption{Summary of the results with Gas Sensor Array Drift ($\times10^{13}$)}
	\label{TabResultsD14}
	\small
	\resizebox{!}{\tableheight}{
		\begin{tabular}{|l|l|l|llllll|llllll|llllll|}
			\hline
			\multicolumn{1}{|c|}{\multirow{3}{*}{$k$}} & \multicolumn{1}{c|}{\multirow{3}{*}{$f^*$}} & \multicolumn{1}{c|}{\multirow{3}{*}{$\overline{f}$}} & \multicolumn{6}{c|}{HPClust-inner} & \multicolumn{6}{c|}{HPClust-competitive} & \multicolumn{6}{c|}{HPClust-cooperative} \\ \cline{4-21}
			\multicolumn{1}{|c|}{} & \multicolumn{1}{c|}{} & \multicolumn{1}{c|}{} & \multicolumn{2}{c|}{$\varepsilon$} & \multicolumn{2}{c|}{$\overline{t}$} & \multicolumn{2}{c|}{$t$} & \multicolumn{2}{c|}{$\varepsilon$} & \multicolumn{2}{c|}{$\overline{t}$} & \multicolumn{2}{c|}{$t$} & \multicolumn{2}{c|}{$\varepsilon$} & \multicolumn{2}{c|}{$\overline{t}$} & \multicolumn{2}{c|}{$t$} \\ \cline{4-21}
			\multicolumn{1}{|c|}{} & \multicolumn{1}{c|}{} & \multicolumn{1}{c|}{} & \multicolumn{1}{c|}{med} & \multicolumn{1}{c|}{std} & \multicolumn{1}{c|}{med} & \multicolumn{1}{c|}{std} & \multicolumn{1}{c|}{med} & \multicolumn{1}{c|}{std} & \multicolumn{1}{c|}{med} & \multicolumn{1}{c|}{std} & \multicolumn{1}{c|}{med} & \multicolumn{1}{c|}{std} & \multicolumn{1}{c|}{med} & \multicolumn{1}{c|}{std} & \multicolumn{1}{c|}{med} & \multicolumn{1}{c|}{std} & \multicolumn{1}{c|}{med} & \multicolumn{1}{c|}{std} & \multicolumn{1}{c|}{med} & \multicolumn{1}{c|}{std} \\ \hline
			2 & 7.91186 & 4.78811 & 0.088 & 0.064 & 0.46 & 0.394 & 0.844 & 0.552 & 0.127 & 0.091 & 0.438 & 0.614 & 1.283 & 0.614 & 0.203 & 0.084 & 0.462 & 0.576 & 1.118 & 0.573 \\
			3 & 5.02412 & 3.01636 & 0.133 & 9.866 & 0.627 & 0.548 & 0.896 & 0.572 & 0.194 & 0.11 & 0.44 & 0.398 & 0.924 & 0.582 & 0.255 & 0.109 & 0.631 & 0.445 & 1.119 & 0.582 \\
			5 & 3.22394 & 2.03175 & 6.935 & 3.508 & 0.727 & 0.638 & 0.797 & 0.572 & 7.107 & 3.625 & 0.718 & 0.48 & 1.151 & 0.544 & 0.178 & 1.763 & 0.589 & 0.446 & 1.145 & 0.525 \\
			10 & 1.65524 & 1.06767 & 3.155 & 2.805 & 0.169 & 0.367 & 0.797 & 0.527 & 0.434 & 1.262 & 0.325 & 0.167 & 1.193 & 0.46 & 0.274 & 1.403 & 0.285 & 0.214 & 1.128 & 0.562 \\
			15 & 1.13801 & 0.74507 & 4.665 & 3.27 & 0.192 & 0.362 & 1.024 & 0.552 & 0.202 & 1.306 & 0.453 & 0.199 & 1.269 & 0.492 & 0.418 & 1.634 & 0.43 & 0.095 & 1.307 & 0.445 \\
			20 & 0.87916 & 0.56988 & 3.129 & 2.638 & 0.717 & 0.544 & 1.319 & 0.604 & 1.84 & 0.792 & 1.023 & 0.415 & 1.488 & 0.422 & 2.375 & 1.02 & 0.853 & 0.342 & 1.488 & 0.396 \\
			25 & 0.72274 & 0.47044 & 4.598 & 1.838 & 0.396 & 0.568 & 1.008 & 0.548 & 2.526 & 0.797 & 1.42 & 0.45 & 1.685 & 0.492 & 2.684 & 0.897 & 1.165 & 0.357 & 1.528 & 0.406 \\
			\hline
			\multicolumn{3}{|c|}{Mean:} & \textbf{3.243} & & \textbf{0.47} & & \textbf{0.955} & & \textbf{1.776} & & \textbf{0.688} & & \textbf{1.285} & & \textbf{0.912} & & \textbf{0.631} & & \textbf{1.262} & \\ \hline
		\end{tabular}
	}
	
	\bigskip
	
	\small
	\resizebox{!}{\tableheight}{
		\begin{tabular}{|l|l|l|llllll|llllll|llllll|}
			\hline
			\multicolumn{1}{|c|}{\multirow{3}{*}{$k$}} & \multicolumn{1}{c|}{\multirow{3}{*}{$f^*$}} & \multicolumn{1}{c|}{\multirow{3}{*}{$\overline{f}$}} & \multicolumn{6}{c|}{HPClust-hybrid} & \multicolumn{6}{c|}{Forgy K-means} & \multicolumn{6}{c|}{PBK-BDC} \\ \cline{4-21}
			\multicolumn{1}{|c|}{} & \multicolumn{1}{c|}{} & \multicolumn{1}{c|}{} & \multicolumn{2}{c|}{$\varepsilon$} & \multicolumn{2}{c|}{$\overline{t}$} & \multicolumn{2}{c|}{$t$} & \multicolumn{2}{c|}{$\varepsilon$} & \multicolumn{2}{c|}{$\overline{t}$} & \multicolumn{2}{c|}{$t$} & \multicolumn{2}{c|}{$\varepsilon$} & \multicolumn{2}{c|}{$\overline{t}$} & \multicolumn{2}{c|}{$t$} \\ \cline{4-21}
			\multicolumn{1}{|c|}{} & \multicolumn{1}{c|}{} & \multicolumn{1}{c|}{} & \multicolumn{1}{c|}{med} & \multicolumn{1}{c|}{std} & \multicolumn{1}{c|}{med} & \multicolumn{1}{c|}{std} & \multicolumn{1}{c|}{med} & \multicolumn{1}{c|}{std} & \multicolumn{1}{c|}{med} & \multicolumn{1}{c|}{std} & \multicolumn{1}{c|}{med} & \multicolumn{1}{c|}{std} & \multicolumn{1}{c|}{med} & \multicolumn{1}{c|}{std} & \multicolumn{1}{c|}{med} & \multicolumn{1}{c|}{std} & \multicolumn{1}{c|}{med} & \multicolumn{1}{c|}{std} & \multicolumn{1}{c|}{med} & \multicolumn{1}{c|}{std} \\ \hline
			2 & 7.91186 & 4.78811 & 0.192 & 0.1 & 0.394 & 0.516 & 0.952 & 0.654 & -0.0 & 0.0 & -- & -- & 0.053 & 0.007 & 0.029 & 0.048 & -- & -- & 0.028 & 0.009 \\
			3 & 5.02412 & 3.01636 & 0.245 & 0.11 & 0.451 & 0.523 & 1.044 & 0.566 & -0.001 & 0.0 & -- & -- & 0.104 & 0.016 & 0.03 & 0.034 & -- & -- & 0.088 & 0.024 \\
			5 & 3.22394 & 2.03175 & 0.366 & 3.35 & 0.817 & 0.525 & 1.245 & 0.557 & 8.108 & 0.387 & -- & -- & 0.247 & 0.064 & 8.156 & 0.394 & -- & -- & 0.118 & 0.03 \\
			10 & 1.65524 & 1.06767 & 0.232 & 0.696 & 0.287 & 0.21 & 1.592 & 0.468 & 37.905 & 17.254 & -- & -- & 0.595 & 0.415 & 41.32 & 13.482 & -- & -- & 0.371 & 0.23 \\
			15 & 1.13801 & 0.74507 & -0.175 & 1.603 & 0.46 & 0.184 & 1.887 & 0.59 & 27.472 & 10.362 & -- & -- & 1.653 & 0.679 & 30.525 & 24.91 & -- & -- & 0.704 & 0.319 \\
			20 & 0.87916 & 0.56988 & 2.03 & 1.033 & 1.132 & 0.391 & 1.673 & 0.48 & 45.732 & 7.967 & -- & -- & 1.684 & 0.509 & 45.904 & 6.855 & -- & -- & 0.982 & 0.472 \\
			25 & 0.72274 & 0.47044 & 2.639 & 0.954 & 1.54 & 0.425 & 1.951 & 0.446 & 50.936 & 12.149 & -- & -- & 2.544 & 1.197 & 52.691 & 14.414 & -- & -- & 1.796 & 0.664 \\
			\hline
			\multicolumn{3}{|c|}{Mean:} & \textbf{0.79} & & \textbf{0.726} & & \textbf{1.478} & & \textbf{24.307} & & \textbf{--} & & \textbf{0.983} & & \textbf{25.522} & & \textbf{--} & & \textbf{0.584} & \\ \hline
		\end{tabular}
	}
	
	\bigskip
	
	\caption{Clustering details with Gas Sensor Array Drift}
	\label{TabDetailsD14}
	\resizebox{!}{\tableheight}{
		\begin{tabular}{|l|l|llll|llll|llll|lllll|l|l|}
			\hline
			\multicolumn{1}{|c|}{\multirow{2}{*}{$k$}} & \multicolumn{1}{c|}{\multirow{2}{*}{$n_{exec}$}} & \multicolumn{4}{c|}{HPClust-inner} & \multicolumn{4}{c|}{HPClust-competitive} & \multicolumn{4}{c|}{HPClust-cooperative} & \multicolumn{5}{c|}{HPClust-hybrid} & \multicolumn{1}{c|}{Forgy K-means} & \multicolumn{1}{c|}{PBK-BDC} \\ \cline{3-21}
			\multicolumn{1}{|c|}{} & \multicolumn{1}{c|}{} & \multicolumn{1}{c|}{$s$} & \multicolumn{1}{c|}{$n_{s}$} & \multicolumn{1}{c|}{$T$} & \multicolumn{1}{c|}{$n_{d}$} & \multicolumn{1}{c|}{$s$} & \multicolumn{1}{c|}{$n_{s}$} & \multicolumn{1}{c|}{$T$} & \multicolumn{1}{c|}{$n_{d}$} & \multicolumn{1}{c|}{$s$} & \multicolumn{1}{c|}{$n_{s}$} & \multicolumn{1}{c|}{$T$} & \multicolumn{1}{c|}{$n_{d}$} & \multicolumn{1}{c|}{$s$} & \multicolumn{1}{c|}{$n_{s}$} & \multicolumn{1}{c|}{$T_1$} & \multicolumn{1}{c|}{$T_2$} & \multicolumn{1}{c|}{$n_{d}$} & \multicolumn{1}{c|}{$n_{d}$} & \multicolumn{1}{c|}{$n_{d}$} \\
			\hline
			2 & 30 & 9000 & 140 & 2.0 & 2.3E+07 & 9000 & 810 & 2.0 & 9.0E+07 & 9000 & 606 & 2.0 & 9.2E+07 & 9000 & 642 & 1.0 & 1.0 & 9.2E+07 & 5.0E+05 & 2.8E+05 \\
			3 & 30 & 9000 & 128 & 2.0 & 2.9E+07 & 9000 & 424 & 2.0 & 9.6E+07 & 9000 & 468 & 2.0 & 9.9E+07 & 9000 & 544 & 0.867 & 1.133 & 1.0E+08 & 1.0E+06 & 6.2E+05 \\
			5 & 30 & 9000 & 84 & 2.0 & 3.8E+07 & 9000 & 300 & 2.0 & 1.0E+08 & 9000 & 292 & 2.0 & 1.0E+08 & 9000 & 292 & 0.867 & 1.133 & 9.8E+07 & 2.5E+06 & 1.3E+06 \\
			10 & 30 & 9000 & 46 & 2.0 & 5.5E+07 & 9000 & 102 & 2.0 & 1.1E+08 & 9000 & 106 & 2.0 & 1.1E+08 & 9000 & 144 & 1.4 & 0.6 & 1.0E+08 & 6.1E+06 & 3.9E+06 \\
			15 & 30 & 9000 & 42 & 2.0 & 5.8E+07 & 9000 & 65 & 2.0 & 1.1E+08 & 9000 & 65 & 2.0 & 1.1E+08 & 9000 & 96 & 1.867 & 0.133 & 1.0E+08 & 1.6E+07 & 7.6E+06 \\
			20 & 30 & 9000 & 40 & 2.0 & 6.2E+07 & 9000 & 44 & 2.0 & 1.1E+08 & 9000 & 38 & 2.0 & 1.1E+08 & 9000 & 30 & 0.2 & 1.8 & 7.9E+07 & 1.8E+07 & 1.1E+07 \\
			25 & 30 & 9000 & 20 & 2.0 & 6.5E+07 & 9000 & 29 & 2.0 & 1.1E+08 & 9000 & 30 & 2.0 & 1.1E+08 & 9000 & 16 & 0.8 & 1.2 & 7.3E+07 & 2.7E+07 & 1.9E+07 \\
			\hline
		\end{tabular}
	}
	
\end{table}

\newpage

%%%%%%%%%%%%%%%%%%%%%%%%%%%%%%%%%%%%%%%%%%%%%%%%%%%%%%%%%%%%%%%%%%%%%%%%%%%%%%%%%%%%%%%
%  END: Gas Sensor Array Drift
%%%%%%%%%%%%%%%%%%%%%%%%%%%%%%%%%%%%%%%%%%%%%%%%%%%%%%%%%%%%%%%%%%%%%%%%%%%%%%%%%%%%%%%

\newpage

%%%%%%%%%%%%%%%%%%%%%%%%%%%%%%%%%%%%%%%%%%%%%%%%%%%%%%%%%%%%%%%%%%%%%%%%%%%%%%%%%%%%%%%
%  START: 3D Road Network
%%%%%%%%%%%%%%%%%%%%%%%%%%%%%%%%%%%%%%%%%%%%%%%%%%%%%%%%%%%%%%%%%%%%%%%%%%%%%%%%%%%%%%%
\subsection{3D Road Network}
Dimensions: $m$ = 434874, $n$ = 3.
\par
Description: 3D road network from Denmark with highly accurate elevation information which contains longitude, latitude and altitude for each road segment or edge in the graph. Usually this data set used in eco-routing and fuel/Co2-estimation routing algorithms.

\begin{table}[!htbp]
	\centering
	
	\caption{Summary of the results with 3D Road Network ($\times10^{6}$)}
	\label{TabResultsD15}
	\small
	\resizebox{!}{\tableheight}{
		\begin{tabular}{|l|l|l|llllll|llllll|llllll|}
			\hline
			\multicolumn{1}{|c|}{\multirow{3}{*}{$k$}} & \multicolumn{1}{c|}{\multirow{3}{*}{$f^*$}} & \multicolumn{1}{c|}{\multirow{3}{*}{$\overline{f}$}} & \multicolumn{6}{c|}{HPClust-inner} & \multicolumn{6}{c|}{HPClust-competitive} & \multicolumn{6}{c|}{HPClust-cooperative} \\ \cline{4-21}
			\multicolumn{1}{|c|}{} & \multicolumn{1}{c|}{} & \multicolumn{1}{c|}{} & \multicolumn{2}{c|}{$\varepsilon$} & \multicolumn{2}{c|}{$\overline{t}$} & \multicolumn{2}{c|}{$t$} & \multicolumn{2}{c|}{$\varepsilon$} & \multicolumn{2}{c|}{$\overline{t}$} & \multicolumn{2}{c|}{$t$} & \multicolumn{2}{c|}{$\varepsilon$} & \multicolumn{2}{c|}{$\overline{t}$} & \multicolumn{2}{c|}{$t$} \\ \cline{4-21}
			\multicolumn{1}{|c|}{} & \multicolumn{1}{c|}{} & \multicolumn{1}{c|}{} & \multicolumn{1}{c|}{med} & \multicolumn{1}{c|}{std} & \multicolumn{1}{c|}{med} & \multicolumn{1}{c|}{std} & \multicolumn{1}{c|}{med} & \multicolumn{1}{c|}{std} & \multicolumn{1}{c|}{med} & \multicolumn{1}{c|}{std} & \multicolumn{1}{c|}{med} & \multicolumn{1}{c|}{std} & \multicolumn{1}{c|}{med} & \multicolumn{1}{c|}{std} & \multicolumn{1}{c|}{med} & \multicolumn{1}{c|}{std} & \multicolumn{1}{c|}{med} & \multicolumn{1}{c|}{std} & \multicolumn{1}{c|}{med} & \multicolumn{1}{c|}{std} \\ \hline
			2 & 49.13298 & 11.15303 & 0.004 & 0.006 & 0.257 & 0.123 & 0.265 & 0.125 & 0.005 & 0.015 & 0.195 & 0.151 & 0.304 & 0.149 & 0.008 & 0.005 & 0.177 & 0.107 & 0.228 & 0.131 \\
			3 & 22.77818 & 5.1707 & 0.005 & 0.007 & 0.19 & 0.133 & 0.318 & 0.141 & 0.011 & 0.011 & 0.15 & 0.102 & 0.285 & 0.145 & 0.015 & 0.015 & 0.13 & 0.093 & 0.263 & 0.145 \\
			5 & 8.82574 & 1.99891 & 0.02 & 0.014 & 0.182 & 0.146 & 0.294 & 0.136 & 0.021 & 0.021 & 0.249 & 0.115 & 0.341 & 0.121 & 0.018 & 0.024 & 0.182 & 0.099 & 0.346 & 0.139 \\
			10 & 2.56661 & 0.58256 & 0.167 & 0.116 & 0.167 & 0.113 & 0.234 & 0.122 & 0.159 & 0.103 & 0.301 & 0.137 & 0.418 & 0.142 & 0.164 & 0.185 & 0.315 & 0.099 & 0.419 & 0.121 \\
			15 & 1.27069 & 0.28889 & 0.334 & 0.38 & 0.276 & 0.104 & 0.382 & 0.144 & 0.223 & 0.377 & 0.502 & 0.346 & 0.504 & 0.332 & 0.343 & 0.331 & 0.442 & 0.198 & 0.503 & 0.193 \\
			20 & 0.80865 & 0.18573 & 1.243 & 0.823 & 0.287 & 0.109 & 0.343 & 0.098 & 0.542 & 0.652 & 0.46 & 0.168 & 0.591 & 0.184 & 0.382 & 0.644 & 0.463 & 0.208 & 0.541 & 0.191 \\
			25 & 0.59259 & 0.13625 & 1.038 & 0.84 & 0.242 & 0.13 & 0.405 & 0.118 & 0.557 & 0.489 & 0.603 & 0.264 & 0.757 & 0.351 & 0.588 & 0.487 & 0.603 & 0.255 & 0.755 & 0.276 \\
			\hline
			\multicolumn{3}{|c|}{Mean:} & \textbf{0.402} & & \textbf{0.229} & & \textbf{0.32} & & \textbf{0.217} & & \textbf{0.351} & & \textbf{0.457} & & \textbf{0.217} & & \textbf{0.33} & & \textbf{0.436} & \\ \hline
		\end{tabular}
	}
	
	\bigskip
	
	\small
	\resizebox{!}{\tableheight}{
		\begin{tabular}{|l|l|l|llllll|llllll|llllll|}
			\hline
			\multicolumn{1}{|c|}{\multirow{3}{*}{$k$}} & \multicolumn{1}{c|}{\multirow{3}{*}{$f^*$}} & \multicolumn{1}{c|}{\multirow{3}{*}{$\overline{f}$}} & \multicolumn{6}{c|}{HPClust-hybrid} & \multicolumn{6}{c|}{Forgy K-means} & \multicolumn{6}{c|}{PBK-BDC} \\ \cline{4-21}
			\multicolumn{1}{|c|}{} & \multicolumn{1}{c|}{} & \multicolumn{1}{c|}{} & \multicolumn{2}{c|}{$\varepsilon$} & \multicolumn{2}{c|}{$\overline{t}$} & \multicolumn{2}{c|}{$t$} & \multicolumn{2}{c|}{$\varepsilon$} & \multicolumn{2}{c|}{$\overline{t}$} & \multicolumn{2}{c|}{$t$} & \multicolumn{2}{c|}{$\varepsilon$} & \multicolumn{2}{c|}{$\overline{t}$} & \multicolumn{2}{c|}{$t$} \\ \cline{4-21}
			\multicolumn{1}{|c|}{} & \multicolumn{1}{c|}{} & \multicolumn{1}{c|}{} & \multicolumn{1}{c|}{med} & \multicolumn{1}{c|}{std} & \multicolumn{1}{c|}{med} & \multicolumn{1}{c|}{std} & \multicolumn{1}{c|}{med} & \multicolumn{1}{c|}{std} & \multicolumn{1}{c|}{med} & \multicolumn{1}{c|}{std} & \multicolumn{1}{c|}{med} & \multicolumn{1}{c|}{std} & \multicolumn{1}{c|}{med} & \multicolumn{1}{c|}{std} & \multicolumn{1}{c|}{med} & \multicolumn{1}{c|}{std} & \multicolumn{1}{c|}{med} & \multicolumn{1}{c|}{std} & \multicolumn{1}{c|}{med} & \multicolumn{1}{c|}{std} \\ \hline
			2 & 49.13298 & 11.15303 & 0.01 & 0.011 & 0.157 & 0.115 & 0.195 & 0.154 & 0.0 & 0.0 & -- & -- & 0.113 & 0.033 & 0.0 & 0.0 & -- & -- & 0.062 & 0.007 \\
			3 & 22.77818 & 5.1707 & 0.012 & 0.012 & 0.123 & 0.082 & 0.233 & 0.121 & 0.0 & 0.0 & -- & -- & 0.187 & 0.065 & 0.0 & 77.393 & -- & -- & 0.106 & 0.009 \\
			5 & 8.82574 & 1.99891 & 0.031 & 0.027 & 0.215 & 0.106 & 0.309 & 0.129 & 0.0 & 0.0 & -- & -- & 0.517 & 0.087 & 77.246 & 43.749 & -- & -- & 0.268 & 0.034 \\
			10 & 2.56661 & 0.58256 & 0.227 & 0.176 & 0.546 & 0.188 & 0.503 & 0.196 & 0.008 & 0.0 & -- & -- & 5.994 & 0.261 & 62.553 & 44.418 & -- & -- & 1.802 & 0.34 \\
			15 & 1.27069 & 0.28889 & 0.224 & 0.25 & 0.455 & 0.443 & 0.501 & 0.48 & 0.002 & 0.0 & -- & -- & 7.558 & 0.788 & 57.087 & 42.217 & -- & -- & 2.768 & 0.357 \\
			20 & 0.80865 & 0.18573 & 0.468 & 0.501 & 0.43 & 0.315 & 0.797 & 0.418 & 0.005 & 0.0 & -- & -- & 25.175 & 1.893 & 42.58 & 23.013 & -- & -- & 4.995 & 0.876 \\
			25 & 0.59259 & 0.13625 & 0.523 & 0.481 & 0.685 & 0.483 & 0.874 & 0.6 & 1.615 & 0.25 & -- & -- & 24.839 & 1.744 & 45.092 & 33.97 & -- & -- & 6.186 & 0.732 \\
			\hline
			\multicolumn{3}{|c|}{Mean:} & \textbf{0.214} & & \textbf{0.373} & & \textbf{0.487} & & \textbf{0.233} & & \textbf{--} & & \textbf{9.198} & & \textbf{40.651} & & \textbf{--} & & \textbf{2.313} & \\ \hline
		\end{tabular}
	}
	
	\bigskip
	
	\caption{Clustering details with 3D Road Network}
	\label{TabDetailsD15}
	\resizebox{!}{\tableheight}{
		\begin{tabular}{|l|l|llll|llll|llll|lllll|l|l|}
			\hline
			\multicolumn{1}{|c|}{\multirow{2}{*}{$k$}} & \multicolumn{1}{c|}{\multirow{2}{*}{$n_{exec}$}} & \multicolumn{4}{c|}{HPClust-inner} & \multicolumn{4}{c|}{HPClust-competitive} & \multicolumn{4}{c|}{HPClust-cooperative} & \multicolumn{5}{c|}{HPClust-hybrid} & \multicolumn{1}{c|}{Forgy K-means} & \multicolumn{1}{c|}{PBK-BDC} \\ \cline{3-21}
			\multicolumn{1}{|c|}{} & \multicolumn{1}{c|}{} & \multicolumn{1}{c|}{$s$} & \multicolumn{1}{c|}{$n_{s}$} & \multicolumn{1}{c|}{$T$} & \multicolumn{1}{c|}{$n_{d}$} & \multicolumn{1}{c|}{$s$} & \multicolumn{1}{c|}{$n_{s}$} & \multicolumn{1}{c|}{$T$} & \multicolumn{1}{c|}{$n_{d}$} & \multicolumn{1}{c|}{$s$} & \multicolumn{1}{c|}{$n_{s}$} & \multicolumn{1}{c|}{$T$} & \multicolumn{1}{c|}{$n_{d}$} & \multicolumn{1}{c|}{$s$} & \multicolumn{1}{c|}{$n_{s}$} & \multicolumn{1}{c|}{$T_1$} & \multicolumn{1}{c|}{$T_2$} & \multicolumn{1}{c|}{$n_{d}$} & \multicolumn{1}{c|}{$n_{d}$} & \multicolumn{1}{c|}{$n_{d}$} \\
			\hline
			2 & 40 & 100000 & 17 & 0.5 & 1.8E+07 & 100000 & 100 & 0.5 & 9.6E+07 & 100000 & 72 & 0.5 & 9.6E+07 & 100000 & 57 & 0.033 & 0.467 & 1.0E+08 & 2.1E+07 & 1.8E+07 \\
			3 & 40 & 100000 & 15 & 0.5 & 2.6E+07 & 100000 & 84 & 0.5 & 1.6E+08 & 100000 & 78 & 0.5 & 1.6E+08 & 100000 & 68 & 0.467 & 0.033 & 1.6E+08 & 4.2E+07 & 4.2E+07 \\
			5 & 40 & 100000 & 14 & 0.5 & 5.7E+07 & 100000 & 74 & 0.5 & 2.8E+08 & 100000 & 71 & 0.5 & 2.7E+08 & 100000 & 52 & 0.117 & 0.383 & 2.7E+08 & 1.5E+08 & 1.3E+08 \\
			10 & 40 & 100000 & 8 & 0.5 & 1.7E+08 & 100000 & 26 & 0.5 & 6.0E+08 & 100000 & 24 & 0.5 & 5.6E+08 & 100000 & 10 & 0.15 & 0.35 & 4.3E+08 & 2.2E+09 & 1.2E+09 \\
			15 & 40 & 100000 & 6 & 0.5 & 2.6E+08 & 100000 & 8 & 0.5 & 8.5E+08 & 100000 & 10 & 0.5 & 8.1E+08 & 100000 & 11 & 0.45 & 0.05 & 7.9E+08 & 3.0E+09 & 2.4E+09 \\
			20 & 40 & 100000 & 6 & 0.5 & 3.6E+08 & 100000 & 6 & 0.5 & 9.5E+08 & 100000 & 6 & 0.5 & 9.9E+08 & 100000 & 6 & 0.333 & 0.167 & 9.7E+08 & 1.1E+10 & 5.0E+09 \\
			25 & 40 & 100000 & 5 & 0.5 & 3.6E+08 & 100000 & 5 & 0.5 & 1.2E+09 & 100000 & 5 & 0.5 & 1.3E+09 & 100000 & 5 & 0.317 & 0.183 & 1.2E+09 & 1.1E+10 & 7.8E+09 \\
			\hline
		\end{tabular}
	}
	
\end{table}

\newpage

%%%%%%%%%%%%%%%%%%%%%%%%%%%%%%%%%%%%%%%%%%%%%%%%%%%%%%%%%%%%%%%%%%%%%%%%%%%%%%%%%%%%%%%
%  END: 3D Road Network
%%%%%%%%%%%%%%%%%%%%%%%%%%%%%%%%%%%%%%%%%%%%%%%%%%%%%%%%%%%%%%%%%%%%%%%%%%%%%%%%%%%%%%%

%%%%%%%%%%%%%%%%%%%%%%%%%%%%%%%%%%%%%%%%%%%%%%%%%%%%%%%%%%%%%%%%%%%%%%%%%%%%%%%%%%%%%%%
%  START: Skin Segmentation
%%%%%%%%%%%%%%%%%%%%%%%%%%%%%%%%%%%%%%%%%%%%%%%%%%%%%%%%%%%%%%%%%%%%%%%%%%%%%%%%%%%%%%%
\subsection{Skin Segmentation}
Dimensions: $m$ = 245057, $n$ = 3.
\par
Description: Skin and Nonskin dataset is generated using skin textures from face images of diversity of age, gender, and race people and constructed over B, G, R color space.

\begin{table}[!htbp]
	\centering
	
	\caption{Summary of the results with Skin Segmentation ($\times10^{9}$)}
	\label{TabResultsD16}
	\small
	\resizebox{!}{\tableheight}{
		\begin{tabular}{|l|l|l|llllll|llllll|llllll|}
			\hline
			\multicolumn{1}{|c|}{\multirow{3}{*}{$k$}} & \multicolumn{1}{c|}{\multirow{3}{*}{$f^*$}} & \multicolumn{1}{c|}{\multirow{3}{*}{$\overline{f}$}} & \multicolumn{6}{c|}{HPClust-inner} & \multicolumn{6}{c|}{HPClust-competitive} & \multicolumn{6}{c|}{HPClust-cooperative} \\ \cline{4-21}
			\multicolumn{1}{|c|}{} & \multicolumn{1}{c|}{} & \multicolumn{1}{c|}{} & \multicolumn{2}{c|}{$\varepsilon$} & \multicolumn{2}{c|}{$\overline{t}$} & \multicolumn{2}{c|}{$t$} & \multicolumn{2}{c|}{$\varepsilon$} & \multicolumn{2}{c|}{$\overline{t}$} & \multicolumn{2}{c|}{$t$} & \multicolumn{2}{c|}{$\varepsilon$} & \multicolumn{2}{c|}{$\overline{t}$} & \multicolumn{2}{c|}{$t$} \\ \cline{4-21}
			\multicolumn{1}{|c|}{} & \multicolumn{1}{c|}{} & \multicolumn{1}{c|}{} & \multicolumn{1}{c|}{med} & \multicolumn{1}{c|}{std} & \multicolumn{1}{c|}{med} & \multicolumn{1}{c|}{std} & \multicolumn{1}{c|}{med} & \multicolumn{1}{c|}{std} & \multicolumn{1}{c|}{med} & \multicolumn{1}{c|}{std} & \multicolumn{1}{c|}{med} & \multicolumn{1}{c|}{std} & \multicolumn{1}{c|}{med} & \multicolumn{1}{c|}{std} & \multicolumn{1}{c|}{med} & \multicolumn{1}{c|}{std} & \multicolumn{1}{c|}{med} & \multicolumn{1}{c|}{std} & \multicolumn{1}{c|}{med} & \multicolumn{1}{c|}{std} \\ \hline
			2 & 1.32236 & 0.04216 & 0.031 & 0.019 & 0.105 & 0.045 & 0.097 & 0.054 & 0.035 & 0.013 & 0.04 & 0.026 & 0.105 & 0.055 & 0.034 & 0.022 & 0.028 & 0.041 & 0.106 & 0.05 \\
			3 & 0.89362 & 0.02822 & 0.054 & 0.032 & 0.058 & 0.052 & 0.084 & 0.06 & 0.043 & 0.024 & 0.066 & 0.035 & 0.099 & 0.049 & 0.038 & 0.03 & 0.038 & 0.046 & 0.107 & 0.065 \\
			5 & 0.50203 & 0.0161 & 0.124 & 2.491 & 0.048 & 0.046 & 0.134 & 0.053 & 0.073 & 0.586 & 0.018 & 0.013 & 0.143 & 0.056 & 0.078 & 0.815 & 0.018 & 0.025 & 0.104 & 0.051 \\
			10 & 0.25121 & 0.00817 & 6.804 & 5.439 & 0.039 & 0.075 & 0.113 & 0.061 & 0.212 & 1.399 & 0.026 & 0.016 & 0.142 & 0.053 & 0.247 & 2.335 & 0.023 & 0.037 & 0.136 & 0.064 \\
			15 & 0.16964 & 0.00544 & 3.665 & 2.287 & 0.064 & 0.044 & 0.128 & 0.059 & 1.201 & 1.962 & 0.046 & 0.029 & 0.142 & 0.042 & 0.734 & 2.771 & 0.038 & 0.032 & 0.12 & 0.053 \\
			20 & 0.12615 & 0.004 & 4.366 & 2.928 & 0.11 & 0.055 & 0.126 & 0.054 & 2.311 & 1.567 & 0.092 & 0.051 & 0.138 & 0.051 & 2.202 & 2.534 & 0.07 & 0.034 & 0.121 & 0.047 \\
			25 & 0.10228 & 0.00335 & 5.333 & 2.735 & 0.067 & 0.035 & 0.104 & 0.051 & 3.485 & 1.754 & 0.052 & 0.016 & 0.15 & 0.051 & 4.461 & 1.755 & 0.056 & 0.034 & 0.155 & 0.046 \\
			\hline
			\multicolumn{3}{|c|}{Mean:} & \textbf{2.911} & & \textbf{0.07} & & \textbf{0.112} & & \textbf{1.052} & & \textbf{0.049} & & \textbf{0.131} & & \textbf{1.113} & & \textbf{0.039} & & \textbf{0.121} & \\ \hline
		\end{tabular}
	}
	
	\bigskip
	
	\small
	\resizebox{!}{\tableheight}{
		\begin{tabular}{|l|l|l|llllll|llllll|llllll|}
			\hline
			\multicolumn{1}{|c|}{\multirow{3}{*}{$k$}} & \multicolumn{1}{c|}{\multirow{3}{*}{$f^*$}} & \multicolumn{1}{c|}{\multirow{3}{*}{$\overline{f}$}} & \multicolumn{6}{c|}{HPClust-hybrid} & \multicolumn{6}{c|}{Forgy K-means} & \multicolumn{6}{c|}{PBK-BDC} \\ \cline{4-21}
			\multicolumn{1}{|c|}{} & \multicolumn{1}{c|}{} & \multicolumn{1}{c|}{} & \multicolumn{2}{c|}{$\varepsilon$} & \multicolumn{2}{c|}{$\overline{t}$} & \multicolumn{2}{c|}{$t$} & \multicolumn{2}{c|}{$\varepsilon$} & \multicolumn{2}{c|}{$\overline{t}$} & \multicolumn{2}{c|}{$t$} & \multicolumn{2}{c|}{$\varepsilon$} & \multicolumn{2}{c|}{$\overline{t}$} & \multicolumn{2}{c|}{$t$} \\ \cline{4-21}
			\multicolumn{1}{|c|}{} & \multicolumn{1}{c|}{} & \multicolumn{1}{c|}{} & \multicolumn{1}{c|}{med} & \multicolumn{1}{c|}{std} & \multicolumn{1}{c|}{med} & \multicolumn{1}{c|}{std} & \multicolumn{1}{c|}{med} & \multicolumn{1}{c|}{std} & \multicolumn{1}{c|}{med} & \multicolumn{1}{c|}{std} & \multicolumn{1}{c|}{med} & \multicolumn{1}{c|}{std} & \multicolumn{1}{c|}{med} & \multicolumn{1}{c|}{std} & \multicolumn{1}{c|}{med} & \multicolumn{1}{c|}{std} & \multicolumn{1}{c|}{med} & \multicolumn{1}{c|}{std} & \multicolumn{1}{c|}{med} & \multicolumn{1}{c|}{std} \\ \hline
			2 & 1.32236 & 0.04216 & 0.036 & 0.022 & 0.034 & 0.04 & 0.144 & 0.057 & -0.0 & 0.0 & -- & -- & 0.042 & 0.008 & -0.0 & 0.007 & -- & -- & 0.014 & 0.001 \\
			3 & 0.89362 & 0.02822 & 0.069 & 0.031 & 0.041 & 0.042 & 0.106 & 0.054 & -0.001 & 0.0 & -- & -- & 0.081 & 0.032 & 0.003 & 59.847 & -- & -- & 0.025 & 0.003 \\
			5 & 0.50203 & 0.0161 & 0.092 & 0.298 & 0.021 & 0.022 & 0.124 & 0.062 & 1.65 & 6.344 & -- & -- & 0.117 & 0.036 & 18.075 & 22.33 & -- & -- & 0.036 & 0.003 \\
			10 & 0.25121 & 0.00817 & 0.202 & 2.214 & 0.029 & 0.03 & 0.176 & 0.059 & 9.122 & 7.003 & -- & -- & 0.219 & 0.062 & 26.085 & 8.432 & -- & -- & 0.062 & 0.007 \\
			15 & 0.16964 & 0.00544 & 0.888 & 2.013 & 0.04 & 0.039 & 0.152 & 0.051 & 13.463 & 7.936 & -- & -- & 0.389 & 0.187 & 29.36 & 13.524 & -- & -- & 0.104 & 0.011 \\
			20 & 0.12615 & 0.004 & 2.121 & 1.667 & 0.089 & 0.04 & 0.151 & 0.033 & 16.816 & 7.379 & -- & -- & 0.548 & 0.224 & 34.997 & 18.226 & -- & -- & 0.145 & 0.018 \\
			25 & 0.10228 & 0.00335 & 3.727 & 1.445 & 0.061 & 0.023 & 0.164 & 0.049 & 22.066 & 6.921 & -- & -- & 0.698 & 0.237 & 35.345 & 16.71 & -- & -- & 0.182 & 0.023 \\
			\hline
			\multicolumn{3}{|c|}{Mean:} & \textbf{1.019} & & \textbf{0.045} & & \textbf{0.145} & & \textbf{9.016} & & \textbf{--} & & \textbf{0.299} & & \textbf{20.552} & & \textbf{--} & & \textbf{0.081} & \\ \hline
		\end{tabular}
	}
	
	\bigskip
	
	\caption{Clustering details with Skin Segmentation}
	\label{TabDetailsD16}
	\resizebox{!}{\tableheight}{
		\begin{tabular}{|l|l|llll|llll|llll|lllll|l|l|}
			\hline
			\multicolumn{1}{|c|}{\multirow{2}{*}{$k$}} & \multicolumn{1}{c|}{\multirow{2}{*}{$n_{exec}$}} & \multicolumn{4}{c|}{HPClust-inner} & \multicolumn{4}{c|}{HPClust-competitive} & \multicolumn{4}{c|}{HPClust-cooperative} & \multicolumn{5}{c|}{HPClust-hybrid} & \multicolumn{1}{c|}{Forgy K-means} & \multicolumn{1}{c|}{PBK-BDC} \\ \cline{3-21}
			\multicolumn{1}{|c|}{} & \multicolumn{1}{c|}{} & \multicolumn{1}{c|}{$s$} & \multicolumn{1}{c|}{$n_{s}$} & \multicolumn{1}{c|}{$T$} & \multicolumn{1}{c|}{$n_{d}$} & \multicolumn{1}{c|}{$s$} & \multicolumn{1}{c|}{$n_{s}$} & \multicolumn{1}{c|}{$T$} & \multicolumn{1}{c|}{$n_{d}$} & \multicolumn{1}{c|}{$s$} & \multicolumn{1}{c|}{$n_{s}$} & \multicolumn{1}{c|}{$T$} & \multicolumn{1}{c|}{$n_{d}$} & \multicolumn{1}{c|}{$s$} & \multicolumn{1}{c|}{$n_{s}$} & \multicolumn{1}{c|}{$T_1$} & \multicolumn{1}{c|}{$T_2$} & \multicolumn{1}{c|}{$n_{d}$} & \multicolumn{1}{c|}{$n_{d}$} & \multicolumn{1}{c|}{$n_{d}$} \\
			\hline
			2 & 30 & 8000 & 10 & 0.2 & 1.6E+06 & 8000 & 89 & 0.2 & 9.8E+06 & 8000 & 80 & 0.2 & 8.8E+06 & 8000 & 114 & 0.047 & 0.153 & 9.4E+06 & 6.9E+06 & 5.4E+06 \\
			3 & 30 & 8000 & 10 & 0.2 & 3.5E+06 & 8000 & 75 & 0.2 & 1.8E+07 & 8000 & 82 & 0.2 & 1.7E+07 & 8000 & 76 & 0.033 & 0.167 & 1.8E+07 & 1.8E+07 & 1.7E+07 \\
			5 & 30 & 8000 & 16 & 0.2 & 5.3E+06 & 8000 & 111 & 0.2 & 2.8E+07 & 8000 & 78 & 0.2 & 2.5E+07 & 8000 & 88 & 0.153 & 0.047 & 2.7E+07 & 2.7E+07 & 2.4E+07 \\
			10 & 30 & 8000 & 8 & 0.2 & 1.1E+07 & 8000 & 81 & 0.2 & 6.3E+07 & 8000 & 90 & 0.2 & 5.9E+07 & 8000 & 96 & 0.127 & 0.073 & 6.0E+07 & 7.6E+07 & 6.5E+07 \\
			15 & 30 & 8000 & 14 & 0.2 & 2.4E+07 & 8000 & 64 & 0.2 & 1.1E+08 & 8000 & 53 & 0.2 & 1.0E+08 & 8000 & 72 & 0.153 & 0.047 & 1.1E+08 & 1.5E+08 & 1.3E+08 \\
			20 & 30 & 8000 & 11 & 0.2 & 3.0E+07 & 8000 & 48 & 0.2 & 1.5E+08 & 8000 & 44 & 0.2 & 1.3E+08 & 8000 & 56 & 0.053 & 0.147 & 1.3E+08 & 2.3E+08 & 1.8E+08 \\
			25 & 30 & 8000 & 10 & 0.2 & 4.6E+07 & 8000 & 44 & 0.2 & 1.5E+08 & 8000 & 48 & 0.2 & 1.5E+08 & 8000 & 48 & 0.16 & 0.04 & 1.5E+08 & 2.8E+08 & 2.4E+08 \\
			\hline
		\end{tabular}
	}
	
\end{table}

\newpage

%%%%%%%%%%%%%%%%%%%%%%%%%%%%%%%%%%%%%%%%%%%%%%%%%%%%%%%%%%%%%%%%%%%%%%%%%%%%%%%%%%%%%%%
%  END: Skin Segmentation
%%%%%%%%%%%%%%%%%%%%%%%%%%%%%%%%%%%%%%%%%%%%%%%%%%%%%%%%%%%%%%%%%%%%%%%%%%%%%%%%%%%%%%%

\newpage

%%%%%%%%%%%%%%%%%%%%%%%%%%%%%%%%%%%%%%%%%%%%%%%%%%%%%%%%%%%%%%%%%%%%%%%%%%%%%%%%%%%%%%%
%  START: KEGG Metabolic Relation Network (Directed)
%%%%%%%%%%%%%%%%%%%%%%%%%%%%%%%%%%%%%%%%%%%%%%%%%%%%%%%%%%%%%%%%%%%%%%%%%%%%%%%%%%%%%%%
\subsection{KEGG Metabolic Relation Network (Directed)}
Dimensions: $m$ = 53413, $n$ = 20.
\par
Description:

\begin{table}[!htbp]
	\centering
	
	\caption{Summary of the results with KEGG Metabolic Relation Network (Directed) ($\times10^{8}$)}
	\label{TabResultsD17}
	\small
	\resizebox{!}{\tableheight}{
		\begin{tabular}{|l|l|l|llllll|llllll|llllll|}
			\hline
			\multicolumn{1}{|c|}{\multirow{3}{*}{$k$}} & \multicolumn{1}{c|}{\multirow{3}{*}{$f^*$}} & \multicolumn{1}{c|}{\multirow{3}{*}{$\overline{f}$}} & \multicolumn{6}{c|}{HPClust-inner} & \multicolumn{6}{c|}{HPClust-competitive} & \multicolumn{6}{c|}{HPClust-cooperative} \\ \cline{4-21}
			\multicolumn{1}{|c|}{} & \multicolumn{1}{c|}{} & \multicolumn{1}{c|}{} & \multicolumn{2}{c|}{$\varepsilon$} & \multicolumn{2}{c|}{$\overline{t}$} & \multicolumn{2}{c|}{$t$} & \multicolumn{2}{c|}{$\varepsilon$} & \multicolumn{2}{c|}{$\overline{t}$} & \multicolumn{2}{c|}{$t$} & \multicolumn{2}{c|}{$\varepsilon$} & \multicolumn{2}{c|}{$\overline{t}$} & \multicolumn{2}{c|}{$t$} \\ \cline{4-21}
			\multicolumn{1}{|c|}{} & \multicolumn{1}{c|}{} & \multicolumn{1}{c|}{} & \multicolumn{1}{c|}{med} & \multicolumn{1}{c|}{std} & \multicolumn{1}{c|}{med} & \multicolumn{1}{c|}{std} & \multicolumn{1}{c|}{med} & \multicolumn{1}{c|}{std} & \multicolumn{1}{c|}{med} & \multicolumn{1}{c|}{std} & \multicolumn{1}{c|}{med} & \multicolumn{1}{c|}{std} & \multicolumn{1}{c|}{med} & \multicolumn{1}{c|}{std} & \multicolumn{1}{c|}{med} & \multicolumn{1}{c|}{std} & \multicolumn{1}{c|}{med} & \multicolumn{1}{c|}{std} & \multicolumn{1}{c|}{med} & \multicolumn{1}{c|}{std} \\ \hline
			2 & 11.3853 & 11.29955 & 0.0 & 8.626 & 0.246 & 0.336 & 0.434 & 0.282 & 0.0 & 0.115 & 0.277 & 0.242 & 0.655 & 0.273 & 0.24 & 0.115 & 0.112 & 0.116 & 0.384 & 0.289 \\
			3 & 4.9006 & 4.84007 & 0.001 & 27.183 & 0.296 & 0.183 & 0.486 & 0.226 & 0.559 & 0.242 & 0.201 & 0.179 & 0.4 & 0.252 & 0.559 & 0.277 & 0.226 & 0.235 & 0.69 & 0.272 \\
			5 & 1.88367 & 1.86304 & 0.005 & 0.315 & 0.521 & 0.276 & 0.585 & 0.292 & 0.016 & 0.708 & 0.321 & 0.196 & 0.499 & 0.238 & 0.014 & 0.707 & 0.381 & 0.198 & 0.557 & 0.232 \\
			10 & 0.60513 & 0.61753 & 0.07 & 7.977 & 0.077 & 0.226 & 0.681 & 0.307 & 0.022 & 1.556 & 0.269 & 0.083 & 0.683 & 0.174 & 0.041 & 0.024 & 0.254 & 0.17 & 0.643 & 0.244 \\
			15 & 0.35393 & 0.35466 & 4.554 & 6.115 & 0.591 & 0.25 & 0.538 & 0.182 & -0.418 & 0.998 & 0.451 & 0.196 & 0.87 & 0.223 & -0.491 & 2.633 & 0.387 & 0.164 & 0.853 & 0.198 \\
			20 & 0.25027 & 0.25131 & 2.103 & 6.812 & 0.152 & 0.267 & 0.76 & 0.3 & 0.149 & 0.63 & 0.799 & 0.198 & 1.006 & 0.195 & 0.433 & 0.795 & 0.792 & 0.213 & 0.966 & 0.272 \\
			25 & 0.19289 & 0.19795 & 4.091 & 2.5 & 0.217 & 0.155 & 0.545 & 0.217 & 1.372 & 1.097 & 0.914 & 0.313 & 1.143 & 0.284 & 1.64 & 5.875 & 0.818 & 0.233 & 1.064 & 0.236 \\
			\hline
			\multicolumn{3}{|c|}{Mean:} & \textbf{1.546} & & \textbf{0.3} & & \textbf{0.575} & & \textbf{0.243} & & \textbf{0.462} & & \textbf{0.751} & & \textbf{0.348} & & \textbf{0.424} & & \textbf{0.737} & \\ \hline
		\end{tabular}
	}
	
	\bigskip
	
	\small
	\resizebox{!}{\tableheight}{
		\begin{tabular}{|l|l|l|llllll|llllll|llllll|}
			\hline
			\multicolumn{1}{|c|}{\multirow{3}{*}{$k$}} & \multicolumn{1}{c|}{\multirow{3}{*}{$f^*$}} & \multicolumn{1}{c|}{\multirow{3}{*}{$\overline{f}$}} & \multicolumn{6}{c|}{HPClust-hybrid} & \multicolumn{6}{c|}{Forgy K-means} & \multicolumn{6}{c|}{PBK-BDC} \\ \cline{4-21}
			\multicolumn{1}{|c|}{} & \multicolumn{1}{c|}{} & \multicolumn{1}{c|}{} & \multicolumn{2}{c|}{$\varepsilon$} & \multicolumn{2}{c|}{$\overline{t}$} & \multicolumn{2}{c|}{$t$} & \multicolumn{2}{c|}{$\varepsilon$} & \multicolumn{2}{c|}{$\overline{t}$} & \multicolumn{2}{c|}{$t$} & \multicolumn{2}{c|}{$\varepsilon$} & \multicolumn{2}{c|}{$\overline{t}$} & \multicolumn{2}{c|}{$t$} \\ \cline{4-21}
			\multicolumn{1}{|c|}{} & \multicolumn{1}{c|}{} & \multicolumn{1}{c|}{} & \multicolumn{1}{c|}{med} & \multicolumn{1}{c|}{std} & \multicolumn{1}{c|}{med} & \multicolumn{1}{c|}{std} & \multicolumn{1}{c|}{med} & \multicolumn{1}{c|}{std} & \multicolumn{1}{c|}{med} & \multicolumn{1}{c|}{std} & \multicolumn{1}{c|}{med} & \multicolumn{1}{c|}{std} & \multicolumn{1}{c|}{med} & \multicolumn{1}{c|}{std} & \multicolumn{1}{c|}{med} & \multicolumn{1}{c|}{std} & \multicolumn{1}{c|}{med} & \multicolumn{1}{c|}{std} & \multicolumn{1}{c|}{med} & \multicolumn{1}{c|}{std} \\ \hline
			2 & 11.3853 & 11.29955 & 0.0 & 0.11 & 0.249 & 0.126 & 0.477 & 0.233 & 18.854 & 0.0 & -- & -- & 0.041 & 0.004 & 18.854 & 0.003 & -- & -- & 0.033 & 0.006 \\
			3 & 4.9006 & 4.84007 & 0.559 & 0.256 & 0.226 & 0.218 & 0.373 & 0.297 & 124.789 & 0.0 & -- & -- & 0.08 & 0.023 & 124.789 & 9.606 & -- & -- & 0.075 & 0.008 \\
			5 & 1.88367 & 1.86304 & 0.072 & 0.715 & 0.376 & 0.249 & 0.731 & 0.205 & 0.0 & 9.787 & -- & -- & 0.201 & 0.042 & 0.0 & 0.001 & -- & -- & 0.189 & 0.07 \\
			10 & 0.60513 & 0.61753 & 0.041 & 8.926 & 0.276 & 0.056 & 0.766 & 0.194 & 36.81 & 3.376 & -- & -- & 0.607 & 0.158 & 36.81 & 3.067 & -- & -- & 0.582 & 0.039 \\
			15 & 0.35393 & 0.35466 & -0.359 & 1.22 & 0.504 & 0.186 & 0.935 & 0.211 & 96.641 & 4.224 & -- & -- & 1.873 & 0.168 & 97.957 & 4.103 & -- & -- & 1.69 & 0.245 \\
			20 & 0.25027 & 0.25131 & 0.15 & 25.29 & 0.631 & 0.364 & 1.26 & 0.303 & 162.301 & 4.756 & -- & -- & 3.433 & 0.856 & 162.039 & 3.883 & -- & -- & 3.784 & 0.849 \\
			25 & 0.19289 & 0.19795 & 1.312 & 0.908 & 0.814 & 0.224 & 1.421 & 0.32 & 230.281 & 6.699 & -- & -- & 5.045 & 0.638 & 223.96 & 5.88 & -- & -- & 5.15 & 0.602 \\
			\hline
			\multicolumn{3}{|c|}{Mean:} & \textbf{0.254} & & \textbf{0.44} & & \textbf{0.852} & & \textbf{95.668} & & \textbf{--} & & \textbf{1.611} & & \textbf{94.916} & & \textbf{--} & & \textbf{1.643} & \\ \hline
		\end{tabular}
	}
	
	\bigskip
	
	\caption{Clustering details with KEGG Metabolic Relation Network (Directed)}
	\label{TabDetailsD17}
	\resizebox{!}{\tableheight}{
		\begin{tabular}{|l|l|llll|llll|llll|lllll|l|l|}
			\hline
			\multicolumn{1}{|c|}{\multirow{2}{*}{$k$}} & \multicolumn{1}{c|}{\multirow{2}{*}{$n_{exec}$}} & \multicolumn{4}{c|}{HPClust-inner} & \multicolumn{4}{c|}{HPClust-competitive} & \multicolumn{4}{c|}{HPClust-cooperative} & \multicolumn{5}{c|}{HPClust-hybrid} & \multicolumn{1}{c|}{Forgy K-means} & \multicolumn{1}{c|}{PBK-BDC} \\ \cline{3-21}
			\multicolumn{1}{|c|}{} & \multicolumn{1}{c|}{} & \multicolumn{1}{c|}{$s$} & \multicolumn{1}{c|}{$n_{s}$} & \multicolumn{1}{c|}{$T$} & \multicolumn{1}{c|}{$n_{d}$} & \multicolumn{1}{c|}{$s$} & \multicolumn{1}{c|}{$n_{s}$} & \multicolumn{1}{c|}{$T$} & \multicolumn{1}{c|}{$n_{d}$} & \multicolumn{1}{c|}{$s$} & \multicolumn{1}{c|}{$n_{s}$} & \multicolumn{1}{c|}{$T$} & \multicolumn{1}{c|}{$n_{d}$} & \multicolumn{1}{c|}{$s$} & \multicolumn{1}{c|}{$n_{s}$} & \multicolumn{1}{c|}{$T_1$} & \multicolumn{1}{c|}{$T_2$} & \multicolumn{1}{c|}{$n_{d}$} & \multicolumn{1}{c|}{$n_{d}$} & \multicolumn{1}{c|}{$n_{d}$} \\
			\hline
			2 & 20 & 53350 & 54 & 1.0 & 3.1E+07 & 53350 & 586 & 1.0 & 2.0E+08 & 53350 & 348 & 1.0 & 2.0E+08 & 53350 & 377 & 0.767 & 0.233 & 1.9E+08 & 1.9E+06 & 2.0E+06 \\
			3 & 20 & 53350 & 62 & 1.0 & 4.4E+07 & 53350 & 222 & 1.0 & 2.3E+08 & 53350 & 412 & 1.0 & 2.2E+08 & 53350 & 206 & 0.967 & 0.033 & 2.3E+08 & 5.4E+06 & 5.4E+06 \\
			5 & 20 & 53350 & 64 & 1.0 & 7.4E+07 & 53350 & 144 & 1.0 & 2.6E+08 & 53350 & 190 & 1.0 & 2.8E+08 & 53350 & 260 & 0.333 & 0.667 & 2.7E+08 & 1.5E+07 & 1.5E+07 \\
			10 & 20 & 53350 & 52 & 1.0 & 1.1E+08 & 53350 & 77 & 1.0 & 2.8E+08 & 53350 & 73 & 1.0 & 3.0E+08 & 53350 & 108 & 0.867 & 0.133 & 2.7E+08 & 5.4E+07 & 5.3E+07 \\
			15 & 20 & 53350 & 34 & 1.0 & 1.3E+08 & 53350 & 62 & 1.0 & 2.9E+08 & 53350 & 56 & 1.0 & 3.0E+08 & 53350 & 35 & 0.6 & 0.4 & 2.3E+08 & 1.7E+08 & 1.5E+08 \\
			20 & 20 & 53350 & 34 & 1.0 & 1.6E+08 & 53350 & 31 & 1.0 & 3.0E+08 & 53350 & 27 & 1.0 & 2.9E+08 & 53350 & 13 & 0.1 & 0.9 & 2.3E+08 & 3.2E+08 & 3.5E+08 \\
			25 & 20 & 53350 & 18 & 1.0 & 1.7E+08 & 53350 & 14 & 1.0 & 3.1E+08 & 53350 & 12 & 1.0 & 3.1E+08 & 53350 & 10 & 0.033 & 0.967 & 2.8E+08 & 4.6E+08 & 4.8E+08 \\
			\hline
		\end{tabular}
	}
	
\end{table}

\newpage

%%%%%%%%%%%%%%%%%%%%%%%%%%%%%%%%%%%%%%%%%%%%%%%%%%%%%%%%%%%%%%%%%%%%%%%%%%%%%%%%%%%%%%%
%  END: KEGG Metabolic Relation Network (Directed)
%%%%%%%%%%%%%%%%%%%%%%%%%%%%%%%%%%%%%%%%%%%%%%%%%%%%%%%%%%%%%%%%%%%%%%%%%%%%%%%%%%%%%%%

%%%%%%%%%%%%%%%%%%%%%%%%%%%%%%%%%%%%%%%%%%%%%%%%%%%%%%%%%%%%%%%%%%%%%%%%%%%%%%%%%%%%%%%
%  START: Shuttle Control
%%%%%%%%%%%%%%%%%%%%%%%%%%%%%%%%%%%%%%%%%%%%%%%%%%%%%%%%%%%%%%%%%%%%%%%%%%%%%%%%%%%%%%%
\subsection{Shuttle Control}
Dimensions: $m$ = 58000, $n$ = 9.
\par
Description: each entity in the dataset contains several shuttle control attributes.

\begin{table}[!htbp]
	\centering
	
	\caption{Summary of the results with Shuttle Control ($\times10^{8}$)}
	\label{TabResultsD18}
	\small
	\resizebox{!}{\tableheight}{
		\begin{tabular}{|l|l|l|llllll|llllll|llllll|}
			\hline
			\multicolumn{1}{|c|}{\multirow{3}{*}{$k$}} & \multicolumn{1}{c|}{\multirow{3}{*}{$f^*$}} & \multicolumn{1}{c|}{\multirow{3}{*}{$\overline{f}$}} & \multicolumn{6}{c|}{HPClust-inner} & \multicolumn{6}{c|}{HPClust-competitive} & \multicolumn{6}{c|}{HPClust-cooperative} \\ \cline{4-21}
			\multicolumn{1}{|c|}{} & \multicolumn{1}{c|}{} & \multicolumn{1}{c|}{} & \multicolumn{2}{c|}{$\varepsilon$} & \multicolumn{2}{c|}{$\overline{t}$} & \multicolumn{2}{c|}{$t$} & \multicolumn{2}{c|}{$\varepsilon$} & \multicolumn{2}{c|}{$\overline{t}$} & \multicolumn{2}{c|}{$t$} & \multicolumn{2}{c|}{$\varepsilon$} & \multicolumn{2}{c|}{$\overline{t}$} & \multicolumn{2}{c|}{$t$} \\ \cline{4-21}
			\multicolumn{1}{|c|}{} & \multicolumn{1}{c|}{} & \multicolumn{1}{c|}{} & \multicolumn{1}{c|}{med} & \multicolumn{1}{c|}{std} & \multicolumn{1}{c|}{med} & \multicolumn{1}{c|}{std} & \multicolumn{1}{c|}{med} & \multicolumn{1}{c|}{std} & \multicolumn{1}{c|}{med} & \multicolumn{1}{c|}{std} & \multicolumn{1}{c|}{med} & \multicolumn{1}{c|}{std} & \multicolumn{1}{c|}{med} & \multicolumn{1}{c|}{std} & \multicolumn{1}{c|}{med} & \multicolumn{1}{c|}{std} & \multicolumn{1}{c|}{med} & \multicolumn{1}{c|}{std} & \multicolumn{1}{c|}{med} & \multicolumn{1}{c|}{std} \\ \hline
			2 & 21.34329 & 19.86153 & 5.043 & 12.056 & 0.724 & 0.333 & 0.716 & 0.364 & 5.043 & 1.082 & 0.22 & 0.357 & 1.0 & 0.34 & 0.0 & 0.744 & 0.093 & 0.067 & 0.821 & 0.37 \\
			3 & 10.85415 & 10.49161 & 0.28 & 29.163 & 0.532 & 0.425 & 0.852 & 0.398 & 3.658 & 1.293 & 0.418 & 0.338 & 1.02 & 0.48 & 3.658 & 1.613 & 0.418 & 0.308 & 0.854 & 0.311 \\
			4 & 8.8691 & 8.62423 & 0.32 & 4.741 & 0.816 & 0.342 & 0.695 & 0.397 & 0.343 & 7.623 & 0.793 & 0.417 & 0.879 & 0.447 & 0.0 & 0.075 & 0.307 & 0.361 & 0.687 & 0.446 \\
			5 & 7.24479 & 7.28912 & 1.484 & 7.359 & 0.017 & 0.037 & 0.73 & 0.456 & 0.178 & 7.394 & 0.034 & 0.077 & 0.757 & 0.403 & 0.392 & 0.21 & 0.033 & 0.005 & 0.714 & 0.429 \\
			10 & 2.83216 & 2.99551 & 8.736 & 21.835 & 0.148 & 0.337 & 0.859 & 0.422 & 1.623 & 2.889 & 0.082 & 0.011 & 0.944 & 0.438 & 0.671 & 0.475 & 0.081 & 0.012 & 0.412 & 0.399 \\
			15 & 1.53154 & 1.69671 & 16.164 & 8.425 & 0.053 & 0.289 & 0.883 & 0.411 & 5.617 & 2.582 & 0.149 & 0.022 & 0.738 & 0.363 & 5.814 & 2.605 & 0.146 & 0.016 & 1.054 & 0.427 \\
			20 & 1.06012 & 1.07621 & 3.493 & 7.041 & 0.181 & 0.409 & 0.952 & 0.41 & -0.758 & 3.626 & 0.225 & 0.045 & 1.102 & 0.419 & -1.494 & 2.123 & 0.21 & 0.079 & 1.04 & 0.437 \\
			25 & 0.77978 & 0.79776 & 9.944 & 4.377 & 0.083 & 0.0 & 0.688 & 0.394 & 2.84 & 3.246 & 0.378 & 0.361 & 1.212 & 0.35 & 3.339 & 3.844 & 0.387 & 0.311 & 1.173 & 0.268 \\
			\hline
			\multicolumn{3}{|c|}{Mean:} & \textbf{5.683} & & \textbf{0.319} & & \textbf{0.797} & & \textbf{2.318} & & \textbf{0.287} & & \textbf{0.957} & & \textbf{1.548} & & \textbf{0.209} & & \textbf{0.844} & \\ \hline
		\end{tabular}
	}
	
	\bigskip
	
	\small
	\resizebox{!}{\tableheight}{
		\begin{tabular}{|l|l|l|llllll|llllll|llllll|}
			\hline
			\multicolumn{1}{|c|}{\multirow{3}{*}{$k$}} & \multicolumn{1}{c|}{\multirow{3}{*}{$f^*$}} & \multicolumn{1}{c|}{\multirow{3}{*}{$\overline{f}$}} & \multicolumn{6}{c|}{HPClust-hybrid} & \multicolumn{6}{c|}{Forgy K-means} & \multicolumn{6}{c|}{PBK-BDC} \\ \cline{4-21}
			\multicolumn{1}{|c|}{} & \multicolumn{1}{c|}{} & \multicolumn{1}{c|}{} & \multicolumn{2}{c|}{$\varepsilon$} & \multicolumn{2}{c|}{$\overline{t}$} & \multicolumn{2}{c|}{$t$} & \multicolumn{2}{c|}{$\varepsilon$} & \multicolumn{2}{c|}{$\overline{t}$} & \multicolumn{2}{c|}{$t$} & \multicolumn{2}{c|}{$\varepsilon$} & \multicolumn{2}{c|}{$\overline{t}$} & \multicolumn{2}{c|}{$t$} \\ \cline{4-21}
			\multicolumn{1}{|c|}{} & \multicolumn{1}{c|}{} & \multicolumn{1}{c|}{} & \multicolumn{1}{c|}{med} & \multicolumn{1}{c|}{std} & \multicolumn{1}{c|}{med} & \multicolumn{1}{c|}{std} & \multicolumn{1}{c|}{med} & \multicolumn{1}{c|}{std} & \multicolumn{1}{c|}{med} & \multicolumn{1}{c|}{std} & \multicolumn{1}{c|}{med} & \multicolumn{1}{c|}{std} & \multicolumn{1}{c|}{med} & \multicolumn{1}{c|}{std} & \multicolumn{1}{c|}{med} & \multicolumn{1}{c|}{std} & \multicolumn{1}{c|}{med} & \multicolumn{1}{c|}{std} & \multicolumn{1}{c|}{med} & \multicolumn{1}{c|}{std} \\ \hline
			2 & 21.34329 & 19.86153 & 1.86 & 2.186 & 0.115 & 0.269 & 0.916 & 0.412 & 51.112 & 11.489 & -- & -- & 0.022 & 0.007 & 51.112 & 0.025 & -- & -- & 0.026 & 0.006 \\
			3 & 10.85415 & 10.49161 & 3.658 & 1.368 & 0.245 & 0.183 & 0.553 & 0.373 & 100.557 & 38.781 & -- & -- & 0.07 & 0.053 & 100.558 & 44.75 & -- & -- & 0.041 & 0.039 \\
			4 & 8.8691 & 8.62423 & 0.0 & 7.064 & 0.204 & 0.374 & 0.36 & 0.463 & 143.415 & 59.951 & -- & -- & 0.057 & 0.032 & 143.415 & 50.661 & -- & -- & 0.043 & 0.019 \\
			5 & 7.24479 & 7.28912 & 0.178 & 5.719 & 0.032 & 0.003 & 0.767 & 0.352 & 38.691 & 61.282 & -- & -- & 0.058 & 0.018 & 38.774 & 47.006 & -- & -- & 0.074 & 0.022 \\
			10 & 2.83216 & 2.99551 & 0.692 & 0.98 & 0.084 & 0.008 & 0.884 & 0.3 & 135.103 & 37.214 & -- & -- & 0.137 & 0.079 & 135.73 & 31.725 & -- & -- & 0.177 & 0.07 \\
			15 & 1.53154 & 1.69671 & 3.768 & 2.945 & 0.145 & 0.016 & 1.193 & 0.401 & 225.668 & 38.69 & -- & -- & 0.228 & 0.102 & 243.615 & 42.423 & -- & -- & 0.213 & 0.051 \\
			20 & 1.06012 & 1.07621 & 0.017 & 2.21 & 0.226 & 0.256 & 1.054 & 0.4 & 324.175 & 40.507 & -- & -- & 0.308 & 0.078 & 284.79 & 24.232 & -- & -- & 0.275 & 0.09 \\
			25 & 0.77978 & 0.79776 & 4.719 & 2.264 & 0.654 & 0.328 & 1.21 & 0.254 & 391.295 & 22.132 & -- & -- & 0.674 & 0.23 & 396.372 & 19.782 & -- & -- & 0.555 & 0.232 \\
			\hline
			\multicolumn{3}{|c|}{Mean:} & \textbf{1.862} & & \textbf{0.213} & & \textbf{0.867} & & \textbf{176.252} & & \textbf{--} & & \textbf{0.194} & & \textbf{174.296} & & \textbf{--} & & \textbf{0.176} & \\ \hline
		\end{tabular}
	}
	
	\bigskip
	
	\caption{Clustering details with Shuttle Control}
	\label{TabDetailsD18}
	\resizebox{!}{\tableheight}{
		\begin{tabular}{|l|l|llll|llll|llll|lllll|l|l|}
			\hline
			\multicolumn{1}{|c|}{\multirow{2}{*}{$k$}} & \multicolumn{1}{c|}{\multirow{2}{*}{$n_{exec}$}} & \multicolumn{4}{c|}{HPClust-inner} & \multicolumn{4}{c|}{HPClust-competitive} & \multicolumn{4}{c|}{HPClust-cooperative} & \multicolumn{5}{c|}{HPClust-hybrid} & \multicolumn{1}{c|}{Forgy K-means} & \multicolumn{1}{c|}{PBK-BDC} \\ \cline{3-21}
			\multicolumn{1}{|c|}{} & \multicolumn{1}{c|}{} & \multicolumn{1}{c|}{$s$} & \multicolumn{1}{c|}{$n_{s}$} & \multicolumn{1}{c|}{$T$} & \multicolumn{1}{c|}{$n_{d}$} & \multicolumn{1}{c|}{$s$} & \multicolumn{1}{c|}{$n_{s}$} & \multicolumn{1}{c|}{$T$} & \multicolumn{1}{c|}{$n_{d}$} & \multicolumn{1}{c|}{$s$} & \multicolumn{1}{c|}{$n_{s}$} & \multicolumn{1}{c|}{$T$} & \multicolumn{1}{c|}{$n_{d}$} & \multicolumn{1}{c|}{$s$} & \multicolumn{1}{c|}{$n_{s}$} & \multicolumn{1}{c|}{$T_1$} & \multicolumn{1}{c|}{$T_2$} & \multicolumn{1}{c|}{$n_{d}$} & \multicolumn{1}{c|}{$n_{d}$} & \multicolumn{1}{c|}{$n_{d}$} \\
			\hline
			2 & 15 & 57950 & 200 & 1.5 & 9.1E+07 & 57950 & 1378 & 1.5 & 5.2E+08 & 57950 & 1169 & 1.5 & 5.3E+08 & 57950 & 1370 & 0.75 & 0.75 & 5.3E+08 & 2.8E+06 & 3.2E+06 \\
			3 & 15 & 57950 & 175 & 1.5 & 1.1E+08 & 57950 & 953 & 1.5 & 6.0E+08 & 57950 & 1008 & 1.5 & 6.3E+08 & 57950 & 618 & 0.65 & 0.85 & 6.2E+08 & 6.1E+06 & 5.9E+06 \\
			4 & 15 & 57950 & 139 & 1.5 & 1.4E+08 & 57950 & 835 & 1.5 & 7.2E+08 & 57950 & 681 & 1.5 & 6.8E+08 & 57950 & 338 & 0.6 & 0.9 & 7.1E+08 & 7.9E+06 & 7.2E+06 \\
			5 & 15 & 57950 & 145 & 1.5 & 1.8E+08 & 57950 & 600 & 1.5 & 7.6E+08 & 57950 & 568 & 1.5 & 7.8E+08 & 57950 & 643 & 0.6 & 0.9 & 8.0E+08 & 9.3E+06 & 8.1E+06 \\
			10 & 15 & 57950 & 106 & 1.5 & 2.8E+08 & 57950 & 432 & 1.5 & 9.6E+08 & 57950 & 172 & 1.5 & 9.2E+08 & 57950 & 416 & 0.1 & 1.4 & 9.0E+08 & 2.8E+07 & 3.6E+07 \\
			15 & 15 & 57950 & 105 & 1.5 & 3.8E+08 & 57950 & 204 & 1.5 & 9.4E+08 & 57950 & 310 & 1.5 & 1.0E+09 & 57950 & 336 & 1.1 & 0.4 & 9.6E+08 & 4.9E+07 & 4.5E+07 \\
			20 & 15 & 57950 & 80 & 1.5 & 4.5E+08 & 57950 & 186 & 1.5 & 9.8E+08 & 57950 & 157 & 1.5 & 9.7E+08 & 57950 & 146 & 0.6 & 0.9 & 9.7E+08 & 6.4E+07 & 5.9E+07 \\
			25 & 15 & 57950 & 56 & 1.5 & 5.0E+08 & 57950 & 157 & 1.5 & 9.8E+08 & 57950 & 135 & 1.5 & 1.0E+09 & 57950 & 118 & 0.35 & 1.15 & 9.1E+08 & 1.0E+08 & 1.2E+08 \\
			\hline
		\end{tabular}
	}
	
\end{table}

\newpage

%%%%%%%%%%%%%%%%%%%%%%%%%%%%%%%%%%%%%%%%%%%%%%%%%%%%%%%%%%%%%%%%%%%%%%%%%%%%%%%%%%%%%%%
%  END: Shuttle Control
%%%%%%%%%%%%%%%%%%%%%%%%%%%%%%%%%%%%%%%%%%%%%%%%%%%%%%%%%%%%%%%%%%%%%%%%%%%%%%%%%%%%%%%

\newpage

%%%%%%%%%%%%%%%%%%%%%%%%%%%%%%%%%%%%%%%%%%%%%%%%%%%%%%%%%%%%%%%%%%%%%%%%%%%%%%%%%%%%%%%
%  START: Shuttle Control (normalized)
%%%%%%%%%%%%%%%%%%%%%%%%%%%%%%%%%%%%%%%%%%%%%%%%%%%%%%%%%%%%%%%%%%%%%%%%%%%%%%%%%%%%%%%
\subsection{Shuttle Control (normalized)}
Dimensions: $m$ = 58000, $n$ = 9.
\par
Description: each entity in the dataset contains several shuttle control attributes. Min-max scaling was used for normalization of data set values for better clusterization.

\begin{table}[!htbp]
	\centering
	
	\caption{Summary of the results with Shuttle Control (normalized) ($\times10^{1}$)}
	\label{TabResultsD19}
	\small
	\resizebox{!}{\tableheight}{
		\begin{tabular}{|l|l|l|llllll|llllll|llllll|}
			\hline
			\multicolumn{1}{|c|}{\multirow{3}{*}{$k$}} & \multicolumn{1}{c|}{\multirow{3}{*}{$f^*$}} & \multicolumn{1}{c|}{\multirow{3}{*}{$\overline{f}$}} & \multicolumn{6}{c|}{HPClust-inner} & \multicolumn{6}{c|}{HPClust-competitive} & \multicolumn{6}{c|}{HPClust-cooperative} \\ \cline{4-21}
			\multicolumn{1}{|c|}{} & \multicolumn{1}{c|}{} & \multicolumn{1}{c|}{} & \multicolumn{2}{c|}{$\varepsilon$} & \multicolumn{2}{c|}{$\overline{t}$} & \multicolumn{2}{c|}{$t$} & \multicolumn{2}{c|}{$\varepsilon$} & \multicolumn{2}{c|}{$\overline{t}$} & \multicolumn{2}{c|}{$t$} & \multicolumn{2}{c|}{$\varepsilon$} & \multicolumn{2}{c|}{$\overline{t}$} & \multicolumn{2}{c|}{$t$} \\ \cline{4-21}
			\multicolumn{1}{|c|}{} & \multicolumn{1}{c|}{} & \multicolumn{1}{c|}{} & \multicolumn{1}{c|}{med} & \multicolumn{1}{c|}{std} & \multicolumn{1}{c|}{med} & \multicolumn{1}{c|}{std} & \multicolumn{1}{c|}{med} & \multicolumn{1}{c|}{std} & \multicolumn{1}{c|}{med} & \multicolumn{1}{c|}{std} & \multicolumn{1}{c|}{med} & \multicolumn{1}{c|}{std} & \multicolumn{1}{c|}{med} & \multicolumn{1}{c|}{std} & \multicolumn{1}{c|}{med} & \multicolumn{1}{c|}{std} & \multicolumn{1}{c|}{med} & \multicolumn{1}{c|}{std} & \multicolumn{1}{c|}{med} & \multicolumn{1}{c|}{std} \\ \hline
			2 & 104.41601 & 3.33677 & 0.106 & 0.211 & 0.244 & 0.103 & 0.218 & 0.101 & 0.184 & 0.167 & 0.068 & 0.065 & 0.164 & 0.103 & 0.245 & 0.102 & 0.112 & 0.078 & 0.182 & 0.122 \\
			3 & 73.28769 & 2.33445 & 0.697 & 0.868 & 0.132 & 0.065 & 0.135 & 0.129 & 0.675 & 0.297 & 0.029 & 0.049 & 0.18 & 0.111 & 0.514 & 0.519 & 0.039 & 0.036 & 0.153 & 0.1 \\
			4 & 50.076 & 1.5748 & 0.781 & 12.197 & 0.212 & 0.116 & 0.242 & 0.116 & 0.675 & 0.508 & 0.019 & 0.03 & 0.204 & 0.111 & 0.46 & 0.347 & 0.023 & 0.026 & 0.132 & 0.109 \\
			5 & 39.78043 & 1.24889 & 1.301 & 1.451 & 0.057 & 0.068 & 0.189 & 0.134 & 1.224 & 0.875 & 0.023 & 0.02 & 0.248 & 0.121 & 1.679 & 0.823 & 0.019 & 0.013 & 0.149 & 0.116 \\
			10 & 15.04997 & 0.44476 & 2.315 & 11.582 & 0.215 & 0.147 & 0.245 & 0.123 & 0.824 & 0.969 & 0.143 & 0.1 & 0.297 & 0.088 & 2.23 & 0.96 & 0.057 & 0.114 & 0.267 & 0.095 \\
			15 & 9.81804 & 0.28928 & 5.001 & 3.919 & 0.066 & 0.094 & 0.25 & 0.111 & 3.02 & 1.72 & 0.042 & 0.018 & 0.217 & 0.096 & 2.906 & 1.421 & 0.027 & 0.025 & 0.26 & 0.116 \\
			20 & 7.233 & 0.19874 & 6.611 & 3.444 & 0.114 & 0.099 & 0.243 & 0.102 & 2.84 & 1.499 & 0.062 & 0.049 & 0.244 & 0.102 & 4.49 & 2.5 & 0.051 & 0.043 & 0.245 & 0.118 \\
			25 & 5.86461 & 0.15054 & 5.645 & 3.749 & 0.14 & 0.13 & 0.207 & 0.126 & 4.909 & 1.212 & 0.14 & 0.076 & 0.255 & 0.108 & 5.227 & 1.257 & 0.094 & 0.088 & 0.246 & 0.101 \\
			\hline
			\multicolumn{3}{|c|}{Mean:} & \textbf{2.807} & & \textbf{0.147} & & \textbf{0.216} & & \textbf{1.794} & & \textbf{0.066} & & \textbf{0.226} & & \textbf{2.219} & & \textbf{0.053} & & \textbf{0.204} & \\ \hline
		\end{tabular}
	}
	
	\bigskip
	
	\small
	\resizebox{!}{\tableheight}{
		\begin{tabular}{|l|l|l|llllll|llllll|llllll|}
			\hline
			\multicolumn{1}{|c|}{\multirow{3}{*}{$k$}} & \multicolumn{1}{c|}{\multirow{3}{*}{$f^*$}} & \multicolumn{1}{c|}{\multirow{3}{*}{$\overline{f}$}} & \multicolumn{6}{c|}{HPClust-hybrid} & \multicolumn{6}{c|}{Forgy K-means} & \multicolumn{6}{c|}{PBK-BDC} \\ \cline{4-21}
			\multicolumn{1}{|c|}{} & \multicolumn{1}{c|}{} & \multicolumn{1}{c|}{} & \multicolumn{2}{c|}{$\varepsilon$} & \multicolumn{2}{c|}{$\overline{t}$} & \multicolumn{2}{c|}{$t$} & \multicolumn{2}{c|}{$\varepsilon$} & \multicolumn{2}{c|}{$\overline{t}$} & \multicolumn{2}{c|}{$t$} & \multicolumn{2}{c|}{$\varepsilon$} & \multicolumn{2}{c|}{$\overline{t}$} & \multicolumn{2}{c|}{$t$} \\ \cline{4-21}
			\multicolumn{1}{|c|}{} & \multicolumn{1}{c|}{} & \multicolumn{1}{c|}{} & \multicolumn{1}{c|}{med} & \multicolumn{1}{c|}{std} & \multicolumn{1}{c|}{med} & \multicolumn{1}{c|}{std} & \multicolumn{1}{c|}{med} & \multicolumn{1}{c|}{std} & \multicolumn{1}{c|}{med} & \multicolumn{1}{c|}{std} & \multicolumn{1}{c|}{med} & \multicolumn{1}{c|}{std} & \multicolumn{1}{c|}{med} & \multicolumn{1}{c|}{std} & \multicolumn{1}{c|}{med} & \multicolumn{1}{c|}{std} & \multicolumn{1}{c|}{med} & \multicolumn{1}{c|}{std} & \multicolumn{1}{c|}{med} & \multicolumn{1}{c|}{std} \\ \hline
			2 & 104.41601 & 3.33677 & 0.231 & 0.153 & 0.148 & 0.084 & 0.221 & 0.124 & 14.732 & 10.172 & -- & -- & 0.018 & 0.005 & 7.998 & 2.641 & -- & -- & 0.006 & 0.001 \\
			3 & 73.28769 & 2.33445 & 0.505 & 0.279 & 0.07 & 0.091 & 0.305 & 0.107 & 1.765 & 12.847 & -- & -- & 0.027 & 0.007 & 14.401 & 16.099 & -- & -- & 0.008 & 0.001 \\
			4 & 50.076 & 1.5748 & 0.514 & 0.415 & 0.025 & 0.058 & 0.208 & 0.124 & 0.0 & 6.86 & -- & -- & 0.032 & 0.006 & 36.908 & 23.315 & -- & -- & 0.008 & 0.001 \\
			5 & 39.78043 & 1.24889 & 1.166 & 0.975 & 0.027 & 0.04 & 0.261 & 0.102 & 0.826 & 4.343 & -- & -- & 0.061 & 0.024 & 18.537 & 18.539 & -- & -- & 0.011 & 0.002 \\
			10 & 15.04997 & 0.44476 & 0.621 & 1.146 & 0.091 & 0.062 & 0.239 & 0.088 & 47.02 & 19.014 & -- & -- & 0.077 & 0.028 & 51.611 & 26.149 & -- & -- & 0.018 & 0.002 \\
			15 & 9.81804 & 0.28928 & 2.955 & 1.34 & 0.029 & 0.012 & 0.284 & 0.089 & 21.544 & 37.916 & -- & -- & 0.092 & 0.061 & 33.001 & 42.832 & -- & -- & 0.028 & 0.003 \\
			20 & 7.233 & 0.19874 & 2.245 & 1.949 & 0.07 & 0.062 & 0.332 & 0.094 & 22.889 & 57.432 & -- & -- & 0.162 & 0.079 & 41.776 & 47.828 & -- & -- & 0.035 & 0.005 \\
			25 & 5.86461 & 0.15054 & 4.658 & 1.881 & 0.123 & 0.078 & 0.223 & 0.111 & 23.942 & 59.568 & -- & -- & 0.237 & 0.087 & 51.317 & 75.254 & -- & -- & 0.044 & 0.005 \\
			\hline
			\multicolumn{3}{|c|}{Mean:} & \textbf{1.612} & & \textbf{0.073} & & \textbf{0.259} & & \textbf{16.59} & & \textbf{--} & & \textbf{0.088} & & \textbf{31.943} & & \textbf{--} & & \textbf{0.02} & \\ \hline
		\end{tabular}
	}
	
	\bigskip
	
	\caption{Clustering details with Shuttle Control (normalized)}
	\label{TabDetailsD19}
	\resizebox{!}{\tableheight}{
		\begin{tabular}{|l|l|llll|llll|llll|lllll|l|l|}
			\hline
			\multicolumn{1}{|c|}{\multirow{2}{*}{$k$}} & \multicolumn{1}{c|}{\multirow{2}{*}{$n_{exec}$}} & \multicolumn{4}{c|}{HPClust-inner} & \multicolumn{4}{c|}{HPClust-competitive} & \multicolumn{4}{c|}{HPClust-cooperative} & \multicolumn{5}{c|}{HPClust-hybrid} & \multicolumn{1}{c|}{Forgy K-means} & \multicolumn{1}{c|}{PBK-BDC} \\ \cline{3-21}
			\multicolumn{1}{|c|}{} & \multicolumn{1}{c|}{} & \multicolumn{1}{c|}{$s$} & \multicolumn{1}{c|}{$n_{s}$} & \multicolumn{1}{c|}{$T$} & \multicolumn{1}{c|}{$n_{d}$} & \multicolumn{1}{c|}{$s$} & \multicolumn{1}{c|}{$n_{s}$} & \multicolumn{1}{c|}{$T$} & \multicolumn{1}{c|}{$n_{d}$} & \multicolumn{1}{c|}{$s$} & \multicolumn{1}{c|}{$n_{s}$} & \multicolumn{1}{c|}{$T$} & \multicolumn{1}{c|}{$n_{d}$} & \multicolumn{1}{c|}{$s$} & \multicolumn{1}{c|}{$n_{s}$} & \multicolumn{1}{c|}{$T_1$} & \multicolumn{1}{c|}{$T_2$} & \multicolumn{1}{c|}{$n_{d}$} & \multicolumn{1}{c|}{$n_{d}$} & \multicolumn{1}{c|}{$n_{d}$} \\
			\hline
			2 & 20 & 2000 & 98 & 0.4 & 1.6E+06 & 2000 & 624 & 0.4 & 1.3E+07 & 2000 & 664 & 0.4 & 1.3E+07 & 2000 & 806 & 0.04 & 0.36 & 1.3E+07 & 1.6E+06 & 1.4E+06 \\
			3 & 20 & 2000 & 74 & 0.4 & 5.0E+06 & 2000 & 598 & 0.4 & 3.2E+07 & 2000 & 541 & 0.4 & 3.6E+07 & 2000 & 1066 & 0.28 & 0.12 & 3.4E+07 & 2.2E+06 & 2.2E+06 \\
			4 & 20 & 2000 & 92 & 0.4 & 5.0E+06 & 2000 & 718 & 0.4 & 4.3E+07 & 2000 & 456 & 0.4 & 4.1E+07 & 2000 & 686 & 0.24 & 0.16 & 4.1E+07 & 3.5E+06 & 3.0E+06 \\
			5 & 20 & 2000 & 74 & 0.4 & 8.4E+06 & 2000 & 716 & 0.4 & 5.8E+07 & 2000 & 468 & 0.4 & 5.4E+07 & 2000 & 786 & 0.187 & 0.213 & 5.8E+07 & 5.9E+06 & 4.6E+06 \\
			10 & 20 & 2000 & 74 & 0.4 & 1.3E+07 & 2000 & 718 & 0.4 & 1.0E+08 & 2000 & 676 & 0.4 & 1.0E+08 & 2000 & 595 & 0.16 & 0.24 & 1.0E+08 & 1.1E+07 & 1.2E+07 \\
			15 & 20 & 2000 & 70 & 0.4 & 2.6E+07 & 2000 & 358 & 0.4 & 1.5E+08 & 2000 & 436 & 0.4 & 1.4E+08 & 2000 & 493 & 0.32 & 0.08 & 1.5E+08 & 1.8E+07 & 2.0E+07 \\
			20 & 20 & 2000 & 62 & 0.4 & 3.7E+07 & 2000 & 295 & 0.4 & 1.8E+08 & 2000 & 310 & 0.4 & 1.7E+08 & 2000 & 414 & 0.253 & 0.147 & 1.7E+08 & 2.8E+07 & 2.6E+07 \\
			25 & 20 & 2000 & 50 & 0.4 & 4.3E+07 & 2000 & 255 & 0.4 & 1.9E+08 & 2000 & 249 & 0.4 & 1.9E+08 & 2000 & 237 & 0.08 & 0.32 & 1.9E+08 & 5.2E+07 & 3.3E+07 \\
			\hline
		\end{tabular}
	}
	
\end{table}

\newpage

%%%%%%%%%%%%%%%%%%%%%%%%%%%%%%%%%%%%%%%%%%%%%%%%%%%%%%%%%%%%%%%%%%%%%%%%%%%%%%%%%%%%%%%
%  END: Shuttle Control (normalized)
%%%%%%%%%%%%%%%%%%%%%%%%%%%%%%%%%%%%%%%%%%%%%%%%%%%%%%%%%%%%%%%%%%%%%%%%%%%%%%%%%%%%%%%

%%%%%%%%%%%%%%%%%%%%%%%%%%%%%%%%%%%%%%%%%%%%%%%%%%%%%%%%%%%%%%%%%%%%%%%%%%%%%%%%%%%%%%%
%  START: EEG Eye State
%%%%%%%%%%%%%%%%%%%%%%%%%%%%%%%%%%%%%%%%%%%%%%%%%%%%%%%%%%%%%%%%%%%%%%%%%%%%%%%%%%%%%%%
\subsection{EEG Eye State}
Dimensions: $m$ = 14980, $n$ = 14.
\par
Description: the data set consists of 14 electroencephalogram (EEG) values for predicting the corresponding eye state.

\begin{table}[!htbp]
	\centering
	
	\caption{Summary of the results with EEG Eye State ($\times10^{8}$)}
	\label{TabResultsD20}
	\small
	\resizebox{!}{\tableheight}{
		\begin{tabular}{|l|l|l|llllll|llllll|llllll|}
			\hline
			\multicolumn{1}{|c|}{\multirow{3}{*}{$k$}} & \multicolumn{1}{c|}{\multirow{3}{*}{$f^*$}} & \multicolumn{1}{c|}{\multirow{3}{*}{$\overline{f}$}} & \multicolumn{6}{c|}{HPClust-inner} & \multicolumn{6}{c|}{HPClust-competitive} & \multicolumn{6}{c|}{HPClust-cooperative} \\ \cline{4-21}
			\multicolumn{1}{|c|}{} & \multicolumn{1}{c|}{} & \multicolumn{1}{c|}{} & \multicolumn{2}{c|}{$\varepsilon$} & \multicolumn{2}{c|}{$\overline{t}$} & \multicolumn{2}{c|}{$t$} & \multicolumn{2}{c|}{$\varepsilon$} & \multicolumn{2}{c|}{$\overline{t}$} & \multicolumn{2}{c|}{$t$} & \multicolumn{2}{c|}{$\varepsilon$} & \multicolumn{2}{c|}{$\overline{t}$} & \multicolumn{2}{c|}{$t$} \\ \cline{4-21}
			\multicolumn{1}{|c|}{} & \multicolumn{1}{c|}{} & \multicolumn{1}{c|}{} & \multicolumn{1}{c|}{med} & \multicolumn{1}{c|}{std} & \multicolumn{1}{c|}{med} & \multicolumn{1}{c|}{std} & \multicolumn{1}{c|}{med} & \multicolumn{1}{c|}{std} & \multicolumn{1}{c|}{med} & \multicolumn{1}{c|}{std} & \multicolumn{1}{c|}{med} & \multicolumn{1}{c|}{std} & \multicolumn{1}{c|}{med} & \multicolumn{1}{c|}{std} & \multicolumn{1}{c|}{med} & \multicolumn{1}{c|}{std} & \multicolumn{1}{c|}{med} & \multicolumn{1}{c|}{std} & \multicolumn{1}{c|}{med} & \multicolumn{1}{c|}{std} \\ \hline
			2 & 7845.09934 & 8178.13658 & 4.245 & 4.728 & 0.661 & 0.358 & 0.719 & 0.359 & 4.247 & 0.002 & 0.411 & 0.21 & 0.706 & 0.374 & 4.246 & 0.002 & 0.183 & 0.199 & 0.995 & 0.37 \\
			3 & 1833.88058 & 1833.87892 & 0.0 & 0.003 & 0.486 & 0.355 & 0.638 & 0.347 & 0.0 & 0.003 & 0.246 & 0.254 & 0.855 & 0.379 & 0.0 & 0.003 & 0.42 & 0.313 & 0.755 & 0.375 \\
			4 & 2.23605 & 2.23431 & 0.0 & 0.001 & 0.629 & 0.383 & 0.678 & 0.428 & 0.002 & 0.001 & 0.352 & 0.307 & 0.563 & 0.474 & 0.0 & 0.001 & 0.206 & 0.276 & 0.615 & 0.363 \\
			5 & 1.33858 & 1.33703 & -0.0 & 14.651 & 0.669 & 0.35 & 0.508 & 0.339 & -0.0 & 120196.81 & 0.276 & 0.354 & 0.583 & 0.433 & -0.0 & 0.0 & 0.1 & 0.221 & 0.668 & 0.481 \\
			10 & 0.4531 & 0.4527 & 0.001 & 0.554 & 0.679 & 0.363 & 0.865 & 0.366 & -0.004 & 88058.848 & 0.612 & 0.347 & 1.088 & 0.469 & -0.005 & 0.005 & 0.397 & 0.307 & 0.95 & 0.406 \\
			15 & 0.34653 & 0.34837 & 0.622 & 0.502 & 0.032 & 0.015 & 0.498 & 0.425 & 0.055 & 0.143 & 0.113 & 0.037 & 1.079 & 0.324 & 0.135 & 0.126 & 0.111 & 0.19 & 0.857 & 0.351 \\
			20 & 0.28986 & 0.29175 & 0.785 & 0.717 & 0.064 & 0.367 & 0.98 & 0.345 & 0.02 & 0.133 & 0.216 & 0.055 & 1.089 & 0.321 & 0.06 & 0.205 & 0.193 & 0.063 & 0.887 & 0.39 \\
			25 & 0.25989 & 0.26088 & 0.636 & 0.604 & 0.204 & 0.274 & 1.099 & 0.377 & 0.156 & 0.095 & 0.222 & 0.049 & 0.87 & 0.353 & 0.137 & 0.082 & 0.225 & 0.042 & 0.874 & 0.303 \\
			\hline
			\multicolumn{3}{|c|}{Mean:} & \textbf{0.786} & & \textbf{0.428} & & \textbf{0.748} & & \textbf{0.559} & & \textbf{0.306} & & \textbf{0.854} & & \textbf{0.571} & & \textbf{0.229} & & \textbf{0.825} & \\ \hline
		\end{tabular}
	}
	
	\bigskip
	
	\small
	\resizebox{!}{\tableheight}{
		\begin{tabular}{|l|l|l|llllll|llllll|llllll|}
			\hline
			\multicolumn{1}{|c|}{\multirow{3}{*}{$k$}} & \multicolumn{1}{c|}{\multirow{3}{*}{$f^*$}} & \multicolumn{1}{c|}{\multirow{3}{*}{$\overline{f}$}} & \multicolumn{6}{c|}{HPClust-hybrid} & \multicolumn{6}{c|}{Forgy K-means} & \multicolumn{6}{c|}{PBK-BDC} \\ \cline{4-21}
			\multicolumn{1}{|c|}{} & \multicolumn{1}{c|}{} & \multicolumn{1}{c|}{} & \multicolumn{2}{c|}{$\varepsilon$} & \multicolumn{2}{c|}{$\overline{t}$} & \multicolumn{2}{c|}{$t$} & \multicolumn{2}{c|}{$\varepsilon$} & \multicolumn{2}{c|}{$\overline{t}$} & \multicolumn{2}{c|}{$t$} & \multicolumn{2}{c|}{$\varepsilon$} & \multicolumn{2}{c|}{$\overline{t}$} & \multicolumn{2}{c|}{$t$} \\ \cline{4-21}
			\multicolumn{1}{|c|}{} & \multicolumn{1}{c|}{} & \multicolumn{1}{c|}{} & \multicolumn{1}{c|}{med} & \multicolumn{1}{c|}{std} & \multicolumn{1}{c|}{med} & \multicolumn{1}{c|}{std} & \multicolumn{1}{c|}{med} & \multicolumn{1}{c|}{std} & \multicolumn{1}{c|}{med} & \multicolumn{1}{c|}{std} & \multicolumn{1}{c|}{med} & \multicolumn{1}{c|}{std} & \multicolumn{1}{c|}{med} & \multicolumn{1}{c|}{std} & \multicolumn{1}{c|}{med} & \multicolumn{1}{c|}{std} & \multicolumn{1}{c|}{med} & \multicolumn{1}{c|}{std} & \multicolumn{1}{c|}{med} & \multicolumn{1}{c|}{std} \\ \hline
			2 & 7845.09934 & 8178.13658 & 4.247 & 3.973 & 0.179 & 0.249 & 0.842 & 0.435 & -0.0 & 1.274 & -- & -- & 0.003 & 0.0 & -0.0 & 16.348 & -- & -- & 0.004 & 0.001 \\
			3 & 1833.88058 & 1833.87892 & 0.0 & 0.003 & 0.245 & 0.27 & 0.773 & 0.42 & 227.909 & 98.687 & -- & -- & 0.004 & 0.001 & 227.909 & 49.672 & -- & -- & 0.005 & 0.001 \\
			4 & 2.23605 & 2.23431 & 0.0 & 0.001 & 0.151 & 0.218 & 0.798 & 0.466 & 268809.803 & 133731.189 & -- & -- & 0.021 & 0.009 & 268809.803 & 128214.104 & -- & -- & 0.019 & 0.007 \\
			5 & 1.33858 & 1.33703 & -0.0 & 6.519 & 0.161 & 0.213 & 1.077 & 0.422 & 449091.754 & 223405.448 & -- & -- & 0.029 & 0.006 & 449091.754 & 205786.243 & -- & -- & 0.028 & 0.005 \\
			10 & 0.4531 & 0.4527 & -0.002 & 0.006 & 0.561 & 0.31 & 0.96 & 0.32 & 1326681.022 & 632723.737 & -- & -- & 0.074 & 0.035 & 1326681.023 & 607938.794 & -- & -- & 0.079 & 0.019 \\
			15 & 0.34653 & 0.34837 & 0.058 & 0.097 & 0.103 & 0.051 & 1.077 & 0.347 & 1734685.672 & 751140.757 & -- & -- & 0.197 & 0.06 & 1.077 & 849818.798 & -- & -- & 0.145 & 0.054 \\
			20 & 0.28986 & 0.29175 & 0.025 & 0.031 & 0.192 & 0.055 & 1.105 & 0.372 & 2073832.95 & 989155.199 & -- & -- & 0.375 & 0.117 & 2073833.29 & 989155.218 & -- & -- & 0.448 & 0.143 \\
			25 & 0.25989 & 0.26088 & 0.109 & 0.13 & 0.202 & 0.07 & 1.211 & 0.312 & 1156493.228 & 1156492.007 & -- & -- & 0.371 & 0.117 & 2312984.49 & 1059942.517 & -- & -- & 0.358 & 0.088 \\
			\hline
			\multicolumn{3}{|c|}{Mean:} & \textbf{0.555} & & \textbf{0.224} & & \textbf{0.98} & & \textbf{876227.792} & & \textbf{--} & & \textbf{0.134} & & \textbf{803953.668} & & \textbf{--} & & \textbf{0.136} & \\ \hline
		\end{tabular}
	}
	
	\bigskip
	
	\caption{Clustering details with EEG Eye State}
	\label{TabDetailsD20}
	\resizebox{!}{\tableheight}{
		\begin{tabular}{|l|l|llll|llll|llll|lllll|l|l|}
			\hline
			\multicolumn{1}{|c|}{\multirow{2}{*}{$k$}} & \multicolumn{1}{c|}{\multirow{2}{*}{$n_{exec}$}} & \multicolumn{4}{c|}{HPClust-inner} & \multicolumn{4}{c|}{HPClust-competitive} & \multicolumn{4}{c|}{HPClust-cooperative} & \multicolumn{5}{c|}{HPClust-hybrid} & \multicolumn{1}{c|}{Forgy K-means} & \multicolumn{1}{c|}{PBK-BDC} \\ \cline{3-21}
			\multicolumn{1}{|c|}{} & \multicolumn{1}{c|}{} & \multicolumn{1}{c|}{$s$} & \multicolumn{1}{c|}{$n_{s}$} & \multicolumn{1}{c|}{$T$} & \multicolumn{1}{c|}{$n_{d}$} & \multicolumn{1}{c|}{$s$} & \multicolumn{1}{c|}{$n_{s}$} & \multicolumn{1}{c|}{$T$} & \multicolumn{1}{c|}{$n_{d}$} & \multicolumn{1}{c|}{$s$} & \multicolumn{1}{c|}{$n_{s}$} & \multicolumn{1}{c|}{$T$} & \multicolumn{1}{c|}{$n_{d}$} & \multicolumn{1}{c|}{$s$} & \multicolumn{1}{c|}{$n_{s}$} & \multicolumn{1}{c|}{$T_1$} & \multicolumn{1}{c|}{$T_2$} & \multicolumn{1}{c|}{$n_{d}$} & \multicolumn{1}{c|}{$n_{d}$} & \multicolumn{1}{c|}{$n_{d}$} \\
			\hline
			2 & 20 & 14979 & 414 & 1.5 & 5.8E+07 & 14979 & 2898 & 1.5 & 4.0E+08 & 14979 & 4582 & 1.5 & 4.0E+08 & 14979 & 4036 & 0.7 & 0.8 & 4.2E+08 & 1.5E+05 & 1.8E+05 \\
			3 & 20 & 14979 & 368 & 1.5 & 8.7E+07 & 14979 & 3046 & 1.5 & 4.9E+08 & 14979 & 2518 & 1.5 & 4.8E+08 & 14979 & 2674 & 0.05 & 1.45 & 5.0E+08 & 3.6E+05 & 4.0E+05 \\
			4 & 20 & 14979 & 342 & 1.5 & 9.9E+07 & 14979 & 1678 & 1.5 & 5.7E+08 & 14979 & 1809 & 1.5 & 5.6E+08 & 14979 & 2296 & 0.3 & 1.2 & 5.5E+08 & 1.5E+06 & 1.7E+06 \\
			5 & 20 & 14979 & 266 & 1.5 & 1.2E+08 & 14979 & 1426 & 1.5 & 5.7E+08 & 14979 & 1598 & 1.5 & 5.8E+08 & 14979 & 2723 & 0.85 & 0.65 & 5.8E+08 & 3.3E+06 & 3.0E+06 \\
			10 & 20 & 14979 & 308 & 1.5 & 1.5E+08 & 14979 & 1563 & 1.5 & 6.9E+08 & 14979 & 1400 & 1.5 & 7.0E+08 & 14979 & 1357 & 0.85 & 0.65 & 6.9E+08 & 9.2E+06 & 1.0E+07 \\
			15 & 20 & 14979 & 174 & 1.5 & 2.6E+08 & 14979 & 935 & 1.5 & 7.2E+08 & 14979 & 742 & 1.5 & 7.0E+08 & 14979 & 996 & 0.35 & 1.15 & 7.3E+08 & 2.4E+07 & 1.9E+07 \\
			20 & 20 & 14979 & 298 & 1.5 & 3.3E+08 & 14979 & 678 & 1.5 & 7.5E+08 & 14979 & 558 & 1.5 & 7.2E+08 & 14979 & 753 & 1.4 & 0.1 & 7.4E+08 & 3.5E+07 & 4.3E+07 \\
			25 & 20 & 14979 & 286 & 1.5 & 3.7E+08 & 14979 & 370 & 1.5 & 7.3E+08 & 14979 & 408 & 1.5 & 7.4E+08 & 14979 & 490 & 0.2 & 1.3 & 6.7E+08 & 5.0E+07 & 4.8E+07 \\
			\hline
		\end{tabular}
	}
	
\end{table}

\newpage

%%%%%%%%%%%%%%%%%%%%%%%%%%%%%%%%%%%%%%%%%%%%%%%%%%%%%%%%%%%%%%%%%%%%%%%%%%%%%%%%%%%%%%%
%  END: EEG Eye State
%%%%%%%%%%%%%%%%%%%%%%%%%%%%%%%%%%%%%%%%%%%%%%%%%%%%%%%%%%%%%%%%%%%%%%%%%%%%%%%%%%%%%%%

\newpage

%%%%%%%%%%%%%%%%%%%%%%%%%%%%%%%%%%%%%%%%%%%%%%%%%%%%%%%%%%%%%%%%%%%%%%%%%%%%%%%%%%%%%%%
%  START: EEG Eye State (normalized)
%%%%%%%%%%%%%%%%%%%%%%%%%%%%%%%%%%%%%%%%%%%%%%%%%%%%%%%%%%%%%%%%%%%%%%%%%%%%%%%%%%%%%%%
\subsection{EEG Eye State (normalized)}
Dimensions: $m$ = 14980, $n$ = 14.
\par
Description: the data set consists of 14 electroencephalogram (EEG) values for predicting the corresponding eye state. Min-max scaling was used for normalization of data set values for better clusterization.

\begin{table}[!htbp]
	\centering
	
	\caption{Summary of the results with EEG Eye State (normalized) ($\times10^{1}$)}
	\label{TabResultsD21}
	\small
	\resizebox{!}{\tableheight}{
		\begin{tabular}{|l|l|l|llllll|llllll|llllll|}
			\hline
			\multicolumn{1}{|c|}{\multirow{3}{*}{$k$}} & \multicolumn{1}{c|}{\multirow{3}{*}{$f^*$}} & \multicolumn{1}{c|}{\multirow{3}{*}{$\overline{f}$}} & \multicolumn{6}{c|}{HPClust-inner} & \multicolumn{6}{c|}{HPClust-competitive} & \multicolumn{6}{c|}{HPClust-cooperative} \\ \cline{4-21}
			\multicolumn{1}{|c|}{} & \multicolumn{1}{c|}{} & \multicolumn{1}{c|}{} & \multicolumn{2}{c|}{$\varepsilon$} & \multicolumn{2}{c|}{$\overline{t}$} & \multicolumn{2}{c|}{$t$} & \multicolumn{2}{c|}{$\varepsilon$} & \multicolumn{2}{c|}{$\overline{t}$} & \multicolumn{2}{c|}{$t$} & \multicolumn{2}{c|}{$\varepsilon$} & \multicolumn{2}{c|}{$\overline{t}$} & \multicolumn{2}{c|}{$t$} \\ \cline{4-21}
			\multicolumn{1}{|c|}{} & \multicolumn{1}{c|}{} & \multicolumn{1}{c|}{} & \multicolumn{1}{c|}{med} & \multicolumn{1}{c|}{std} & \multicolumn{1}{c|}{med} & \multicolumn{1}{c|}{std} & \multicolumn{1}{c|}{med} & \multicolumn{1}{c|}{std} & \multicolumn{1}{c|}{med} & \multicolumn{1}{c|}{std} & \multicolumn{1}{c|}{med} & \multicolumn{1}{c|}{std} & \multicolumn{1}{c|}{med} & \multicolumn{1}{c|}{std} & \multicolumn{1}{c|}{med} & \multicolumn{1}{c|}{std} & \multicolumn{1}{c|}{med} & \multicolumn{1}{c|}{std} & \multicolumn{1}{c|}{med} & \multicolumn{1}{c|}{std} \\ \hline
			2 & 1.15267 & 1.15216 & 6.104 & 9.638 & 0.258 & 0.307 & 0.574 & 0.311 & 0.002 & 8.681 & 0.193 & 0.199 & 0.581 & 0.287 & 0.002 & 0.001 & 0.139 & 0.152 & 0.599 & 0.274 \\
			3 & 0.82423 & 0.87097 & 5.716 & 13.325 & 0.005 & 0.278 & 0.551 & 0.267 & 0.001 & 9.655 & 0.009 & 0.005 & 0.331 & 0.29 & 0.001 & 1.026 & 0.009 & 0.008 & 0.482 & 0.278 \\
			4 & 0.5429 & 0.57038 & 5.15 & 14.019 & 0.005 & 0.192 & 0.507 & 0.297 & 0.001 & 10.29 & 0.012 & 0.001 & 0.438 & 0.259 & 0.001 & 0.001 & 0.011 & 0.001 & 0.549 & 0.301 \\
			5 & 0.28952 & 0.28903 & 0.002 & 33.997 & 0.413 & 0.331 & 0.504 & 0.315 & 0.002 & 15.033 & 0.161 & 0.163 & 0.339 & 0.255 & 0.002 & 0.0 & 0.195 & 0.186 & 0.472 & 0.281 \\
			10 & 0.10269 & 0.10335 & 0.707 & 0.479 & 0.029 & 0.19 & 0.601 & 0.303 & -0.003 & 67.68 & 0.064 & 0.015 & 0.449 & 0.294 & -0.004 & 0.126 & 0.059 & 0.014 & 0.671 & 0.29 \\
			15 & 0.07469 & 0.07479 & 0.2 & 0.789 & 0.05 & 0.24 & 0.606 & 0.254 & 0.036 & 0.053 & 0.134 & 0.049 & 0.712 & 0.209 & 0.052 & 0.066 & 0.139 & 0.045 & 0.495 & 0.276 \\
			20 & 0.06125 & 0.06154 & 0.457 & 0.629 & 0.059 & 0.077 & 0.566 & 0.313 & 0.177 & 0.146 & 0.166 & 0.06 & 0.654 & 0.222 & 0.205 & 0.167 & 0.166 & 0.054 & 0.623 & 0.211 \\
			25 & 0.05385 & 0.0543 & 0.873 & 0.774 & 0.065 & 0.186 & 0.575 & 0.267 & -0.154 & 0.152 & 0.224 & 0.048 & 0.777 & 0.192 & -0.151 & 80.458 & 0.201 & 0.044 & 0.637 & 0.196 \\
			\hline
			\multicolumn{3}{|c|}{Mean:} & \textbf{2.401} & & \textbf{0.11} & & \textbf{0.56} & & \textbf{0.008} & & \textbf{0.12} & & \textbf{0.535} & & \textbf{0.014} & & \textbf{0.115} & & \textbf{0.566} & \\ \hline
		\end{tabular}
	}
	
	\bigskip
	
	\small
	\resizebox{!}{\tableheight}{
		\begin{tabular}{|l|l|l|llllll|llllll|llllll|}
			\hline
			\multicolumn{1}{|c|}{\multirow{3}{*}{$k$}} & \multicolumn{1}{c|}{\multirow{3}{*}{$f^*$}} & \multicolumn{1}{c|}{\multirow{3}{*}{$\overline{f}$}} & \multicolumn{6}{c|}{HPClust-hybrid} & \multicolumn{6}{c|}{Forgy K-means} & \multicolumn{6}{c|}{PBK-BDC} \\ \cline{4-21}
			\multicolumn{1}{|c|}{} & \multicolumn{1}{c|}{} & \multicolumn{1}{c|}{} & \multicolumn{2}{c|}{$\varepsilon$} & \multicolumn{2}{c|}{$\overline{t}$} & \multicolumn{2}{c|}{$t$} & \multicolumn{2}{c|}{$\varepsilon$} & \multicolumn{2}{c|}{$\overline{t}$} & \multicolumn{2}{c|}{$t$} & \multicolumn{2}{c|}{$\varepsilon$} & \multicolumn{2}{c|}{$\overline{t}$} & \multicolumn{2}{c|}{$t$} \\ \cline{4-21}
			\multicolumn{1}{|c|}{} & \multicolumn{1}{c|}{} & \multicolumn{1}{c|}{} & \multicolumn{1}{c|}{med} & \multicolumn{1}{c|}{std} & \multicolumn{1}{c|}{med} & \multicolumn{1}{c|}{std} & \multicolumn{1}{c|}{med} & \multicolumn{1}{c|}{std} & \multicolumn{1}{c|}{med} & \multicolumn{1}{c|}{std} & \multicolumn{1}{c|}{med} & \multicolumn{1}{c|}{std} & \multicolumn{1}{c|}{med} & \multicolumn{1}{c|}{std} & \multicolumn{1}{c|}{med} & \multicolumn{1}{c|}{std} & \multicolumn{1}{c|}{med} & \multicolumn{1}{c|}{std} & \multicolumn{1}{c|}{med} & \multicolumn{1}{c|}{std} \\ \hline
			2 & 1.15267 & 1.15216 & 0.001 & 2.823 & 0.258 & 0.256 & 0.578 & 0.275 & 25.398 & 0.011 & -- & -- & 0.016 & 0.003 & 25.398 & 0.0 & -- & -- & 0.018 & 0.003 \\
			3 & 0.82423 & 0.87097 & 0.001 & 7.024 & 0.009 & 0.001 & 0.435 & 0.25 & 69.038 & 0.035 & -- & -- & 0.017 & 0.004 & 69.038 & 4.986 & -- & -- & 0.015 & 0.003 \\
			4 & 0.5429 & 0.57038 & 0.001 & 6.519 & 0.011 & 0.001 & 0.56 & 0.248 & 152.474 & 0.048 & -- & -- & 0.022 & 0.007 & 152.479 & 0.049 & -- & -- & 0.02 & 0.007 \\
			5 & 0.28952 & 0.28903 & 0.002 & 13.447 & 0.295 & 0.235 & 0.561 & 0.315 & 367.097 & 32.271 & -- & -- & 0.033 & 0.011 & 367.097 & 24.212 & -- & -- & 0.036 & 0.009 \\
			10 & 0.10269 & 0.10335 & -0.004 & 0.131 & 0.063 & 0.051 & 0.789 & 0.265 & 633.846 & 193.064 & -- & -- & 0.116 & 0.037 & 879.525 & 132.438 & -- & -- & 0.098 & 0.038 \\
			15 & 0.07469 & 0.07479 & 0.037 & 0.068 & 0.138 & 0.146 & 0.688 & 0.194 & 853.035 & 256.91 & -- & -- & 0.179 & 0.105 & 853.015 & 297.335 & -- & -- & 0.154 & 0.062 \\
			20 & 0.06125 & 0.06154 & 0.226 & 0.146 & 0.176 & 0.034 & 0.781 & 0.239 & 1044.241 & 312.789 & -- & -- & 0.301 & 0.151 & 1044.477 & 285.929 & -- & -- & 0.242 & 0.092 \\
			25 & 0.05385 & 0.0543 & -0.122 & 0.197 & 0.21 & 0.047 & 0.925 & 0.236 & 1190.906 & 385.599 & -- & -- & 0.44 & 0.156 & 1190.787 & 365.14 & -- & -- & 0.303 & 0.119 \\
			\hline
			\multicolumn{3}{|c|}{Mean:} & \textbf{0.018} & & \textbf{0.145} & & \textbf{0.665} & & \textbf{542.004} & & \textbf{--} & & \textbf{0.141} & & \textbf{572.727} & & \textbf{--} & & \textbf{0.111} & \\ \hline
		\end{tabular}
	}
	
	\bigskip
	
	\caption{Clustering details with EEG Eye State (normalized)}
	\label{TabDetailsD21}
	\resizebox{!}{\tableheight}{
		\begin{tabular}{|l|l|llll|llll|llll|lllll|l|l|}
			\hline
			\multicolumn{1}{|c|}{\multirow{2}{*}{$k$}} & \multicolumn{1}{c|}{\multirow{2}{*}{$n_{exec}$}} & \multicolumn{4}{c|}{HPClust-inner} & \multicolumn{4}{c|}{HPClust-competitive} & \multicolumn{4}{c|}{HPClust-cooperative} & \multicolumn{5}{c|}{HPClust-hybrid} & \multicolumn{1}{c|}{Forgy K-means} & \multicolumn{1}{c|}{PBK-BDC} \\ \cline{3-21}
			\multicolumn{1}{|c|}{} & \multicolumn{1}{c|}{} & \multicolumn{1}{c|}{$s$} & \multicolumn{1}{c|}{$n_{s}$} & \multicolumn{1}{c|}{$T$} & \multicolumn{1}{c|}{$n_{d}$} & \multicolumn{1}{c|}{$s$} & \multicolumn{1}{c|}{$n_{s}$} & \multicolumn{1}{c|}{$T$} & \multicolumn{1}{c|}{$n_{d}$} & \multicolumn{1}{c|}{$s$} & \multicolumn{1}{c|}{$n_{s}$} & \multicolumn{1}{c|}{$T$} & \multicolumn{1}{c|}{$n_{d}$} & \multicolumn{1}{c|}{$s$} & \multicolumn{1}{c|}{$n_{s}$} & \multicolumn{1}{c|}{$T_1$} & \multicolumn{1}{c|}{$T_2$} & \multicolumn{1}{c|}{$n_{d}$} & \multicolumn{1}{c|}{$n_{d}$} & \multicolumn{1}{c|}{$n_{d}$} \\
			\hline
			2 & 30 & 14979 & 337 & 1.0 & 3.6E+07 & 14979 & 2470 & 1.0 & 2.6E+08 & 14979 & 2653 & 1.0 & 2.6E+08 & 14979 & 2379 & 0.233 & 0.767 & 2.6E+08 & 9.4E+05 & 1.0E+06 \\
			3 & 30 & 14979 & 314 & 1.0 & 5.2E+07 & 14979 & 1104 & 1.0 & 3.1E+08 & 14979 & 1702 & 1.0 & 3.1E+08 & 14979 & 1378 & 0.533 & 0.467 & 3.0E+08 & 1.3E+06 & 1.5E+06 \\
			4 & 30 & 14979 & 252 & 1.0 & 7.2E+07 & 14979 & 1234 & 1.0 & 3.5E+08 & 14979 & 1438 & 1.0 & 3.5E+08 & 14979 & 1590 & 0.933 & 0.067 & 3.4E+08 & 2.3E+06 & 2.0E+06 \\
			5 & 30 & 14979 & 224 & 1.0 & 8.3E+07 & 14979 & 770 & 1.0 & 3.7E+08 & 14979 & 1136 & 1.0 & 3.8E+08 & 14979 & 1370 & 0.733 & 0.267 & 3.7E+08 & 3.7E+06 & 3.9E+06 \\
			10 & 30 & 14979 & 192 & 1.0 & 1.2E+08 & 14979 & 560 & 1.0 & 4.3E+08 & 14979 & 862 & 1.0 & 4.3E+08 & 14979 & 1040 & 0.5 & 0.5 & 4.3E+08 & 1.4E+07 & 1.3E+07 \\
			15 & 30 & 14979 & 166 & 1.0 & 1.7E+08 & 14979 & 602 & 1.0 & 4.6E+08 & 14979 & 358 & 1.0 & 4.5E+08 & 14979 & 531 & 0.333 & 0.667 & 4.7E+08 & 2.0E+07 & 2.0E+07 \\
			20 & 30 & 14979 & 152 & 1.0 & 2.0E+08 & 14979 & 361 & 1.0 & 4.6E+08 & 14979 & 318 & 1.0 & 4.6E+08 & 14979 & 442 & 0.733 & 0.267 & 4.7E+08 & 3.1E+07 & 3.3E+07 \\
			25 & 30 & 14979 & 174 & 1.0 & 2.5E+08 & 14979 & 282 & 1.0 & 4.6E+08 & 14979 & 228 & 1.0 & 4.7E+08 & 14979 & 415 & 0.933 & 0.067 & 4.7E+08 & 4.3E+07 & 4.1E+07 \\
			\hline
		\end{tabular}
	}
	
\end{table}

\newpage

%%%%%%%%%%%%%%%%%%%%%%%%%%%%%%%%%%%%%%%%%%%%%%%%%%%%%%%%%%%%%%%%%%%%%%%%%%%%%%%%%%%%%%%
%  END: EEG Eye State (normalized)
%%%%%%%%%%%%%%%%%%%%%%%%%%%%%%%%%%%%%%%%%%%%%%%%%%%%%%%%%%%%%%%%%%%%%%%%%%%%%%%%%%%%%%%

%%%%%%%%%%%%%%%%%%%%%%%%%%%%%%%%%%%%%%%%%%%%%%%%%%%%%%%%%%%%%%%%%%%%%%%%%%%%%%%%%%%%%%%
%  START: Pla85900
%%%%%%%%%%%%%%%%%%%%%%%%%%%%%%%%%%%%%%%%%%%%%%%%%%%%%%%%%%%%%%%%%%%%%%%%%%%%%%%%%%%%%%%
\subsection{Pla85900}
Dimensions: $m$ = 85900, $n$ = 2.
\par
Description: a data set contains cities coordinates for traveling salesman problem.

\begin{table}[!htbp]
	\centering
	
	\caption{Summary of the results with Pla85900 ($\times10^{15}$)}
	\label{TabResultsD22}
	\small
	\resizebox{!}{\tableheight}{
		\begin{tabular}{|l|l|l|llllll|llllll|llllll|}
			\hline
			\multicolumn{1}{|c|}{\multirow{3}{*}{$k$}} & \multicolumn{1}{c|}{\multirow{3}{*}{$f^*$}} & \multicolumn{1}{c|}{\multirow{3}{*}{$\overline{f}$}} & \multicolumn{6}{c|}{HPClust-inner} & \multicolumn{6}{c|}{HPClust-competitive} & \multicolumn{6}{c|}{HPClust-cooperative} \\ \cline{4-21}
			\multicolumn{1}{|c|}{} & \multicolumn{1}{c|}{} & \multicolumn{1}{c|}{} & \multicolumn{2}{c|}{$\varepsilon$} & \multicolumn{2}{c|}{$\overline{t}$} & \multicolumn{2}{c|}{$t$} & \multicolumn{2}{c|}{$\varepsilon$} & \multicolumn{2}{c|}{$\overline{t}$} & \multicolumn{2}{c|}{$t$} & \multicolumn{2}{c|}{$\varepsilon$} & \multicolumn{2}{c|}{$\overline{t}$} & \multicolumn{2}{c|}{$t$} \\ \cline{4-21}
			\multicolumn{1}{|c|}{} & \multicolumn{1}{c|}{} & \multicolumn{1}{c|}{} & \multicolumn{1}{c|}{med} & \multicolumn{1}{c|}{std} & \multicolumn{1}{c|}{med} & \multicolumn{1}{c|}{std} & \multicolumn{1}{c|}{med} & \multicolumn{1}{c|}{std} & \multicolumn{1}{c|}{med} & \multicolumn{1}{c|}{std} & \multicolumn{1}{c|}{med} & \multicolumn{1}{c|}{std} & \multicolumn{1}{c|}{med} & \multicolumn{1}{c|}{std} & \multicolumn{1}{c|}{med} & \multicolumn{1}{c|}{std} & \multicolumn{1}{c|}{med} & \multicolumn{1}{c|}{std} & \multicolumn{1}{c|}{med} & \multicolumn{1}{c|}{std} \\ \hline
			2 & 3.74908 & 0.60031 & 0.054 & 0.718 & 0.59 & 0.405 & 0.776 & 0.44 & 0.013 & 0.021 & 0.19 & 0.276 & 0.603 & 0.429 & 0.011 & 0.224 & 0.108 & 0.17 & 0.73 & 0.44 \\
			3 & 2.28057 & 0.36407 & 0.026 & 0.036 & 0.809 & 0.508 & 0.941 & 0.444 & 0.024 & 0.021 & 0.259 & 0.307 & 0.552 & 0.428 & 0.025 & 0.032 & 0.274 & 0.233 & 0.735 & 0.391 \\
			5 & 1.33972 & 0.21512 & 0.09 & 0.751 & 0.292 & 0.34 & 0.821 & 0.438 & 0.046 & 0.029 & 0.099 & 0.083 & 0.683 & 0.42 & 0.051 & 0.305 & 0.051 & 0.331 & 0.718 & 0.427 \\
			10 & 0.68294 & 0.10944 & 0.587 & 0.371 & 0.85 & 0.459 & 0.802 & 0.468 & 0.111 & 0.148 & 0.221 & 0.297 & 0.923 & 0.48 & 0.151 & 0.317 & 0.146 & 0.274 & 0.844 & 0.433 \\
			15 & 0.46029 & 0.07355 & 0.291 & 0.476 & 0.557 & 0.504 & 0.919 & 0.433 & 0.251 & 0.153 & 0.539 & 0.349 & 0.769 & 0.36 & 0.268 & 0.183 & 0.335 & 0.398 & 0.979 & 0.438 \\
			20 & 0.34988 & 0.05595 & 0.656 & 0.422 & 0.545 & 0.379 & 0.833 & 0.421 & 0.316 & 0.254 & 0.507 & 0.36 & 0.971 & 0.411 & 0.304 & 0.338 & 0.3 & 0.191 & 0.655 & 0.38 \\
			25 & 0.28259 & 0.04518 & 0.89 & 0.318 & 0.853 & 0.383 & 0.884 & 0.373 & 0.617 & 0.281 & 0.618 & 0.392 & 0.806 & 0.408 & 0.763 & 0.455 & 0.432 & 0.304 & 0.711 & 0.402 \\
			\hline
			\multicolumn{3}{|c|}{Mean:} & \textbf{0.371} & & \textbf{0.642} & & \textbf{0.854} & & \textbf{0.197} & & \textbf{0.348} & & \textbf{0.758} & & \textbf{0.225} & & \textbf{0.235} & & \textbf{0.767} & \\ \hline
		\end{tabular}
	}
	
	\bigskip
	
	\small
	\resizebox{!}{\tableheight}{
		\begin{tabular}{|l|l|l|llllll|llllll|llllll|}
			\hline
			\multicolumn{1}{|c|}{\multirow{3}{*}{$k$}} & \multicolumn{1}{c|}{\multirow{3}{*}{$f^*$}} & \multicolumn{1}{c|}{\multirow{3}{*}{$\overline{f}$}} & \multicolumn{6}{c|}{HPClust-hybrid} & \multicolumn{6}{c|}{Forgy K-means} & \multicolumn{6}{c|}{PBK-BDC} \\ \cline{4-21}
			\multicolumn{1}{|c|}{} & \multicolumn{1}{c|}{} & \multicolumn{1}{c|}{} & \multicolumn{2}{c|}{$\varepsilon$} & \multicolumn{2}{c|}{$\overline{t}$} & \multicolumn{2}{c|}{$t$} & \multicolumn{2}{c|}{$\varepsilon$} & \multicolumn{2}{c|}{$\overline{t}$} & \multicolumn{2}{c|}{$t$} & \multicolumn{2}{c|}{$\varepsilon$} & \multicolumn{2}{c|}{$\overline{t}$} & \multicolumn{2}{c|}{$t$} \\ \cline{4-21}
			\multicolumn{1}{|c|}{} & \multicolumn{1}{c|}{} & \multicolumn{1}{c|}{} & \multicolumn{1}{c|}{med} & \multicolumn{1}{c|}{std} & \multicolumn{1}{c|}{med} & \multicolumn{1}{c|}{std} & \multicolumn{1}{c|}{med} & \multicolumn{1}{c|}{std} & \multicolumn{1}{c|}{med} & \multicolumn{1}{c|}{std} & \multicolumn{1}{c|}{med} & \multicolumn{1}{c|}{std} & \multicolumn{1}{c|}{med} & \multicolumn{1}{c|}{std} & \multicolumn{1}{c|}{med} & \multicolumn{1}{c|}{std} & \multicolumn{1}{c|}{med} & \multicolumn{1}{c|}{std} & \multicolumn{1}{c|}{med} & \multicolumn{1}{c|}{std} \\ \hline
			2 & 3.74908 & 0.60031 & 0.016 & 0.014 & 0.139 & 0.224 & 0.988 & 0.428 & 0.0 & 0.686 & -- & -- & 0.024 & 0.007 & 6.458 & 2.618 & -- & -- & 0.01 & 0.002 \\
			3 & 2.28057 & 0.36407 & 0.024 & 0.029 & 0.287 & 0.353 & 0.671 & 0.41 & 0.0 & 0.0 & -- & -- & 0.078 & 0.025 & 0.005 & 20.855 & -- & -- & 0.022 & 0.006 \\
			5 & 1.33972 & 0.21512 & 0.036 & 0.027 & 0.091 & 0.089 & 1.106 & 0.456 & 0.407 & 1.133 & -- & -- & 0.082 & 0.06 & 6.719 & 5.906 & -- & -- & 0.027 & 0.012 \\
			10 & 0.68294 & 0.10944 & 0.136 & 0.186 & 0.273 & 0.318 & 1.022 & 0.348 & 0.42 & 0.774 & -- & -- & 0.201 & 0.087 & 14.084 & 10.514 & -- & -- & 0.067 & 0.017 \\
			15 & 0.46029 & 0.07355 & 0.226 & 0.143 & 0.555 & 0.281 & 1.004 & 0.337 & 0.495 & 0.806 & -- & -- & 0.313 & 0.156 & 17.409 & 9.89 & -- & -- & 0.098 & 0.022 \\
			20 & 0.34988 & 0.05595 & 0.331 & 0.126 & 0.469 & 0.305 & 0.728 & 0.394 & 0.45 & 0.601 & -- & -- & 0.453 & 0.213 & 15.152 & 8.883 & -- & -- & 0.125 & 0.043 \\
			25 & 0.28259 & 0.04518 & 0.618 & 0.299 & 0.645 & 0.359 & 0.996 & 0.369 & 0.932 & 0.495 & -- & -- & 0.697 & 0.229 & 13.672 & 7.007 & -- & -- & 0.163 & 0.038 \\
			\hline
			\multicolumn{3}{|c|}{Mean:} & \textbf{0.198} & & \textbf{0.351} & & \textbf{0.931} & & \textbf{0.386} & & \textbf{--} & & \textbf{0.264} & & \textbf{10.5} & & \textbf{--} & & \textbf{0.073} & \\ \hline
		\end{tabular}
	}
	
	\bigskip
	
	\caption{Clustering details with Pla85900}
	\label{TabDetailsD22}
	\resizebox{!}{\tableheight}{
		\begin{tabular}{|l|l|llll|llll|llll|lllll|l|l|}
			\hline
			\multicolumn{1}{|c|}{\multirow{2}{*}{$k$}} & \multicolumn{1}{c|}{\multirow{2}{*}{$n_{exec}$}} & \multicolumn{4}{c|}{HPClust-inner} & \multicolumn{4}{c|}{HPClust-competitive} & \multicolumn{4}{c|}{HPClust-cooperative} & \multicolumn{5}{c|}{HPClust-hybrid} & \multicolumn{1}{c|}{Forgy K-means} & \multicolumn{1}{c|}{PBK-BDC} \\ \cline{3-21}
			\multicolumn{1}{|c|}{} & \multicolumn{1}{c|}{} & \multicolumn{1}{c|}{$s$} & \multicolumn{1}{c|}{$n_{s}$} & \multicolumn{1}{c|}{$T$} & \multicolumn{1}{c|}{$n_{d}$} & \multicolumn{1}{c|}{$s$} & \multicolumn{1}{c|}{$n_{s}$} & \multicolumn{1}{c|}{$T$} & \multicolumn{1}{c|}{$n_{d}$} & \multicolumn{1}{c|}{$s$} & \multicolumn{1}{c|}{$n_{s}$} & \multicolumn{1}{c|}{$T$} & \multicolumn{1}{c|}{$n_{d}$} & \multicolumn{1}{c|}{$s$} & \multicolumn{1}{c|}{$n_{s}$} & \multicolumn{1}{c|}{$T_1$} & \multicolumn{1}{c|}{$T_2$} & \multicolumn{1}{c|}{$n_{d}$} & \multicolumn{1}{c|}{$n_{d}$} & \multicolumn{1}{c|}{$n_{d}$} \\
			\hline
			2 & 40 & 14000 & 237 & 1.5 & 4.4E+07 & 14000 & 1470 & 1.5 & 3.4E+08 & 14000 & 1718 & 1.5 & 3.2E+08 & 14000 & 2442 & 1.0 & 0.5 & 3.4E+08 & 5.1E+06 & 4.1E+06 \\
			3 & 40 & 14000 & 240 & 1.5 & 8.1E+07 & 14000 & 1190 & 1.5 & 5.9E+08 & 14000 & 1476 & 1.5 & 6.0E+08 & 14000 & 1413 & 1.4 & 0.1 & 6.0E+08 & 1.6E+07 & 1.1E+07 \\
			5 & 40 & 14000 & 217 & 1.5 & 1.4E+08 & 14000 & 1238 & 1.5 & 9.3E+08 & 14000 & 1272 & 1.5 & 9.1E+08 & 14000 & 1943 & 1.05 & 0.45 & 9.1E+08 & 1.9E+07 & 1.6E+07 \\
			10 & 40 & 14000 & 186 & 1.5 & 3.7E+08 & 14000 & 1000 & 1.5 & 1.7E+09 & 14000 & 907 & 1.5 & 1.7E+09 & 14000 & 1098 & 1.2 & 0.3 & 1.7E+09 & 7.6E+07 & 5.8E+07 \\
			15 & 40 & 14000 & 159 & 1.5 & 5.6E+08 & 14000 & 498 & 1.5 & 2.4E+09 & 14000 & 748 & 1.5 & 2.3E+09 & 14000 & 728 & 0.6 & 0.9 & 2.3E+09 & 1.4E+08 & 1.0E+08 \\
			20 & 40 & 14000 & 117 & 1.5 & 7.6E+08 & 14000 & 482 & 1.5 & 2.7E+09 & 14000 & 336 & 1.5 & 2.7E+09 & 14000 & 359 & 1.4 & 0.1 & 2.7E+09 & 2.2E+08 & 1.4E+08 \\
			25 & 40 & 14000 & 110 & 1.5 & 9.3E+08 & 14000 & 270 & 1.5 & 2.8E+09 & 14000 & 264 & 1.5 & 2.9E+09 & 14000 & 324 & 1.15 & 0.35 & 2.8E+09 & 3.5E+08 & 1.9E+08 \\
			\hline
		\end{tabular}
	}
	
\end{table}

\newpage

%%%%%%%%%%%%%%%%%%%%%%%%%%%%%%%%%%%%%%%%%%%%%%%%%%%%%%%%%%%%%%%%%%%%%%%%%%%%%%%%%%%%%%%
%  END: Pla85900
%%%%%%%%%%%%%%%%%%%%%%%%%%%%%%%%%%%%%%%%%%%%%%%%%%%%%%%%%%%%%%%%%%%%%%%%%%%%%%%%%%%%%%%

\newpage

%%%%%%%%%%%%%%%%%%%%%%%%%%%%%%%%%%%%%%%%%%%%%%%%%%%%%%%%%%%%%%%%%%%%%%%%%%%%%%%%%%%%%%%
%  START: D15112
%%%%%%%%%%%%%%%%%%%%%%%%%%%%%%%%%%%%%%%%%%%%%%%%%%%%%%%%%%%%%%%%%%%%%%%%%%%%%%%%%%%%%%%
\subsection{D15112}
Dimensions: $m$ = 15112, $n$ = 2.
\par
Description: a data set with German cities coordinates for travelling salesman problem.

\begin{table}[!htbp]
	\centering
	
	\caption{Summary of the results with D15112 ($\times10^{11}$)}
	\label{TabResultsD23}
	\small
	\resizebox{!}{\tableheight}{
		\begin{tabular}{|l|l|l|llllll|llllll|llllll|}
			\hline
			\multicolumn{1}{|c|}{\multirow{3}{*}{$k$}} & \multicolumn{1}{c|}{\multirow{3}{*}{$f^*$}} & \multicolumn{1}{c|}{\multirow{3}{*}{$\overline{f}$}} & \multicolumn{6}{c|}{HPClust-inner} & \multicolumn{6}{c|}{HPClust-competitive} & \multicolumn{6}{c|}{HPClust-cooperative} \\ \cline{4-21}
			\multicolumn{1}{|c|}{} & \multicolumn{1}{c|}{} & \multicolumn{1}{c|}{} & \multicolumn{2}{c|}{$\varepsilon$} & \multicolumn{2}{c|}{$\overline{t}$} & \multicolumn{2}{c|}{$t$} & \multicolumn{2}{c|}{$\varepsilon$} & \multicolumn{2}{c|}{$\overline{t}$} & \multicolumn{2}{c|}{$t$} & \multicolumn{2}{c|}{$\varepsilon$} & \multicolumn{2}{c|}{$\overline{t}$} & \multicolumn{2}{c|}{$t$} \\ \cline{4-21}
			\multicolumn{1}{|c|}{} & \multicolumn{1}{c|}{} & \multicolumn{1}{c|}{} & \multicolumn{1}{c|}{med} & \multicolumn{1}{c|}{std} & \multicolumn{1}{c|}{med} & \multicolumn{1}{c|}{std} & \multicolumn{1}{c|}{med} & \multicolumn{1}{c|}{std} & \multicolumn{1}{c|}{med} & \multicolumn{1}{c|}{std} & \multicolumn{1}{c|}{med} & \multicolumn{1}{c|}{std} & \multicolumn{1}{c|}{med} & \multicolumn{1}{c|}{std} & \multicolumn{1}{c|}{med} & \multicolumn{1}{c|}{std} & \multicolumn{1}{c|}{med} & \multicolumn{1}{c|}{std} & \multicolumn{1}{c|}{med} & \multicolumn{1}{c|}{std} \\ \hline
			2 & 3.68403 & 1.91227 & 0.011 & 0.012 & 0.84 & 0.495 & 0.84 & 0.44 & 0.014 & 0.014 & 0.218 & 0.268 & 0.602 & 0.222 & 0.02 & 0.023 & 0.258 & 0.231 & 0.745 & 0.256 \\
			3 & 2.5324 & 1.30699 & 0.021 & 0.023 & 0.289 & 0.54 & 1.009 & 0.425 & 0.023 & 0.019 & 0.277 & 0.238 & 0.45 & 0.299 & 0.036 & 0.027 & 0.205 & 0.166 & 0.619 & 0.379 \\
			5 & 1.32707 & 0.68683 & 0.041 & 0.023 & 0.507 & 0.402 & 0.907 & 0.427 & 0.034 & 0.022 & 0.07 & 0.123 & 0.801 & 0.399 & 0.045 & 0.02 & 0.178 & 0.159 & 0.777 & 0.329 \\
			10 & 0.64491 & 0.33574 & 0.734 & 1.319 & 0.495 & 0.411 & 0.649 & 0.458 & 0.118 & 0.145 & 0.104 & 0.205 & 0.973 & 0.434 & 0.098 & 0.278 & 0.15 & 0.356 & 1.158 & 0.315 \\
			15 & 0.43136 & 0.22393 & 0.776 & 0.79 & 0.205 & 0.127 & 0.546 & 0.389 & 0.235 & 0.091 & 0.247 & 0.194 & 0.365 & 0.369 & 0.309 & 0.2 & 0.163 & 0.323 & 0.596 & 0.444 \\
			20 & 0.32177 & 0.16878 & 0.888 & 0.619 & 0.214 & 0.171 & 0.558 & 0.449 & 0.28 & 0.144 & 0.098 & 0.081 & 1.023 & 0.398 & 0.626 & 0.497 & 0.063 & 0.041 & 0.623 & 0.4 \\
			25 & 0.25308 & 0.13159 & 0.868 & 0.851 & 0.396 & 0.516 & 0.623 & 0.409 & 0.306 & 0.206 & 0.487 & 0.257 & 0.626 & 0.338 & 0.867 & 0.432 & 0.675 & 0.342 & 0.945 & 0.361 \\
			\hline
			\multicolumn{3}{|c|}{Mean:} & \textbf{0.477} & & \textbf{0.421} & & \textbf{0.733} & & \textbf{0.144} & & \textbf{0.214} & & \textbf{0.691} & & \textbf{0.286} & & \textbf{0.242} & & \textbf{0.78} & \\ \hline
		\end{tabular}
	}
	
	\bigskip
	
	\small
	\resizebox{!}{\tableheight}{
		\begin{tabular}{|l|l|l|llllll|llllll|llllll|}
			\hline
			\multicolumn{1}{|c|}{\multirow{3}{*}{$k$}} & \multicolumn{1}{c|}{\multirow{3}{*}{$f^*$}} & \multicolumn{1}{c|}{\multirow{3}{*}{$\overline{f}$}} & \multicolumn{6}{c|}{HPClust-hybrid} & \multicolumn{6}{c|}{Forgy K-means} & \multicolumn{6}{c|}{PBK-BDC} \\ \cline{4-21}
			\multicolumn{1}{|c|}{} & \multicolumn{1}{c|}{} & \multicolumn{1}{c|}{} & \multicolumn{2}{c|}{$\varepsilon$} & \multicolumn{2}{c|}{$\overline{t}$} & \multicolumn{2}{c|}{$t$} & \multicolumn{2}{c|}{$\varepsilon$} & \multicolumn{2}{c|}{$\overline{t}$} & \multicolumn{2}{c|}{$t$} & \multicolumn{2}{c|}{$\varepsilon$} & \multicolumn{2}{c|}{$\overline{t}$} & \multicolumn{2}{c|}{$t$} \\ \cline{4-21}
			\multicolumn{1}{|c|}{} & \multicolumn{1}{c|}{} & \multicolumn{1}{c|}{} & \multicolumn{1}{c|}{med} & \multicolumn{1}{c|}{std} & \multicolumn{1}{c|}{med} & \multicolumn{1}{c|}{std} & \multicolumn{1}{c|}{med} & \multicolumn{1}{c|}{std} & \multicolumn{1}{c|}{med} & \multicolumn{1}{c|}{std} & \multicolumn{1}{c|}{med} & \multicolumn{1}{c|}{std} & \multicolumn{1}{c|}{med} & \multicolumn{1}{c|}{std} & \multicolumn{1}{c|}{med} & \multicolumn{1}{c|}{std} & \multicolumn{1}{c|}{med} & \multicolumn{1}{c|}{std} & \multicolumn{1}{c|}{med} & \multicolumn{1}{c|}{std} \\ \hline
			2 & 3.68403 & 1.91227 & 0.021 & 0.012 & 0.293 & 0.276 & 0.909 & 0.428 & 0.0 & 0.0 & -- & -- & 0.003 & 0.0 & 0.013 & 0.008 & -- & -- & 0.003 & 0.0 \\
			3 & 2.5324 & 1.30699 & 0.03 & 0.04 & 0.342 & 0.24 & 0.799 & 0.341 & 0.001 & 0.0 & -- & -- & 0.007 & 0.002 & 0.038 & 0.084 & -- & -- & 0.004 & 0.001 \\
			5 & 1.32707 & 0.68683 & 0.052 & 0.029 & 0.254 & 0.22 & 1.129 & 0.346 & -0.0 & 7.357 & -- & -- & 0.005 & 0.002 & 0.048 & 4.148 & -- & -- & 0.004 & 0.001 \\
			10 & 0.64491 & 0.33574 & 0.1 & 0.033 & 0.126 & 0.214 & 1.1 & 0.296 & 1.411 & 1.559 & -- & -- & 0.032 & 0.02 & 0.955 & 1.46 & -- & -- & 0.018 & 0.006 \\
			15 & 0.43136 & 0.22393 & 0.283 & 0.147 & 0.381 & 0.372 & 0.663 & 0.467 & 2.788 & 1.452 & -- & -- & 0.045 & 0.013 & 2.639 & 1.792 & -- & -- & 0.015 & 0.005 \\
			20 & 0.32177 & 0.16878 & 0.3 & 0.155 & 0.071 & 0.049 & 0.818 & 0.377 & 1.635 & 2.513 & -- & -- & 0.05 & 0.014 & 3.321 & 2.902 & -- & -- & 0.019 & 0.006 \\
			25 & 0.25308 & 0.13159 & 0.297 & 0.339 & 0.226 & 0.293 & 0.91 & 0.432 & 2.208 & 1.762 & -- & -- & 0.093 & 0.037 & 2.838 & 1.386 & -- & -- & 0.04 & 0.012 \\
			\hline
			\multicolumn{3}{|c|}{Mean:} & \textbf{0.155} & & \textbf{0.242} & & \textbf{0.904} & & \textbf{1.149} & & \textbf{--} & & \textbf{0.033} & & \textbf{1.407} & & \textbf{--} & & \textbf{0.015} & \\ \hline
		\end{tabular}
	}
	
	\bigskip
	
	\caption{Clustering details with D15112}
	\label{TabDetailsD23}
	\resizebox{!}{\tableheight}{
		\begin{tabular}{|l|l|llll|llll|llll|lllll|l|l|}
			\hline
			\multicolumn{1}{|c|}{\multirow{2}{*}{$k$}} & \multicolumn{1}{c|}{\multirow{2}{*}{$n_{exec}$}} & \multicolumn{4}{c|}{HPClust-inner} & \multicolumn{4}{c|}{HPClust-competitive} & \multicolumn{4}{c|}{HPClust-cooperative} & \multicolumn{5}{c|}{HPClust-hybrid} & \multicolumn{1}{c|}{Forgy K-means} & \multicolumn{1}{c|}{PBK-BDC} \\ \cline{3-21}
			\multicolumn{1}{|c|}{} & \multicolumn{1}{c|}{} & \multicolumn{1}{c|}{$s$} & \multicolumn{1}{c|}{$n_{s}$} & \multicolumn{1}{c|}{$T$} & \multicolumn{1}{c|}{$n_{d}$} & \multicolumn{1}{c|}{$s$} & \multicolumn{1}{c|}{$n_{s}$} & \multicolumn{1}{c|}{$T$} & \multicolumn{1}{c|}{$n_{d}$} & \multicolumn{1}{c|}{$s$} & \multicolumn{1}{c|}{$n_{s}$} & \multicolumn{1}{c|}{$T$} & \multicolumn{1}{c|}{$n_{d}$} & \multicolumn{1}{c|}{$s$} & \multicolumn{1}{c|}{$n_{s}$} & \multicolumn{1}{c|}{$T_1$} & \multicolumn{1}{c|}{$T_2$} & \multicolumn{1}{c|}{$n_{d}$} & \multicolumn{1}{c|}{$n_{d}$} & \multicolumn{1}{c|}{$n_{d}$} \\
			\hline
			2 & 15 & 8000 & 1083 & 1.5 & 8.7E+07 & 8000 & 6286 & 1.5 & 4.6E+08 & 8000 & 7270 & 1.5 & 4.7E+08 & 8000 & 9539 & 0.95 & 0.55 & 7.3E+08 & 4.8E+05 & 2.4E+05 \\
			3 & 15 & 8000 & 1184 & 1.5 & 1.5E+08 & 8000 & 3347 & 1.5 & 8.5E+08 & 8000 & 4976 & 1.5 & 8.3E+08 & 8000 & 6275 & 0.7 & 0.8 & 1.1E+09 & 1.9E+06 & 9.3E+05 \\
			5 & 15 & 8000 & 759 & 1.5 & 2.0E+08 & 8000 & 5774 & 1.5 & 1.3E+09 & 8000 & 5444 & 1.5 & 1.4E+09 & 8000 & 7672 & 0.7 & 0.8 & 1.4E+09 & 1.5E+06 & 9.2E+05 \\
			10 & 15 & 8000 & 392 & 1.5 & 4.5E+08 & 8000 & 3398 & 1.5 & 2.3E+09 & 8000 & 3959 & 1.5 & 2.3E+09 & 8000 & 3865 & 0.9 & 0.6 & 2.3E+09 & 8.9E+06 & 3.8E+06 \\
			15 & 15 & 8000 & 304 & 1.5 & 6.2E+08 & 8000 & 622 & 1.5 & 2.7E+09 & 8000 & 1352 & 1.5 & 2.9E+09 & 8000 & 1290 & 1.35 & 0.15 & 2.6E+09 & 1.5E+07 & 6.9E+06 \\
			20 & 15 & 8000 & 231 & 1.5 & 8.9E+08 & 8000 & 1400 & 1.5 & 2.9E+09 & 8000 & 1015 & 1.5 & 3.1E+09 & 8000 & 1087 & 1.35 & 0.15 & 3.0E+09 & 2.5E+07 & 9.6E+06 \\
			25 & 15 & 8000 & 207 & 1.5 & 9.7E+08 & 8000 & 615 & 1.5 & 3.0E+09 & 8000 & 1099 & 1.5 & 3.1E+09 & 8000 & 964 & 0.85 & 0.65 & 3.1E+09 & 2.6E+07 & 1.3E+07 \\
			\hline
		\end{tabular}
	}
	
\end{table}

\newpage

%%%%%%%%%%%%%%%%%%%%%%%%%%%%%%%%%%%%%%%%%%%%%%%%%%%%%%%%%%%%%%%%%%%%%%%%%%%%%%%%%%%%%%%
%  END: D15112
%%%%%%%%%%%%%%%%%%%%%%%%%%%%%%%%%%%%%%%%%%%%%%%%%%%%%%%%%%%%%%%%%%%%%%%%%%%%%%%%%%%%%%%

\newpage

\begin{figure}[htb]
	\centering
	\begin{subfigure}[b]{0.32\linewidth}
		\centering
		\begin{tikzpicture}[scale=0.5]
			\begin{axis}[xlabel={No of clusters}, ylabel={No of dist. func. eval.}, legend pos=north west, legend style={nodes={scale=0.5, transform shape}}]
				\addplot[plotStyle1] coordinates {
					(2, 36975232.0)
					(3, 54438848.0)
					(5, 79734080.0)
					(10, 121772160.0)
					(15, 141890240.0)
					(20, 165528320.0)
					(25, 188526400.0)
				};
				\addlegendentry{HPClust-inner}
				\addplot[plotStyle2] coordinates {
					(2, 156143232.0)
					(3, 202598848.0)
					(5, 253526080.0)
					(10, 346924160.0)
					(15, 372802240.0)
					(20, 376280320.0)
					(25, 371278400.0)
				};
				\addlegendentry{HPClust-competitive}
				\addplot[plotStyle3] coordinates {
					(2, 154159232.0)
					(3, 196070848.0)
					(5, 259766080.0)
					(10, 336684160.0)
					(15, 363682240.0)
					(20, 378840320.0)
					(25, 372878400.0)
				};
				\addlegendentry{HPClust-cooperative}
				\addplot[plotStyle4] coordinates {
					(2, 162863232.0)
					(3, 200870848.0)
					(5, 257366080.0)
					(10, 333164160.0)
					(15, 346882240.0)
					(20, 327000320.0)
					(25, 338478400.0)
				};
				\addlegendentry{HPClust-hybrid}
				\addplot[plotStyle5] coordinates {
					(2, 14390784.0)
					(3, 55764288.0)
					(5, 125919360.0)
					(10, 689558400.0)
					(15, 1268187840.0)
					(20, 2050686720.0)
					(25, 2938118400.0)
				};
				\addlegendentry{Forgy K-means}
				\addplot[plotStyle6] coordinates {
					(2, 12719376.0)
					(3, 48935334.0)
					(5, 102999880.0)
					(10, 422323360.0)
					(15, 791426640.0)
					(20, 1032202720.0)
					(25, 1503880400.0)
				};
				\addlegendentry{PBK-BDC}
			\end{axis}
		\end{tikzpicture}
		\caption{CORD-19 Embeddings}
		\label{Fig1Exp2Ds1}
	\end{subfigure}%
	\begin{subfigure}[b]{0.32\linewidth}
		\centering
		\begin{tikzpicture}[scale=0.5]
			\begin{axis}[xlabel={No of clusters}, ylabel={No of dist. func. eval.}, legend pos=north west, legend style={nodes={scale=0.5, transform shape}}]
				\addplot[plotStyle1] coordinates {
					(2, 32136000.0)
					(3, 55564000.0)
					(5, 95572000.0)
					(10, 195112000.0)
					(15, 305212000.0)
					(20, 441552000.0)
					(25, 572772000.0)
				};
				\addlegendentry{HPClust-inner}
				\addplot[plotStyle2] coordinates {
					(2, 106632000.0)
					(3, 201548000.0)
					(5, 375956000.0)
					(10, 795816000.0)
					(15, 1330556000.0)
					(20, 1906256000.0)
					(25, 2389476000.0)
				};
				\addlegendentry{HPClust-competitive}
				\addplot[plotStyle3] coordinates {
					(2, 99976000.0)
					(3, 186252000.0)
					(5, 363220000.0)
					(10, 743592000.0)
					(15, 1157756000.0)
					(20, 1791056000.0)
					(25, 2203876000.0)
				};
				\addlegendentry{HPClust-cooperative}
				\addplot[plotStyle4] coordinates {
					(2, 104712000.0)
					(3, 198476000.0)
					(5, 368916000.0)
					(10, 772136000.0)
					(15, 1257596000.0)
					(20, 1807696000.0)
					(25, 2347876000.0)
				};
				\addlegendentry{HPClust-hybrid}
				\addplot[plotStyle5] coordinates {
					(2, 567000000.0)
					(3, 1354500000.0)
					(5, 4305000000.0)
					(10, 23835000000.0)
					(15, 44257500000.0)
					(20, 70350000000.0)
					(25, 78487500000.0)
				};
				\addlegendentry{Forgy K-means}
				\addplot[plotStyle6] coordinates {
					(2, 400649968.0)
					(3, 1110158952.0)
					(5, 2707564600.0)
					(10, 10780987600.0)
					(15, 22025893500.0)
					(20, 29437027200.0)
					(25, 36516530000.0)
				};
				\addlegendentry{PBK-BDC}
			\end{axis}
		\end{tikzpicture}
		\caption{HEPMASS}
		\label{Fig1Exp2Ds2}
	\end{subfigure}%
	\begin{subfigure}[b]{0.32\linewidth}
		\centering
		\begin{tikzpicture}[scale=0.5]
			\begin{axis}[xlabel={No of clusters}, ylabel={No of dist. func. eval.}, legend pos=north west, legend style={nodes={scale=0.5, transform shape}}]
				\addplot[plotStyle1] coordinates {
					(2, 5924570.0)
					(3, 8658855.0)
					(5, 15009425.0)
					(10, 34560850.0)
					(15, 55087275.0)
					(20, 77773700.0)
					(25, 93920125.0)
				};
				\addlegendentry{HPClust-inner}
				\addplot[plotStyle2] coordinates {
					(2, 11258570.0)
					(3, 16602855.0)
					(5, 31770425.0)
					(10, 74556850.0)
					(15, 124033275.0)
					(20, 169789700.0)
					(25, 204836125.0)
				};
				\addlegendentry{HPClust-competitive}
				\addplot[plotStyle3] coordinates {
					(2, 10928570.0)
					(3, 16650855.0)
					(5, 28683425.0)
					(10, 70278850.0)
					(15, 115303275.0)
					(20, 153289700.0)
					(25, 195086125.0)
				};
				\addlegendentry{HPClust-cooperative}
				\addplot[plotStyle4] coordinates {
					(2, 11090570.0)
					(3, 16656855.0)
					(5, 29850425.0)
					(10, 78816850.0)
					(15, 118138275.0)
					(20, 152869700.0)
					(25, 202361125.0)
				};
				\addlegendentry{HPClust-hybrid}
				\addplot[plotStyle5] coordinates {
					(2, 14749710.0)
					(3, 36874275.0)
					(5, 362597037.5)
					(10, 946439725.0)
					(15, 1714653787.5)
					(20, 2704113500.0)
					(25, 4117627375.0)
				};
				\addlegendentry{Forgy K-means}
				\addplot[plotStyle6] coordinates {
					(2, 18887842.0)
					(3, 44126799.0)
					(5, 148740012.5)
					(10, 508262125.0)
					(15, 899585722.5)
					(20, 1312795670.0)
					(25, 1695252212.5)
				};
				\addlegendentry{PBK-BDC}
			\end{axis}
		\end{tikzpicture}
		\caption{US Census Data 1990}
		\label{Fig1Exp2Ds3}
	\end{subfigure}%
	\vskip\baselineskip%
	\begin{subfigure}[b]{0.32\linewidth}
		\centering
		\begin{tikzpicture}[scale=0.5]
			\begin{axis}[xlabel={No of clusters}, ylabel={No of dist. func. eval.}, legend pos=north west, legend style={nodes={scale=0.5, transform shape}}]
				\addplot[plotStyle1] coordinates {
					(2, 1087000.0)
					(3, 1520500.0)
					(5, 2047500.0)
					(10, 2915000.0)
					(15, 3932500.0)
					(20, 4450000.0)
					(25, 6067500.0)
				};
				\addlegendentry{HPClust-inner}
				\addplot[plotStyle2] coordinates {
					(2, 3367000.0)
					(3, 4020500.0)
					(5, 6307500.0)
					(10, 16075000.0)
					(15, 25242500.0)
					(20, 34710000.0)
					(25, 44177500.0)
				};
				\addlegendentry{HPClust-competitive}
				\addplot[plotStyle3] coordinates {
					(2, 3387000.0)
					(3, 4050500.0)
					(5, 6407500.0)
					(10, 16175000.0)
					(15, 24942500.0)
					(20, 33310000.0)
					(25, 43927500.0)
				};
				\addlegendentry{HPClust-cooperative}
				\addplot[plotStyle4] coordinates {
					(2, 3307000.0)
					(3, 3870500.0)
					(5, 6407500.0)
					(10, 16575000.0)
					(15, 24042500.0)
					(20, 35110000.0)
					(25, 44427500.0)
				};
				\addlegendentry{HPClust-hybrid}
				\addplot[plotStyle5] coordinates {
					(2, 702000.0)
					(3, 1579500.0)
					(5, 6750000.0)
					(10, 9720000.0)
					(15, 12757500.0)
					(20, 23220000.0)
					(25, 24975000.0)
				};
				\addlegendentry{Forgy K-means}
				\addplot[plotStyle6] coordinates {
					(2, 667008.0)
					(3, 1630518.0)
					(5, 3167550.0)
					(10, 6835200.0)
					(15, 9502950.0)
					(20, 12870800.0)
					(25, 15088750.0)
				};
				\addlegendentry{PBK-BDC}
			\end{axis}
		\end{tikzpicture}
		\caption{Gisette}
		\label{Fig1Exp2Ds4}
	\end{subfigure}%
	\begin{subfigure}[b]{0.32\linewidth}
		\centering
		\begin{tikzpicture}[scale=0.5]
			\begin{axis}[xlabel={No of clusters}, ylabel={No of dist. func. eval.}, legend pos=north west, legend style={nodes={scale=0.5, transform shape}}]
				\addplot[plotStyle1] coordinates {
					(2, 19923148.0)
					(3, 26893722.0)
					(5, 37765870.0)
					(10, 54693740.0)
					(15, 60446610.0)
					(20, 63799480.0)
					(25, 66402350.0)
				};
				\addlegendentry{HPClust-inner}
				\addplot[plotStyle2] coordinates {
					(2, 87111148.0)
					(3, 97495722.0)
					(5, 109036870.0)
					(10, 117429740.0)
					(15, 119627610.0)
					(20, 122935480.0)
					(25, 123993350.0)
				};
				\addlegendentry{HPClust-competitive}
				\addplot[plotStyle3] coordinates {
					(2, 86007148.0)
					(3, 97126722.0)
					(5, 109561870.0)
					(10, 119319740.0)
					(15, 118142610.0)
					(20, 120955480.0)
					(25, 119793350.0)
				};
				\addlegendentry{HPClust-cooperative}
				\addplot[plotStyle4] coordinates {
					(2, 84657148.0)
					(3, 96109722.0)
					(5, 107416870.0)
					(10, 113349740.0)
					(15, 105812610.0)
					(20, 100255480.0)
					(25, 88068350.0)
				};
				\addlegendentry{HPClust-hybrid}
				\addplot[plotStyle5] coordinates {
					(2, 4902404.0)
					(3, 9112077.0)
					(5, 24245585.0)
					(10, 122560100.0)
					(15, 299739375.0)
					(20, 332510880.0)
					(25, 522212600.0)
				};
				\addlegendentry{Forgy K-means}
				\addplot[plotStyle6] coordinates {
					(2, 3333318.0)
					(3, 8015334.0)
					(5, 17139782.5)
					(10, 62613590.0)
					(15, 104878335.0)
					(20, 149869680.0)
					(25, 193200912.5)
				};
				\addlegendentry{PBK-BDC}
			\end{axis}
		\end{tikzpicture}
		\caption{Music Analysis}
		\label{Fig1Exp2Ds5}
	\end{subfigure}%
	\begin{subfigure}[b]{0.32\linewidth}
		\centering
		\begin{tikzpicture}[scale=0.5]
			\begin{axis}[xlabel={No of clusters}, ylabel={No of dist. func. eval.}, legend pos=north west, legend style={nodes={scale=0.5, transform shape}}]
				\addplot[plotStyle1] coordinates {
					(2, 41059502.0)
					(3, 68029253.0)
					(5, 91056755.0)
					(10, 144145510.0)
					(15, 182674265.0)
					(20, 187603020.0)
					(25, 198131775.0)
				};
				\addlegendentry{HPClust-inner}
				\addplot[plotStyle2] coordinates {
					(2, 211411502.0)
					(3, 260445253.0)
					(5, 297752755.0)
					(10, 325361510.0)
					(15, 355770265.0)
					(20, 381699020.0)
					(25, 447947775.0)
				};
				\addlegendentry{HPClust-competitive}
				\addplot[plotStyle3] coordinates {
					(2, 223843502.0)
					(3, 269181253.0)
					(5, 303912755.0)
					(10, 313601510.0)
					(15, 344010265.0)
					(20, 402979020.0)
					(25, 446547775.0)
				};
				\addlegendentry{HPClust-cooperative}
				\addplot[plotStyle4] coordinates {
					(2, 215219502.0)
					(3, 257421253.0)
					(5, 282352755.0)
					(10, 204961510.0)
					(15, 313770265.0)
					(20, 369379020.0)
					(25, 432547775.0)
				};
				\addlegendentry{HPClust-hybrid}
				\addplot[plotStyle5] coordinates {
					(2, 6704546.0)
					(3, 20113638.0)
					(5, 57571645.0)
					(10, 246319190.0)
					(15, 647134440.0)
					(20, 746245120.0)
					(25, 1002038125.0)
				};
				\addlegendentry{Forgy K-means}
				\addplot[plotStyle6] coordinates {
					(2, 5555518.0)
					(3, 16229307.0)
					(5, 39368855.0)
					(10, 173377910.0)
					(15, 355829865.0)
					(20, 490122220.0)
					(25, 700847525.0)
				};
				\addlegendentry{PBK-BDC}
			\end{axis}
		\end{tikzpicture}
		\caption{Protein Homology}
		\label{Fig1Exp2Ds6}
	\end{subfigure}%
	\vskip\baselineskip%
	\begin{subfigure}[b]{0.32\linewidth}
		\centering
		\begin{tikzpicture}[scale=0.5]
			\begin{axis}[xlabel={No of clusters}, ylabel={No of dist. func. eval.}, legend pos=north west, legend style={nodes={scale=0.5, transform shape}}]
				\addplot[plotStyle1] coordinates {
					(2, 54860128.0)
					(3, 71500192.0)
					(5, 105690320.0)
					(10, 168740640.0)
					(15, 208390960.0)
					(20, 231141280.0)
					(25, 249991600.0)
				};
				\addlegendentry{HPClust-inner}
				\addplot[plotStyle2] coordinates {
					(2, 206180128.0)
					(3, 250640192.0)
					(5, 309270320.0)
					(10, 382720640.0)
					(15, 579020960.0)
					(20, 822121280.0)
					(25, 1106171600.0)
				};
				\addlegendentry{HPClust-competitive}
				\addplot[plotStyle3] coordinates {
					(2, 208260128.0)
					(3, 255710192.0)
					(5, 307320320.0)
					(10, 374920640.0)
					(15, 590720960.0)
					(20, 850721280.0)
					(25, 992421600.0)
				};
				\addlegendentry{HPClust-cooperative}
				\addplot[plotStyle4] coordinates {
					(2, 194740128.0)
					(3, 247520192.0)
					(5, 223470320.0)
					(10, 338520640.0)
					(15, 621920960.0)
					(20, 845521280.0)
					(25, 1135421600.0)
				};
				\addlegendentry{HPClust-hybrid}
				\addplot[plotStyle5] coordinates {
					(2, 2341152.0)
					(3, 8584224.0)
					(5, 44872080.0)
					(10, 412302880.0)
					(15, 544317840.0)
					(20, 756972480.0)
					(25, 949467200.0)
				};
				\addlegendentry{Forgy K-means}
				\addplot[plotStyle6] coordinates {
					(2, 4680136.0)
					(3, 8190210.0)
					(5, 48100370.0)
					(10, 375700840.0)
					(15, 485551410.0)
					(20, 782602080.0)
					(25, 932752850.0)
				};
				\addlegendentry{PBK-BDC}
			\end{axis}
		\end{tikzpicture}
		\caption{MiniBooNE Particle Identification}
		\label{Fig1Exp2Ds7}
	\end{subfigure}%
	\begin{subfigure}[b]{0.32\linewidth}
		\centering
		\begin{tikzpicture}[scale=0.5]
			\begin{axis}[xlabel={No of clusters}, ylabel={No of dist. func. eval.}, legend pos=north west, legend style={nodes={scale=0.5, transform shape}}]
				\addplot[plotStyle1] coordinates {
					(2, 7532128.0)
					(3, 15720192.0)
					(5, 22736320.0)
					(10, 55096640.0)
					(15, 74916960.0)
					(20, 82977280.0)
					(25, 87527600.0)
				};
				\addlegendentry{HPClust-inner}
				\addplot[plotStyle2] coordinates {
					(2, 42848128.0)
					(3, 66978192.0)
					(5, 90308320.0)
					(10, 119128640.0)
					(15, 131178960.0)
					(20, 134889280.0)
					(25, 142259600.0)
				};
				\addlegendentry{HPClust-competitive}
				\addplot[plotStyle3] coordinates {
					(2, 44660128.0)
					(3, 69588192.0)
					(5, 91268320.0)
					(10, 126508640.0)
					(15, 129558960.0)
					(20, 130569280.0)
					(25, 144809600.0)
				};
				\addlegendentry{HPClust-cooperative}
				\addplot[plotStyle4] coordinates {
					(2, 42524128.0)
					(3, 68652192.0)
					(5, 89408320.0)
					(10, 108808640.0)
					(15, 121458960.0)
					(20, 101769280.0)
					(25, 140009600.0)
				};
				\addlegendentry{HPClust-hybrid}
				\addplot[plotStyle5] coordinates {
					(2, 4422176.0)
					(3, 9949896.0)
					(5, 23411520.0)
					(10, 122260160.0)
					(15, 307276200.0)
					(20, 417505440.0)
					(25, 614552400.0)
				};
				\addlegendentry{Forgy K-means}
				\addplot[plotStyle6] coordinates {
					(2, 4136208.0)
					(3, 7212417.0)
					(5, 17901320.0)
					(10, 85966140.0)
					(15, 152808960.0)
					(20, 231707280.0)
					(25, 302870212.5)
				};
				\addlegendentry{PBK-BDC}
			\end{axis}
		\end{tikzpicture}
		\caption{MiniBooNE Particle Identification (normalized)}
		\label{Fig1Exp2Ds8}
	\end{subfigure}%
	\begin{subfigure}[b]{0.32\linewidth}
		\centering
		\begin{tikzpicture}[scale=0.5]
			\begin{axis}[xlabel={No of clusters}, ylabel={No of dist. func. eval.}, legend pos=north west, legend style={nodes={scale=0.5, transform shape}}]
				\addplot[plotStyle1] coordinates {
					(2, 13850268.0)
					(3, 18663402.0)
					(5, 27371670.0)
					(10, 49907340.0)
					(15, 60383010.0)
					(20, 65638680.0)
					(25, 70954350.0)
				};
				\addlegendentry{HPClust-inner}
				\addplot[plotStyle2] coordinates {
					(2, 60806268.0)
					(3, 72009402.0)
					(5, 91283670.0)
					(10, 104399340.0)
					(15, 110525010.0)
					(20, 116470680.0)
					(25, 118336350.0)
				};
				\addlegendentry{HPClust-competitive}
				\addplot[plotStyle3] coordinates {
					(2, 57446268.0)
					(3, 76005402.0)
					(5, 87773670.0)
					(10, 108059340.0)
					(15, 108095010.0)
					(20, 113230680.0)
					(25, 121486350.0)
				};
				\addlegendentry{HPClust-cooperative}
				\addplot[plotStyle4] coordinates {
					(2, 56354268.0)
					(3, 72513402.0)
					(5, 89093670.0)
					(10, 106199340.0)
					(15, 79475010.0)
					(20, 109990680.0)
					(25, 115486350.0)
				};
				\addlegendentry{HPClust-hybrid}
				\addplot[plotStyle5] coordinates {
					(2, 3490494.0)
					(3, 5363442.0)
					(5, 19155150.0)
					(10, 56614110.0)
					(15, 153879705.0)
					(20, 261361380.0)
					(25, 416092425.0)
				};
				\addlegendentry{Forgy K-means}
				\addplot[plotStyle6] coordinates {
					(2, 3122338.0)
					(3, 4827591.0)
					(5, 14856195.0)
					(10, 48375540.0)
					(15, 98755672.5)
					(20, 164332080.0)
					(25, 210054600.0)
				};
				\addlegendentry{PBK-BDC}
			\end{axis}
		\end{tikzpicture}
		\caption{MFCCs for Speech Emotion Recognition}
		\label{Fig1Exp2Ds9}
	\end{subfigure}%
	\caption{Number of distance evaluations, 1}
\end{figure}
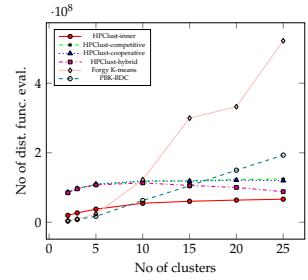
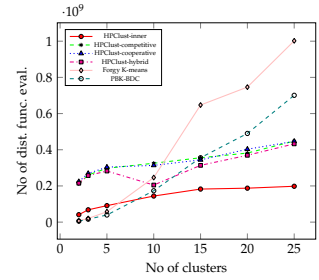
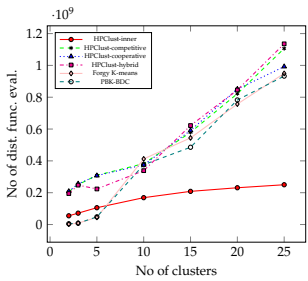
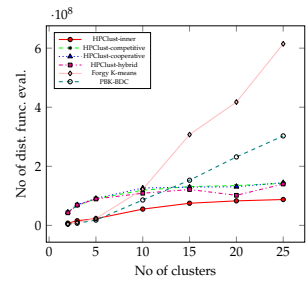
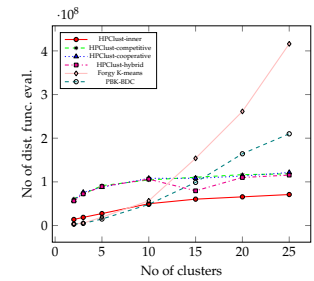
\begin{figure}[htb]
	\centering
	\begin{subfigure}[b]{0.32\linewidth}
		\centering
		\begin{tikzpicture}[scale=0.5]
			\begin{axis}[xlabel={No of clusters}, ylabel={No of dist. func. eval.}, legend pos=north west, legend style={nodes={scale=0.5, transform shape}}]
				\addplot[plotStyle1] coordinates {
					(2, 12503594.0)
					(3, 14211391.0)
					(5, 17930985.0)
					(10, 25109970.0)
					(15, 28488955.0)
					(20, 30947940.0)
					(25, 32586925.0)
				};
				\addlegendentry{HPClust-inner}
				\addplot[plotStyle2] coordinates {
					(2, 49295594.0)
					(3, 54043391.0)
					(5, 56414985.0)
					(10, 58693970.0)
					(15, 59032955.0)
					(20, 57851940.0)
					(25, 55630925.0)
				};
				\addlegendentry{HPClust-competitive}
				\addplot[plotStyle3] coordinates {
					(2, 50735594.0)
					(3, 53755391.0)
					(5, 55814985.0)
					(10, 59093970.0)
					(15, 59092955.0)
					(20, 58971940.0)
					(25, 57630925.0)
				};
				\addlegendentry{HPClust-cooperative}
				\addplot[plotStyle4] coordinates {
					(2, 50495594.0)
					(3, 52951391.0)
					(5, 54954985.0)
					(10, 57373970.0)
					(15, 47632955.0)
					(20, 56571940.0)
					(25, 45030925.0)
				};
				\addlegendentry{HPClust-hybrid}
				\addplot[plotStyle5] coordinates {
					(2, 171534.0)
					(3, 491211.0)
					(5, 1052595.0)
					(10, 2495040.0)
					(15, 2923875.0)
					(20, 5613840.0)
					(25, 7797000.0)
				};
				\addlegendentry{Forgy K-means}
				\addplot[plotStyle6] coordinates {
					(2, 111602.0)
					(3, 287409.0)
					(5, 479035.0)
					(10, 1078170.0)
					(15, 1917405.0)
					(20, 2396740.0)
					(25, 3196175.0)
				};
				\addlegendentry{PBK-BDC}
			\end{axis}
		\end{tikzpicture}
		\caption{ISOLET}
		\label{Fig1Exp2Ds10}
	\end{subfigure}%
	\begin{subfigure}[b]{0.32\linewidth}
		\centering
		\begin{tikzpicture}[scale=0.5]
			\begin{axis}[xlabel={No of clusters}, ylabel={No of dist. func. eval.}, legend pos=north west, legend style={nodes={scale=0.5, transform shape}}]
				\addplot[plotStyle1] coordinates {
					(2, 19363518.0)
					(3, 26032527.0)
					(5, 39809295.0)
					(10, 57213090.0)
					(15, 73593135.0)
					(20, 77103180.0)
					(25, 83245725.0)
				};
				\addlegendentry{HPClust-inner}
				\addplot[plotStyle2] coordinates {
					(2, 85176018.0)
					(3, 98046027.0)
					(5, 109921545.0)
					(10, 129811590.0)
					(15, 222387885.0)
					(20, 278109180.0)
					(25, 366882975.0)
				};
				\addlegendentry{HPClust-competitive}
				\addplot[plotStyle3] coordinates {
					(2, 85000518.0)
					(3, 99801027.0)
					(5, 113431545.0)
					(10, 125716590.0)
					(15, 232040385.0)
					(20, 288639180.0)
					(25, 347870475.0)
				};
				\addlegendentry{HPClust-cooperative}
				\addplot[plotStyle4] coordinates {
					(2, 83947518.0)
					(3, 94974777.0)
					(5, 103925295.0)
					(10, 118404090.0)
					(15, 221071635.0)
					(20, 274599180.0)
					(25, 363226725.0)
				};
				\addlegendentry{HPClust-hybrid}
				\addplot[plotStyle5] coordinates {
					(2, 3861594.0)
					(3, 13515579.0)
					(5, 15943702.5)
					(10, 57338820.0)
					(15, 65822625.0)
					(20, 128134710.0)
					(25, 218677387.5)
				};
				\addlegendentry{Forgy K-means}
				\addplot[plotStyle6] coordinates {
					(2, 3861026.0)
					(3, 14654295.0)
					(5, 16818845.0)
					(10, 59085290.0)
					(15, 58793085.0)
					(20, 117585980.0)
					(25, 212795225.0)
				};
				\addlegendentry{PBK-BDC}
			\end{axis}
		\end{tikzpicture}
		\caption{Sensorless Drive Diagnosis}
		\label{Fig1Exp2Ds11}
	\end{subfigure}%
	\begin{subfigure}[b]{0.32\linewidth}
		\centering
		\begin{tikzpicture}[scale=0.5]
			\begin{axis}[xlabel={No of clusters}, ylabel={No of dist. func. eval.}, legend pos=north west, legend style={nodes={scale=0.5, transform shape}}]
				\addplot[plotStyle1] coordinates {
					(2, 2217018.0)
					(3, 3439277.0)
					(5, 5413045.0)
					(10, 9695590.0)
					(15, 15360635.0)
					(20, 17543180.0)
					(25, 20574475.0)
				};
				\addlegendentry{HPClust-inner}
				\addplot[plotStyle2] coordinates {
					(2, 14043518.0)
					(3, 19334527.0)
					(5, 25882795.0)
					(10, 33516590.0)
					(15, 37256635.0)
					(20, 39474180.0)
					(25, 40344225.0)
				};
				\addlegendentry{HPClust-competitive}
				\addplot[plotStyle3] coordinates {
					(2, 14197518.0)
					(3, 19660027.0)
					(5, 25672795.0)
					(10, 32029090.0)
					(15, 36915385.0)
					(20, 38004180.0)
					(25, 39425475.0)
				};
				\addlegendentry{HPClust-cooperative}
				\addplot[plotStyle4] coordinates {
					(2, 14215018.0)
					(3, 19129777.0)
					(5, 25244045.0)
					(10, 32536590.0)
					(15, 35996635.0)
					(20, 34049180.0)
					(25, 36144225.0)
				};
				\addlegendentry{HPClust-hybrid}
				\addplot[plotStyle5] coordinates {
					(2, 1170180.0)
					(3, 2720668.5)
					(5, 5850900.0)
					(10, 19015425.0)
					(15, 32033677.5)
					(20, 55583550.0)
					(25, 62165812.5)
				};
				\addlegendentry{Forgy K-means}
				\addplot[plotStyle6] coordinates {
					(2, 1051678.0)
					(3, 2569887.0)
					(5, 5386845.0)
					(10, 13827090.0)
					(15, 22511985.0)
					(20, 29001780.0)
					(25, 37155137.5)
				};
				\addlegendentry{PBK-BDC}
			\end{axis}
		\end{tikzpicture}
		\caption{Sensorless Drive Diagnosis (normalized)}
		\label{Fig1Exp2Ds12}
	\end{subfigure}%
	\vskip\baselineskip%
	\begin{subfigure}[b]{0.32\linewidth}
		\centering
		\begin{tikzpicture}[scale=0.5]
			\begin{axis}[xlabel={No of clusters}, ylabel={No of dist. func. eval.}, legend pos=north west, legend style={nodes={scale=0.5, transform shape}}]
				\addplot[plotStyle1] coordinates {
					(2, 8699288.0)
					(3, 12713932.0)
					(5, 20303220.0)
					(10, 30226440.0)
					(15, 42499660.0)
					(20, 47272880.0)
					(25, 51346100.0)
				};
				\addlegendentry{HPClust-inner}
				\addplot[plotStyle2] coordinates {
					(2, 44529288.0)
					(3, 58728932.0)
					(5, 65838220.0)
					(10, 72786440.0)
					(15, 76709660.0)
					(20, 81532880.0)
					(25, 84081100.0)
				};
				\addlegendentry{HPClust-competitive}
				\addplot[plotStyle3] coordinates {
					(2, 44379288.0)
					(3, 61503932.0)
					(5, 67863220.0)
					(10, 74936440.0)
					(15, 78734660.0)
					(20, 79832880.0)
					(25, 84956100.0)
				};
				\addlegendentry{HPClust-cooperative}
				\addplot[plotStyle4] coordinates {
					(2, 44599288.0)
					(3, 60363932.0)
					(5, 65863220.0)
					(10, 66186440.0)
					(15, 57359660.0)
					(20, 81132880.0)
					(25, 82331100.0)
				};
				\addlegendentry{HPClust-hybrid}
				\addplot[plotStyle5] coordinates {
					(2, 555016.0)
					(3, 4876212.0)
					(5, 6045710.0)
					(10, 23984620.0)
					(15, 48167460.0)
					(20, 91181200.0)
					(25, 208626550.0)
				};
				\addlegendentry{Forgy K-means}
				\addplot[plotStyle6] coordinates {
					(2, 439312.0)
					(3, 2008986.0)
					(5, 4148445.0)
					(10, 19347490.0)
					(15, 34047697.5)
					(20, 66299480.0)
					(25, 81627350.0)
				};
				\addlegendentry{PBK-BDC}
			\end{axis}
		\end{tikzpicture}
		\caption{Online News Popularity}
		\label{Fig1Exp2Ds13}
	\end{subfigure}%
	\begin{subfigure}[b]{0.32\linewidth}
		\centering
		\begin{tikzpicture}[scale=0.5]
			\begin{axis}[xlabel={No of clusters}, ylabel={No of dist. func. eval.}, legend pos=north west, legend style={nodes={scale=0.5, transform shape}}]
				\addplot[plotStyle1] coordinates {
					(2, 22572820.0)
					(3, 28589730.0)
					(5, 38301550.0)
					(10, 54616100.0)
					(15, 58105650.0)
					(20, 62360200.0)
					(25, 64679750.0)
				};
				\addlegendentry{HPClust-inner}
				\addplot[plotStyle2] coordinates {
					(2, 90018820.0)
					(3, 96341730.0)
					(5, 103065550.0)
					(10, 106285100.0)
					(15, 107322150.0)
					(20, 106964200.0)
					(25, 106741250.0)
				};
				\addlegendentry{HPClust-competitive}
				\addplot[plotStyle3] coordinates {
					(2, 92115820.0)
					(3, 98636730.0)
					(5, 100523050.0)
					(10, 107860100.0)
					(15, 106984650.0)
					(20, 110114200.0)
					(25, 111016250.0)
				};
				\addlegendentry{HPClust-cooperative}
				\addplot[plotStyle4] coordinates {
					(2, 92124820.0)
					(3, 101053230.0)
					(5, 97890550.0)
					(10, 101245100.0)
					(15, 100437150.0)
					(20, 79154200.0)
					(25, 72991250.0)
				};
				\addlegendentry{HPClust-hybrid}
				\addplot[plotStyle5] coordinates {
					(2, 500760.0)
					(3, 1001520.0)
					(5, 2503800.0)
					(10, 6120400.0)
					(15, 15961725.0)
					(20, 17804800.0)
					(25, 27124500.0)
				};
				\addlegendentry{Forgy K-means}
				\addplot[plotStyle6] coordinates {
					(2, 279828.0)
					(3, 622248.0)
					(5, 1262100.0)
					(10, 3874300.0)
					(15, 7634100.0)
					(20, 10719000.0)
					(25, 19361500.0)
				};
				\addlegendentry{PBK-BDC}
			\end{axis}
		\end{tikzpicture}
		\caption{Gas Sensor Array Drift}
		\label{Fig1Exp2Ds14}
	\end{subfigure}%
	\begin{subfigure}[b]{0.32\linewidth}
		\centering
		\begin{tikzpicture}[scale=0.5]
			\begin{axis}[xlabel={No of clusters}, ylabel={No of dist. func. eval.}, legend pos=north west, legend style={nodes={scale=0.5, transform shape}}]
				\addplot[plotStyle1] coordinates {
					(2, 17669748.0)
					(3, 26004622.0)
					(5, 56974370.0)
					(10, 166148740.0)
					(15, 262073110.0)
					(20, 364497480.0)
					(25, 360671850.0)
				};
				\addlegendentry{HPClust-inner}
				\addplot[plotStyle2] coordinates {
					(2, 96069748.0)
					(3, 156904622.0)
					(5, 278824370.0)
					(10, 600248740.0)
					(15, 848673110.0)
					(20, 954097480.0)
					(25, 1248021850.0)
				};
				\addlegendentry{HPClust-competitive}
				\addplot[plotStyle3] coordinates {
					(2, 96469748.0)
					(3, 161104622.0)
					(5, 273824370.0)
					(10, 560248740.0)
					(15, 806673110.0)
					(20, 986097480.0)
					(25, 1254271850.0)
				};
				\addlegendentry{HPClust-cooperative}
				\addplot[plotStyle4] coordinates {
					(2, 101369748.0)
					(3, 155254622.0)
					(5, 267574370.0)
					(10, 429248740.0)
					(15, 787173110.0)
					(20, 973097480.0)
					(25, 1239271850.0)
				};
				\addlegendentry{HPClust-hybrid}
				\addplot[plotStyle5] coordinates {
					(2, 20873952.0)
					(3, 41747904.0)
					(5, 147857160.0)
					(10, 2217857400.0)
					(15, 2974538160.0)
					(20, 10545694500.0)
					(25, 10654413000.0)
				};
				\addlegendentry{Forgy K-means}
				\addplot[plotStyle6] coordinates {
					(2, 17669788.0)
					(3, 42104730.0)
					(5, 130174670.0)
					(10, 1210850340.0)
					(15, 2394528510.0)
					(20, 5048704680.0)
					(25, 7772134350.0)
				};
				\addlegendentry{PBK-BDC}
			\end{axis}
		\end{tikzpicture}
		\caption{3D Road Network}
		\label{Fig1Exp2Ds15}
	\end{subfigure}%
	\vskip\baselineskip%
	\begin{subfigure}[b]{0.32\linewidth}
		\centering
		\begin{tikzpicture}[scale=0.5]
			\begin{axis}[xlabel={No of clusters}, ylabel={No of dist. func. eval.}, legend pos=north west, legend style={nodes={scale=0.5, transform shape}}]
				\addplot[plotStyle1] coordinates {
					(2, 1586114.0)
					(3, 3467171.0)
					(5, 5309285.0)
					(10, 10834570.0)
					(15, 24419855.0)
					(20, 29525140.0)
					(25, 45810425.0)
				};
				\addlegendentry{HPClust-inner}
				\addplot[plotStyle2] coordinates {
					(2, 9786114.0)
					(3, 18151171.0)
					(5, 27737285.0)
					(10, 62682570.0)
					(15, 108247855.0)
					(20, 146693140.0)
					(25, 152398425.0)
				};
				\addlegendentry{HPClust-competitive}
				\addplot[plotStyle3] coordinates {
					(2, 8762114.0)
					(3, 16871171.0)
					(5, 25453285.0)
					(10, 59122570.0)
					(15, 101887855.0)
					(20, 133093140.0)
					(25, 145498425.0)
				};
				\addlegendentry{HPClust-cooperative}
				\addplot[plotStyle4] coordinates {
					(2, 9442114.0)
					(3, 17767171.0)
					(5, 26757285.0)
					(10, 60202570.0)
					(15, 108247855.0)
					(20, 132693140.0)
					(25, 147298425.0)
				};
				\addlegendentry{HPClust-hybrid}
				\addplot[plotStyle5] coordinates {
					(2, 6861596.0)
					(3, 17644104.0)
					(5, 26956270.0)
					(10, 75967670.0)
					(15, 154385910.0)
					(20, 225452440.0)
					(25, 284878762.5)
				};
				\addlegendentry{Forgy K-means}
				\addplot[plotStyle6] coordinates {
					(2, 5418474.0)
					(3, 16923981.0)
					(5, 24168660.0)
					(10, 65432975.0)
					(15, 125476200.0)
					(20, 181678090.0)
					(25, 236621687.5)
				};
				\addlegendentry{PBK-BDC}
			\end{axis}
		\end{tikzpicture}
		\caption{Skin Segmentation}
		\label{Fig1Exp2Ds16}
	\end{subfigure}%
	\begin{subfigure}[b]{0.32\linewidth}
		\centering
		\begin{tikzpicture}[scale=0.5]
			\begin{axis}[xlabel={No of clusters}, ylabel={No of dist. func. eval.}, legend pos=north west, legend style={nodes={scale=0.5, transform shape}}]
				\addplot[plotStyle1] coordinates {
					(2, 31263226.0)
					(3, 44227339.0)
					(5, 73916740.0)
					(10, 110861930.0)
					(15, 131935495.0)
					(20, 158877560.0)
					(25, 167947375.0)
				};
				\addlegendentry{HPClust-inner}
				\addplot[plotStyle2] coordinates {
					(2, 196274776.0)
					(3, 228258164.0)
					(5, 262428965.0)
					(10, 284836280.0)
					(15, 292839095.0)
					(20, 297374160.0)
					(25, 313246100.0)
				};
				\addlegendentry{HPClust-competitive}
				\addplot[plotStyle3] coordinates {
					(2, 203316976.0)
					(3, 222336314.0)
					(5, 275899840.0)
					(10, 300307780.0)
					(15, 298440845.0)
					(20, 292572660.0)
					(25, 311245475.0)
				};
				\addlegendentry{HPClust-cooperative}
				\addplot[plotStyle4] coordinates {
					(2, 190726376.0)
					(3, 225937439.0)
					(5, 269897965.0)
					(10, 271498780.0)
					(15, 226018220.0)
					(20, 225351660.0)
					(25, 283903600.0)
				};
				\addlegendentry{HPClust-hybrid}
				\addplot[plotStyle5] coordinates {
					(2, 1922868.0)
					(3, 5448126.0)
					(5, 15356237.5)
					(10, 54481260.0)
					(15, 167850352.5)
					(20, 318875610.0)
					(25, 462022450.0)
				};
				\addlegendentry{Forgy K-means}
				\addplot[plotStyle6] coordinates {
					(2, 2027434.0)
					(3, 5441907.0)
					(5, 15205115.0)
					(10, 52550580.0)
					(15, 154449645.0)
					(20, 353179060.0)
					(25, 480819700.0)
				};
				\addlegendentry{PBK-BDC}
			\end{axis}
		\end{tikzpicture}
		\caption{KEGG Metabolic Relation Network (Directed)}
		\label{Fig1Exp2Ds17}
	\end{subfigure}%
	\begin{subfigure}[b]{0.32\linewidth}
		\centering
		\begin{tikzpicture}[scale=0.5]
			\begin{axis}[xlabel={No of clusters}, ylabel={No of dist. func. eval.}, legend pos=north west, legend style={nodes={scale=0.5, transform shape}}]
				\addplot[plotStyle1] coordinates {
					(2, 91213400.0)
					(3, 111669800.0)
					(4, 140818700.0)
					(5, 176631850.0)
					(10, 279203600.0)
					(15, 380616350.0)
					(20, 449577100.0)
					(25, 502601600.0)
				};
				\addlegendentry{HPClust-inner}
				\addplot[plotStyle2] coordinates {
					(2, 517841300.0)
					(3, 601463200.0)
					(4, 720202800.0)
					(5, 761405300.0)
					(10, 958145800.0)
					(15, 936994300.0)
					(20, 981905800.0)
					(25, 978429050.0)
				};
				\addlegendentry{HPClust-competitive}
				\addplot[plotStyle3] coordinates {
					(2, 525258900.0)
					(3, 627540700.0)
					(4, 677319800.0)
					(5, 780528800.0)
					(10, 920478300.0)
					(15, 996103300.0)
					(20, 967997800.0)
					(25, 1020442800.0)
				};
				\addlegendentry{HPClust-cooperative}
				\addplot[plotStyle4] coordinates {
					(2, 534762700.0)
					(3, 621455950.0)
					(4, 709540000.0)
					(5, 801390800.0)
					(10, 902513800.0)
					(15, 956117800.0)
					(20, 969156800.0)
					(25, 907440300.0)
				};
				\addlegendentry{HPClust-hybrid}
				\addplot[plotStyle5] coordinates {
					(2, 2784000.0)
					(3, 6090000.0)
					(4, 7888000.0)
					(5, 9280000.0)
					(10, 27840000.0)
					(15, 48720000.0)
					(20, 63800000.0)
					(25, 104400000.0)
				};
				\addlegendentry{Forgy K-means}
				\addplot[plotStyle6] coordinates {
					(2, 3245308.0)
					(3, 5911068.0)
					(4, 7186032.0)
					(5, 8113300.0)
					(10, 35929700.0)
					(15, 45202200.0)
					(20, 59110800.0)
					(25, 117351250.0)
				};
				\addlegendentry{PBK-BDC}
			\end{axis}
		\end{tikzpicture}
		\caption{Shuttle Control}
		\label{Fig1Exp2Ds18}
	\end{subfigure}%
	\caption{Number of distance evaluations, 2}
\end{figure}
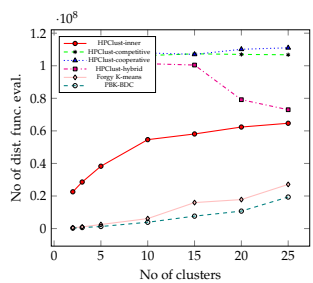
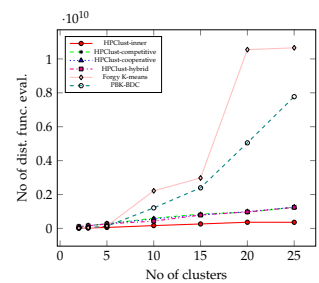
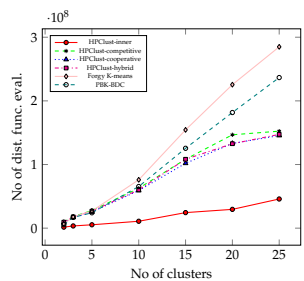
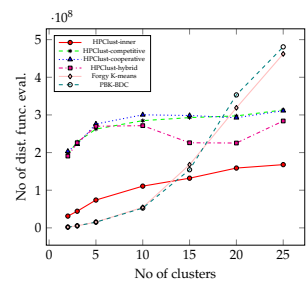
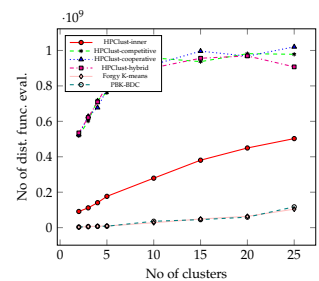
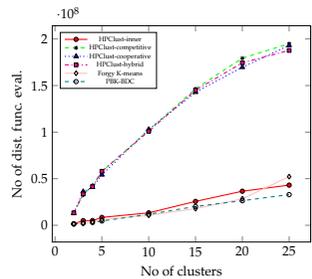
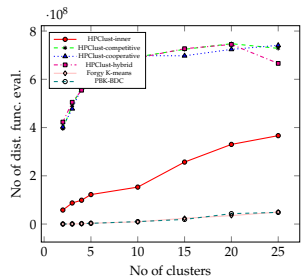
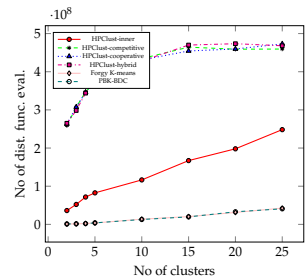
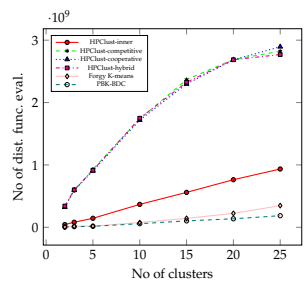
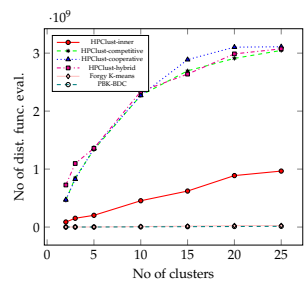
\begin{figure}[htb]
	\centering
	\begin{subfigure}[b]{0.32\linewidth}
		\centering
		\begin{tikzpicture}[scale=0.5]
			\begin{axis}[xlabel={No of clusters}, ylabel={No of dist. func. eval.}, legend pos=north west, legend style={nodes={scale=0.5, transform shape}}]
				\addplot[plotStyle1] coordinates {
					(2, 1586000.0)
					(3, 5021000.0)
					(4, 5032000.0)
					(5, 8421000.0)
					(10, 13326000.0)
					(15, 25631000.0)
					(20, 36556000.0)
					(25, 43103000.0)
				};
				\addlegendentry{HPClust-inner}
				\addplot[plotStyle2] coordinates {
					(2, 13444000.0)
					(3, 32452000.0)
					(4, 42896000.0)
					(5, 57728000.0)
					(10, 102058000.0)
					(15, 146488000.0)
					(20, 179396000.0)
					(25, 194910000.0)
				};
				\addlegendentry{HPClust-competitive}
				\addplot[plotStyle3] coordinates {
					(2, 12960000.0)
					(3, 36178000.0)
					(4, 40868000.0)
					(5, 54218000.0)
					(10, 102958000.0)
					(15, 142873000.0)
					(20, 169668000.0)
					(25, 193118000.0)
				};
				\addlegendentry{HPClust-cooperative}
				\addplot[plotStyle4] coordinates {
					(2, 13318000.0)
					(3, 33670000.0)
					(4, 41436000.0)
					(5, 57983000.0)
					(10, 100888000.0)
					(15, 145048000.0)
					(20, 174424000.0)
					(25, 187680000.0)
				};
				\addlegendentry{HPClust-hybrid}
				\addplot[plotStyle5] coordinates {
					(2, 1624000.0)
					(3, 2175000.0)
					(4, 3480000.0)
					(5, 5945000.0)
					(10, 10730000.0)
					(15, 17835000.0)
					(20, 28420000.0)
					(25, 52200000.0)
				};
				\addlegendentry{Forgy K-means}
				\addplot[plotStyle6] coordinates {
					(2, 1360638.0)
					(3, 2236174.5)
					(4, 3041624.0)
					(5, 4593262.5)
					(10, 11783050.0)
					(15, 20171775.0)
					(20, 26390800.0)
					(25, 32924250.0)
				};
				\addlegendentry{PBK-BDC}
			\end{axis}
		\end{tikzpicture}
		\caption{Shuttle Control (normalized)}
		\label{Fig1Exp2Ds19}
	\end{subfigure}%
	\begin{subfigure}[b]{0.32\linewidth}
		\centering
		\begin{tikzpicture}[scale=0.5]
			\begin{axis}[xlabel={No of clusters}, ylabel={No of dist. func. eval.}, legend pos=north west, legend style={nodes={scale=0.5, transform shape}}]
				\addplot[plotStyle1] coordinates {
					(2, 58088564.0)
					(3, 87327573.0)
					(4, 98831446.0)
					(5, 122198687.0)
					(10, 153130327.0)
					(15, 256897354.5)
					(20, 330107222.0)
					(25, 366393854.5)
				};
				\addlegendentry{HPClust-inner}
				\addplot[plotStyle2] coordinates {
					(2, 395175980.0)
					(3, 490742001.0)
					(4, 567823936.0)
					(5, 571621113.5)
					(10, 693063361.0)
					(15, 724257133.5)
					(20, 747512036.0)
					(25, 728301473.5)
				};
				\addlegendentry{HPClust-competitive}
				\addplot[plotStyle3] coordinates {
					(2, 402845228.0)
					(3, 477845082.0)
					(4, 564228976.0)
					(5, 579410193.5)
					(10, 698156221.0)
					(15, 697407276.0)
					(20, 723995006.0)
					(25, 740846386.0)
				};
				\addlegendentry{HPClust-cooperative}
				\addplot[plotStyle4] coordinates {
					(2, 422707382.0)
					(3, 504807282.0)
					(4, 554762248.0)
					(5, 582555783.5)
					(10, 688569661.0)
					(15, 726841011.0)
					(20, 744516236.0)
					(25, 665951386.0)
				};
				\addlegendentry{HPClust-hybrid}
				\addplot[plotStyle5] coordinates {
					(2, 149800.0)
					(3, 359520.0)
					(4, 1468040.0)
					(5, 3258150.0)
					(10, 9212700.0)
					(15, 23705850.0)
					(20, 35053200.0)
					(25, 49621250.0)
				};
				\addlegendentry{Forgy K-means}
				\addplot[plotStyle6] coordinates {
					(2, 179758.0)
					(3, 404454.0)
					(4, 1737600.0)
					(5, 2995855.0)
					(10, 10111035.0)
					(15, 19211032.5)
					(20, 43140340.0)
					(25, 48308550.0)
				};
				\addlegendentry{PBK-BDC}
			\end{axis}
		\end{tikzpicture}
		\caption{EEG Eye State}
		\label{Fig1Exp2Ds20}
	\end{subfigure}%
	\begin{subfigure}[b]{0.32\linewidth}
		\centering
		\begin{tikzpicture}[scale=0.5]
			\begin{axis}[xlabel={No of clusters}, ylabel={No of dist. func. eval.}, legend pos=north west, legend style={nodes={scale=0.5, transform shape}}]
				\addplot[plotStyle1] coordinates {
					(2, 36219224.0)
					(3, 52231776.0)
					(4, 71689498.0)
					(5, 82579232.0)
					(10, 116506672.0)
					(15, 167135697.0)
					(20, 197992442.0)
					(25, 248246992.0)
				};
				\addlegendentry{HPClust-inner}
				\addplot[plotStyle2] coordinates {
					(2, 259031849.0)
					(3, 307106950.5)
					(4, 347273140.0)
					(5, 372400413.5)
					(10, 434376031.0)
					(15, 464408931.0)
					(20, 459016496.0)
					(25, 459428423.5)
				};
				\addlegendentry{HPClust-competitive}
				\addplot[plotStyle3] coordinates {
					(2, 262911410.0)
					(3, 307443978.0)
					(4, 346014904.0)
					(5, 375358766.0)
					(10, 430855966.0)
					(15, 454410448.5)
					(20, 459615656.0)
					(25, 472909523.5)
				};
				\addlegendentry{HPClust-cooperative}
				\addplot[plotStyle4] coordinates {
					(2, 264888638.0)
					(3, 298501515.0)
					(4, 343947802.0)
					(5, 371389331.0)
					(10, 427860166.0)
					(15, 469913713.5)
					(20, 473396336.0)
					(25, 468041348.5)
				};
				\addlegendentry{HPClust-hybrid}
				\addplot[plotStyle5] coordinates {
					(2, 943740.0)
					(3, 1348200.0)
					(4, 2336880.0)
					(5, 3670100.0)
					(10, 13631800.0)
					(15, 20335350.0)
					(20, 31308200.0)
					(25, 42505750.0)
				};
				\addlegendentry{Forgy K-means}
				\addplot[plotStyle6] coordinates {
					(2, 1018582.0)
					(3, 1482942.0)
					(4, 2037180.0)
					(5, 3932042.5)
					(10, 12807255.0)
					(15, 19660402.5)
					(20, 32954620.0)
					(25, 40819050.0)
				};
				\addlegendentry{PBK-BDC}
			\end{axis}
		\end{tikzpicture}
		\caption{EEG Eye State (normalized)}
		\label{Fig1Exp2Ds21}
	\end{subfigure}%
	\vskip\baselineskip%
	\begin{subfigure}[b]{0.32\linewidth}
		\centering
		\begin{tikzpicture}[scale=0.5]
			\begin{axis}[xlabel={No of clusters}, ylabel={No of dist. func. eval.}, legend pos=north west, legend style={nodes={scale=0.5, transform shape}}]
				\addplot[plotStyle1] coordinates {
					(2, 43795800.0)
					(3, 81226700.0)
					(5, 144391500.0)
					(10, 367281000.0)
					(15, 559755500.0)
					(20, 762870000.0)
					(25, 934519500.0)
				};
				\addlegendentry{HPClust-inner}
				\addplot[plotStyle2] coordinates {
					(2, 337333800.0)
					(3, 587781700.0)
					(5, 930190500.0)
					(10, 1738105000.0)
					(15, 2367554500.0)
					(20, 2690334000.0)
					(25, 2822573500.0)
				};
				\addlegendentry{HPClust-competitive}
				\addplot[plotStyle3] coordinates {
					(2, 322143800.0)
					(3, 595173700.0)
					(5, 905585500.0)
					(10, 1719205000.0)
					(15, 2297729500.0)
					(20, 2683754000.0)
					(25, 2894498500.0)
				};
				\addlegendentry{HPClust-cooperative}
				\addplot[plotStyle4] coordinates {
					(2, 340665800.0)
					(3, 600297700.0)
					(5, 914720500.0)
					(10, 1749095000.0)
					(15, 2325134500.0)
					(20, 2685994000.0)
					(25, 2769548500.0)
				};
				\addlegendentry{HPClust-hybrid}
				\addplot[plotStyle5] coordinates {
					(2, 5068100.0)
					(3, 15848550.0)
					(5, 18683250.0)
					(10, 76451000.0)
					(15, 143667750.0)
					(20, 223340000.0)
					(25, 347895000.0)
				};
				\addlegendentry{Forgy K-means}
				\addplot[plotStyle6] coordinates {
					(2, 4105860.0)
					(3, 11261862.0)
					(5, 16284950.0)
					(10, 57841100.0)
					(15, 102935925.0)
					(20, 136688800.0)
					(25, 185920000.0)
				};
				\addlegendentry{PBK-BDC}
			\end{axis}
		\end{tikzpicture}
		\caption{Pla85900}
		\label{Fig1Exp2Ds22}
	\end{subfigure}%
	\begin{subfigure}[b]{0.32\linewidth}
		\centering
		\begin{tikzpicture}[scale=0.5]
			\begin{axis}[xlabel={No of clusters}, ylabel={No of dist. func. eval.}, legend pos=north west, legend style={nodes={scale=0.5, transform shape}}]
				\addplot[plotStyle1] coordinates {
					(2, 87166224.0)
					(3, 151709336.0)
					(5, 202539560.0)
					(10, 454055120.0)
					(15, 621210680.0)
					(20, 887966240.0)
					(25, 965761800.0)
				};
				\addlegendentry{HPClust-inner}
				\addplot[plotStyle2] coordinates {
					(2, 463198224.0)
					(3, 847381336.0)
					(5, 1337187560.0)
					(10, 2271303120.0)
					(15, 2694938680.0)
					(20, 2909614240.0)
					(25, 3046649800.0)
				};
				\addlegendentry{HPClust-competitive}
				\addplot[plotStyle3] coordinates {
					(2, 469742224.0)
					(3, 826021336.0)
					(5, 1355387560.0)
					(10, 2270903120.0)
					(15, 2887658680.0)
					(20, 3102894240.0)
					(25, 3110249800.0)
				};
				\addlegendentry{HPClust-cooperative}
				\addplot[plotStyle4] coordinates {
					(2, 725598224.0)
					(3, 1096189336.0)
					(5, 1358987560.0)
					(10, 2339143120.0)
					(15, 2638298680.0)
					(20, 2987054240.0)
					(25, 3074449800.0)
				};
				\addlegendentry{HPClust-hybrid}
				\addplot[plotStyle5] coordinates {
					(2, 483584.0)
					(3, 1904112.0)
					(5, 1511200.0)
					(10, 8916080.0)
					(15, 14960880.0)
					(20, 24783680.0)
					(25, 25690400.0)
				};
				\addlegendentry{Forgy K-means}
				\addplot[plotStyle6] coordinates {
					(2, 238232.0)
					(3, 933354.0)
					(5, 915610.0)
					(10, 3831320.0)
					(15, 6947130.0)
					(20, 9583040.0)
					(25, 12979050.0)
				};
				\addlegendentry{PBK-BDC}
			\end{axis}
		\end{tikzpicture}
		\caption{D15112}
		\label{Fig1Exp2Ds23}
	\end{subfigure}%
	\caption{Number of distance evaluations, 3}
\end{figure}
	
\end{landscape}

\end{document}